\pdfoutput=1

\documentclass[11pt]{article}
 \usepackage[preprint]{acl}

\usepackage{times}
\usepackage{latexsym}

\usepackage[T1]{fontenc}

\usepackage[utf8]{inputenc}

\usepackage{microtype}

\usepackage{inconsolata}

\usepackage{graphicx}

\usepackage{enumitem}
\usepackage{tabularx}
\usepackage{xspace}
\usepackage{enumitem}
\usepackage[flushleft]{threeparttable}
\usepackage{makecell}
\usepackage{multirow}
\usepackage{xcolor, colortbl}
\usepackage{subcaption}
\usepackage{tikz}
\usepackage{wrapfig}
\usepackage{textcomp}  %
\usepackage{scalerel}  %
\usepackage{tipa}
\usepackage{pifont}
\usepackage[subtle]{savetrees}
\usepackage[most]{tcolorbox}
\usepackage{listings}
\usepackage{caption}
\usepackage[normalem]{ulem}
\usepackage[utf8]{inputenc}
\usepackage{lettrine}
\usepackage{amsmath}
\usepackage{amsfonts}
\usepackage{graphicx}
\usepackage{array}
\usepackage{tabularx}
\usepackage{bbding}
\usepackage{makecell}
\usepackage{subcaption}
\usepackage{svg}
\usepackage{enumitem}
\usepackage{booktabs}
\usepackage{longtable}
\usepackage[table]{xcolor}
\usepackage{geometry}

\usepackage{booktabs}
\usepackage{multirow}
\usepackage[most]{tcolorbox}

\usepackage{minitoc}
\mtcsettitle{parttoc}{Contents}
\mtcsetrules{parttoc}{on}

\mtcsetfont{parttoc}{section}{\normalsize\bfseries\linespread{0.9}\selectfont}
\mtcsetfont{parttoc}{subsection}{\normalsize\linespread{0.9}\selectfont}

\definecolor{findblue}{RGB}{0, 85, 145}

\newtcolorbox[auto counter]{findingbox}[2][]{%
    colback=gray!6!white,
    colframe=findblue,
    fontupper=\normalsize,
    boxrule=0pt,
    leftrule=2.5pt,
    arc=0pt,
    enhanced,
    top=2pt, bottom=2pt, left=5pt, right=4pt,
    #1
}

\usepackage{xcolor}
\usepackage{listings}

\definecolor{codenum}{HTML}{B8BEC8}
\definecolor{codeplain}{HTML}{24292E}
\definecolor{codekw}{HTML}{D73A49}
\definecolor{codefunc}{HTML}{6F42C1}
\definecolor{codelib}{HTML}{3867D6}
\definecolor{codestr}{HTML}{032F62}
\definecolor{codecomment}{HTML}{6A9955} 
\definecolor{coderule}{HTML}{AEB6C2}

\usepackage{graphicx}
\usepackage{booktabs}
\usepackage{tabularx}
\usepackage{array}

\usepackage{subcaption}
\lstdefinestyle{prettyPythonKernel}{
  language=Python,
  backgroundcolor=\color{white},
  basicstyle=\ttfamily\footnotesize\color{codeplain},
  numbers=left,
  numberstyle=\scriptsize\color{codenum},
  numbersep=10pt,
  xleftmargin=1.8em,
  frame=none,
  columns=fullflexible,
  keepspaces=true,
  showstringspaces=false,
  breaklines=true,
  tabsize=4,
  keywordstyle=\color{codekw},
  commentstyle=\color{codecomment},
  stringstyle=\color{codestr},
  moredelim=[s][\color{codecomment}]{"""}{"""},
  morekeywords={
    def,return,and,or,not,is,if,in,None,True,False
  },
  morekeywords=[2]{
    torch,config,expert_map,topk_ids
  },
  keywordstyle=[2]\color{codelib},
  morekeywords=[3]{
    view,full,numel,moe_align_block_size,
    _prepare_expert_assignment
  },
  keywordstyle=[3]\color{codefunc},
}

\lstdefinestyle{prettyCppKernel}{
  language=C++,
  backgroundcolor=\color{white},
  basicstyle=\ttfamily\footnotesize\color{codeplain},
  numbers=left,
  numberstyle=\scriptsize\color{codenum},
  numbersep=10pt,
  xleftmargin=1.8em,
  frame=none,
  columns=fullflexible,
  keepspaces=true,
  showstringspaces=false,
  breaklines=true,
  tabsize=4,
  keywordstyle=\color{codekw},
  commentstyle=\color{codecomment},
  stringstyle=\color{codestr},
  morekeywords={
    template,typename,bool,int,const,using,inline,
    return,if,else,true,false,nullptr,
    constexpr,extern,static
  },
  morekeywords=[2]{
    __device__,__global__,__host__,__shared__,
    threadIdx,blockIdx,blockDim,gridDim,
    Kernel_traits,Params,Element,ElementAccum,index_t,
    Tensor
  },
  keywordstyle=[2]\color{codelib},
  morekeywords=[3]{
    compute_attn_1rowblock,
    make_tensor,make_gmem_ptr,reinterpret_cast,
    make_shape,make_stride,local_tile,make_coord,
    Shape,Int,q_offset
  },
  keywordstyle=[3]\color{codefunc},
}

\title{How Much Parallelism Is "Free"?\\ A Principle of Near-Free Parallelism for Parallel Decoding}

\author{Minghua He
\thanks{This work was done during the internship at WeChat AI.}$^{1,2}$, Lingzhe Zhang$^{2}$, Yuan Liu$^{1}$, Xiao Zhou$^{1}$, Aiwei Liu\thanks{Corresponding author.}$^{1}$\\
  $^{1}$WeChat AI, Tencent,
  $^{2}$Peking University\\
\texttt{hemh2120@stu.pku.edu.cn, coveliu@tencent.com}
}

\begin{document}
\maketitle

\doparttoc
\faketableofcontents

\begin{abstract}
Parallel decoding improves generation efficiency by processing multiple decode positions within a single decode forward, but reported speedups conflate algorithmic token utilization with the system cost of executing multiple positions.
We isolate the system side by introducing Near-Free Parallelism (NFP), the maximum number of positions executable at near-free latency.
Analyzing Dense FFNs, MoE FFNs, and Attention against an idle-compute baseline, we find that NFP is shaped not by memory-bound resource slack alone, but also by implementation-induced kernel-granularity slack.
Based on these mechanisms, we establish a Near-Free Parallelism principle that predicts the NFP boundary from hardware balance and implementation granularity.
Validation on representative Dense and MoE models---spanning both diffusion and autoregressive decoding---shows that the principle accurately predicts practical NFP boundaries, revealing that the standard idle-compute intuition can over-predict by up to \(23\times\)---offering a system-side budget for parallelism selection and model-system co-design.
\end{abstract}

\section{Introduction}

Autoregressive (AR) decoding is the dominant generation paradigm for large language models \cite{brown2020language, touvron2023llama}, but its token-by-token nature has long been a fundamental bottleneck for inference efficiency. As models scale, each decode step typically produces only a single new token while accessing the full model parameters and KV cache, leaving modern accelerators underutilized in memory-bound regimes \cite{williams2009roofline, pope2023efficiently}. This bottleneck has motivated a growing body of work on decoding architectures that introduce parallelism during generation.

Representative approaches include speculative decoding \cite{leviathan2023fast, chen2023accelerating}, multi-token prediction (MTP) \cite{gloeckle2024better, liu2024deepseek}, and diffusion language models (DLLMs) \cite{sahoo2024simple, nielarge, ye2025dream}. Despite differing in algorithmic design, they share a common systems structure: processing multiple decode positions within a single decode forward. Existing work typically reports aggregate end-to-end speedups, but these numbers entangle two distinct factors: how many useful tokens an algorithm extracts from processed positions, and how cheaply the model-system stack can process those positions. Naively, increasing the number of positions \(N\) should increase computation and data movement. Yet in practical inference systems, latency often remains nearly flat over a non-trivial range of N. This raises a hidden systems question: \textbf{how many decode positions in total can the model-system stack process before latency becomes expensive?}

We call this system-side capacity boundary \textbf{Near-Free Parallelism (NFP)}: the maximum number of decode positions that a decode forward can process while latency remains within a prescribed tolerance. By isolating system-side capacity from algorithmic token utilization, NFP provides a common basis for comparing decoding methods, selecting parallelism levels, and co-designing model architectures with inference systems.

Despite its utility, the NFP boundary remains poorly understood. A common intuition attributes it to idle compute in memory-bound decoding \cite{williams2009roofline, ivanov2021data}---underutilized compute absorbs extra positions at near-zero cost---but this explains only why a near-free regime may arise, not how far it extends or what governs its boundary.
The central question is therefore: \textbf{what determines the NFP boundary, and can it be predicted for a given setting?}

To study this boundary, we abstract parallel decoding methods as a \emph{multi-position decode forward} and formalize the idle-compute intuition as a baseline prediction. We then examine when this baseline succeeds or fails.

We perform a modular study of Dense FFNs, Attention, and MoE FFNs across multiple inference frameworks and hardware platforms. For each module, we ask: Does a near-free regime exist? Is its boundary predicted by idle compute alone? If not, what additional mechanisms govern it?

Our analysis shows that the NFP boundary is not determined by idle compute alone. Instead, it is shaped by two interacting sources of slack: memory-bound resource slack and kernel-granularity slack introduced by implementation details such as padding and tiling, which can shift the realized boundary beyond pure compute-to-memory predictions. 
Based on this joint mechanism, we establish a \textbf{Near-Free Parallelism principle} that predicts the NFP boundary from hardware and execution configurations, distilling empirical observations into systematic prediction.

We validate the principle on representative Dense and MoE models---covering both diffusion and autoregressive decoding paradigms---and show that module-level mechanisms compose into an accurate full-model predictor across hardware platforms. We further use the principle to reason about future hardware trends: as compute grows faster than memory bandwidth, the opportunity space for parallel decoding may enlarge.

In summary, our contributions are as follows:
\begin{itemize}
\item \textbf{A system-side capacity for parallel decoding.} We formulate NFP as the maximum number of decode positions processable at near-free latency, isolating system-side capacity from algorithmic token utilization.
\item \textbf{A joint mechanism beyond idle compute.} We show that NFP is not determined by memory-bound idle compute alone. Implementation-induced slack, especially padding slack in kernel implementations, can substantially shift the realized boundary.
\item \textbf{A predictive NFP principle.} We establish a principle that predicts the NFP boundary from hardware balance and implementation granularity.
\item \textbf{Cross-stack validation.} We validate the principle across representative modules and models under diverse inference frameworks, kernel implementations, and hardware platforms, and analyze future hardware implications.
\end{itemize}
\section{Near-Free Parallelism: Abstraction and Baseline}

\subsection{Parallel Decoding as Multi-Position Decode Forward}

We abstract parallel decoding methods by ignoring how positions are generated, verified, or updated, and focus only on the number of positions processed by one decode forward.

In standard autoregressive decoding, each decode forward processes one position and produces the output for one new token:
\begin{equation}
\ell_t = F_\theta(z_t),
\qquad N = 1.
\end{equation}
Here \(z_t\) denotes the input representation at the current decode position, \(\ell_t\) denotes its output logits, and \(N\) is the number of decode positions processed in this forward pass. Thus, autoregressive decoding is a single-position decode forward.

In contrast, parallel decoding processes multiple decode positions within the same decode forward:
\begin{equation}
\boldsymbol{\ell}_{t:t+N-1} = F_\theta(Z_{t:t+N-1}),
\qquad N > 1.
\end{equation}
Here \(Z_{t:t+N-1}\) denotes the inputs placed at \(N\) decode positions, and \(\boldsymbol{\ell}_{t:t+N-1}\) denotes their corresponding outputs. The shared system-side question is then: \textbf{how does decode latency change as \(N\) increases?}

\subsection{Near-Free Parallelism}

In practice, decode latency often remains nearly flat as the number of positions increases over a non-trivial range. We formalize the boundary of this range as \textbf{Near-Free Parallelism (NFP)}.

Let \(T(N)\) denote the latency of one decode forward when it processes \(N\) decode positions, where \(N=1\) corresponds to standard autoregressive decoding. Because latency is rarely exactly constant, we define near-free execution with respect to a tolerance \(\epsilon\). A forward with \(N\) positions is near-free if its latency remains within \((1+\epsilon)\) of the single-position AR baseline:
\begin{equation}
T(N) \leq (1+\epsilon)T(1).
\end{equation}
The near-free boundary is then the largest number of decode positions that satisfies this condition:
\begin{equation}
N_{\max}(\epsilon)
=
\max \left\{
N:
T(N) \leq (1+\epsilon)T(1)
\right\}.
\end{equation}

This definition is algorithm-independent: it measures only the system-side capacity to execute positions at negligible additional latency cost.

\subsection{Idle-Compute Intuition as a Baseline}
Prior work has suggested idle compute in memory-bound decoding as one explanation for NFP \cite{kim2023full, xia2024unlocking}. We use this intuition as a simple baseline. 
In a standard AR decode step, the model processes only one position while still moving model weights and KV-cache data, leading to latency dominated by memory traffic. In this regime, additional computation can be absorbed by otherwise idle compute units, so increasing the number of decoded positions may have little latency effect.

When \(N\) positions are processed in one forward, let \(C(N)\) and \(B(N)\) denote the resulting computation and data movement, and define arithmetic intensity (AI). We count one fused multiply-add as two floating-point operations, consistent with hardware-reported TFLOP/s and profiler-reported FLOPs.
\begin{equation}
AI(N) = \frac{C(N)}{B(N)}, \qquad
\rho = \frac{\phi}{\beta},
\end{equation}
where \(\phi\) and \(\beta\) are hardware-dependent peak compute throughput and peak memory bandwidth, respectively, and \(\rho\) is the corresponding hardware balance point.
The idle-compute baseline predicts that parallel positions remain near-free until the decode forward reaches this balance point:
\begin{equation}
AI(N_{\mathrm{idle}}) = \rho.
\end{equation}

Thus, \(N_{\mathrm{idle}}\) provides a first-order estimate of the observed near-free parallelism boundary \(N_{\max}\). It captures only the parallelism explained by memory-bound idle compute, excluding implementation effects such as kernel granularity and padding.

\section{Module-Level Analysis of Near-Free Parallelism}

\begin{figure*}[t]
    \centering
    \scalebox{1}[1.0]{%
    \begin{minipage}{\textwidth}
        \centering

        \begin{subfigure}[t]{0.24\textwidth}
            \centering
            \includegraphics[width=\linewidth]{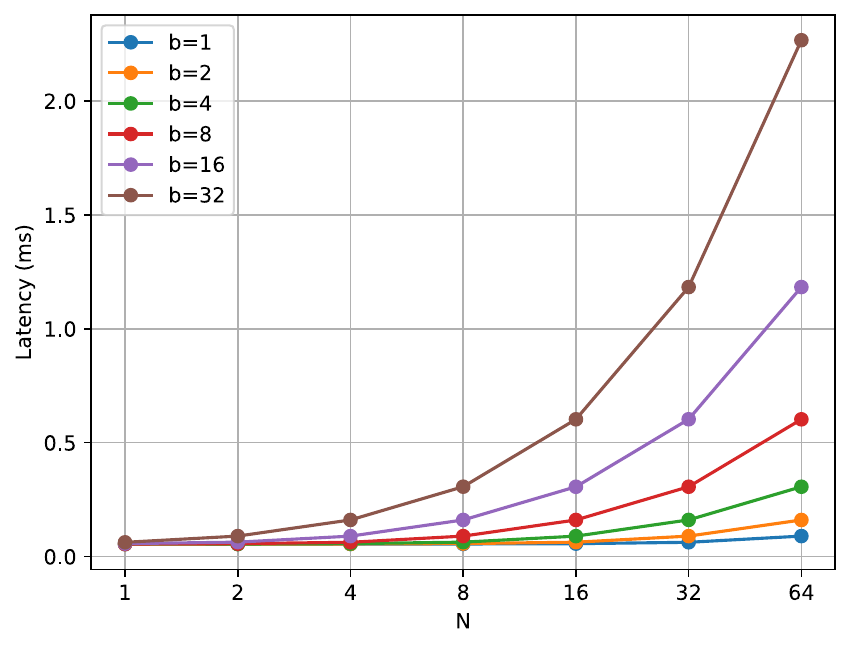}
            \caption{Latency vs. $N$}
            \label{fig:dense-ffn-h20-latency_main}
        \end{subfigure}
        \hfill
        \begin{subfigure}[t]{0.24\textwidth}
            \centering
            \includegraphics[width=\linewidth]{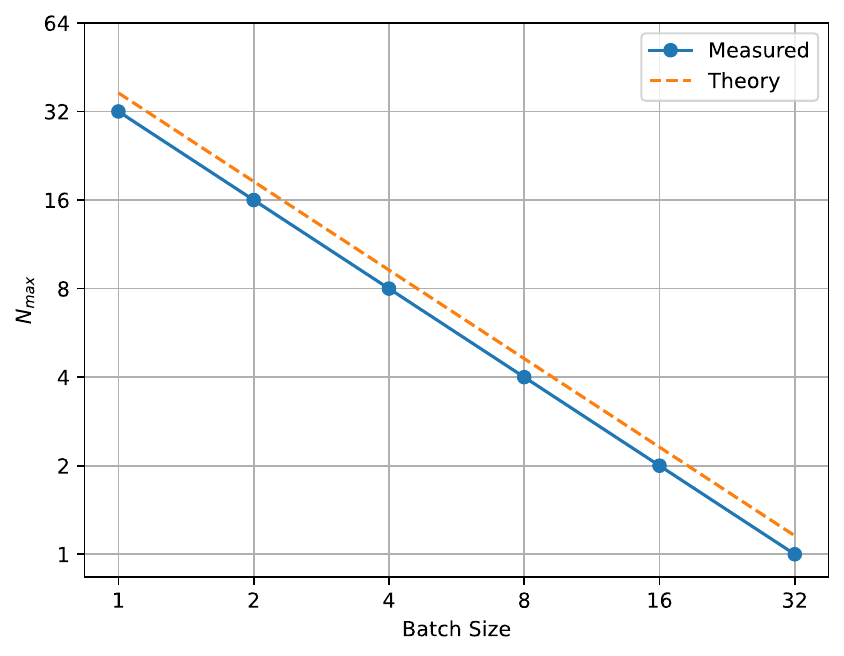}
            \caption{$N_{\max}$ vs. $b$}
            \label{fig:dense-ffn-h20-nmax_main}
        \end{subfigure}
        \hfill
        \begin{subfigure}[t]{0.24\textwidth}
            \centering
            \includegraphics[width=\linewidth,height=1.0\linewidth,keepaspectratio]{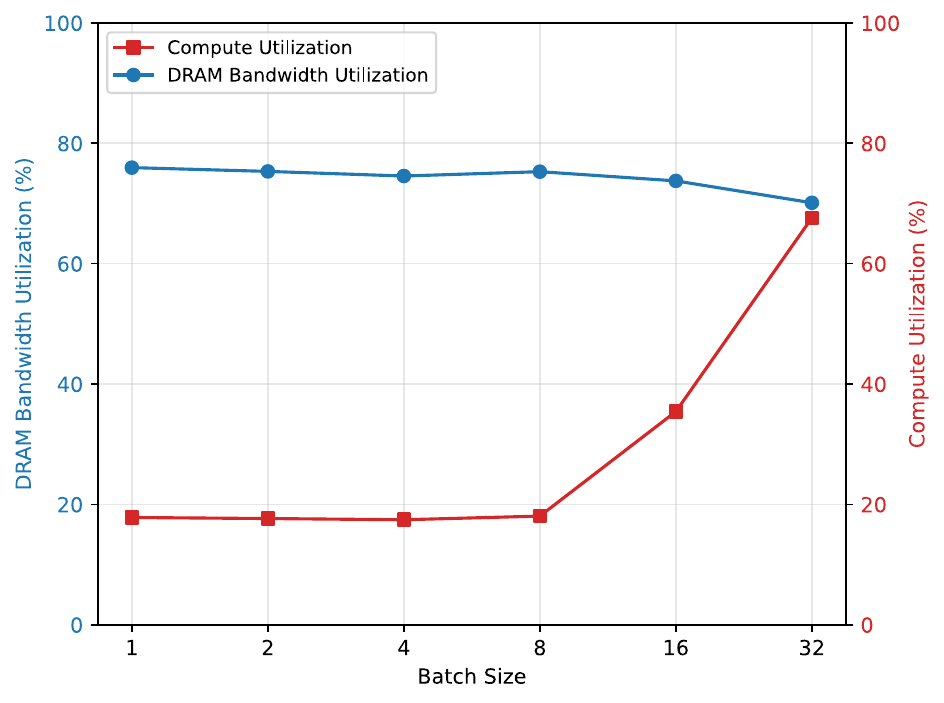}
            \caption{$N=1$ utilization vs. $b$}
            \label{fig:dense-ffn-h20-util-batch_main}
        \end{subfigure}
        \hfill
        \begin{subfigure}[t]{0.24\textwidth}
            \centering
            \includegraphics[width=\linewidth,height=1.0\linewidth,keepaspectratio]{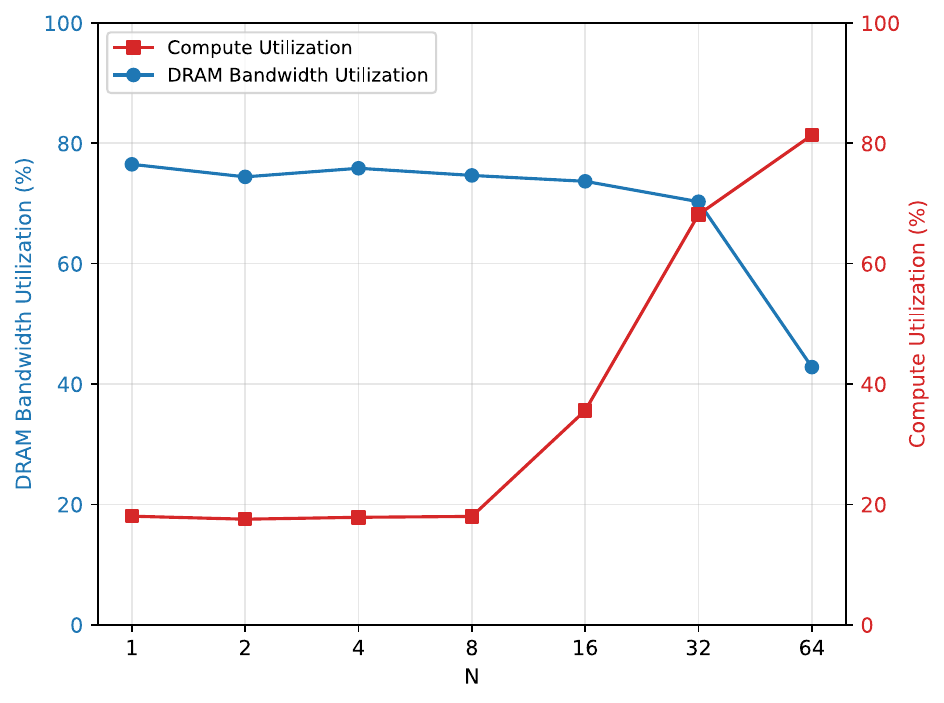}
            \caption{$b=1$}
            \label{fig:dense-ffn-h20-util-b1_main}
        \end{subfigure}

    \end{minipage}%
    }

    \caption{
    Dense FFN evaluation results on \textbf{NVIDIA H20 GPU}.
    }
    \label{fig:dense-ffn-h20_main}
\end{figure*}

\subsection{Study Design}

\textbf{Study Subject.}
In this section, we study NFP at the module level.
Since a full decode forward involves multiple modules with distinct execution characteristics, its overall NFP boundary may conflate different mechanisms.
We therefore isolate three representative module classes in modern LLMs \cite{vaswani2017attention}: Dense FFNs, MoE FFNs, and Attention. 
Dense FFNs capture regular dense matrix computation, MoE FFNs introduce routed expert computation with token grouping and padding effects, and Attention captures KV-cache-dominated memory access.

\textbf{Study Method.}
To enable a systematic comparison across modules, we analyze each module using the same set of research questions (RQ):

\begin{itemize}
    \item \textbf{RQ1}: Does a near-free regime exist as the number of decode positions \(N\) increases?
    \item \textbf{RQ2}: Is the observed boundary predicted by the idle-compute baseline?
    \item \textbf{RQ3}: What factors govern the observed boundary?
\end{itemize}

For each module, we apply a common analysis procedure that connects the theoretical baseline to the measured latency behavior.
We first instantiate the idle-compute baseline by deriving the arithmetic intensity \(AI(N)\), which gives the predicted boundary \(N_{\mathrm{idle}}\).
We then measure the latency curve \(T(N)\) and define the observed NFP boundary \(N_{\max}=N_{\max}(0.2)\) as the largest \(N\) whose latency remains within a 20\% increase over the baseline.\footnote{This tolerance captures a practical low-overhead regime while reducing sensitivity to GPU timing noise. The choice of \(\epsilon\) affects the extracted boundary value but not the mechanism identification, which relies on latency trends, profiling, and source-level evidence independent of the threshold. A full sensitivity sweep across \(\epsilon \in [0.05, 0.30]\) confirms this robustness (Appendix~\ref{app:sensitivity}).}
Except for the load-balanced MoE case, where the baseline is the smallest \(N\) that activates all experts, the baseline is \(T(1)\).
Finally, we compare \(N_{\max}\) with \(N_{\mathrm{idle}}\): consistency between the two suggests that memory-bound idle compute is the dominant mechanism, whereas a mismatch indicates additional mechanism effects.
Detailed experimental configurations, hardware platforms, and implementations for all module experiments are provided in Appendix~\ref{appendix:experimental-setup}.

\subsection{Dense FFN Layers}
For the Dense FFN analysis, we model a standard feed-forward layer as two GEMM operators, corresponding to the up-projection and down-projection, respectively. We omit the activation function in order to isolate the dense matrix-computation behavior that dominates the module-level cost.
Full results are provided in Appendix~\ref{app:dense-ffn-results}.

\subsubsection{Theoretical Baseline Prediction}

We instantiate the idle-compute baseline for Dense FFNs. Let \(b\) denote the batch size, \(N\) the number of decode positions per request, \(d_{\mathrm{model}}\) the hidden dimension, \(d_{\mathrm{ff}}\) the FFN intermediate dimension, and \(s\) the number of bytes per element. Under a weight-traffic-dominated abstraction, a Dense FFN processes \(bN\) positions using two GEMM operators, whose computation and dominant memory traffic are approximated as
\begin{equation}
\begin{aligned}
C_{\mathrm{dense}}(N)
&\approx
4bN d_{\mathrm{model}}d_{\mathrm{ff}}, \\
B_{\mathrm{dense}}(N)
&\approx
2d_{\mathrm{model}}d_{\mathrm{ff}}s .
\end{aligned}
\end{equation}
The key asymmetry is that computation scales with the total number of processed positions \(bN\), while the dominant weight traffic is amortized across positions. The AI is therefore
\begin{equation}
AI_{\mathrm{dense}}(N)
=
\frac{C_{\mathrm{dense}}(N)}{B_{\mathrm{dense}}(N)}
\approx
\frac{2bN}{s}.
\end{equation}
Substituting this intensity into the idle-compute baseline condition yields
\begin{equation}
N_{\mathrm{idle}}^{\mathrm{dense}}
\approx
\frac{\rho s}{2b},
\end{equation}

This result predicts that the NFP boundary of Dense FFNs decreases inversely with batch size. 
Intuitively, increasing batch size consumes the same compute slack as increasing N, thereby reducing the near-free range.

\subsubsection{RQ1: NFP Existence}
To examine whether Dense FFNs exhibit module-level NFP, we sweep the number of decode positions \(N\) under different batch sizes \(b\) and measure the corresponding latency \(T(N)\). As shown in Figure~\ref{fig:dense-ffn-h20-latency_main}, \(T(N)\) remains nearly flat over an initial range of \(N\) and then increases as more positions are processed. This near-free region is larger at smaller batch sizes and shrinks as \(b\) increases.

\subsubsection{RQ2: Idle-Compute Prediction}
To evaluate the idle-compute prediction for Dense FFNs, we compare the measured boundary \(N_{\max}\) with the theoretical estimate \(N_{\mathrm{idle}}^{\mathrm{dense}}\). As shown in Figure~\ref{fig:dense-ffn-h20-nmax_main}, the measured boundary closely follows the predicted trend: as \(b\) increases, \(N_{\max}\) decreases approximately proportionally to \(1/b\). This agreement indicates that, for Dense FFNs, the observed NFP boundary is well explained by the idle-compute baseline.

\subsubsection{RQ3: Boundary Mechanism}
To verify the mechanism behind the Dense FFN boundary, we profile compute utilization and DRAM bandwidth utilization at \(N=1\) and while sweeping \(N\). 
As shown in Figure~\ref{fig:dense-ffn-h20-util-batch_main}--\ref{fig:dense-ffn-h20-util-b1_main}, Dense FFN execution at \(N=1\) is bandwidth-dominated: DRAM bandwidth utilization is high, whereas compute utilization remains low.
Across batch sizes, larger batches increase compute utilization at \(N=1\), thereby reducing the available compute slack.
As \(N\) increases, compute utilization rises while DRAM bandwidth utilization remains nearly unchanged, indicating that increasing N consumes idle compute until execution reaches the compute-bandwidth balance point predicted by the idle-compute baseline.

\begin{findingbox}
    \textbf{Finding~\thetcbcounter:} Dense FFN NFP is governed by idle-compute slack. It scales inversely with batch size.
\end{findingbox}

\begin{figure*}[t]
    \centering
    \scalebox{1}[1.0]{%
    \begin{minipage}{\textwidth}
        \centering

        \begin{subfigure}[t]{0.24\textwidth}
            \centering
            \includegraphics[width=\linewidth]{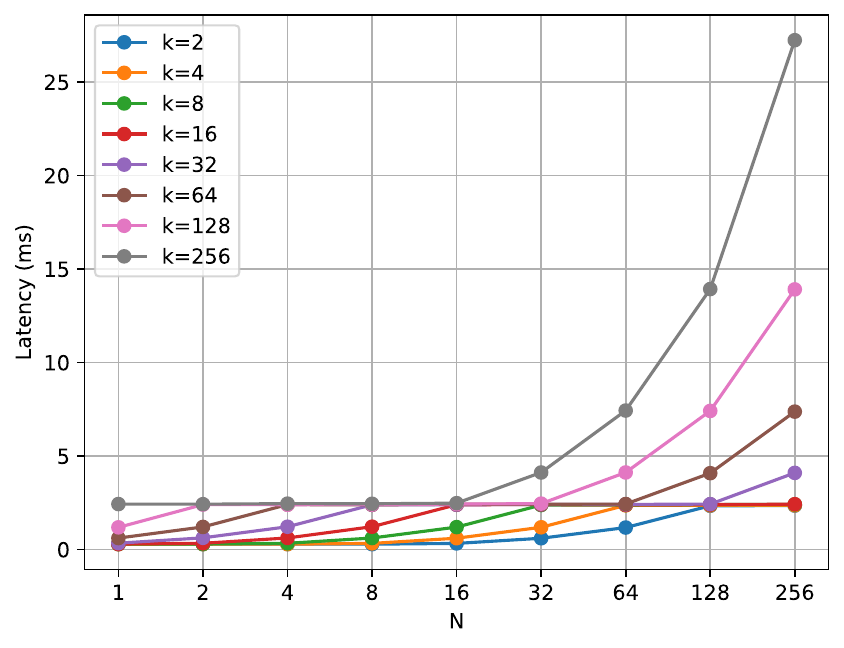}
            \caption{Latency vs. $N$}
            \label{fig:moe-upper-vllm-h20-latency_main}
        \end{subfigure}
        \hfill
        \begin{subfigure}[t]{0.24\textwidth}
            \centering
            \includegraphics[width=\linewidth]{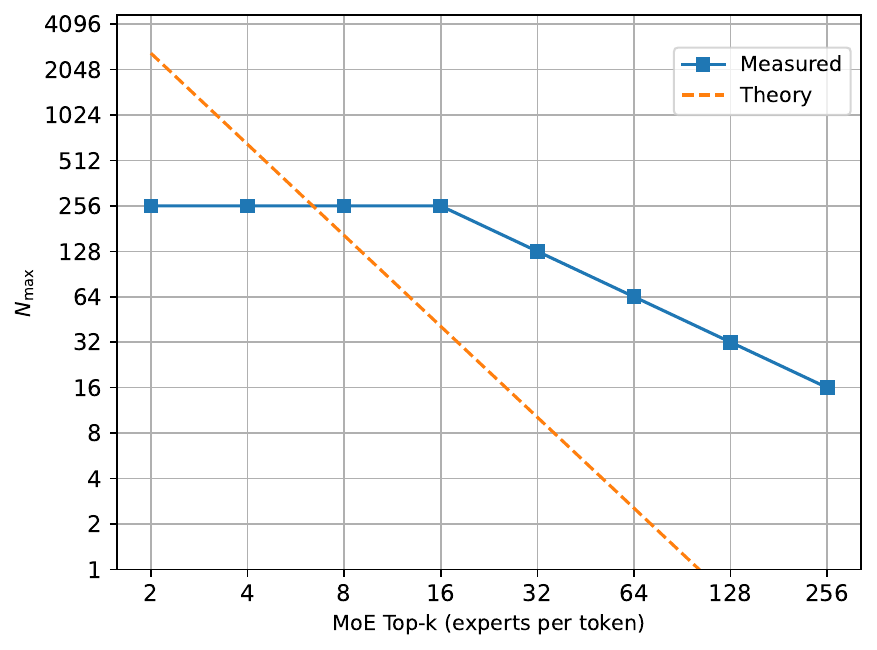}
            \caption{$N_{\max}$ vs. $k$}
            \label{fig:moe-upper-vllm-h20-nmax_main}
        \end{subfigure}
        \hfill
        \begin{subfigure}[t]{0.24\textwidth}
            \centering
            \includegraphics[width=\linewidth,height=0.75\linewidth,keepaspectratio]{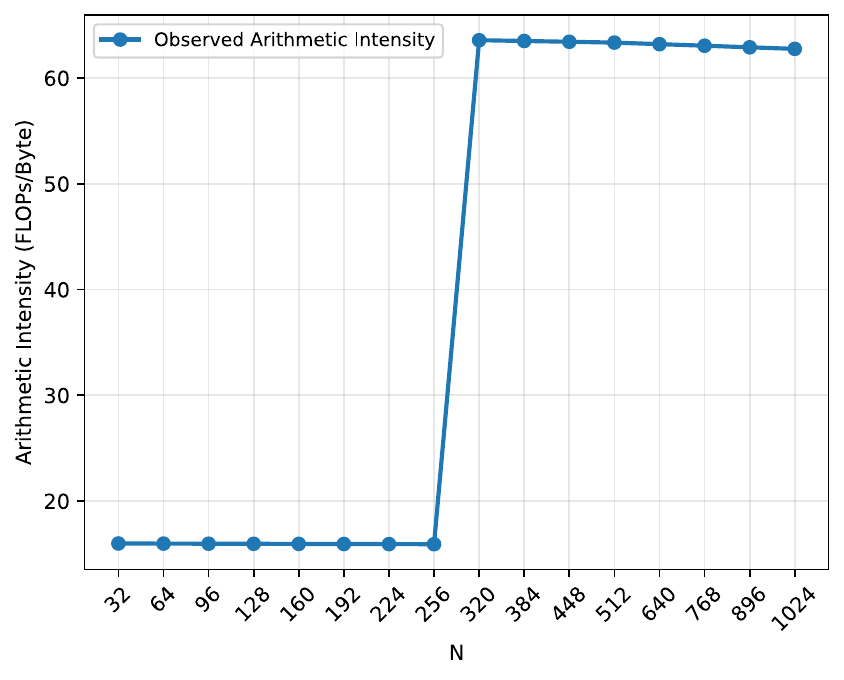}
            \caption{AI, $k=8$}
            \label{fig:moe-upper-vllm-h20-ai-k8_main}
        \end{subfigure}
        \hfill
        \begin{subfigure}[t]{0.24\textwidth}
            \centering
            \includegraphics[width=\linewidth,height=0.75\linewidth,keepaspectratio]{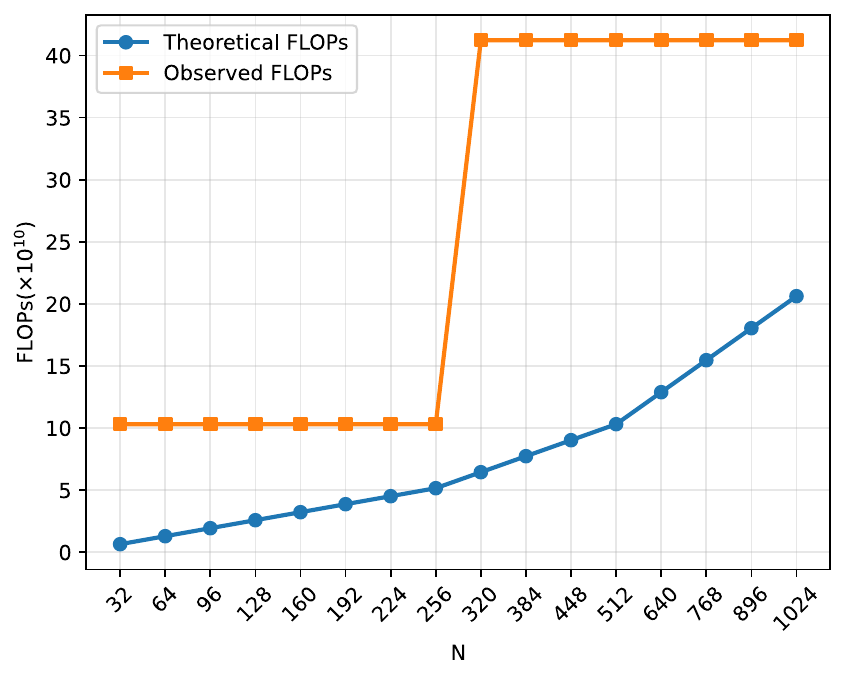}
            \caption{FLOPs, $k=8$}
            \label{fig:moe-upper-vllm-h20-flops-k8_main}
        \end{subfigure}

    \end{minipage}%
    }

    \caption{
    MoE FFN evaluation for load-balanced routing (upper-bound case) with \textbf{vLLM} on \textbf{NVIDIA H20}. Load-skewed (lower-bound) results in Figure~\ref{fig:moe-lower-vllm-h20}.
    }
    \label{fig:moe-upper-vllm-h20_main}
\end{figure*}

\subsection{MoE FFN Layers}
For the MoE FFN analysis, we instantiate a MoE layer \cite{shazeer2017outrageously, jiang2024mixtral} as a routed expert-execution pipeline consisting of token dispatch, expert FFN computation, and weighted combine. 
Each expert is implemented as a standard two-GEMM FFN, and the full layer applies these expert computations to the tokens assigned to each expert through dispatch and combine. 
To capture practical implementation behavior, we use the fused MoE operators from vLLM \cite{kwon2023efficient} and SGLang \cite{zheng2024sglang}, which jointly implement these stages. 

Since real MoE routing is input-dependent and yields nondeterministic expert-load distributions, we analyze two controlled routing cases that expose distinct expert-load behaviors. 
The \textbf{load-balanced} case evenly distributes tokens across experts, whereas the \textbf{load-skewed} case concentrates all tokens on the same \(k\) experts. 
Under load-balanced routing, increasing N from 1 also activates additional experts, adding weight traffic unrelated to parallelism. To isolate the parallelism effect, we define the load-balanced baseline as the smallest N that activates all experts, so that further increases in N only add decode positions without changing the active expert set.
Full results are provided in Appendix~\ref{app:moe-ffn-results}.

\subsubsection{Theoretical Baseline Prediction}

We instantiate the idle-compute baseline for MoE FFNs (full derivation in Appendix~\ref{app:moe-derivation}).
Let \(k\) denote the number of selected experts per token and \(E_{\mathrm{act}}\) the number of distinct experts activated by a routing pattern.
Unlike Dense FFNs, MoE memory traffic includes both active expert weights and routing-induced activation movement.
The idle-compute baseline condition then yields
\begin{equation}
N_{\mathrm{idle}}^{\mathrm{moe}}
\approx
\frac{
2\rho s E_{\mathrm{act}}d_{\mathrm{ff}}
}{
b\left(4kd_{\mathrm{ff}}-\rho s(1+3k+\eta k)\right)
},
\end{equation}
when \(4kd_{\mathrm{ff}}>\rho s(1+3k+\eta k)\); otherwise, the execution remains memory-bound under this abstraction.\footnote{\(\eta=2\) counts per-expert activation accesses in the combine stage.}
This result predicts that the MoE NFP boundary decreases inversely with batch size, and is primarily controlled by \(k\) and \(E_{\mathrm{act}}\): larger \(k\) reduces the boundary, whereas larger \(E_{\mathrm{act}}\) increases it.
Thus, load-balanced routing serves as the upper-bound case and load-skewed routing as the lower-bound case.

\subsubsection{RQ1: NFP Existence}
We sweep \(N\) under both routing cases (\(b=1\)); load-balanced (upper-bound) results are shown in Figure~\ref{fig:moe-upper-vllm-h20_main}, with load-skewed (lower-bound) results in Figure~\ref{fig:moe-lower-vllm-h20}.
As shown in Figure~\ref{fig:moe-upper-vllm-h20-latency_main}, \(T(N)\) remains nearly flat over an initial range of \(N\) and then increases in both cases, confirming module-level NFP in MoE FFNs.

\subsubsection{RQ2: Idle-Compute Prediction}
We compare the measured boundary \(N_{\max}\) with the theoretical estimate \(N_{\mathrm{idle}}^{\mathrm{moe}}\) under both routing cases.
As shown in Figure~\ref{fig:moe-upper-vllm-h20-nmax_main}, the measured boundaries do not consistently follow the idle-compute prediction in either case. 
This mismatch indicates that, unlike Dense FFNs, the MoE boundary requires mechanisms beyond idle compute alone.

\subsubsection{RQ3: Boundary Mechanism}
To identify the mechanisms missing from the idle-compute baseline, we profile the achieved arithmetic intensity while sweeping \(N\).
As shown in Figures~\ref{fig:moe-upper-vllm-h20-ai-k8_main}--\ref{fig:moe-upper-vllm-h20-flops-k8_main}, the measured arithmetic intensity changes in discrete steps rather than smoothly with \(N\), suggesting that the MoE FFN execution is affected by implementation-level kernel granularity.

To further test this hypothesis, we compare the theoretical FLOPs from the baseline model with the actual FLOPs executed at runtime. 
The runtime FLOPs often exceed the theoretical FLOPs and exhibit a similar staircase pattern, indicating that the fused MoE kernels perform additional padded computation beyond the logical expert workload. 
We confirm this by inspecting the vLLM and SGLang fused MoE kernels in Appendix \ref{app:moe-padding}, where tokens assigned to each expert are padded according to framework-defined block or tile sizes before expert GEMM execution.
These results show that MoE FFN NFP arises primarily from implementation-induced kernel-granularity slack: larger N remains near-free within existing padded kernel regions until crossing a granularity boundary.


\begin{figure*}[htbp]
    \centering

    \begin{subfigure}[t]{0.24\textwidth}
        \centering
        \includegraphics[width=\linewidth]{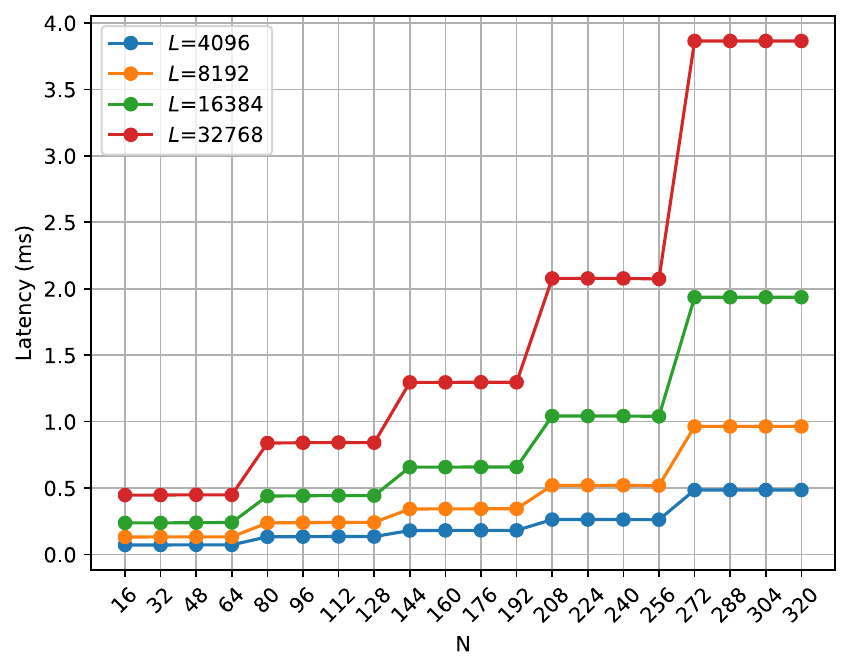}
        \caption{Latency vs. $N$}
        \label{fig:attention-flashatten-h20-latency_main}
    \end{subfigure}
    \hfill
    \begin{subfigure}[t]{0.24\textwidth}
        \centering
        \includegraphics[width=\linewidth]{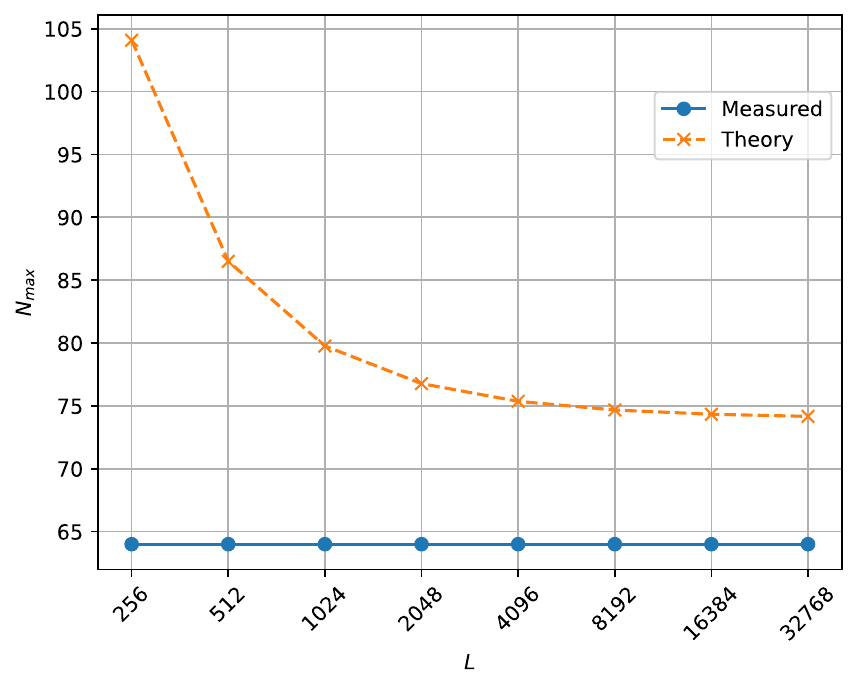}
        \caption{$N_{\max}$ vs. $L$}
        \label{fig:attention-flashatten-h20-nmax_main}
    \end{subfigure}
    \hfill
    \begin{subfigure}[t]{0.24\textwidth}
        \centering
        \includegraphics[width=\linewidth]{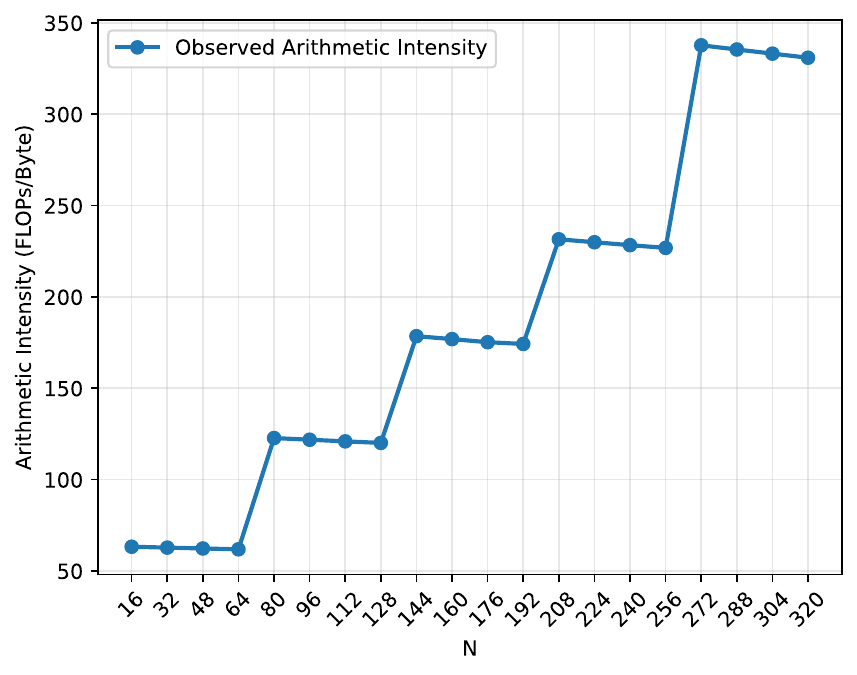}
        \caption{AI, $L=1024$}
        \label{fig:attention-flashatten-h20-ai-l1024_main}
    \end{subfigure}
    \hfill
    \begin{subfigure}[t]{0.24\textwidth}
        \centering
        \includegraphics[width=\linewidth]{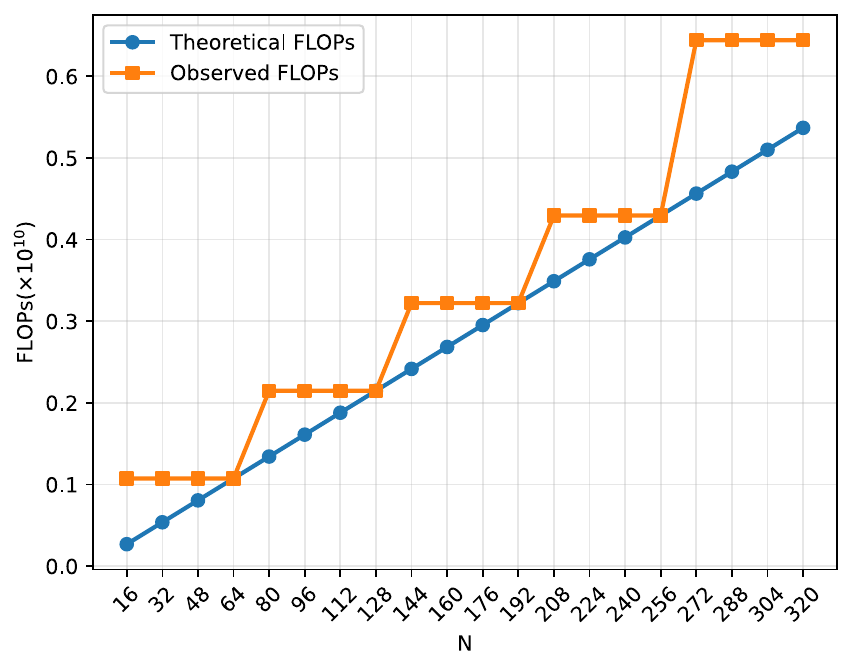}
        \caption{FLOPs, $L=1024$}
        \label{fig:attention-flashatten-h20-flops-l1024_main}
    \end{subfigure}

    \caption{
    Attention evaluation results with \textbf{FlashAttention} on \textbf{NVIDIA H20}.
    }
    \label{fig:attention-flashatten-h20_main}
\end{figure*}

\begin{findingbox}
    \textbf{Finding~\thetcbcounter:} MoE FFN NFP is governed by expert-token padding granularity. Balanced routing gives the upper bound on NFP; skewed routing gives the lower bound.
\end{findingbox}

\subsection{Attention Layers}
For the Attention analysis, we model a decode causal MHA layer over a KV cache, processing \(N\) query positions against a cached sequence of length \(L\), instantiated using FlashAttention \cite{dao2022flashattention} and FlashInfer \cite{ye2025flashinfer} to capture practical backend behavior.
We set \(b=1\) and omit QKV and output projections to isolate KV-cache-dominated module-level cost (full results in Appendix~\ref{app:attention-results}).

\subsubsection{Theoretical Baseline Prediction}

We instantiate the idle-compute baseline for Attention layers (full derivation in Appendix~\ref{app:attn-derivation}).
Let \(L\) denote the cached sequence length.
The dominant computation scales with query-cache interactions \(NL\), while the dominant KV-cache traffic is primarily controlled by \(L\).
The idle-compute baseline condition then yields
\begin{equation}
N_{\mathrm{idle}}^{\mathrm{attn}}
\approx
\begin{cases}
\frac{\rho s L}{2L-\rho s}, & 2L>\rho s, \\
+\infty, & 2L\le \rho s .
\end{cases}
\end{equation}
This predicts that the Attention NFP boundary is primarily governed by \(L\): when \(L\) is small, attention remains memory-bound over a wide range of \(N\); when \(L\) is large, the boundary approaches a finite value.

\subsubsection{RQ1: NFP Existence}
We sweep \(N\) under different cached sequence lengths \(L\).
As shown in Figure~\ref{fig:attention-flashatten-h20-latency_main}, \(T(N)\) exhibits a clear staircase pattern, with latency remaining nearly constant over several consecutive values of \(N\) before jumping to the next level, confirming module-level NFP in Attention layers.

\subsubsection{RQ2: Idle-Compute Prediction}
To evaluate the idle-compute prediction for Attention layers, we compare the measured boundary \(N_{\max}\) with the theoretical estimate \(N_{\mathrm{idle}}^{\mathrm{attn}}\) under different cached sequence lengths \(L\). 
As shown in Figure~\ref{fig:attention-flashatten-h20-nmax_main}, the measured boundary remains nearly constant across \(L\), whereas the idle-compute baseline predicts a length-dependent boundary. 
This mismatch indicates that the Attention NFP boundary is not fully explained by KV-cache memory-bound idle compute alone. 
This motivates examining implementation-level effects in RQ3.

\subsubsection{RQ3: Boundary Mechanism}
We apply the same profiling protocol as for MoE FFNs (Figure~\ref{fig:attention-flashatten-h20-ai-l1024_main}--\ref{fig:attention-flashatten-h20-flops-l1024_main}).
The results show the same pattern: arithmetic intensity and runtime FLOPs change in discrete steps due to query-position padding in FlashAttention and FlashInfer (Appendix~\ref{app:attention-query-tile}).
Thus, Attention NFP is also governed by kernel-granularity slack.
Unlike MoE, the query-tile boundary is reached before idle compute is exhausted, making granularity rather than resource balance the effective constraint on the Attention NFP boundary.

\begin{findingbox}
    \textbf{Finding~\thetcbcounter:} Attention NFP is governed by backend query-tile granularity. It is largely independent of cached sequence length.
\end{findingbox}

\begin{figure*}[htbp]
    \centering

    \begin{subfigure}[t]{0.24\textwidth}
        \centering
        \includegraphics[width=\linewidth]{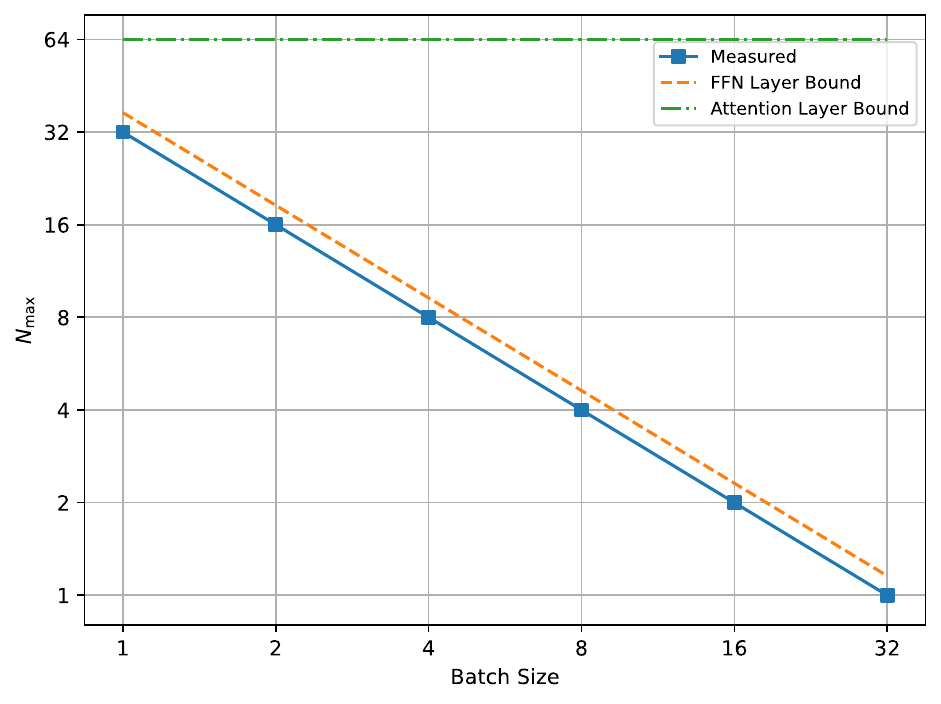}
        \caption{Dense, $L=128$}
        \label{fig:dense-model-h20-seq128_main}
    \end{subfigure}
    \hfill
    \begin{subfigure}[t]{0.24\textwidth}
        \centering
        \includegraphics[width=\linewidth]{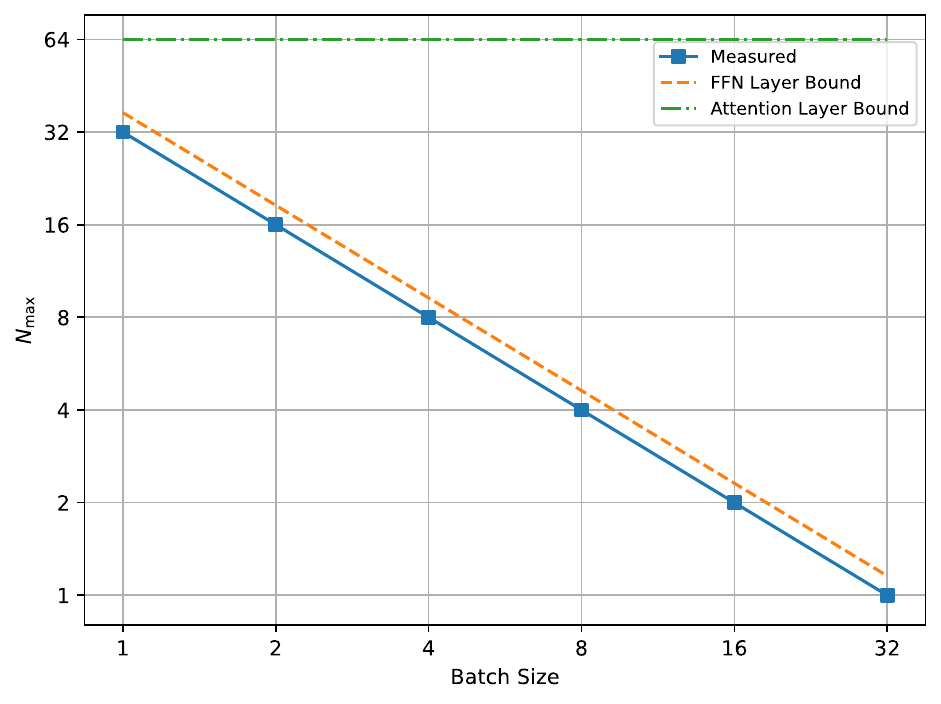}
        \caption{Dense, $L=256$}
        \label{fig:dense-model-h20-seq256_main}
    \end{subfigure}
    \hfill
    \begin{subfigure}[t]{0.24\textwidth}
        \centering
        \includegraphics[width=\linewidth]{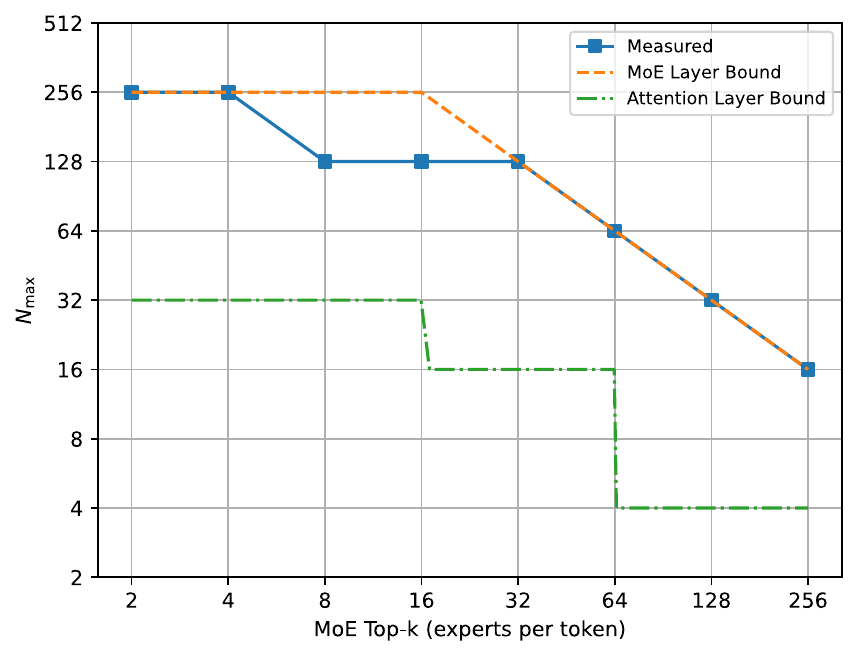}
        \caption{Load-Balanced, $L=256$}
        \label{fig:moe-upper-h20-seq256_main}
    \end{subfigure}
    \hfill
    \begin{subfigure}[t]{0.24\textwidth}
        \centering
        \includegraphics[width=\linewidth]{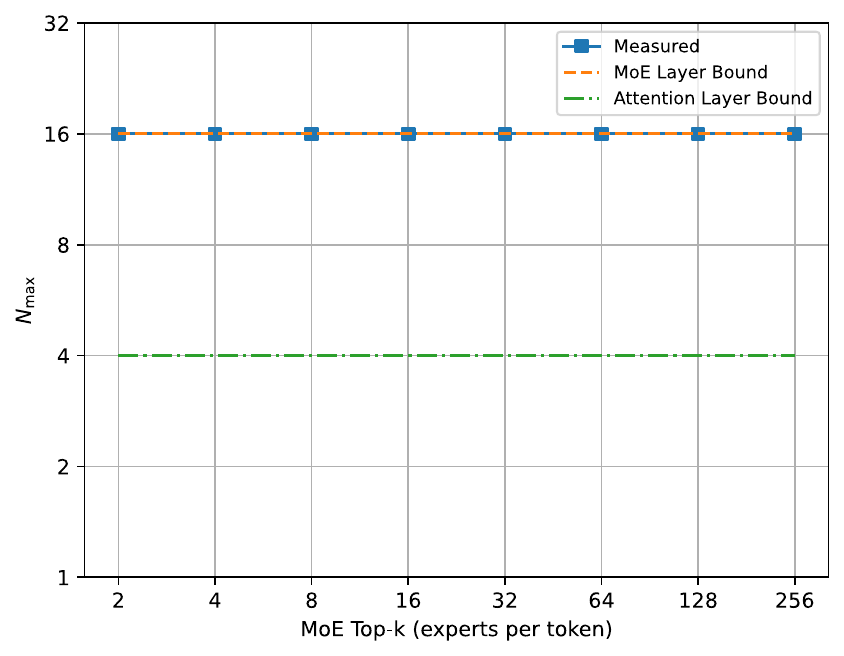}
        \caption{Load-Skewed, $L=256$}
        \label{fig:moe-lower-h20-seq256_main}
    \end{subfigure}

    \caption{
    Full-model NFP principle validation results on \textbf{NVIDIA H20}.
    }
    \label{fig:model_valid_h20_main}
\end{figure*}

\section{Near-Free Parallelism Principle}
\label{sec:nfp_law}

The module-level analysis identifies two NFP mechanisms: memory-bound idle compute for Dense FFNs, and implementation-induced kernel-granularity slack for MoE FFNs and Attention. 
At the model level, we use a first-exiting-module approximation: since modules within each layer execute sequentially without inter-layer overlap in standard inference, a full-model forward remains near-free only while all major modules remain within their module-level near-free regimes.
Thus, the model-level NFP boundary is the minimum of module-level boundaries under a fixed backend and execution configuration.

\subsection{Principle for Dense Models}

For dense models, NFP is jointly constrained by Dense FFNs and Attention. 
The Dense FFN constraint comes from the idle-compute boundary, while the Attention constraint comes from backend-specific query granularity.
Combining these two constraints gives
\begin{equation}
N_{\max}^{\mathrm{dense}}
\approx
\min\left(
\frac{\rho s}{2b},
M_{\mathrm{attn}}
\right).
\end{equation}
where \(M_{\mathrm{attn}}\) denotes the maximum number of query positions absorbable within one backend-specific Attention execution granularity.

\subsection{Principle for MoE Models}

For MoE models, NFP is jointly constrained by MoE FFNs and Attention. 
Let \(E\) be the number of experts, \(k\) the number of selected experts per token, \(M_{\mathrm{moe}}\) the expert-token padding granularity of the MoE implementation, \(M_{\mathrm{attn}}\) the query granularity of the Attention backend, and \(\tau\) the token-count threshold beyond which the MoE backend switches kernel configuration.

In the load-balanced case, expert-token padding slack is distributed across experts. 
The MoE side can absorb roughly \(M_{\mathrm{moe}}E\) padded expert-token slots, while each decode position consumes \(k\) expert-token slots. 
This gives the backend-aware upper-bound boundary
\begin{equation}
N_{\max}^{\mathrm{moe,bal}}
\approx
\min\left(
\frac{M_{\mathrm{moe}}E}{k},
\tau,
M_{\mathrm{attn}}
\right).
\end{equation}
The first term captures aggregate MoE padding capacity, the second term enforces the validity range of the selected MoE backend branch, and the third term captures the Attention backend constraint. 
For the evaluated vLLM and SGLang backend configurations, \(\tau=E\), yielding the specialized form used in our experiments.

In the load-skewed case, tokens concentrate on the same \(k\) experts. 
The MoE boundary is therefore fixed by a single expert-token padding granularity rather than the aggregate slack across all experts:
\begin{equation}
N_{\max}^{\mathrm{moe,skew}}
\approx
\min\left(
M_{\mathrm{moe}},
M_{\mathrm{attn}}
\right).
\end{equation}

Both \(M_{\mathrm{moe}}\) and \(M_{\mathrm{attn}}\) are implementation-dependent and are determined from the backend configurations analyzed in Appendices~\ref{app:moe-padding-rules} and~\ref{app:attention-query-tile-rules}.

\section{Model-Level Validation of Near-Free Parallelism Principle}

We validate the principle on full-model inference using WeDLM-8B \cite{liu2025wedlm} and Qwen3-8B \cite{yang2025qwen3} (Dense, nano-vLLM + FlashAttention) and LLaDA-2.1-mini \cite{bie2026llada2} and Ling-2.0-mini \cite{team2025every} (MoE, SGLang + FlashInfer), covering DLLM and AR settings.
We retain the \(20\%\) latency tolerance and controlled routing patterns from the module-level study.
Detailed configurations are in Appendix~\ref{sec:model_valid_setup}.

\subsection{Dense Model Validation}
As shown in Figures~\ref{fig:dense-model-h20-seq128_main}--\ref{fig:dense-model-h20-seq256_main} and~\ref{fig:dense-ar-qwen3-h800}, the measured boundaries closely match the principle-predicted boundaries across both DLLM and AR settings.
Full cross-platform results are provided in Appendix~\ref{app:dense-model-results}.
This confirms that, for dense models, the module-level Dense FFN and Attention constraints compose into an accurate full-model NFP predictor regardless of the decoding paradigm.

\subsection{MoE Model Validation}
As shown in Figures~\ref{fig:moe-upper-h20-seq256_main}--\ref{fig:moe-lower-h20-seq256_main} and~\ref{fig:moe-upper-ar-ling2-h800}--\ref{fig:moe-lower-ar-ling2-h800}, the measured boundaries follow the predicted trend across both DLLM and AR settings.
Full cross-platform and sequence-length results are provided in Appendix~\ref{app:moe-model-results}.
At shorter sequence lengths, the boundary is primarily limited by the MoE FFN granularity, whereas at longer sequence lengths the limiting factor shifts to the Attention backend.
These results show that the NFP principle captures not only the boundary value but also the dominant limiting module in full MoE model inference, consistent across decoding paradigms.

\section{Implications}
The NFP principle turns near-free parallelism from an empirical observation into a predictive system-side budget for parallel decoding: given a deployment configuration, it estimates how many decode positions can be executed at near-free latency---without predicting end-to-end speedup of any specific algorithm.
First, this budget directly guides parallelism selection---speculative verification length, MTP prediction length, or diffusion-style block size. Table~\ref{tab:practical_lookup} instantiates it as a deployment lookup, showing that the standard idle-compute intuition can over-predict by up to \(23\times\).
Second, NFP enables capacity-normalized evaluation. End-to-end speedup conflates system-side capacity with algorithm-side utilization; measuring NFP separates these factors, distinguishing whether limited speedup stems from insufficient system capacity or underutilized positions (additional implications in Appendix~\ref{app:implications}).
\section{Related Work}

\subsection{Parallel Decoding}

Speculative decoding generates candidate tokens with a draft mechanism and verifies them in parallel using the target model \cite{stern2018blockwise, leviathan2023fast, chen2023accelerating, cai2024medusa, li2024eagle, li2024eagle2, miao2024specinfer}.
Multi-token prediction methods predict multiple positions from one forward pass \cite{gloeckle2024better, liu2024deepseek, fu2024break, kou2024cllms}.
Diffusion language models and masked iterative generation expose parallelism by updating multiple positions simultaneously \cite{austin2021structured, lou2023discrete, sahoo2024simple, gat2024discrete, nielarge, ye2025dream, cheng2025sdar}.
Despite differing mechanisms, all share the same system-level structure: a single decode forward processes multiple decode positions.
This work addresses the shared system-side question---how many positions can a decode forward process at near-free latency---independent of algorithm-specific token acceptance.

\subsection{Performance Modeling for LLM Inference}
Prior work analyzes LLM inference performance at multiple levels, including end-to-end latency and throughput \cite{pope2023efficiently, aminabadi2022deepspeed, sheng2023flexgen, patel2024splitwise}, batching and scheduling effects \cite{yu2022orca, kwon2023efficient, zhong2024distserve, agrawal2024taming}, memory traffic such as model weights and KV-cache access \cite{ivanov2021data}, and operator- or kernel-level execution behavior \cite{dao2022flashattention, dao2024flashattention, ye2025flashinfer}.
Among these, roofline-style resource-balance analysis \cite{williams2009roofline, yuan2024llm} is commonly used to explain whether a workload is limited by compute or memory.

Our work adopts this modeling perspective to examine the mechanism behind near-free parallelism: increasing N can be near-free when it consumes idle compute in memory-bound execution. 
We further show that this resource-balance view is incomplete, since implementation-level granularity, such as padding and tiling, can also determine the realized NFP boundary.
\section{Conclusion}

This paper introduced Near-Free Parallelism (NFP) as a system-side capacity boundary for parallel decoding. 
Rather than treating the latency behavior of multi-position execution as an empirical artifact, we showed that this boundary can be explained and predicted from the interaction between model structure, hardware balance, and implementation granularity.
Our analysis reveals that NFP is governed by two sources of slack: memory-bound idle compute and kernel-granularity slack introduced by practical implementations. 
Building on these mechanisms, we established a Near-Free Parallelism principle and validated that it predicts full-model NFP boundaries across representative Dense and MoE DLLMs, serving stacks, and hardware platforms.

Beyond explaining existing behavior, the NFP principle provides a design signal for future parallel decoding systems. 
It can guide how many positions algorithms should expose, how runtimes and kernels should absorb them efficiently, and where future hardware scaling can expand low-latency parallel capacity. 
Overall, NFP offers a unified systems lens for co-designing decoding algorithms, inference implementations, and hardware around the parallelism that is actually near-free.

\section*{Limitations}

\textbf{Experimental conditions.}
The module-level analysis uses single-GPU execution, small batch sizes, and controlled routing patterns.
These choices isolate the target mechanisms but limit direct coverage: the Dense FFN boundary scales as $1/b$ under larger batches (validated in our sweep), real MoE routing falls between the analyzed balanced and skewed bounds, and multi-GPU settings introduce communication overlap that requires separate modeling.

\textbf{Threshold and implementation versions.}
The NFP boundary is extracted using a fixed 20\% latency tolerance, and the granularity parameters are tied to the evaluated framework versions and BF16 precision.
These choices affect the extracted boundary values but not the mechanism identification; a sensitivity sweep across $\epsilon \in [0.05, 0.30]$ confirms this robustness (Appendix~\ref{app:sensitivity}). Updating to newer backends or precisions requires only re-querying kernel configuration rules.

\textbf{Execution model.}
The current analysis assumes sequential module execution without inter-layer overlap, and does not account for potential effects of kernel fusion or cross-layer pipelining.

\bibliography{custom}

\clearpage
\onecolumn
\appendix

\renewcommand{\thepart}{}
\renewcommand{\partname}{}

\part{Appendix}
\parttoc

\clearpage

\section{Notation and Definitions}

\begin{center}
\renewcommand{\arraystretch}{1.15}
\begin{longtable}{p{0.25\linewidth} p{0.68\linewidth}}
\toprule
\textbf{Symbol} & \textbf{Definition} \\
\midrule
\endfirsthead

\toprule
\textbf{Symbol} & \textbf{Definition} \\
\midrule
\endhead

\endfoot

\bottomrule
\\[-0.2em]
\caption{Notation and definitions used throughout the paper.}
\label{tab:notation} \\
\endlastfoot

\multicolumn{2}{c}{\textbf{Parallel Decoding Notation}} \\
\midrule
\(t\) & Decode position index. \\
\(\theta\) & Model parameters. \\
\(F_\theta(\cdot)\) & Model forward function parameterized by \(\theta\). \\
\(z_t\) & Input representation at decode position \(t\). \\
\(\ell_t\) & Output logits at decode position \(t\). \\
\(Z_{t:t+N-1}\) & Inputs placed at the \(N\) decode positions from \(t\) to \(t+N-1\). \\
\(\boldsymbol{\ell}_{t:t+N-1}\) & Output logits for the \(N\) decode positions from \(t\) to \(t+N-1\). \\
\(N\) & Number of decode positions processed in one forward pass. \(N=1\) corresponds to standard autoregressive decoding. \\
\(T(N)\) & Latency of one decode forward that processes \(N\) decode positions. \\
\(T(1)\) & Latency of the single-position autoregressive baseline. \\
\(\epsilon\) & Latency tolerance used to define near-free execution. \\
\(N_{\max}(\epsilon)\) & Maximum number of decode positions whose latency remains within \((1+\epsilon)T(1)\). \\
\(N_{\max}\) & Observed near-free parallelism boundary. In the experiments, \(N_{\max}=N_{\max}(0.2)\). \\

\midrule
\multicolumn{2}{c}{\textbf{Shared Notation and Idle-Compute Baseline Notation}} \\
\midrule
\(b\) & Batch size. \\
\(d_{\mathrm{model}}\) & Model hidden dimension. \\
\(d_{\mathrm{ff}}\) & FFN intermediate dimension. \\
\(s\) & Number of bytes per element. \\
\(C(N)\) & Computation cost of a decode forward with \(N\) positions. \\
\(B(N)\) & Data movement, or memory traffic, of a decode forward with \(N\) positions. \\
\(AI(N)\) & Arithmetic intensity, defined as \(AI(N)=C(N)/B(N)\). \\
\(\phi\) & peak hardware compute throughput. \\
\(\beta\) & peak hardware memory bandwidth. \\
\(\rho\) & Hardware balance point, defined as \(\rho=\phi/\beta\). \\
\(N_{\mathrm{idle}}\) & Near-free boundary predicted by the idle-compute baseline, obtained from \(AI(N_{\mathrm{idle}})=\rho\). \\

\midrule
\multicolumn{2}{c}{\textbf{Dense FFN Notation}} \\
\midrule
\(C_{\mathrm{dense}}(N)\) & Computation cost of a Dense FFN layer with \(N\) decode positions. \\
\(B_{\mathrm{dense}}(N)\) & Dominant memory traffic of a Dense FFN layer with \(N\) decode positions. \\
\(AI_{\mathrm{dense}}(N)\) & Arithmetic intensity of a Dense FFN layer. \\
\(N_{\mathrm{idle}}^{\mathrm{dense}}\) & Idle-compute-predicted near-free boundary for Dense FFN layers. \\
\(N_{\max}^{\mathrm{dense}}\) & Principle-predicted model-level near-free boundary for dense models. \\
\midrule

\pagebreak[4]

\multicolumn{2}{c}{\textbf{MoE FFN Notation}} \\
\midrule
\(k\) & Number of selected experts per token in top-\(k\) routing. \\
\(E\) & Total number of experts. \\
\(E_{\mathrm{act}}\) & Number of distinct experts activated by a routing pattern. \\
\(\eta\) & Implementation-dependent factor capturing additional read-modify-write overhead in the MoE combine stage. \\
\(C_{\mathrm{moe}}(N)\) & Computation cost of a MoE FFN layer with \(N\) decode positions. \\
\(B_{\mathrm{moe},W}(N)\) & Memory traffic from loading active expert weights. \\
\(B_{\mathrm{moe},A}(N)\) & Routing-induced activation memory traffic from dispatch and combine. \\
\(B_{\mathrm{moe}}(N)\) & Total dominant memory traffic of a MoE FFN layer. \\
\(AI_{\mathrm{moe}}(N)\) & Arithmetic intensity of a MoE FFN layer. \\
\(N_{\mathrm{idle}}^{\mathrm{moe}}\) & Idle-compute-predicted near-free boundary for MoE FFN layers. \\
\(N_{\max}^{\mathrm{moe,bal}}\) & Principle-predicted model-level near-free boundary for MoE models under load-balanced routing. \\
\(N_{\max}^{\mathrm{moe,skew}}\) & Principle-predicted model-level near-free boundary for MoE models under load-skewed routing. \\
\(M_{\mathrm{moe}}\) & Expert-token padding granularity of the MoE implementation. \\
\(\tau\) & Validity bound of the backend branch that selects \(M_{\mathrm{moe}}\). \\

\midrule
\multicolumn{2}{c}{\textbf{Attention Notation}} \\
\midrule
\(L\) & Cached sequence length in attention. \\
\(d_{\mathrm{kv}}\) & KV hidden dimension. For standard multi-head attention, \(d_{\mathrm{kv}}=d_{\mathrm{model}}\). \\
\(C_{\mathrm{attn}}(N)\) & Computation cost of an attention layer with \(N\) query positions. \\
\(B_{\mathrm{attn}}(N)\) & KV-cache-dominated memory traffic of an attention layer with \(N\) query positions. \\
\(AI_{\mathrm{attn}}(N)\) & Arithmetic intensity of an attention layer. \\
\(N_{\mathrm{idle}}^{\mathrm{attn}}\) & Idle-compute-predicted near-free boundary for attention layers. \\
\(M_{\mathrm{attn}}\) & Maximum number of query positions that can be absorbed within one backend-specific attention execution granularity. \\

\end{longtable}
\renewcommand{\arraystretch}{1.0}
\end{center}

\clearpage

\section{Theoretical Baseline Derivations}
\label{app:baseline-derivations}

The Dense FFN idle-compute baseline derivation is presented in the main text. This section provides the full derivations for MoE FFNs and Attention layers.

\subsection{MoE FFN Idle-Compute Baseline Derivation}
\label{app:moe-derivation}

We instantiate the idle-compute baseline for MoE FFNs.
Let \(k\) denote the number of selected experts per token, \(E_{\mathrm{act}}\) the number of distinct experts activated by a routing pattern, and \(\eta=2\) the number of per-expert activation accesses in the combine stage.\footnote{One read of each expert output and one write for weighted accumulation.}
A MoE FFN processes \(bN\) positions, each routed to \(k\) experts.
Since each expert is implemented as a two-GEMM FFN, the total computation is approximated as
\begin{equation}
C_{\mathrm{moe}}(N)
\approx
4bNk d_{\mathrm{model}}d_{\mathrm{ff}} .
\end{equation}

Unlike Dense FFNs, the dominant memory traffic in MoE FFNs consists of both active expert weights and routing-induced activation movement.
The weight traffic is proportional to the number of activated experts,
\begin{equation}
B_{\mathrm{moe},W}(N)
\approx
2E_{\mathrm{act}}d_{\mathrm{model}}d_{\mathrm{ff}}s ,
\end{equation}
while dispatch and combine move activations between tokens and experts.
We approximate this routing-induced activation traffic as
\begin{equation}
B_{\mathrm{moe},A}(N)
\approx
bN d_{\mathrm{model}}s(1+3k+\eta k).
\end{equation}
Combining weight and activation traffic yields the arithmetic intensity
\begin{equation}
AI_{\mathrm{moe}}(N)
\approx
\frac{
4bNk d_{\mathrm{ff}}
}{
s\left(2E_{\mathrm{act}}d_{\mathrm{ff}} + bN(1+3k+\eta k)\right)
}.
\end{equation}
Substituting this intensity into the idle-compute baseline condition yields
\begin{equation}
N_{\mathrm{idle}}^{\mathrm{moe}}
\approx
\frac{
2\rho s E_{\mathrm{act}}d_{\mathrm{ff}}
}{
b\left(4kd_{\mathrm{ff}}-\rho s(1+3k+\eta k)\right)
},
\end{equation}
when \(4kd_{\mathrm{ff}}>\rho s(1+3k+\eta k)\); otherwise, the execution remains memory-bound under this abstraction.

\clearpage

\subsection{Attention Idle-Compute Baseline Derivation}
\label{app:attn-derivation}

We instantiate the idle-compute baseline for Attention layers.
Let \(L\) denote the cached sequence length, \(N\) the number of decode positions per request, \(d_{\mathrm{kv}}\) the KV hidden dimension, and \(s\) the number of bytes per element.
For standard MHA, \(d_{\mathrm{kv}}=d_{\mathrm{model}}\).
Under a KV-cache-dominated abstraction, a decode attention layer processes \(bN\) query positions over a cache of length \(L\).
The dominant computation comes from the attention score computation and value aggregation, while the dominant memory traffic comes from accessing the KV cache:
\begin{equation}
\begin{aligned}
C_{\mathrm{attn}}(N)
&\approx
4bNLd_{\mathrm{kv}}, \\
B_{\mathrm{attn}}(N)
&\approx
2b(L+N)d_{\mathrm{kv}}s .
\end{aligned}
\end{equation}

The key asymmetry is that computation scales with the number of query-cache interactions \(NL\), while the dominant KV-cache traffic is primarily controlled by the cached sequence length.
The AI is therefore
\begin{equation}
AI_{\mathrm{attn}}(N)
=
\frac{C_{\mathrm{attn}}(N)}{B_{\mathrm{attn}}(N)}
\approx
\frac{2NL}{(L+N)s}.
\end{equation}
Substituting this intensity into the idle-compute baseline condition yields
\begin{equation}
N_{\mathrm{idle}}^{\mathrm{attn}}
\approx
\begin{cases}
\frac{\rho s L}{2L-\rho s}, & 2L>\rho s, \\
+\infty, & 2L\le \rho s .
\end{cases}
\end{equation}

This result predicts that the NFP boundary of Attention layers is primarily governed by the cached sequence length \(L\).
When \(L\) is small, the attention core remains memory-bound over a wide range of \(N\).
When \(L\) is sufficiently large, the boundary approaches a finite value determined by the compute-to-memory balance, and increasing N is near-free only until it exhausts the idle compute exposed by KV-cache-dominated execution.

\clearpage

\section{Module-Level Analysis Implementation Details}
\label{appendix:experimental-setup}

\subsection{Overall Settings}

\subsubsection{Experimental Goal and Scope}
The goal is not to benchmark complete models, but to
measure the NFP boundary of individual modules under controlled configurations.
This module-level analysis serves as the first step toward the model-level NFP
principle: by isolating the boundary of each major module, we can later identify the
limiting module and compose module-level constraints into full-model
predictions. We consider the same three module classes as in the main analysis:
Dense FFNs, MoE FFNs, and Attention. The scope of the module-level experiments is summarized as follows:
\begin{itemize}
    \item \textbf{Dense FFN layers}: isolated two-GEMM FFN operators used to
    study resource slack and the idle-compute baseline.
    \item \textbf{MoE FFN layers}: fused MoE operators with controlled routing
    patterns used to study expert-token padding and kernel-granularity slack.
    \item \textbf{Attention layers}: decode attention kernels over a
    pre-allocated KV cache used to study backend-specific query granularity.
\end{itemize}

\subsubsection{Common NFP Measurement Protocol}

For each module and configuration, we vary only the number of decode positions
$N$ and keep all other factors fixed, including the module shape,
hardware platform, inference backend, and numerical precision. We denote by
$T(N)$ the measured latency of one module forward that processes $N$ decode
positions under this fixed configuration. The single-position case $N=1$ is
used as the autoregressive baseline $T(1)$.

Following the definition in the main text, we define a configuration with $N$
decode positions to be near-free if its latency remains within a 20\% increase
over the single-position baseline:
\begin{equation}
    T(N) \leq 1.2 T(1),
\end{equation}
The observed NFP boundary is then extracted as
\begin{equation}
    N_{\max}=N_{\max}(0.2)
    =
    \max \left\{
    N \in \mathcal{N}:
    T(N) \leq 1.2T(1)
    \right\}.
\end{equation}
where $N$ denotes the set of sampled decode-position counts in the
sweep.

This protocol turns the visual notion of a flat latency region into a uniform
measurement rule. The 20\% tolerance captures a practical low-overhead regime
rather than exact latency equality, while reducing sensitivity to GPU timing
noise, kernel launch jitter, and minor backend-dependent latency
discontinuities. We use this threshold only as an operational criterion for
extracting a boundary from each latency curve; the mechanism analysis further
relies on the latency trend, profiling metrics, and comparison with the
idle-compute baseline. We apply the same threshold and extraction rule to all
modules, hardware platforms, and inference backends, so that the measured
boundaries can be compared consistently with the idle-compute baseline and the
model-level NFP principle.

\subsubsection{Latency Measurement Procedure}

For each sampled value of $N$, we measure the steady-state latency of the
corresponding module forward. Before timing, all input tensors, output buffers,
routing tensors, and KV-cache buffers required by the module are pre-allocated
on the GPU, so that the reported latency excludes memory allocation, data
movement from the host, and other setup overheads. Each configuration is first
warmed up for 50 iterations to trigger CUDA context initialization, kernel
loading or compilation, backend autotuning, and cache stabilization.

After warm-up, we perform 10 independent measurement rounds. Each round contains
200 timed iterations of the same module forward under the fixed configuration.
Latency is measured using CUDA events with explicit device synchronization,
which avoids artifacts from asynchronous GPU execution. For each configuration,
we compute the median latency within each measurement round and report the
median across the 10 rounds as $T(N)$. This median-based procedure provides a
stable estimate of the steady-state module latency and reduces the influence of
occasional timing spikes when extracting the NFP boundary.

\subsubsection{Hardware Platforms}

We evaluate all module-level experiments on three GPU platforms: NVIDIA A800,
H800, and H20. These platforms cover different architectures and
compute-to-memory balance points, allowing us to test whether the measured NFP
boundary follows the hardware-dependent resource-balance prediction or is
instead dominated by backend-specific kernel granularity. The same set of
hardware platforms is used for Dense FFN, MoE FFN, and Attention experiments to
ensure that the observed trends are not artifacts of a single GPU generation or
hardware configuration. All experiments are conducted on a single GPU to avoid
potential confounding effects from inter-GPU communication.

Table~\ref{tab:hardware_platforms} summarizes the hardware platforms. We use
$\phi$ to denote the peak FP16 compute throughput and $\beta$ to denote the
peak memory bandwidth. The hardware balance point is computed as
$\rho=\phi/\beta$. All module-level experiments are executed in BF16 precision.
We use the same numerical precision across modules and platforms to avoid
precision-dependent differences in kernel selection and throughput, while
matching a common precision setting for LLM inference.
\begin{table}[htbp]
\centering
\small
\begin{tabular}{lccccc}
\toprule
GPU & Release Date & Architecture & $\phi$ (FP16/BF16 TFLOP/s) & $\beta$ (TB/s) & $\rho=\phi/\beta$ \\
\midrule
H20  & 2024 & Hopper & 148 & 4.0 & 37.0 \\
A800 & 2022 & Ampere & 312 & 2.039 & 153.0 \\
H800 & 2023 & Hopper & 989 & 3.35 & 295.2 \\
\bottomrule
\end{tabular}
\caption{Hardware platforms used in the module-level experiments. $\phi$ denotes
peak FP16/BF16 compute throughput, $\beta$ denotes peak memory bandwidth, and
$\rho=\phi/\beta$ is the compute-to-memory balance point.}
\label{tab:hardware_platforms}
\end{table}

\subsubsection{Profiling Details}

We profile the measured forward region using NVIDIA Nsight Compute in roofline
mode. The profiled region is delimited by CUDA profiler start and stop markers,
and includes all GPU kernels launched by the corresponding module forward.
Kernel-level counters are aggregated into forward-level metrics using kernel
execution time as the weight.

For compute-side profiling, we use dtype-matched tensor-core throughput
counters according to the numerical precision of the experiment, i.e., BF16 or
FP16. The reported compute utilization is obtained by normalizing the
time-weighted achieved tensor-core throughput by the corresponding peak
tensor-core throughput. We also derive the observed runtime FLOPs from the
achieved tensor-core throughput and kernel execution time, normalized by the
number of profiled forward executions.

For memory-side profiling, we use DRAM throughput counters. The observed DRAM
traffic per forward is computed from the measured DRAM byte throughput and
kernel execution time, again normalized by the number of profiled forward
executions. When the profiler directly reports percentage-of-peak DRAM
throughput, we report the time-weighted average as DRAM bandwidth utilization.
The observed arithmetic intensity is computed as the ratio between observed
runtime FLOPs and observed DRAM traffic. These metrics are used to interpret
whether the measured NFP boundary is dominated by resource slack or by
implementation-induced kernel-granularity effects.

\subsubsection{Reproducibility Controls}
We use a fixed random seed, \texttt{torch.manual\_seed(0)}. All measurements are conducted on
pre-allocated GPU tensors and focus only on the measured module forward. Data
loading, CPU preprocessing, graph construction, and other host-side setup
operations are excluded from the measured latency. First-run initialization and
compilation overheads are also excluded through the warm-up procedure described
above. These controls ensure that the reported latency and profiling metrics
reflect the steady-state execution behavior of the module rather than
input-generation or setup overheads.

\clearpage

\subsection{Dense FFN Layers Implementation Details}

We implement the Dense FFN layer as an isolated two-GEMM operator, matching the
module abstraction used in the main analysis. One forward consists of an
up-projection followed by a down-projection. We omit activation functions, bias
terms, and gating branches, so that the measured latency reflects the dense
matrix-computation behavior of the FFN rather than additional elementwise or
routing operations. We set $d_{\mathrm{model}}=4096$ and
$d_{\mathrm{ff}}=9216$, following the hidden size and intermediate size of
LLaDA-2.1-Flash, a representative existing DLLM, to make the microbenchmark
reflect realistic model dimensions.

The two GEMMs are executed as separate \texttt{torch.matmul} operations, using
the underlying cuBLAS/cuBLASLt backend. We do not use fused MLP kernels,
compiler fusion, Triton kernels, or custom CUTLASS kernels in this benchmark.
This design keeps the Dense FFN experiment aligned with the idle-compute
baseline and avoids introducing additional implementation-specific granularity
effects. All tensors are stored in BF16 precision. The
benchmark measures a single FFN layer in isolation, without attention,
normalization, residual connections, or layer stacking.

\subsection{MoE FFN Layers Implementation Details}

We instantiate MoE FFN layers using the fused MoE operators from vLLM v0.9.1
and SGLang v0.5.9. Each MoE forward consists of token dispatch, expert FFN
computation, and weighted combine, following the execution path of practical
MoE inference backends. Each expert is implemented as a two-GEMM FFN, and the
benchmark isolates a single MoE FFN layer without attention, normalization,
residual connections, or layer stacking.

We use controlled top-$k$ routing instead of input-dependent gating, so that the
expert-load distribution is deterministic and comparable across $N$, hardware
platforms, and backends. We use $E=256$ experts and sweep $k$ from 2 to 256
selected experts per token. Each expert uses
$d_{\mathrm{model}}=4096$ and $d_{\mathrm{ff}}=1024$, following the hidden size
and MoE intermediate size of LLaDA-2.1-Flash. All MoE experiments are executed
in BF16 precision using the same logical workload across vLLM and SGLang. We
evaluate both backends on the same set of GPU platforms to obtain cross-stack
evidence and to avoid attributing an implementation- or hardware-specific
artifact to the general MoE NFP mechanism.

\subsubsection{Controlled Routing Patterns}

We evaluate two controlled routing patterns. In the load-balanced case, tokens
are assigned to experts in a deterministic round-robin manner:
\begin{equation}
    \{(i\cdot k + j) \bmod E\}_{j=0}^{k-1}
\end{equation}
for token $i$. With $T=bN$ total tokens, this gives
$E_{\mathrm{act}}=\min(E,Tk)$. Since $b=1$ in the MoE experiments,
$E_{\mathrm{act}}=\min(E,Nk)$. This case spreads tokens across as many experts
as possible.

For the load-balanced case, we do not use $N=1$ as the latency baseline,
because the number of activated experts would still change as $N$ increases.
Instead, for each top-$k$ setting, we first increase $N$ until all experts are
activated and use this smallest saturated point as the baseline:
\begin{equation}
    N_{\mathrm{bal},0}(k)
    =
    \left\lceil \frac{E}{bk} \right\rceil .
\end{equation}
In our experiments, $E=256$ and $b=1$, so
$N_{\mathrm{bal},0}(k)=\lceil 256/k\rceil$ as $k$ is swept from 2 to 256. The
load-balanced NFP boundary is extracted relative to
$T(N_{\mathrm{bal},0}(k))$, so that the measured near-free region reflects
increasing N under a fixed fully activated expert set rather than the
cost of activating more experts.

In the load-skewed case, all tokens are routed to the same $k$ experts, so
$E_{\mathrm{act}}=k$ regardless of $N$. We therefore use the standard
single-position baseline for this case. These two patterns represent the
upper- and lower-bound expert-load cases used to study how expert-token
distribution and fused-kernel padding affect the MoE NFP boundary.

\clearpage

\subsection{Attention Layers Implementation Details}

We benchmark the decode attention core over a pre-allocated KV cache. Each
forward processes $N$ query positions against a cached sequence of length $L$
under causal masking. We omit the QKV projections and output projection, so
that the measured latency reflects the KV-cache attention backend rather than
additional dense GEMM operations.

For simplicity, we instantiate the attention module as standard multi-head
attention (MHA). We use $n_{\mathrm{heads}}=32$ and
$d_{\mathrm{head}}=128$, giving
$d_{\mathrm{model}}=n_{\mathrm{heads}}d_{\mathrm{head}}=4096$. Since the
benchmark uses MHA, the key-value hidden dimension is
$d_{\mathrm{kv}}=d_{\mathrm{model}}=4096$. The same formulation also applies
to GQA or MQA settings, where only the KV width changes to
$d_{\mathrm{kv}}=n_{\mathrm{kv\_heads}}d_{\mathrm{head}}$.

We evaluate two practical decode attention backends: FlashAttention-2 and
FlashInfer. The same logical attention workload is used across both backends
and all hardware platforms. This allows us to attribute differences in the
measured NFP boundary to backend-specific query granularity and kernel
implementation behavior rather than changes in model shape or attention
semantics. All attention experiments are executed in BF16 precision and isolate
a single attention layer without normalization, residual connections, or layer
stacking.

\clearpage

\section{Module-Level Analysis Results}

\subsection{Dense FFN Layers Results}
\label{app:dense-ffn-results}

\begin{figure*}[!p]
    \centering
    \scalebox{1}[1.0]{%
    \begin{minipage}{\textwidth}
        \centering

        \begin{subfigure}[t]{0.24\textwidth}
            \centering
            \includegraphics[width=\linewidth]{figs/Dense_FFN/H20/fig1_latency_vs_N_log2.pdf}
            \caption{Latency vs. $N$}
            \label{fig:dense-ffn-h20-latency}
        \end{subfigure}
        \hfill
        \begin{subfigure}[t]{0.24\textwidth}
            \centering
            \includegraphics[width=\linewidth]{figs/Dense_FFN/H20/fig2_Nmax_vs_b_with_theory.pdf}
            \caption{$N_{\max}$ vs. $b$}
            \label{fig:dense-ffn-h20-nmax}
        \end{subfigure}
        \hfill
        \begin{subfigure}[t]{0.24\textwidth}
            \centering
            \includegraphics[width=\linewidth,height=1.0\linewidth,keepaspectratio]{figs/Dense_FFN/H20/fig3_utilization_vs_batch_roofline.pdf}
            \caption{$N=1$ utilization vs. $b$}
            \label{fig:dense-ffn-h20-util-batch}
        \end{subfigure}
        \hfill
        \begin{subfigure}[t]{0.24\textwidth}
            \centering
            \includegraphics[width=\linewidth,height=1.0\linewidth,keepaspectratio]{figs/Dense_FFN/H20/fig4_utilization_vs_N_b1_roofline.pdf}
            \caption{$b=1$}
            \label{fig:dense-ffn-h20-util-b1}
        \end{subfigure}

        \vspace{0.8em}
        \begin{subfigure}[t]{0.24\textwidth}
            \centering
            \includegraphics[width=\linewidth,height=1.0\linewidth,keepaspectratio]{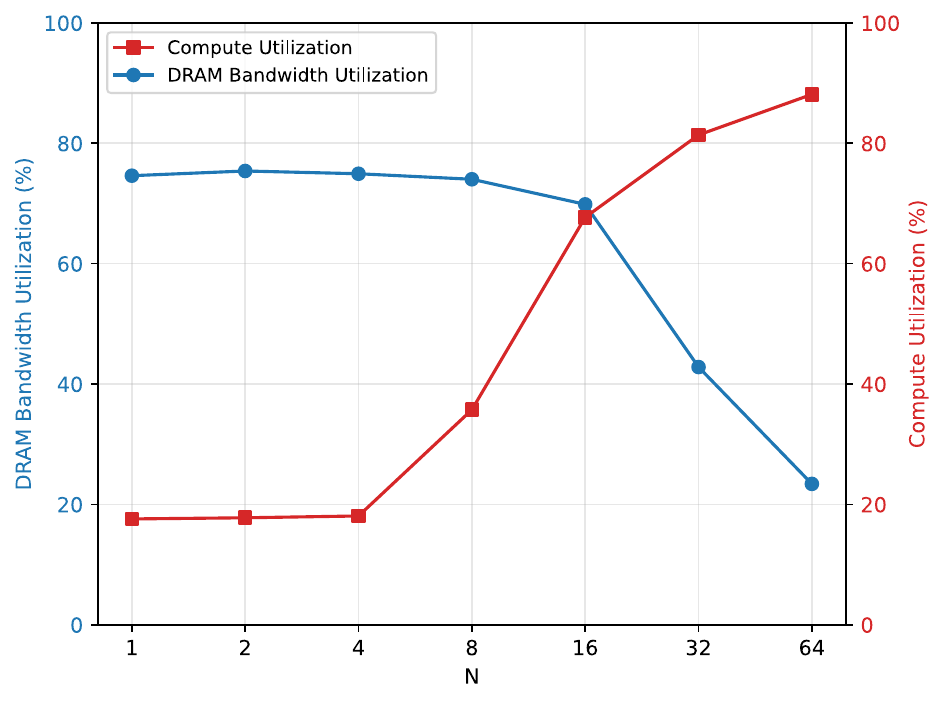}
            \caption{$b=2$}
            \label{fig:dense-ffn-h20-util-b2}
        \end{subfigure}
        \hfill
        \begin{subfigure}[t]{0.24\textwidth}
            \centering
            \includegraphics[width=\linewidth,height=1.0\linewidth,keepaspectratio]{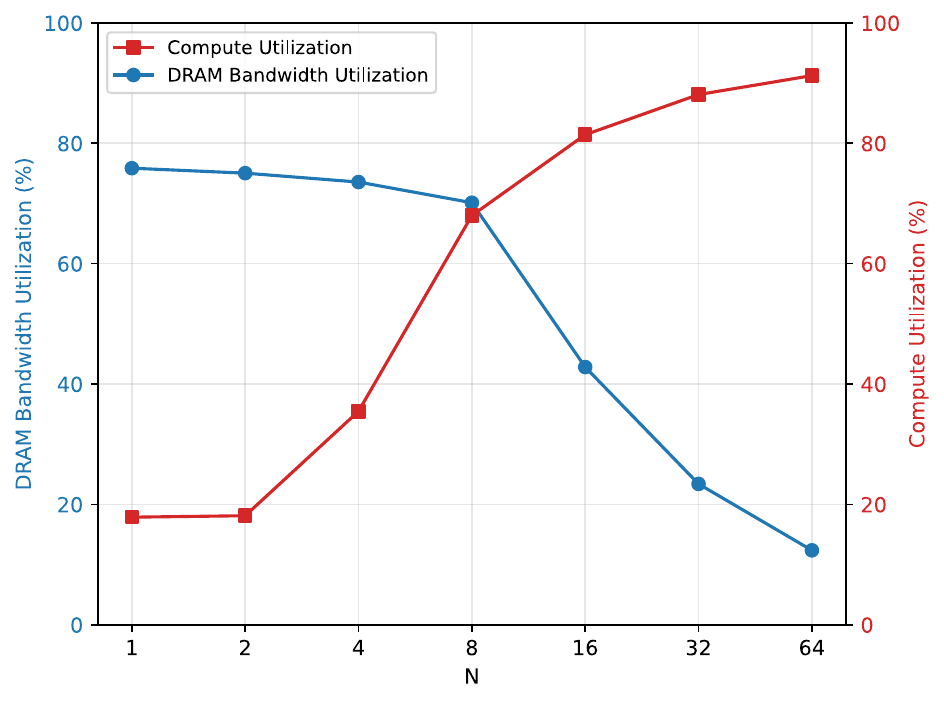}
            \caption{$b=4$}
            \label{fig:dense-ffn-h20-util-b4}
        \end{subfigure}
        \hfill
        \begin{subfigure}[t]{0.24\textwidth}
            \centering
            \includegraphics[width=\linewidth,height=1.0\linewidth,keepaspectratio]{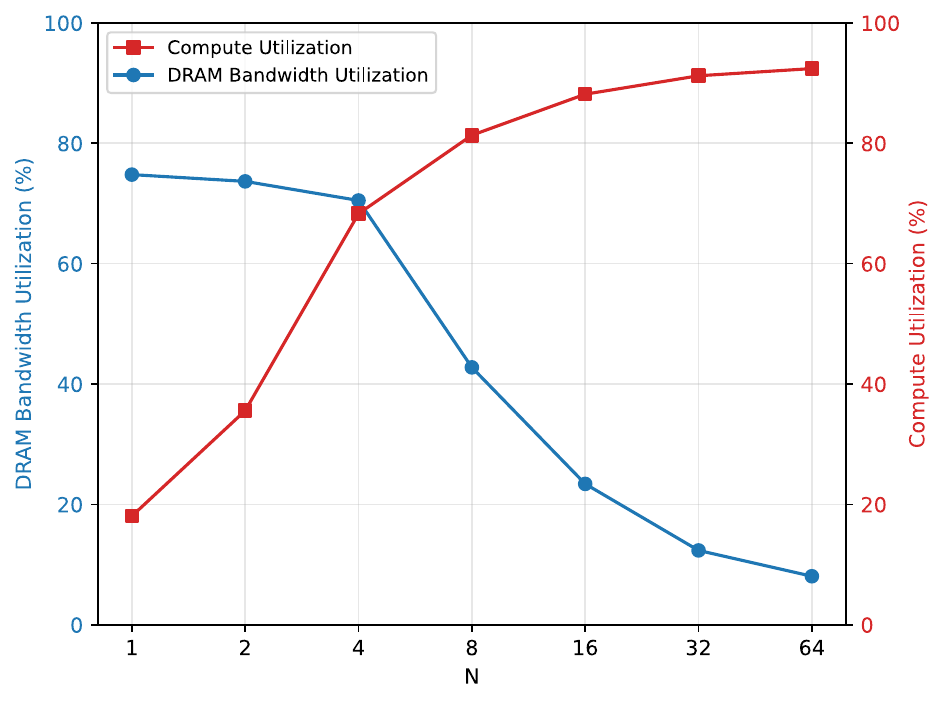}
            \caption{$b=8$}
            \label{fig:dense-ffn-h20-util-b8}
        \end{subfigure}
        \hfill
        \begin{subfigure}[t]{0.24\textwidth}
            \centering
            \includegraphics[width=\linewidth,height=1.0\linewidth,keepaspectratio]{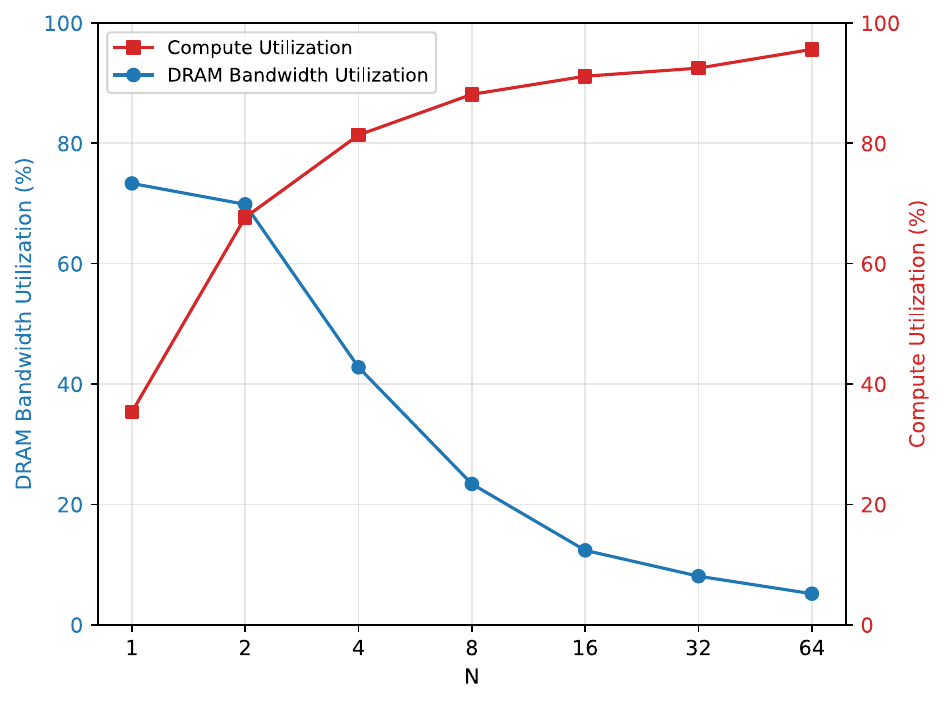}
            \caption{$b=16$}
            \label{fig:dense-ffn-h20-util-b16}
        \end{subfigure}

    \end{minipage}%
    }

    \caption{
    Dense FFN evaluation results on \textbf{NVIDIA H20 GPU}.
    }
    \label{fig:dense-ffn-h20}
\end{figure*}

\begin{figure*}[!p]
    \centering
    \scalebox{1}[1.0]{%
    \begin{minipage}{\textwidth}
        \centering

        \begin{subfigure}[t]{0.24\textwidth}
            \centering
            \includegraphics[width=\linewidth]{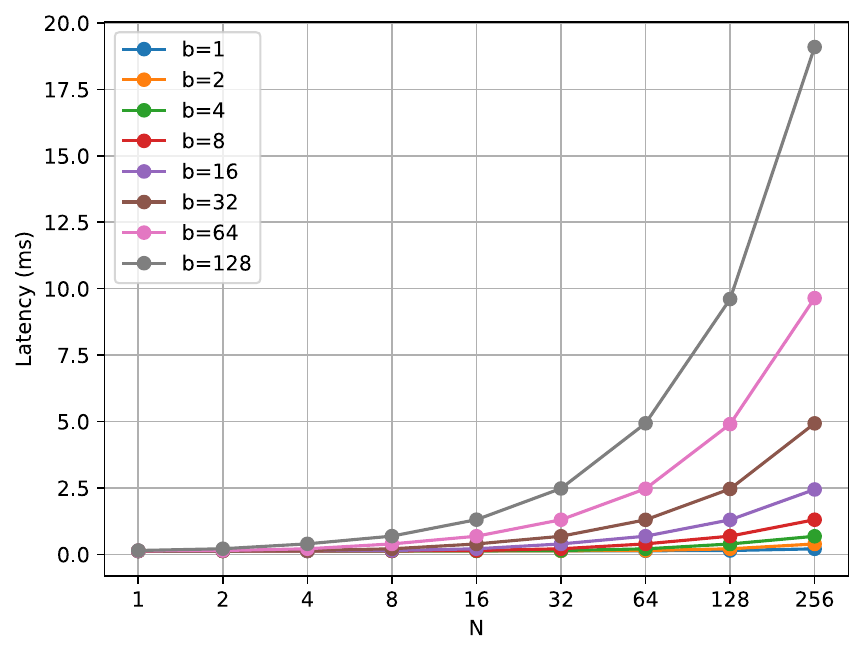}
            \caption{Latency vs. $N$}
            \label{fig:dense-ffn-a800-latency}
        \end{subfigure}
        \hfill
        \begin{subfigure}[t]{0.24\textwidth}
            \centering
            \includegraphics[width=\linewidth]{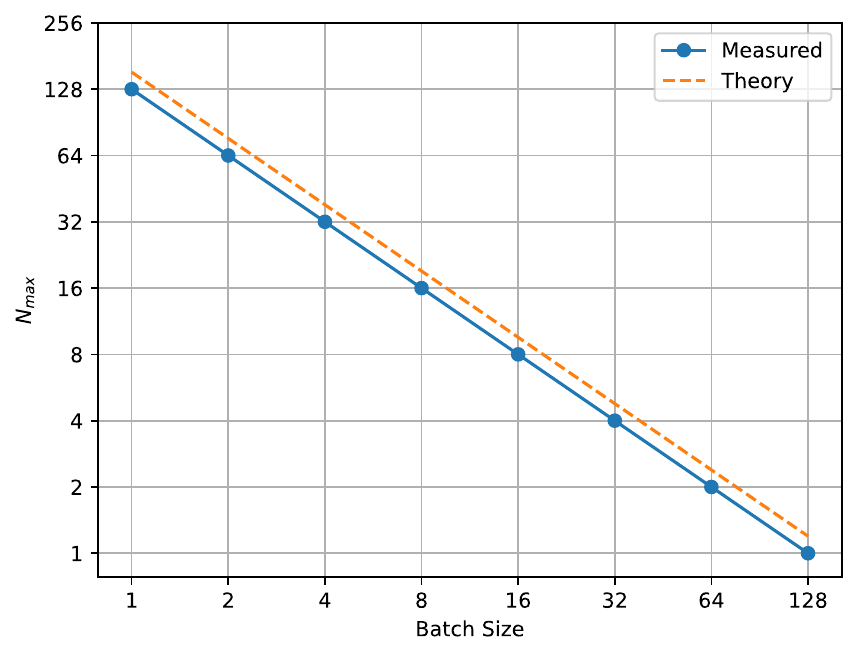}
            \caption{$N_{\max}$ vs. $b$}
            \label{fig:dense-ffn-a800-nmax}
        \end{subfigure}
        \hfill
        \begin{subfigure}[t]{0.24\textwidth}
            \centering
            \includegraphics[width=\linewidth,height=1.0\linewidth,keepaspectratio]{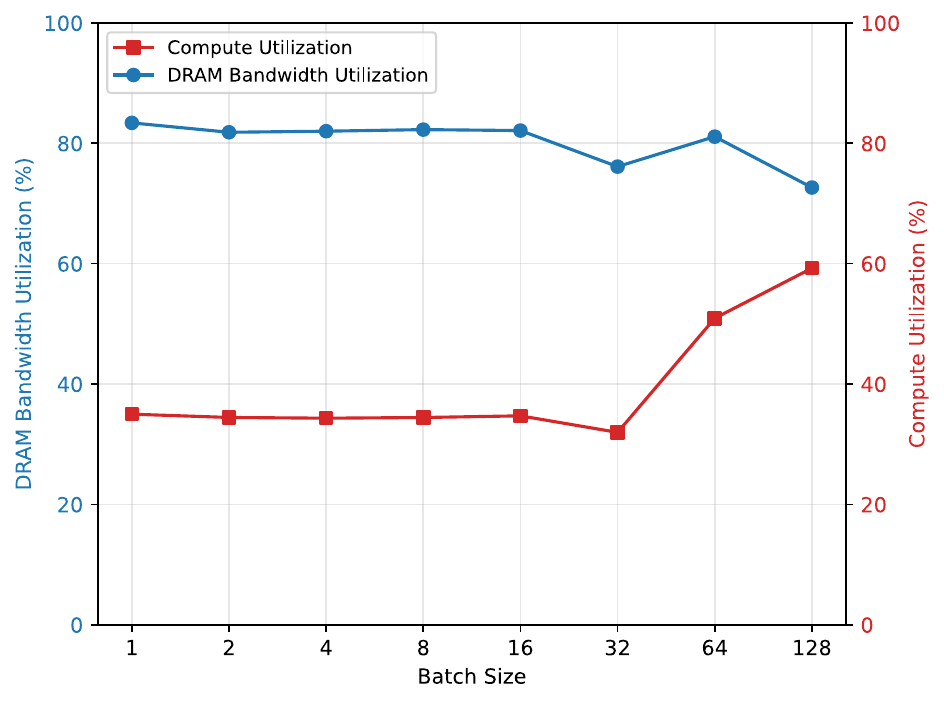}
            \caption{$N=1$ utilization vs. $b$}
            \label{fig:dense-ffn-a800-util-batch}
        \end{subfigure}
        \hfill
        \begin{subfigure}[t]{0.24\textwidth}
            \centering
            \includegraphics[width=\linewidth,height=1.0\linewidth,keepaspectratio]{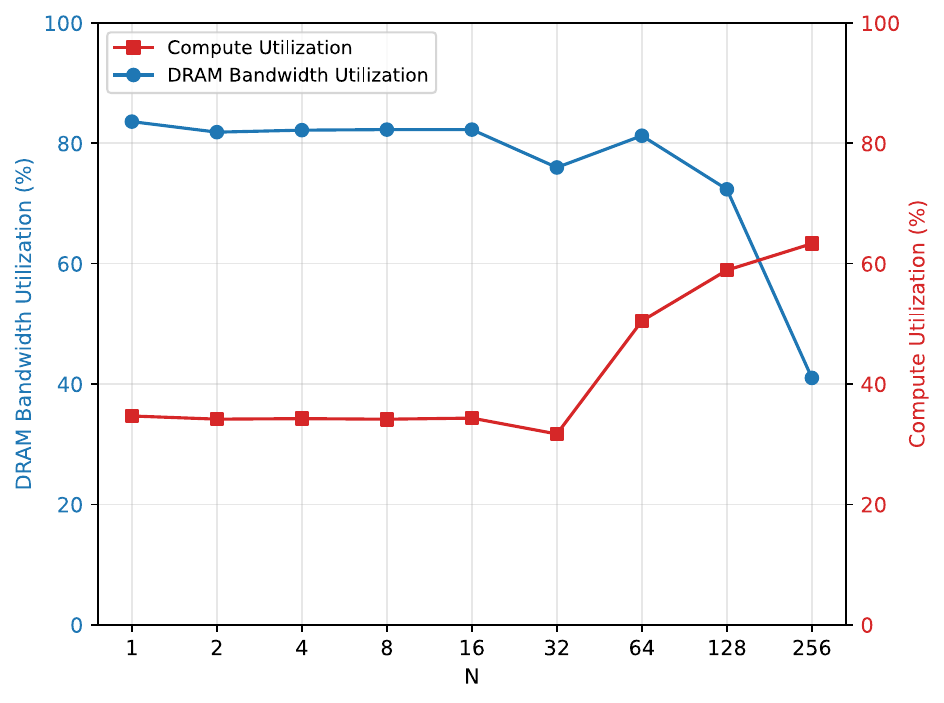}
            \caption{$b=1$}
            \label{fig:dense-ffn-a800-util-b1}
        \end{subfigure}

        \vspace{0.8em}
        \begin{subfigure}[t]{0.24\textwidth}
            \centering
            \includegraphics[width=\linewidth,height=1.0\linewidth,keepaspectratio]{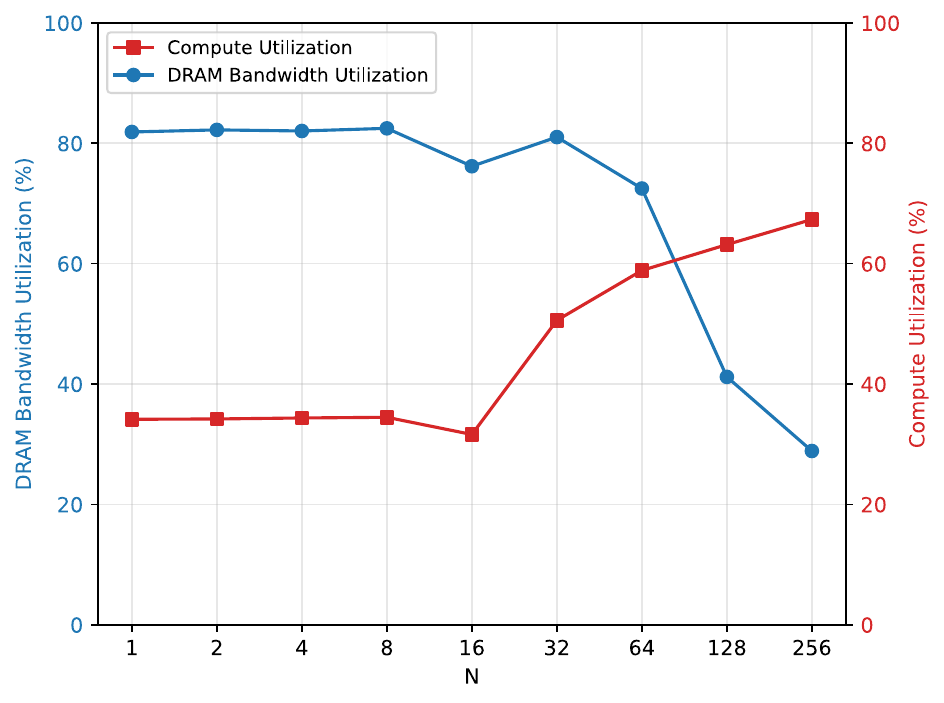}
            \caption{$b=2$}
            \label{fig:dense-ffn-a800-util-b2}
        \end{subfigure}
        \hfill
        \begin{subfigure}[t]{0.24\textwidth}
            \centering
            \includegraphics[width=\linewidth,height=1.0\linewidth,keepaspectratio]{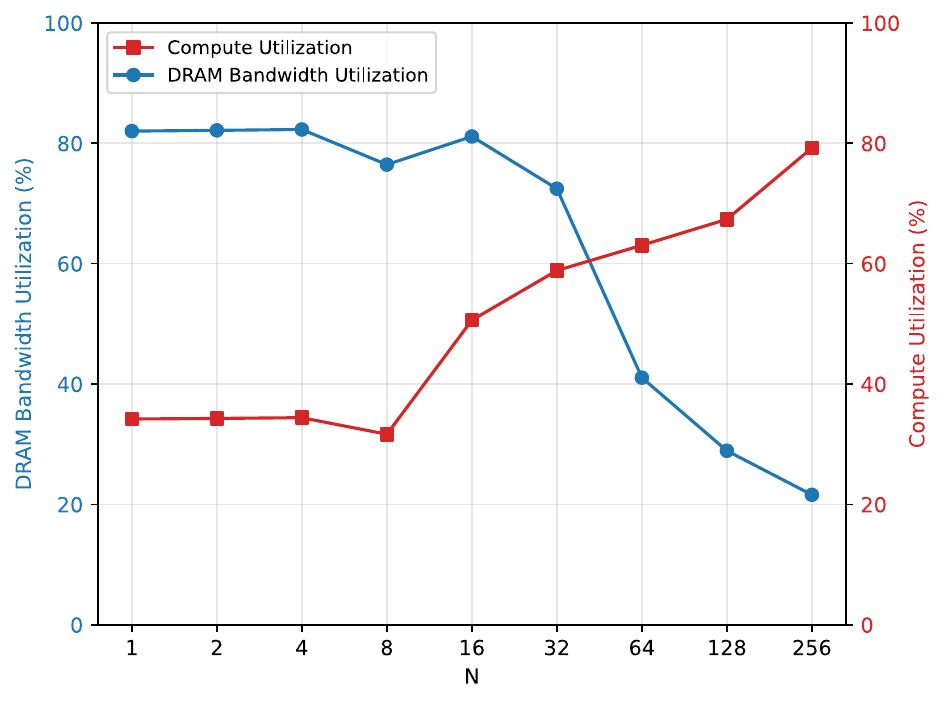}
            \caption{$b=4$}
            \label{fig:dense-ffn-a800-util-b4}
        \end{subfigure}
        \hfill
        \begin{subfigure}[t]{0.24\textwidth}
            \centering
            \includegraphics[width=\linewidth,height=1.0\linewidth,keepaspectratio]{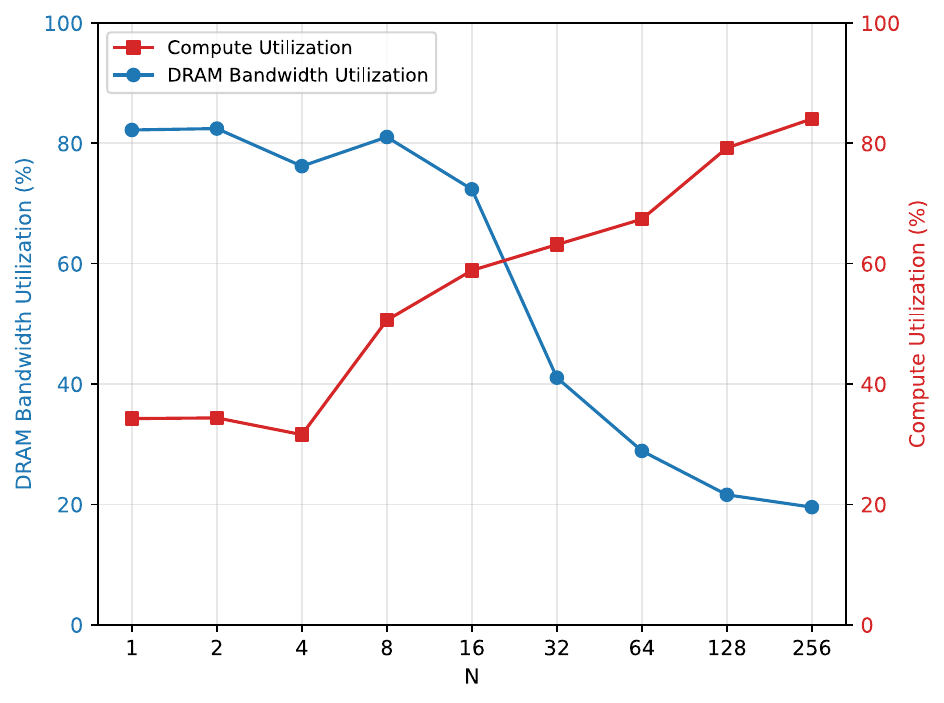}
            \caption{$b=8$}
            \label{fig:dense-ffn-a800-util-b8}
        \end{subfigure}
        \hfill
        \begin{subfigure}[t]{0.24\textwidth}
            \centering
            \includegraphics[width=\linewidth,height=1.0\linewidth,keepaspectratio]{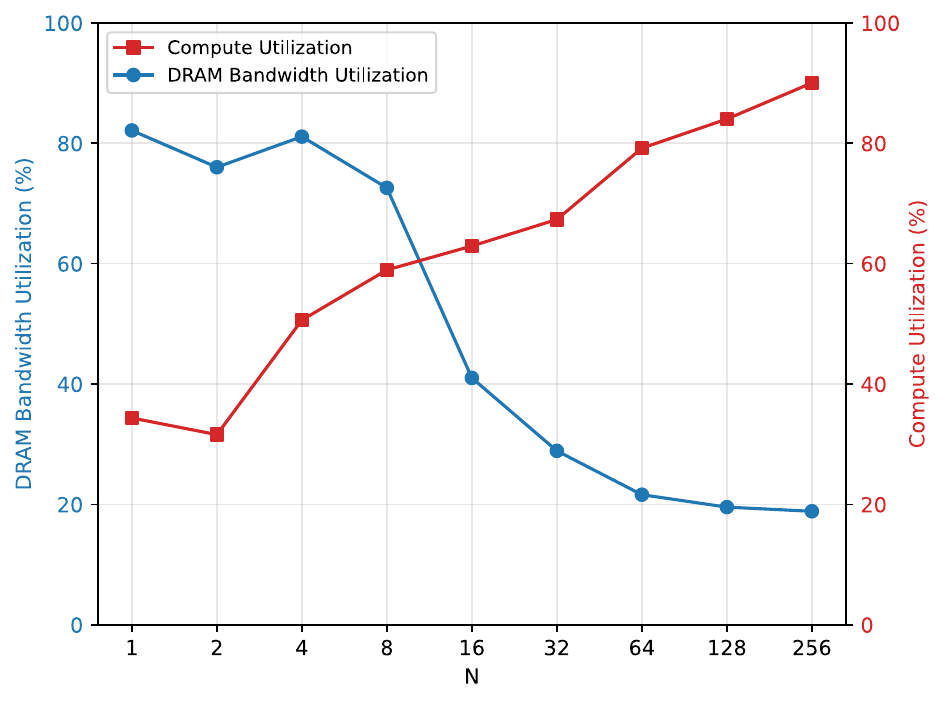}
            \caption{$b=16$}
            \label{fig:dense-ffn-a800-util-b16}
        \end{subfigure}

    \end{minipage}%
    }

    \caption{
    Dense FFN evaluation results on \textbf{NVIDIA A800 GPU}.
    }
    \label{fig:dense-ffn-a800}
\end{figure*}

\begin{figure*}[!p]
    \centering
    \scalebox{1}[1.0]{%
    \begin{minipage}{\textwidth}
        \centering

        \begin{subfigure}[t]{0.24\textwidth}
            \centering
            \includegraphics[width=\linewidth]{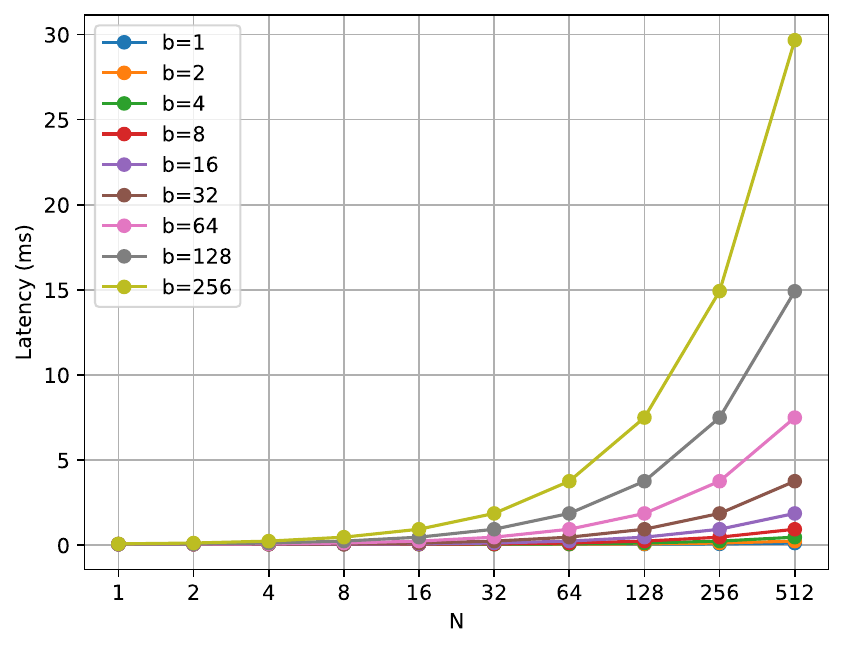}
            \caption{Latency vs. $N$}
            \label{fig:dense-ffn-h800-latency}
        \end{subfigure}
        \hfill
        \begin{subfigure}[t]{0.24\textwidth}
            \centering
            \includegraphics[width=\linewidth]{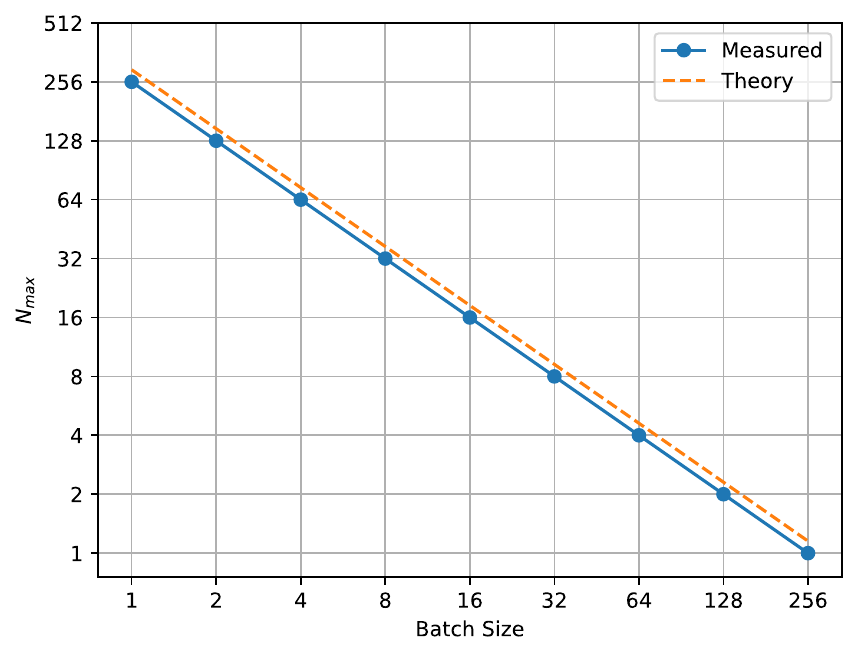}
            \caption{$N_{\max}$ vs. $b$}
            \label{fig:dense-ffn-h800-nmax}
        \end{subfigure}
        \hfill
        \begin{subfigure}[t]{0.24\textwidth}
            \centering
            \includegraphics[width=\linewidth,height=1.0\linewidth,keepaspectratio]{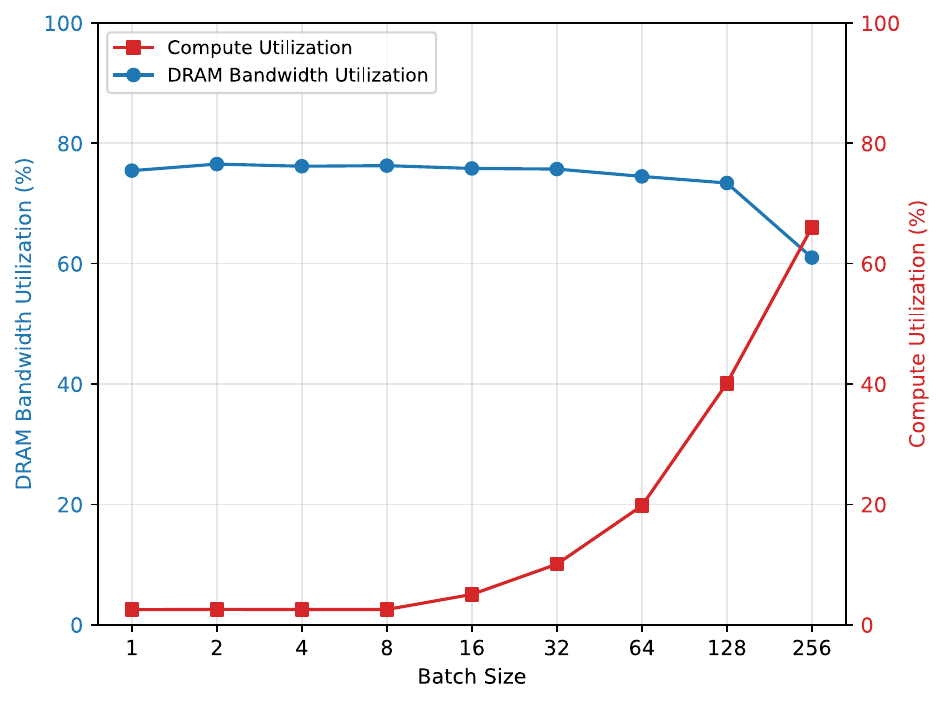}
            \caption{$N=1$ utilization vs. $b$}
            \label{fig:dense-ffn-h800-util-batch}
        \end{subfigure}
        \hfill
        \begin{subfigure}[t]{0.24\textwidth}
            \centering
            \includegraphics[width=\linewidth,height=1.0\linewidth,keepaspectratio]{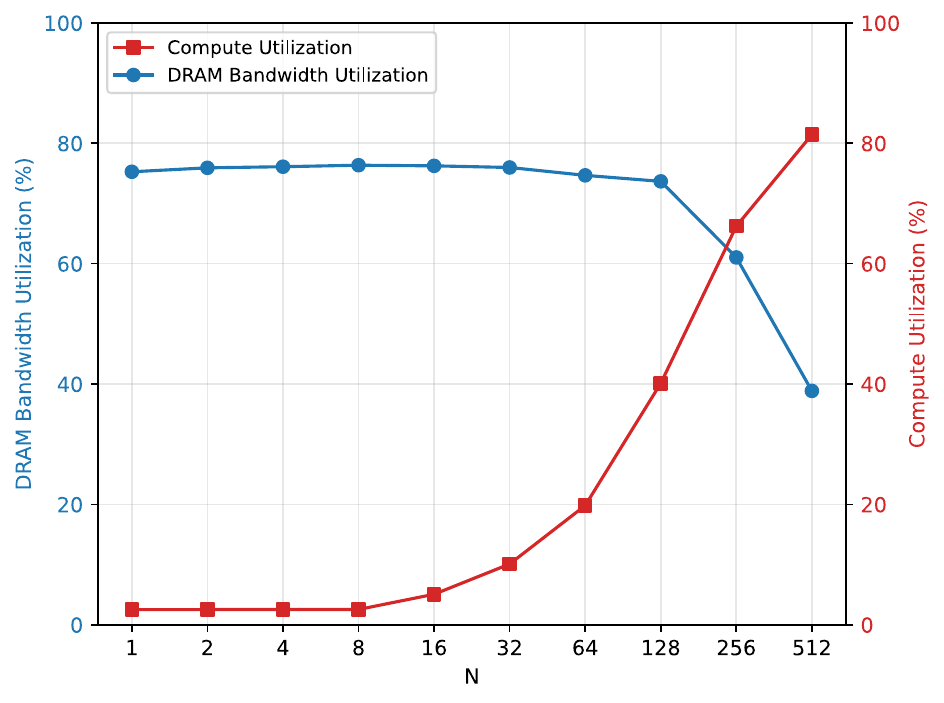}
            \caption{$b=1$}
            \label{fig:dense-ffn-h800-util-b1}
        \end{subfigure}

        \vspace{0.8em}
        \begin{subfigure}[t]{0.24\textwidth}
            \centering
            \includegraphics[width=\linewidth,height=1.0\linewidth,keepaspectratio]{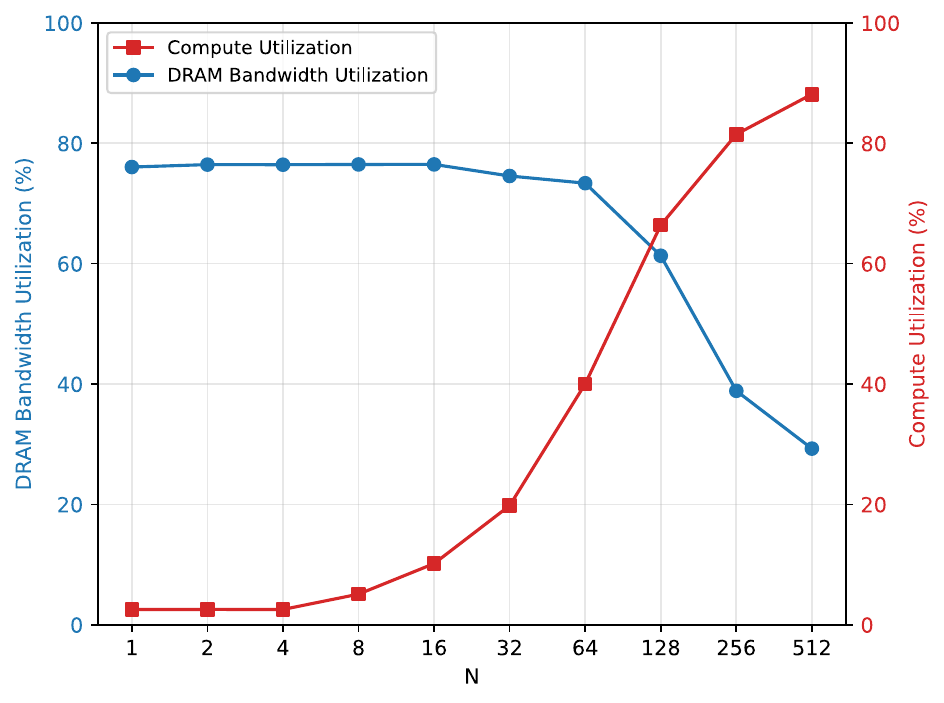}
            \caption{$b=2$}
            \label{fig:dense-ffn-h800-util-b2}
        \end{subfigure}
        \hfill
        \begin{subfigure}[t]{0.24\textwidth}
            \centering
            \includegraphics[width=\linewidth,height=1.0\linewidth,keepaspectratio]{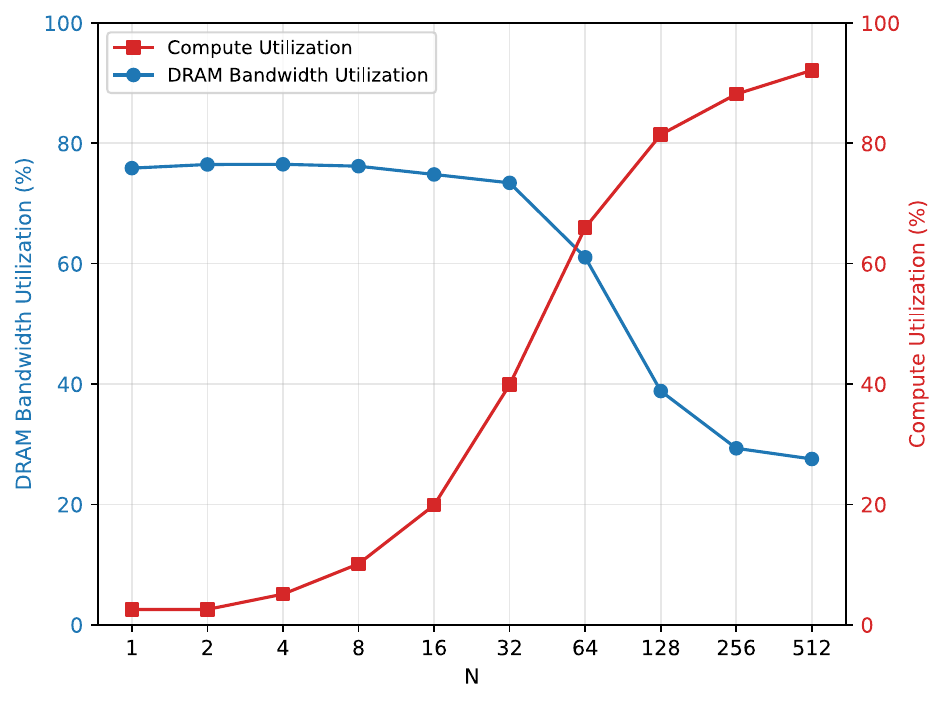}
            \caption{$b=4$}
            \label{fig:dense-ffn-h800-util-b4}
        \end{subfigure}
        \hfill
        \begin{subfigure}[t]{0.24\textwidth}
            \centering
            \includegraphics[width=\linewidth,height=1.0\linewidth,keepaspectratio]{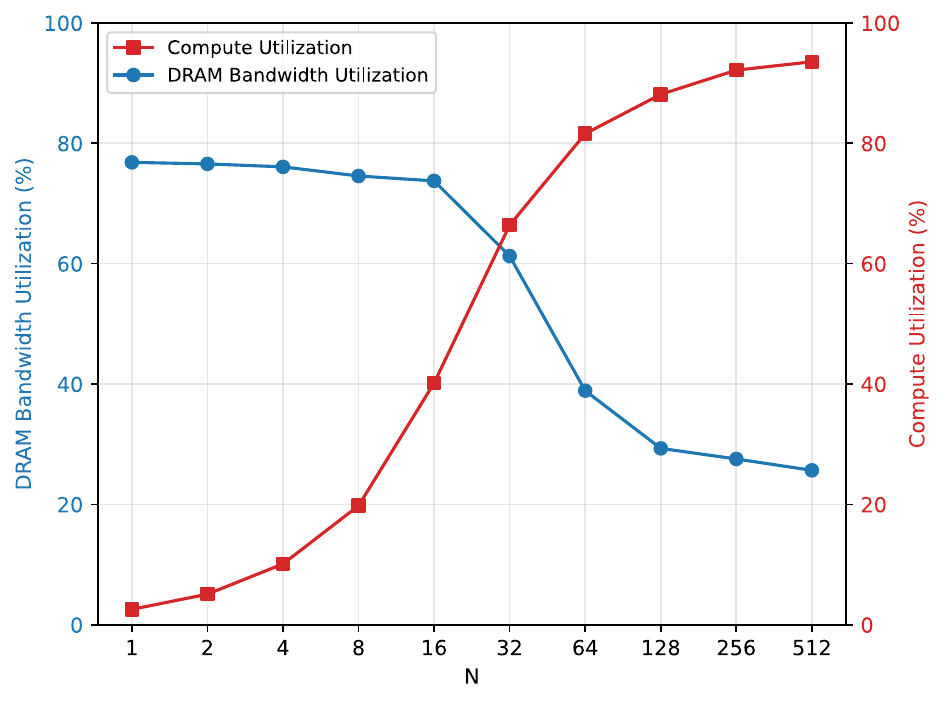}
            \caption{$b=8$}
            \label{fig:dense-ffn-h800-util-b8}
        \end{subfigure}
        \hfill
        \begin{subfigure}[t]{0.24\textwidth}
            \centering
            \includegraphics[width=\linewidth,height=1.0\linewidth,keepaspectratio]{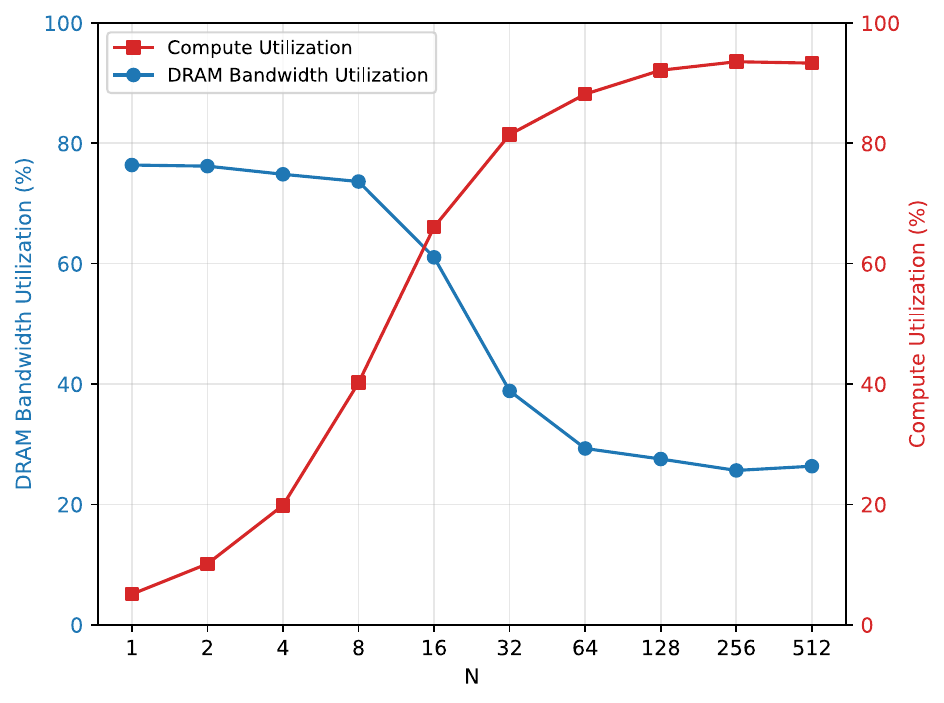}
            \caption{$b=16$}
            \label{fig:dense-ffn-h800-util-b16}
        \end{subfigure}

    \end{minipage}%
    }

    \caption{
    Dense FFN evaluation results on \textbf{NVIDIA H800 GPU}.
    }
    \label{fig:dense-ffn-h800}
\end{figure*}

Figures~\ref{fig:dense-ffn-h20}--\ref{fig:dense-ffn-h800} report the complete Dense FFN module-level results from single-GPU experiments on NVIDIA H20, A800, and H800, respectively. These experiments complement the main-text analysis by validating the Dense FFN idle-compute mechanism on GPU platforms with different compute-to-memory balance points. In each figure, the first subfigure shows the latency curve as the number of decode positions $N$ increases, the second reports the extracted NFP boundary $N_{\max}$ as a function of batch size $b$, the third shows the $N=1$ resource utilization across batch sizes, and the remaining subfigures show the utilization trends when sweeping $N$ for each fixed batch size.

On each evaluated GPU, Dense FFNs exhibit a clear near-free region. For each batch size, latency remains nearly flat over an initial range of $N$ and increases only after the available slack is exhausted. The size of this region is batch-size dependent: smaller batches support a larger near-free range, while larger batches reduce the maximum near-free N. This behavior is consistent across H20, A800, and H800.

The extracted boundaries further match the idle-compute prediction. As shown in Figures~\ref{fig:dense-ffn-h20}\subref{fig:dense-ffn-h20-nmax}, \ref{fig:dense-ffn-a800}\subref{fig:dense-ffn-a800-nmax}, and \ref{fig:dense-ffn-h800}\subref{fig:dense-ffn-h800-nmax}, $N_{\max}$ decreases approximately inversely with $b$, following the predicted trend $N_{\mathrm{idle}}^{\mathrm{dense}} \approx \rho s / 2b$. The absolute boundary differs across GPU platforms because the amount of available compute slack depends on the hardware balance point $\rho$. A GPU with larger $\rho$ exposes more compute capacity relative to memory bandwidth and therefore supports a larger Dense FFN near-free region under the same workload.

The utilization profiles explain this trend. At $N=1$, Dense FFN execution is bandwidth-dominated: DRAM bandwidth utilization is already high, while tensor-core utilization remains comparatively low. As $N$ increases, tensor-core utilization rises, indicating that increasing N consumes otherwise idle compute. In contrast, DRAM bandwidth utilization changes more slowly because the dominant weight traffic is already incurred by the single-position baseline. Increasing the batch size raises the initial compute utilization at $N=1$, leaving less remaining compute slack and therefore reducing $N_{\max}$.

Overall, these single-GPU results confirm that the Dense FFN NFP boundary is primarily governed by memory-bound resource slack rather than implementation-induced kernel granularity. Dense FFNs therefore contribute the hardware-sensitive term $\rho s / 2b$ to the model-level NFP principle.

\clearpage

\vspace{-0.2cm}
\clearpage

\subsection{MoE FFN Layers Results}
\label{app:moe-ffn-results}

This section reports the complete MoE FFN module-level results. These experiments complement the main-text analysis by validating the implementation-induced kernel-granularity mechanism across routing patterns, fused MoE backends, and single-GPU platforms. We consider two controlled routing cases: load-balanced routing, which distributes tokens across experts and exposes the upper-bound MoE NFP behavior, and load-skewed routing, which routes all tokens to the same selected experts and exposes the lower-bound behavior. For each case, we evaluate both vLLM and SGLang fused MoE operators on NVIDIA H20, A800, and H800.

\subsubsection{Load-balanced Routing as Upper Bound}

\begin{figure*}[!p]
    \centering
    \scalebox{1}[1.0]{%
    \begin{minipage}{\textwidth}
        \centering

        \begin{subfigure}[t]{0.24\textwidth}
            \centering
            \includegraphics[width=\linewidth]{figs/MoE_FFN_Upper_vLLM/H20/fig1_total_latency_vs_N_log2.pdf}
            \caption{Latency vs. $N$}
            \label{fig:moe-upper-vllm-h20-latency}
        \end{subfigure}
        \hfill
        \begin{subfigure}[t]{0.24\textwidth}
            \centering
            \includegraphics[width=\linewidth]{figs/MoE_FFN_Upper_vLLM/H20/fig2_Nmax_vs_k_with_theory.pdf}
            \caption{$N_{\max}$ vs. $k$}
            \label{fig:moe-upper-vllm-h20-nmax}
        \end{subfigure}
        \hfill
        \begin{subfigure}[t]{0.24\textwidth}
            \centering
            \includegraphics[width=\linewidth,height=0.75\linewidth,keepaspectratio]{figs/MoE_FFN_Upper_vLLM/H20/fig4_main_kernel_arithmetic_intensity_vs_N_k8_roofline.pdf}
            \caption{AI, $k=8$}
            \label{fig:moe-upper-vllm-h20-ai-k8}
        \end{subfigure}
        \hfill
        \begin{subfigure}[t]{0.24\textwidth}
            \centering
            \includegraphics[width=\linewidth,height=0.75\linewidth,keepaspectratio]{figs/MoE_FFN_Upper_vLLM/H20/fig3_main_kernel_flops_vs_N_k8_roofline.pdf}
            \caption{FLOPs, $k=8$}
            \label{fig:moe-upper-vllm-h20-flops-k8}
        \end{subfigure}

        \vspace{0.8em}
        \begin{subfigure}[t]{0.24\textwidth}
            \centering
            \includegraphics[width=\linewidth,height=0.75\linewidth,keepaspectratio]{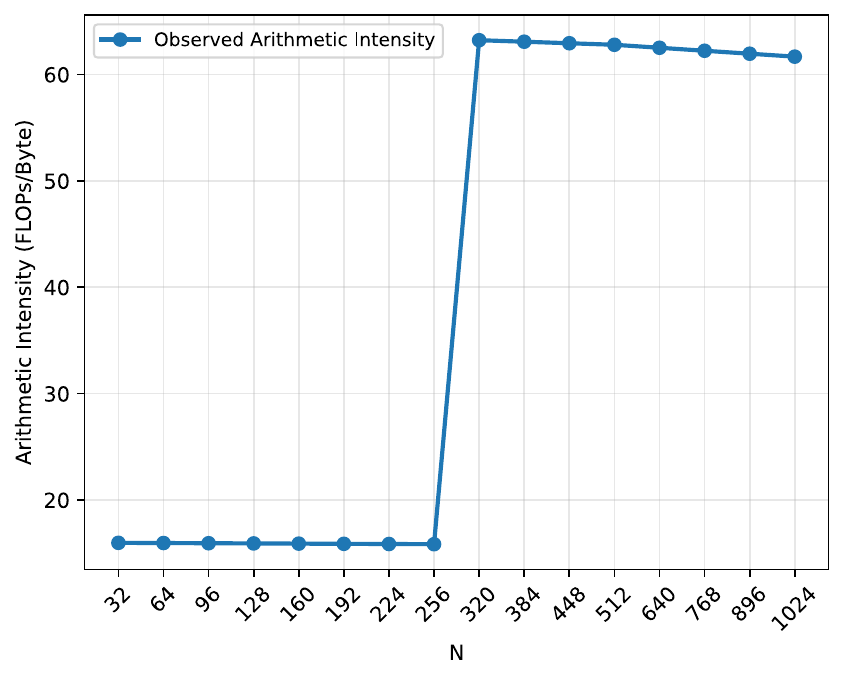}
            \caption{AI, $k=16$}
            \label{fig:moe-upper-vllm-h20-ai-k16}
        \end{subfigure}
        \hfill
        \begin{subfigure}[t]{0.24\textwidth}
            \centering
            \includegraphics[width=\linewidth,height=0.75\linewidth,keepaspectratio]{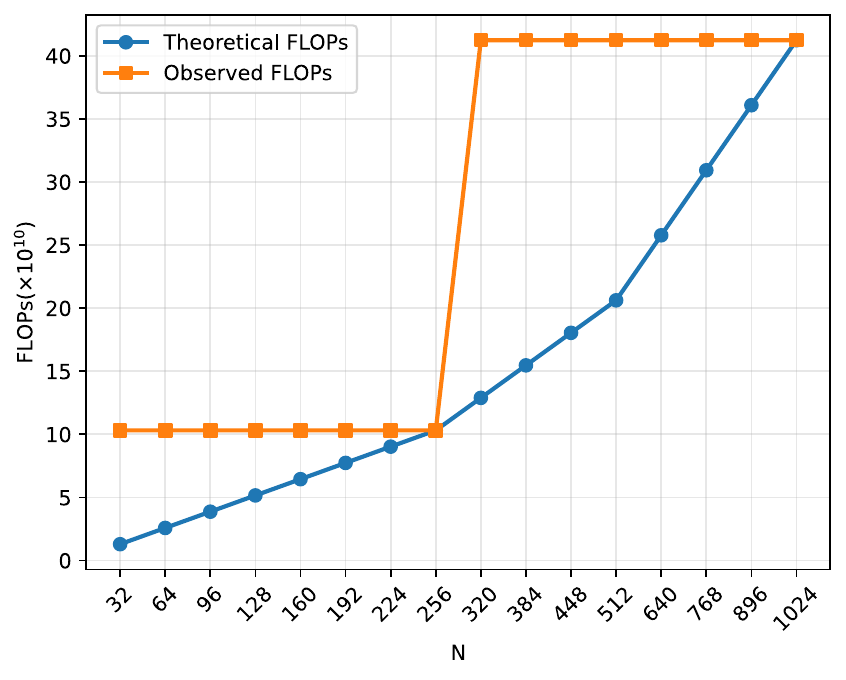}
            \caption{FLOPs, $k=16$}
            \label{fig:moe-upper-vllm-h20-flops-k16}
        \end{subfigure}
        \hfill
        \begin{subfigure}[t]{0.24\textwidth}
            \centering
            \includegraphics[width=\linewidth,height=0.75\linewidth,keepaspectratio]{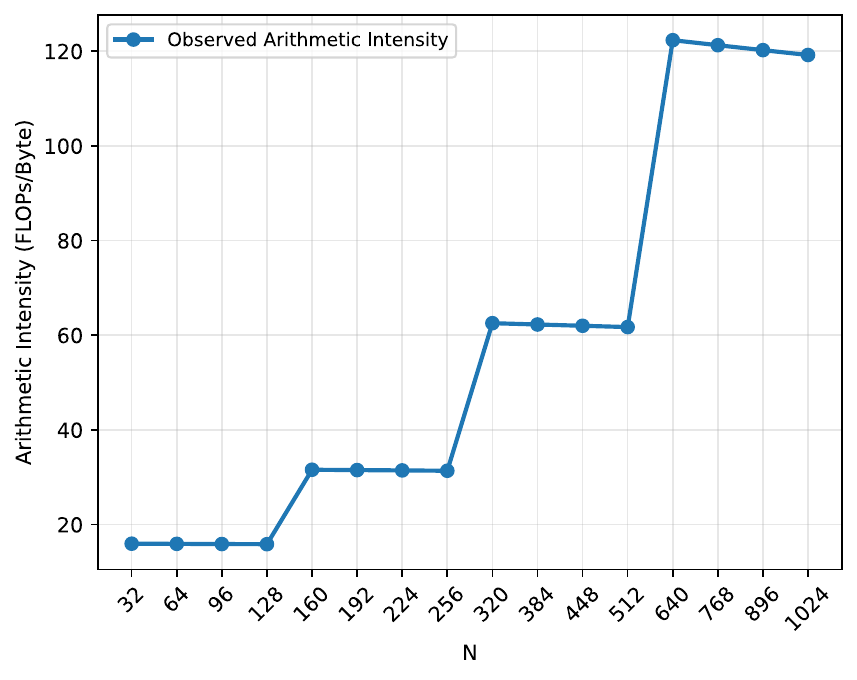}
            \caption{AI, $k=32$}
            \label{fig:moe-upper-vllm-h20-ai-k32}
        \end{subfigure}
        \hfill
        \begin{subfigure}[t]{0.24\textwidth}
            \centering
            \includegraphics[width=\linewidth,height=0.75\linewidth,keepaspectratio]{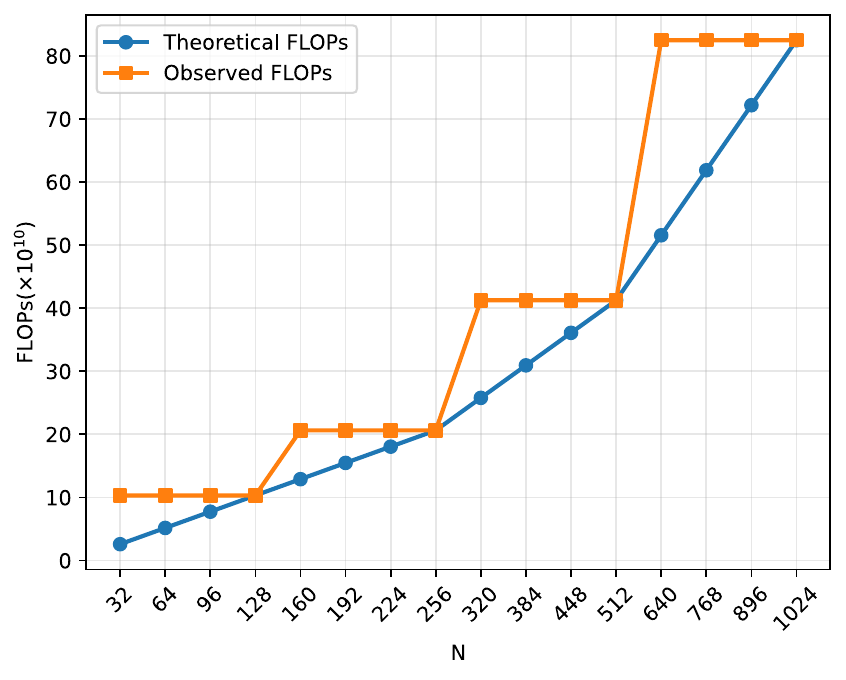}
            \caption{FLOPs, $k=32$}
            \label{fig:moe-upper-vllm-h20-flops-k32}
        \end{subfigure}
    \end{minipage}%
    }

    \caption{
    MoE FFN evaluation for load-balanced routing, the upper-bound case, with \textbf{vLLM} on \textbf{NVIDIA H20}.
    }
    \label{fig:moe-upper-vllm-h20}
\end{figure*}

\begin{figure*}[!p]
    \centering
    \scalebox{1}[1.0]{%
    \begin{minipage}{\textwidth}
        \centering

        \begin{subfigure}[t]{0.24\textwidth}
            \centering
            \includegraphics[width=\linewidth]{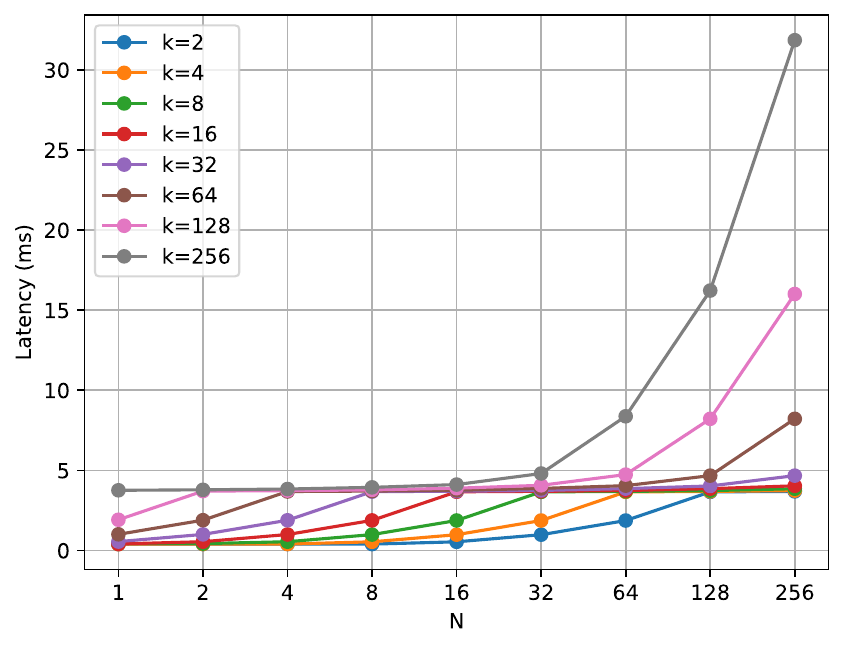}
            \caption{Latency vs. $N$}
            \label{fig:moe-upper-vllm-a800-latency}
        \end{subfigure}
        \hfill
        \begin{subfigure}[t]{0.24\textwidth}
            \centering
            \includegraphics[width=\linewidth]{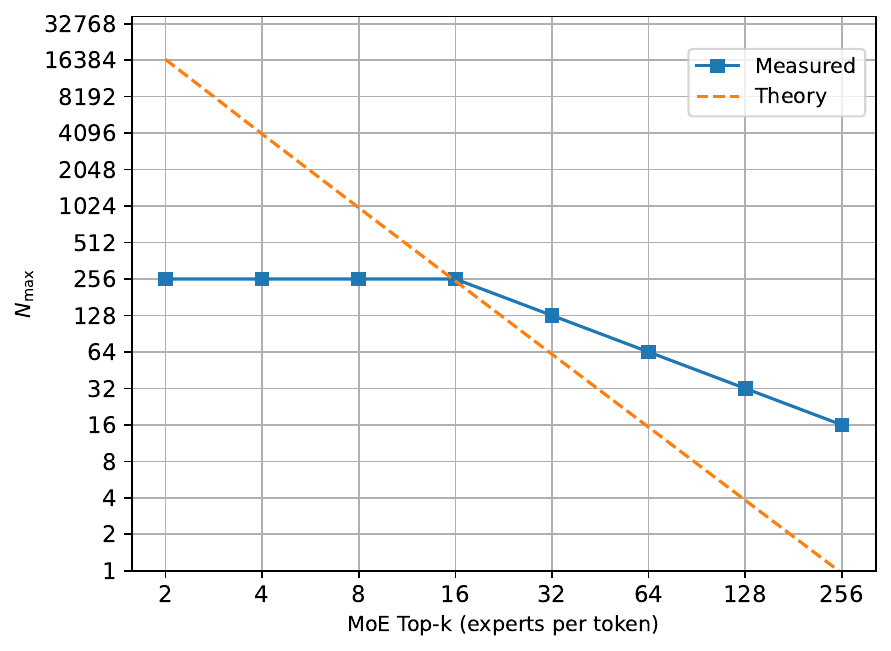}
            \caption{$N_{\max}$ vs. $k$}
            \label{fig:moe-upper-vllm-a800-nmax}
        \end{subfigure}
        \hfill
        \begin{subfigure}[t]{0.24\textwidth}
            \centering
            \includegraphics[width=\linewidth,height=0.75\linewidth,keepaspectratio]{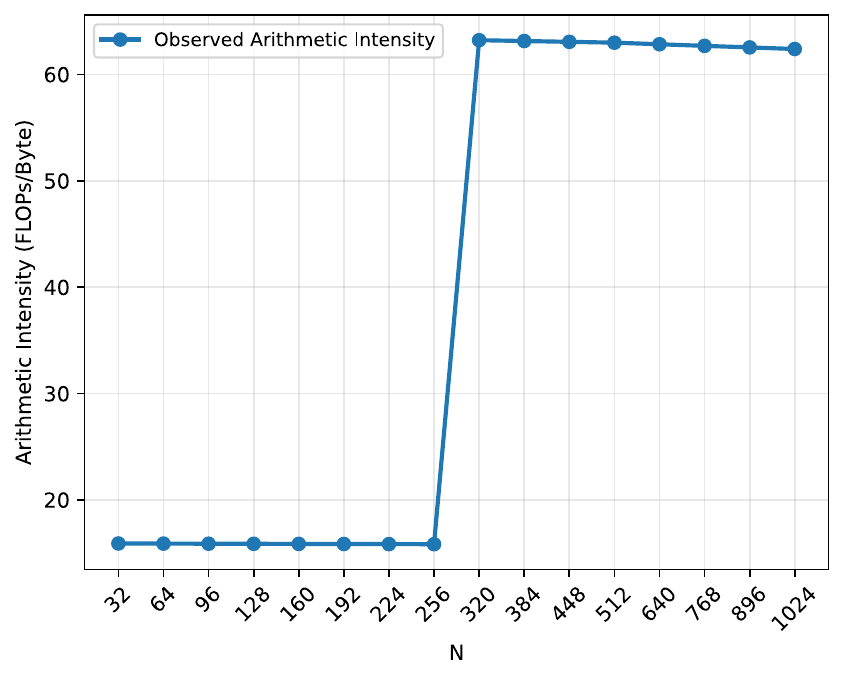}
            \caption{AI, $k=8$}
            \label{fig:moe-upper-vllm-a800-ai-k8}
        \end{subfigure}
        \hfill
        \begin{subfigure}[t]{0.24\textwidth}
            \centering
            \includegraphics[width=\linewidth,height=0.75\linewidth,keepaspectratio]{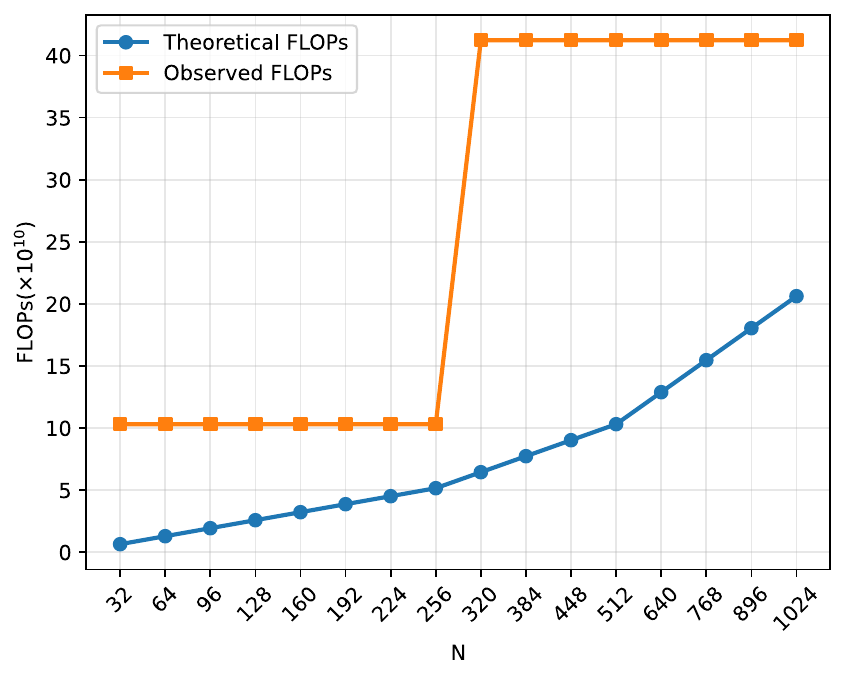}
            \caption{FLOPs, $k=8$}
            \label{fig:moe-upper-vllm-a800-flops-k8}
        \end{subfigure}

        \vspace{0.8em}
        \begin{subfigure}[t]{0.24\textwidth}
            \centering
            \includegraphics[width=\linewidth,height=0.75\linewidth,keepaspectratio]{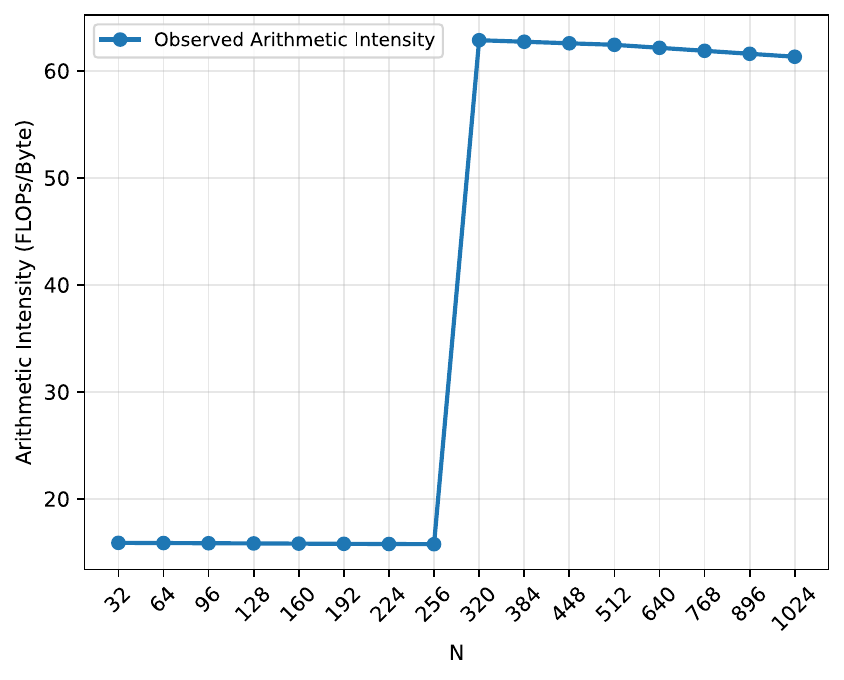}
            \caption{AI, $k=16$}
            \label{fig:moe-upper-vllm-a800-ai-k16}
        \end{subfigure}
        \hfill
        \begin{subfigure}[t]{0.24\textwidth}
            \centering
            \includegraphics[width=\linewidth,height=0.75\linewidth,keepaspectratio]{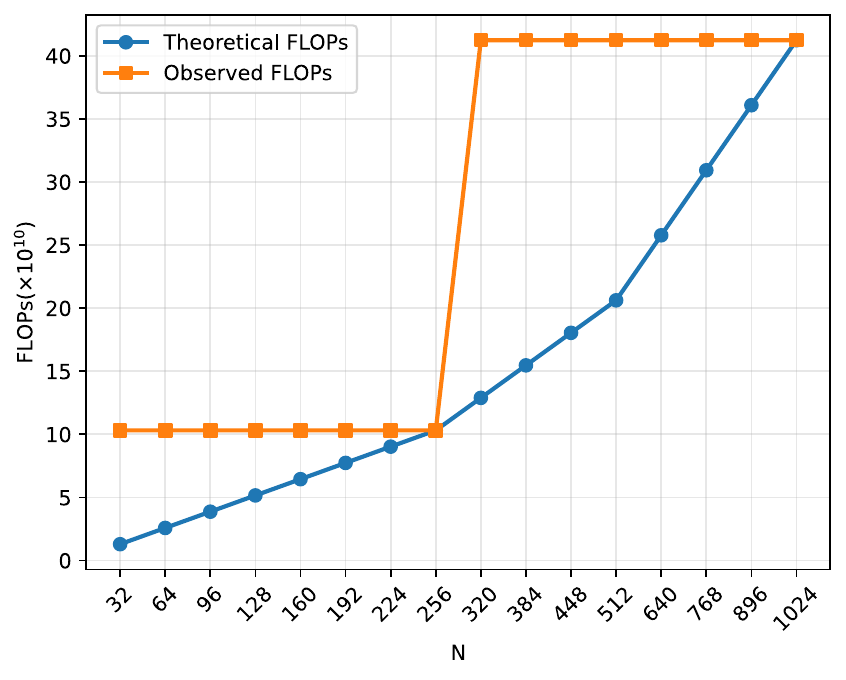}
            \caption{FLOPs, $k=16$}
            \label{fig:moe-upper-vllm-a800-flops-k16}
        \end{subfigure}
        \hfill
        \begin{subfigure}[t]{0.24\textwidth}
            \centering
            \includegraphics[width=\linewidth,height=0.75\linewidth,keepaspectratio]{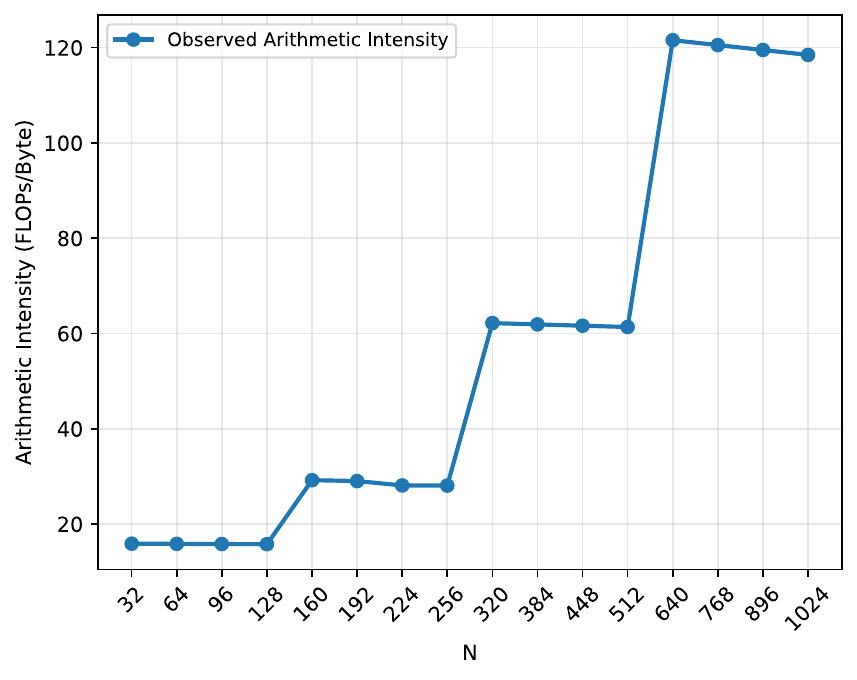}
            \caption{AI, $k=32$}
            \label{fig:moe-upper-vllm-a800-ai-k32}
        \end{subfigure}
        \hfill
        \begin{subfigure}[t]{0.24\textwidth}
            \centering
            \includegraphics[width=\linewidth,height=0.75\linewidth,keepaspectratio]{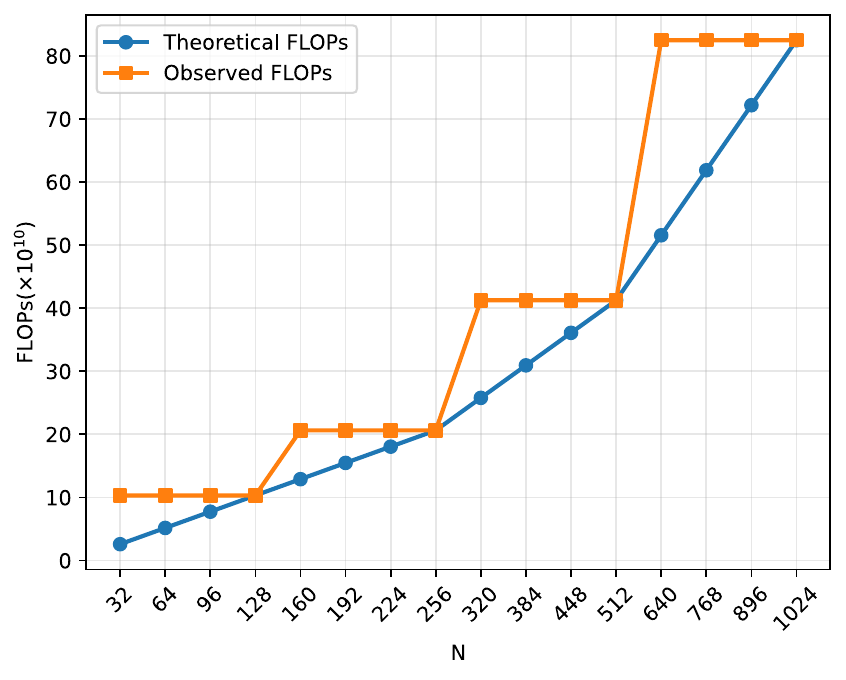}
            \caption{FLOPs, $k=32$}
            \label{fig:moe-upper-vllm-a800-flops-k32}
        \end{subfigure}
    \end{minipage}%
    }

    \caption{
    MoE FFN evaluation for load-balanced routing, the upper-bound case, with \textbf{vLLM} on \textbf{NVIDIA A800}.
    }
    \label{fig:moe-upper-vllm-a800}
\end{figure*}

\begin{figure*}[!p]
    \centering
    \scalebox{1}[1.0]{%
    \begin{minipage}{\textwidth}
        \centering

        \begin{subfigure}[t]{0.24\textwidth}
            \centering
            \includegraphics[width=\linewidth]{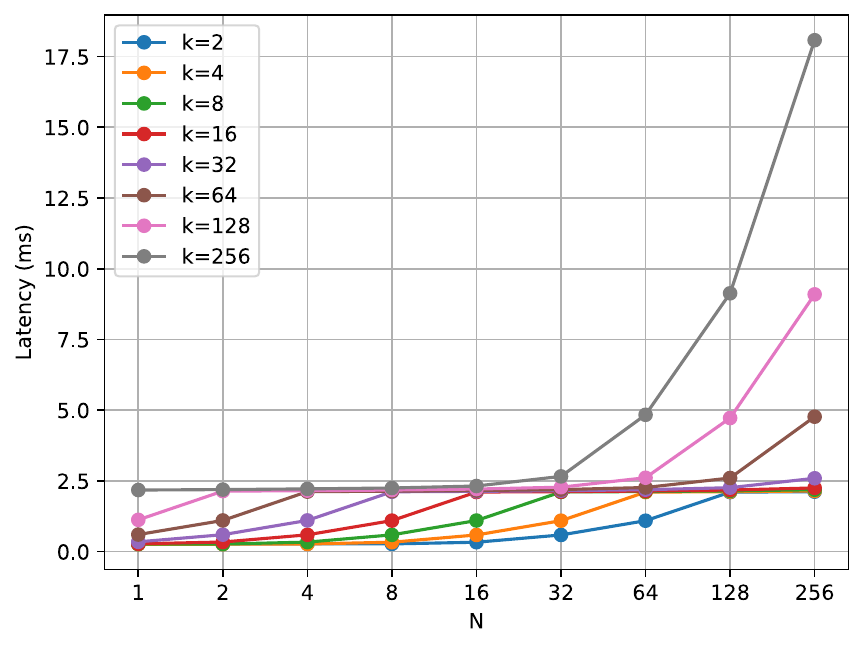}
            \caption{Latency vs. $N$}
            \label{fig:moe-upper-vllm-h800-latency}
        \end{subfigure}
        \hfill
        \begin{subfigure}[t]{0.24\textwidth}
            \centering
            \includegraphics[width=\linewidth]{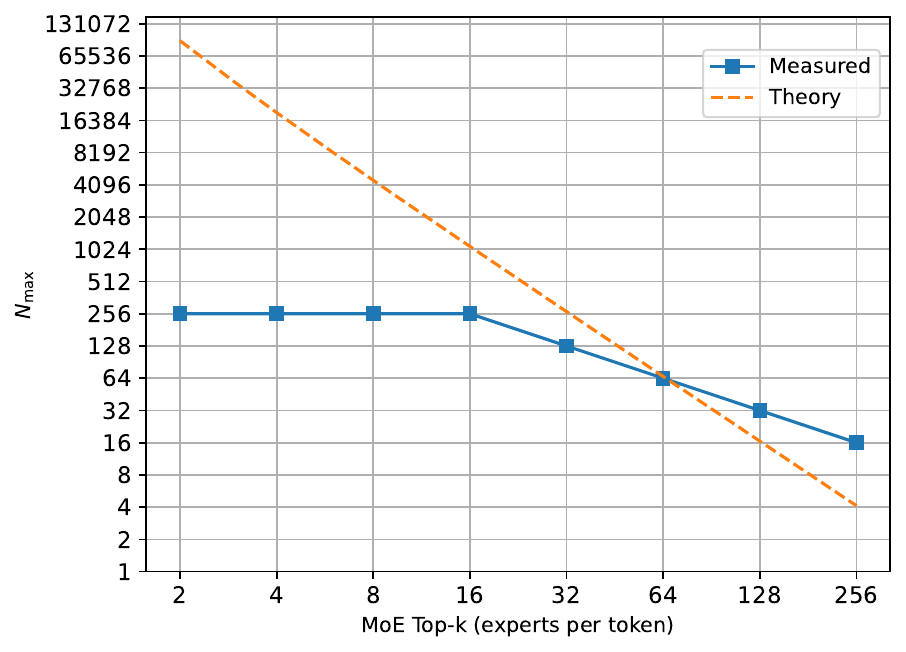}
            \caption{$N_{\max}$ vs. $k$}
            \label{fig:moe-upper-vllm-h800-nmax}
        \end{subfigure}
        \hfill
        \begin{subfigure}[t]{0.24\textwidth}
            \centering
            \includegraphics[width=\linewidth,height=0.75\linewidth,keepaspectratio]{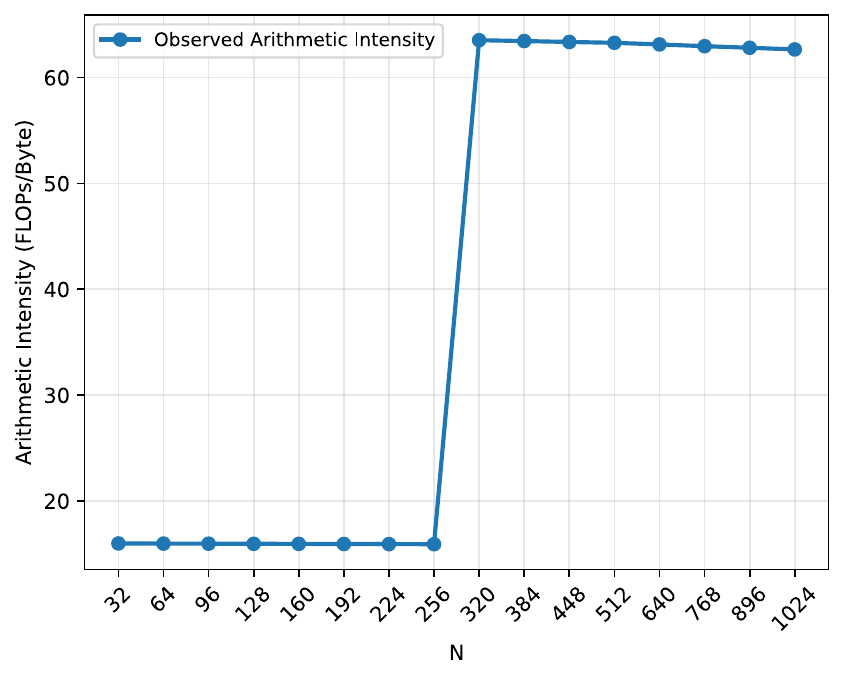}
            \caption{AI, $k=8$}
            \label{fig:moe-upper-vllm-h800-ai-k8}
        \end{subfigure}
        \hfill
        \begin{subfigure}[t]{0.24\textwidth}
            \centering
            \includegraphics[width=\linewidth,height=0.75\linewidth,keepaspectratio]{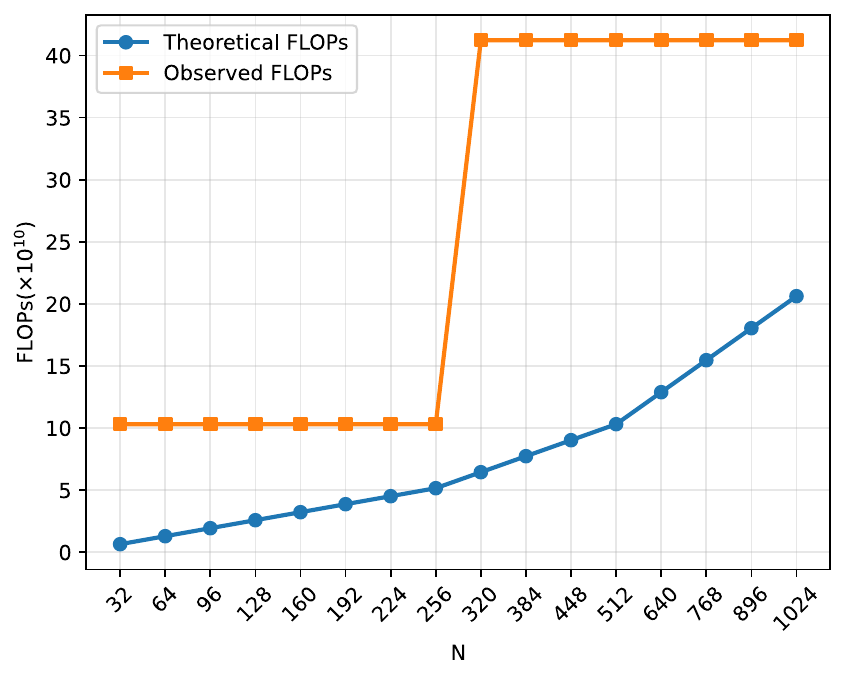}
            \caption{FLOPs, $k=8$}
            \label{fig:moe-upper-vllm-h800-flops-k8}
        \end{subfigure}

        \vspace{0.8em}
        \begin{subfigure}[t]{0.24\textwidth}
            \centering
            \includegraphics[width=\linewidth,height=0.75\linewidth,keepaspectratio]{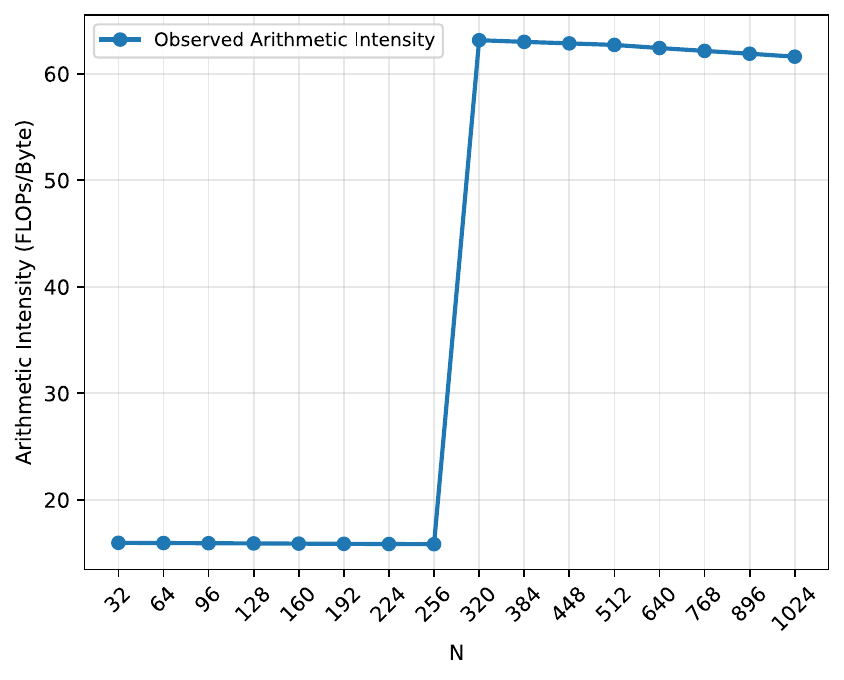}
            \caption{AI, $k=16$}
            \label{fig:moe-upper-vllm-h800-ai-k16}
        \end{subfigure}
        \hfill
        \begin{subfigure}[t]{0.24\textwidth}
            \centering
            \includegraphics[width=\linewidth,height=0.75\linewidth,keepaspectratio]{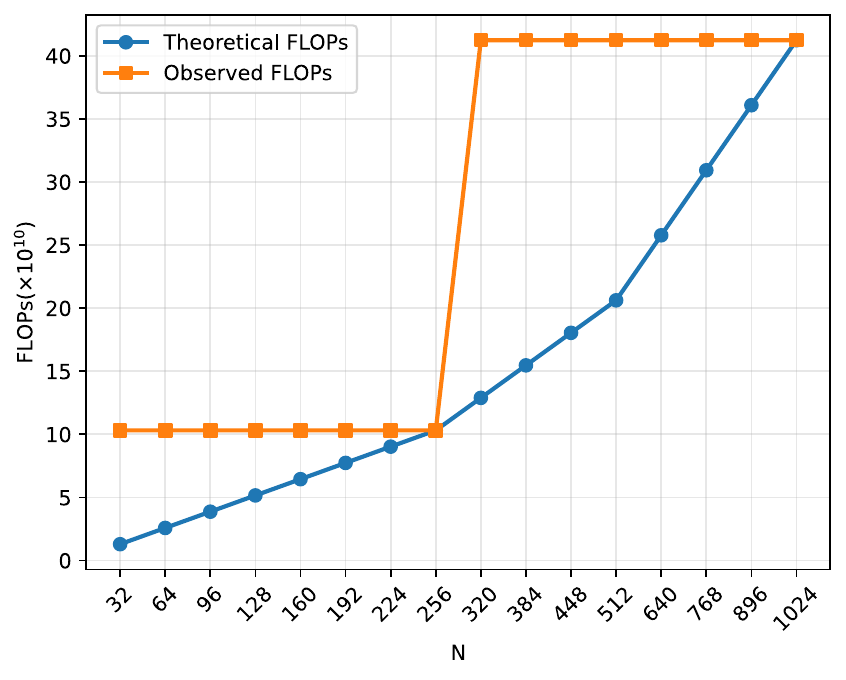}
            \caption{FLOPs, $k=16$}
            \label{fig:moe-upper-vllm-h800-flops-k16}
        \end{subfigure}
        \hfill
        \begin{subfigure}[t]{0.24\textwidth}
            \centering
            \includegraphics[width=\linewidth,height=0.75\linewidth,keepaspectratio]{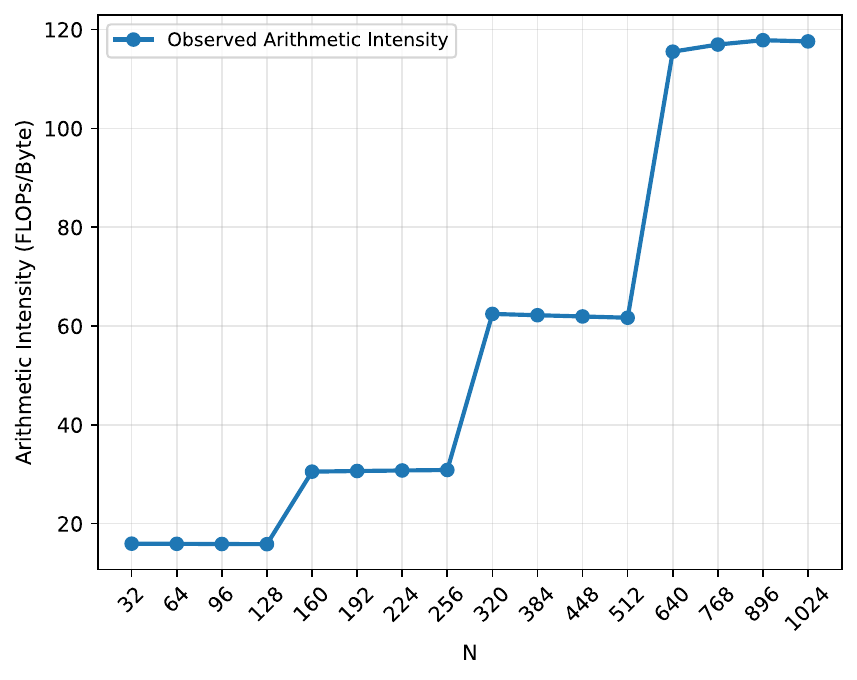}
            \caption{AI, $k=32$}
            \label{fig:moe-upper-vllm-h800-ai-k32}
        \end{subfigure}
        \hfill
        \begin{subfigure}[t]{0.24\textwidth}
            \centering
            \includegraphics[width=\linewidth,height=0.75\linewidth,keepaspectratio]{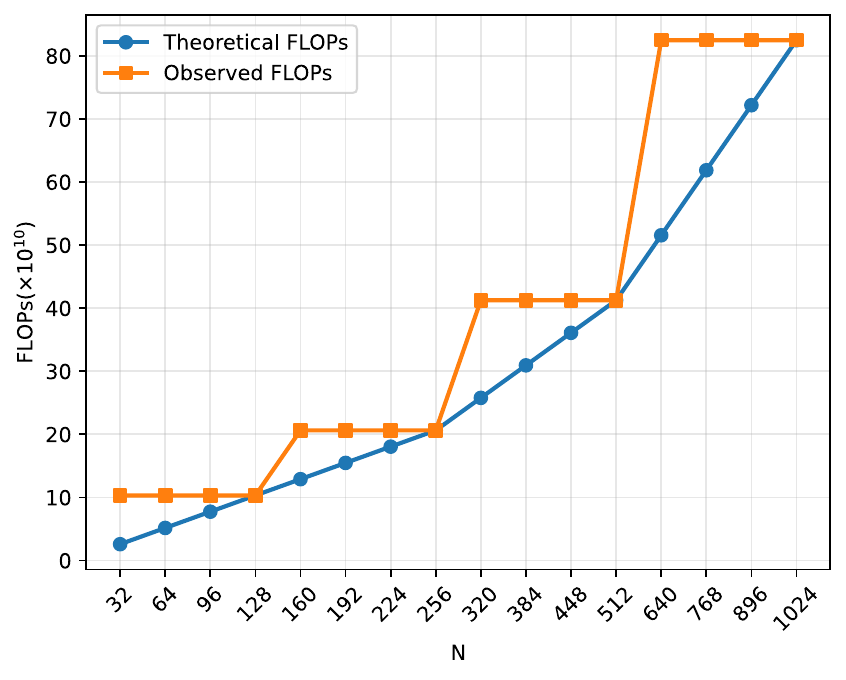}
            \caption{FLOPs, $k=32$}
            \label{fig:moe-upper-vllm-h800-flops-k32}
        \end{subfigure}
    \end{minipage}%
    }

    \caption{
    MoE FFN evaluation for load-balanced routing, the upper-bound case, with \textbf{vLLM} on \textbf{NVIDIA H800}.
    }
    \label{fig:moe-upper-vllm-h800}
\end{figure*}

Figures~\ref{fig:moe-upper-vllm-h20}, \ref{fig:moe-upper-vllm-a800}, and \ref{fig:moe-upper-vllm-h800} report the load-balanced MoE FFN results using the vLLM fused MoE backend. In this setting, tokens are distributed across experts, so the padding slack of fused expert GEMMs is aggregated over many activated experts. Across the evaluated GPU platforms, the latency curves exhibit clear near-free regions followed by discrete jumps, indicating that larger N remains within existing padded expert-token blocks until the next kernel-granularity boundary is crossed.

The extracted boundaries follow the expected upper-bound trend. As shown in Figures~\ref{fig:moe-upper-vllm-h20}\subref{fig:moe-upper-vllm-h20-nmax}, \ref{fig:moe-upper-vllm-a800}\subref{fig:moe-upper-vllm-a800-nmax}, and \ref{fig:moe-upper-vllm-h800}\subref{fig:moe-upper-vllm-h800-nmax}, $N_{\max}$ increases as routing becomes more sparse. This is consistent with the granularity-based upper-bound prediction: smaller $k$ spreads the workload over more available expert-token padding capacity, while larger $k$ consumes more expert computation per token and reduces the maximum near-free N. The arithmetic-intensity and runtime-FLOPs profiles further show staircase-like behavior, confirming that the measured boundary is governed by padded kernel execution rather than a smooth idle-compute transition.

\begin{figure*}[!p]
    \centering
    \scalebox{1}[1.0]{%
    \begin{minipage}{\textwidth}
        \centering

        \begin{subfigure}[t]{0.24\textwidth}
            \centering
            \includegraphics[width=\linewidth]{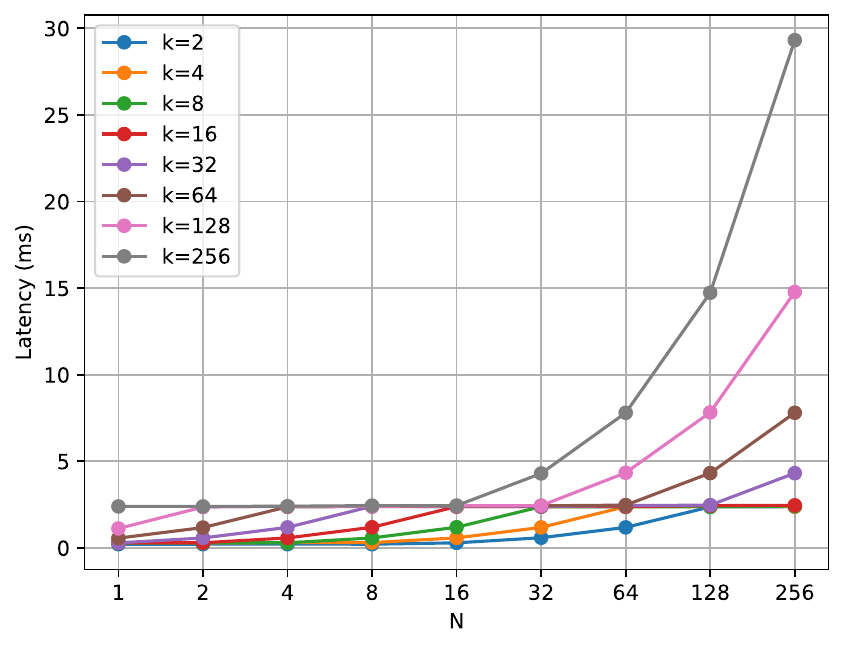}
            \caption{Latency vs. $N$}
            \label{fig:moe-upper-sglang-h20-latency}
        \end{subfigure}
        \hfill
        \begin{subfigure}[t]{0.24\textwidth}
            \centering
            \includegraphics[width=\linewidth]{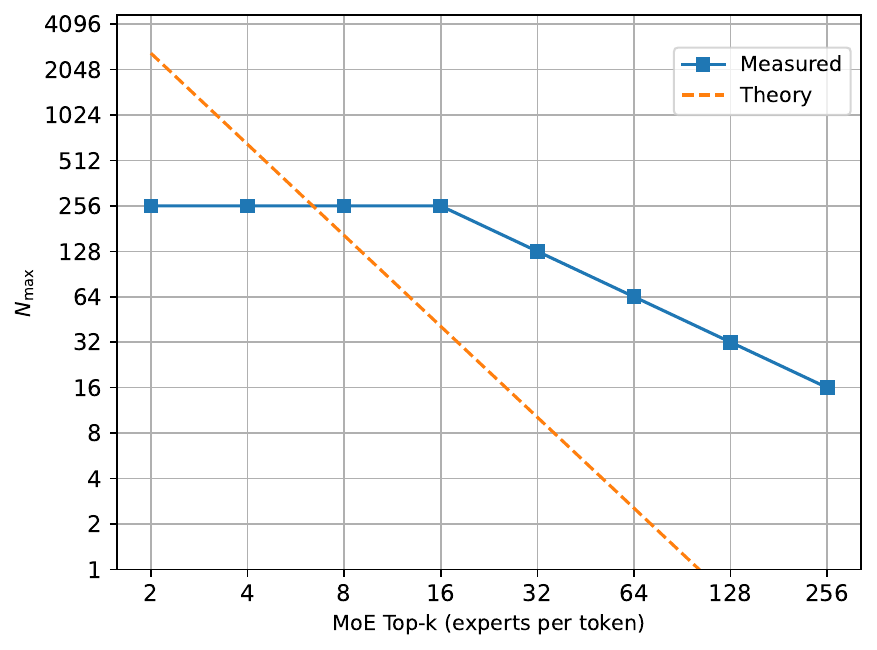}
            \caption{$N_{\max}$ vs. $k$}
            \label{fig:moe-upper-sglang-h20-nmax}
        \end{subfigure}
        \hfill
        \begin{subfigure}[t]{0.24\textwidth}
            \centering
            \includegraphics[width=\linewidth,height=0.75\linewidth,keepaspectratio]{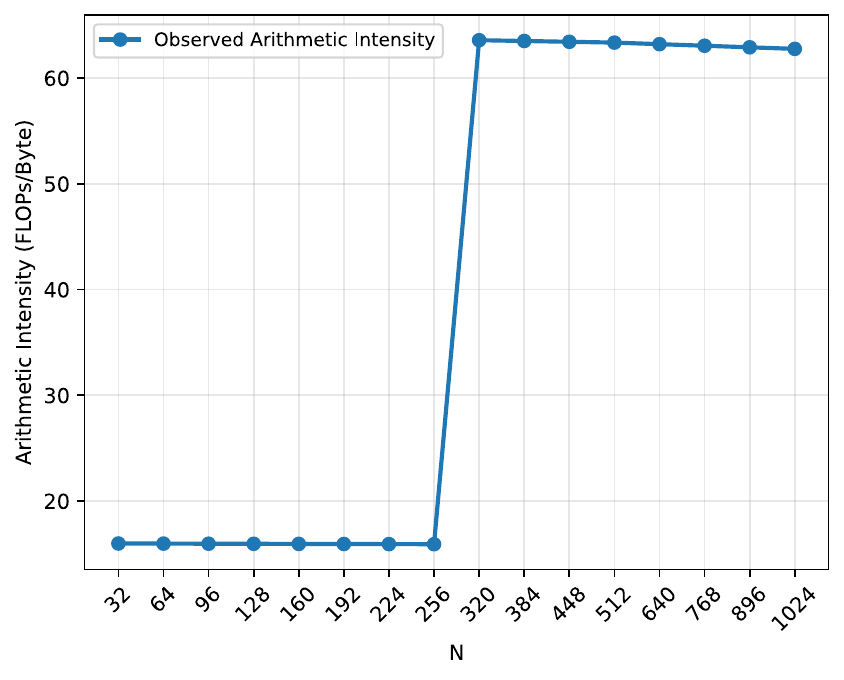}
            \caption{AI, $k=8$}
            \label{fig:moe-upper-sglang-h20-ai-k8}
        \end{subfigure}
        \hfill
        \begin{subfigure}[t]{0.24\textwidth}
            \centering
            \includegraphics[width=\linewidth,height=0.75\linewidth,keepaspectratio]{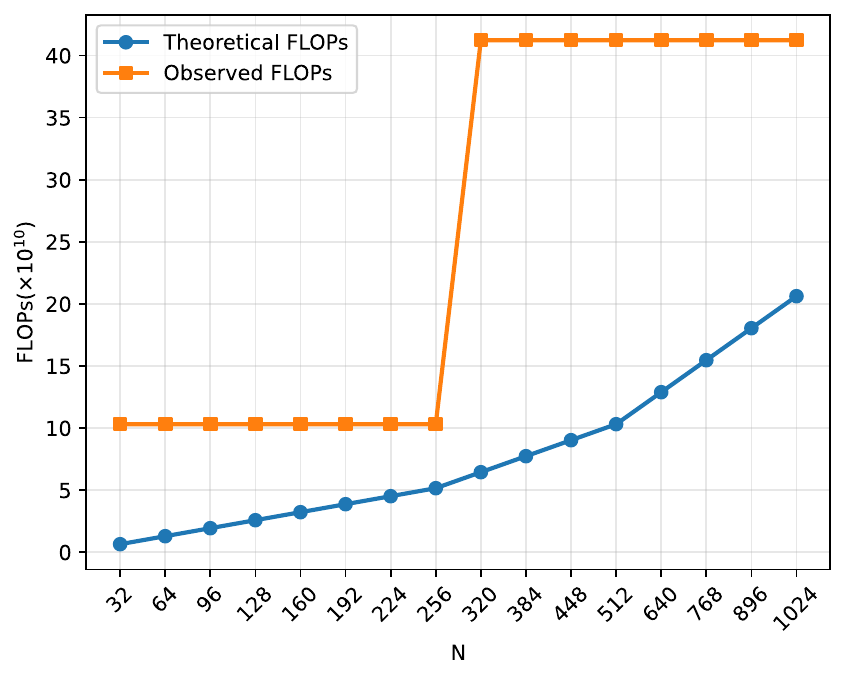}
            \caption{FLOPs, $k=8$}
            \label{fig:moe-upper-sglang-h20-flops-k8}
        \end{subfigure}

        \vspace{0.8em}
        \begin{subfigure}[t]{0.24\textwidth}
            \centering
            \includegraphics[width=\linewidth,height=0.75\linewidth,keepaspectratio]{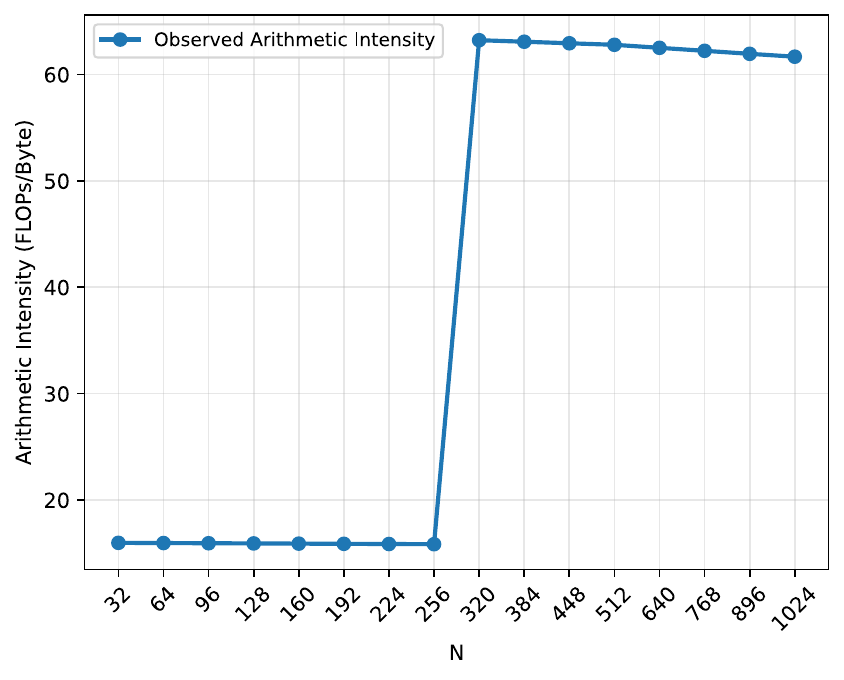}
            \caption{AI, $k=16$}
            \label{fig:moe-upper-sglang-h20-ai-k16}
        \end{subfigure}
        \hfill
        \begin{subfigure}[t]{0.24\textwidth}
            \centering
            \includegraphics[width=\linewidth,height=0.75\linewidth,keepaspectratio]{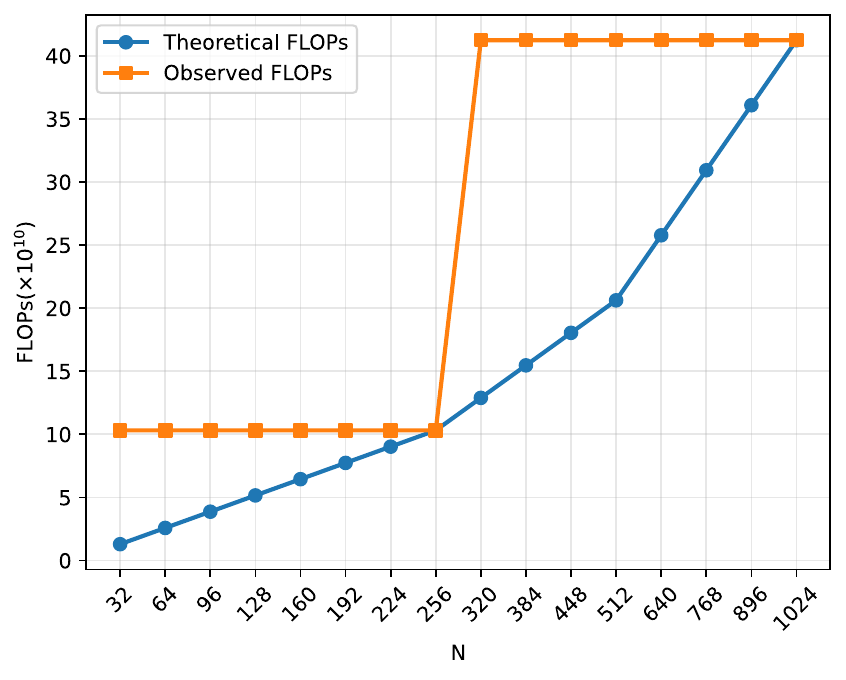}
            \caption{FLOPs, $k=16$}
            \label{fig:moe-upper-sglang-h20-flops-k16}
        \end{subfigure}
        \hfill
        \begin{subfigure}[t]{0.24\textwidth}
            \centering
            \includegraphics[width=\linewidth,height=0.75\linewidth,keepaspectratio]{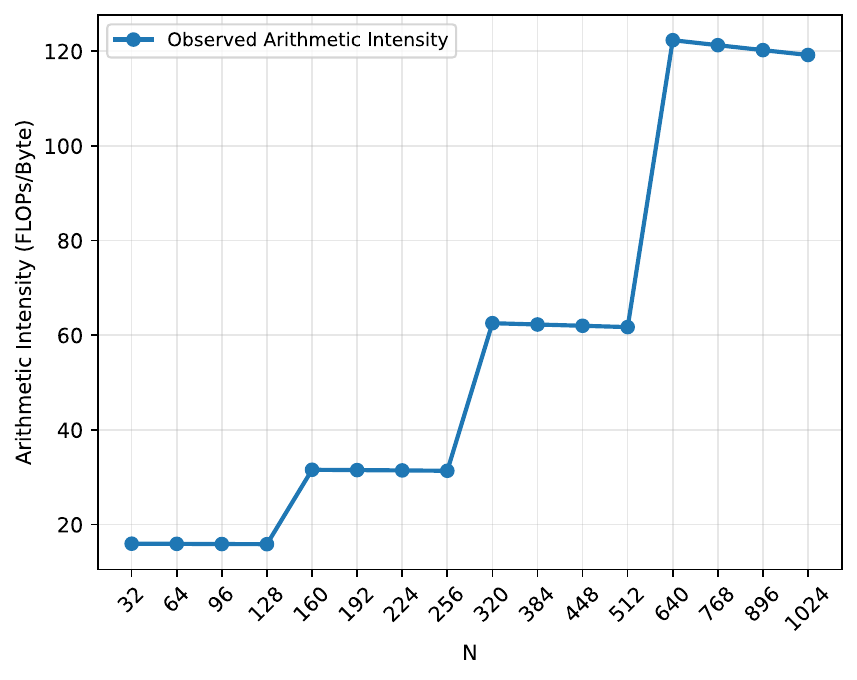}
            \caption{AI, $k=32$}
            \label{fig:moe-upper-sglang-h20-ai-k32}
        \end{subfigure}
        \hfill
        \begin{subfigure}[t]{0.24\textwidth}
            \centering
            \includegraphics[width=\linewidth,height=0.75\linewidth,keepaspectratio]{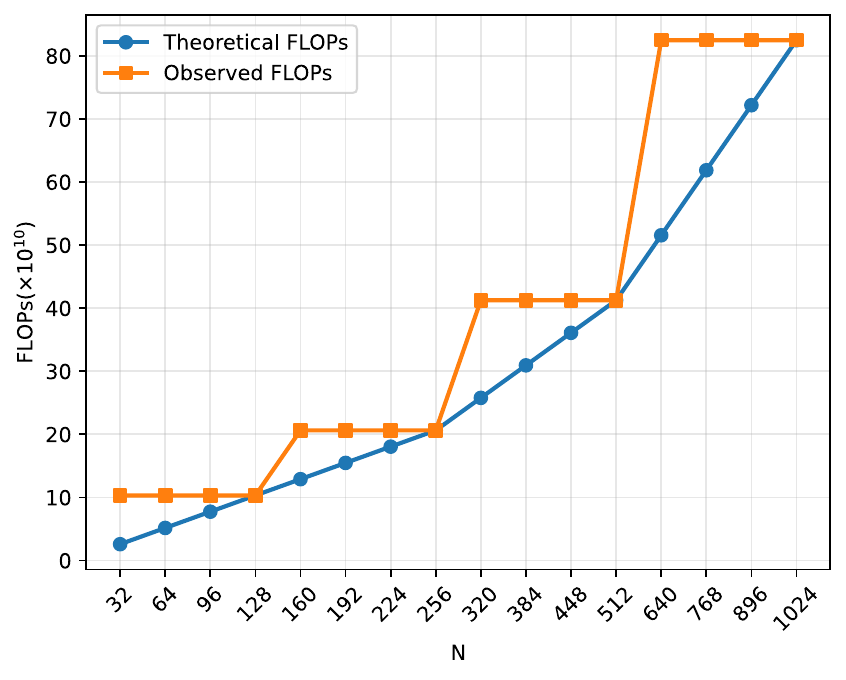}
            \caption{FLOPs, $k=32$}
            \label{fig:moe-upper-sglang-h20-flops-k32}
        \end{subfigure}
    \end{minipage}%
    }

    \caption{
    MoE FFN evaluation for load-balanced routing, the upper-bound case, with \textbf{SGLang} on \textbf{NVIDIA H20}.
    }
    \label{fig:moe-upper-sglang-h20}
\end{figure*}

\begin{figure*}[!p]
    \centering
    \scalebox{1}[1.0]{%
    \begin{minipage}{\textwidth}
        \centering

        \begin{subfigure}[t]{0.24\textwidth}
            \centering
            \includegraphics[width=\linewidth]{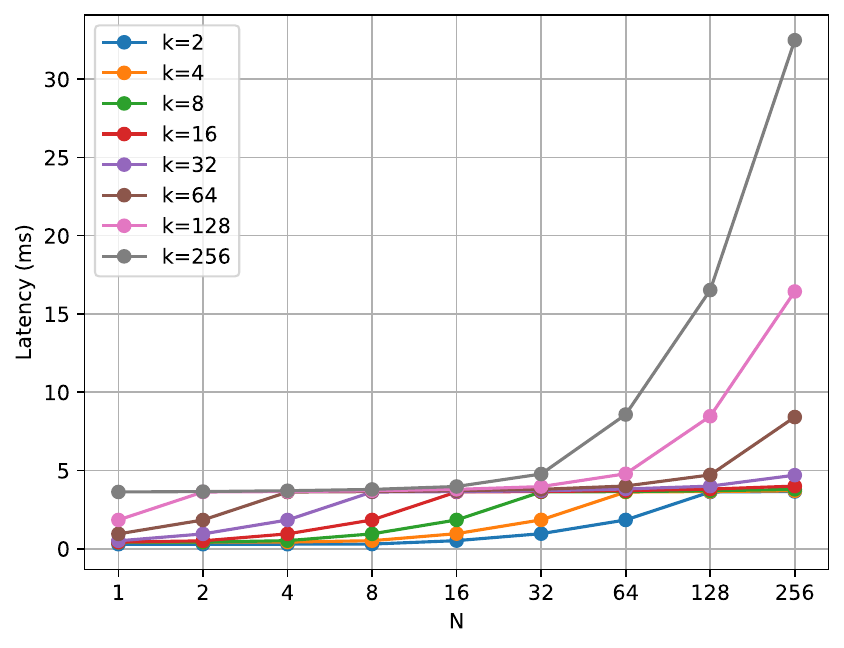}
            \caption{Latency vs. $N$}
            \label{fig:moe-upper-sglang-a800-latency}
        \end{subfigure}
        \hfill
        \begin{subfigure}[t]{0.24\textwidth}
            \centering
            \includegraphics[width=\linewidth]{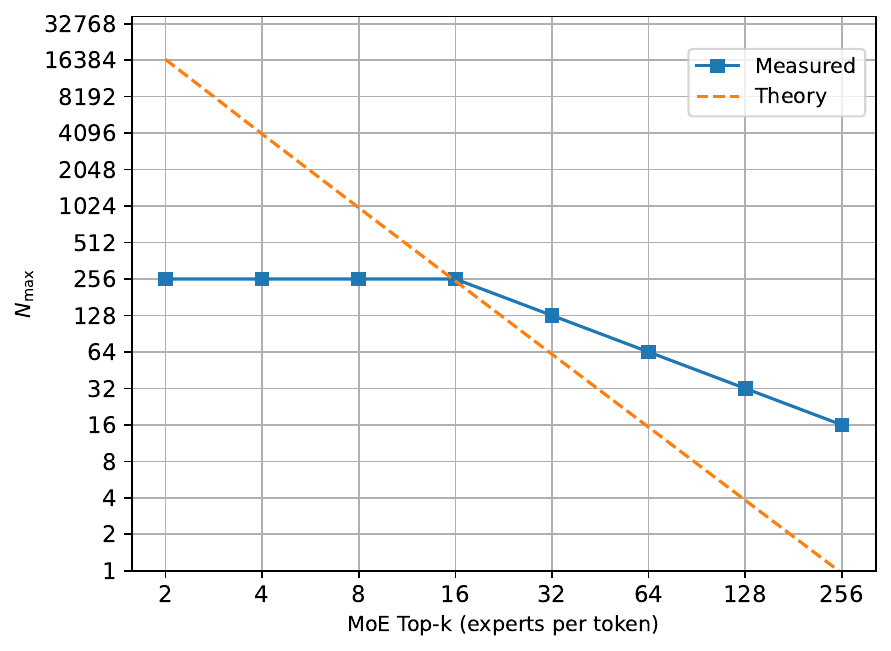}
            \caption{$N_{\max}$ vs. $k$}
            \label{fig:moe-upper-sglang-a800-nmax}
        \end{subfigure}
        \hfill
        \begin{subfigure}[t]{0.24\textwidth}
            \centering
            \includegraphics[width=\linewidth,height=0.75\linewidth,keepaspectratio]{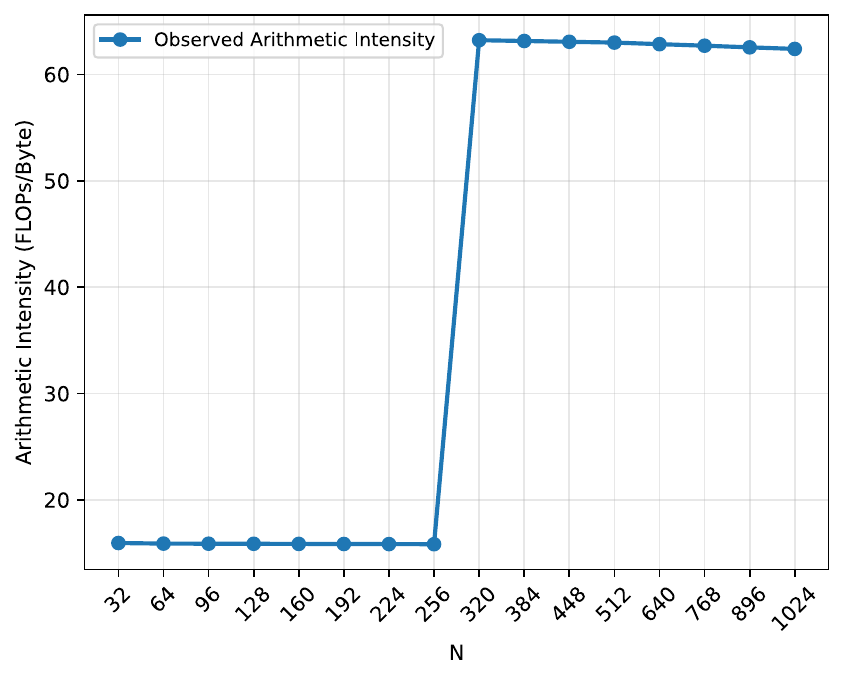}
            \caption{AI, $k=8$}
            \label{fig:moe-upper-sglang-a800-ai-k8}
        \end{subfigure}
        \hfill
        \begin{subfigure}[t]{0.24\textwidth}
            \centering
            \includegraphics[width=\linewidth,height=0.75\linewidth,keepaspectratio]{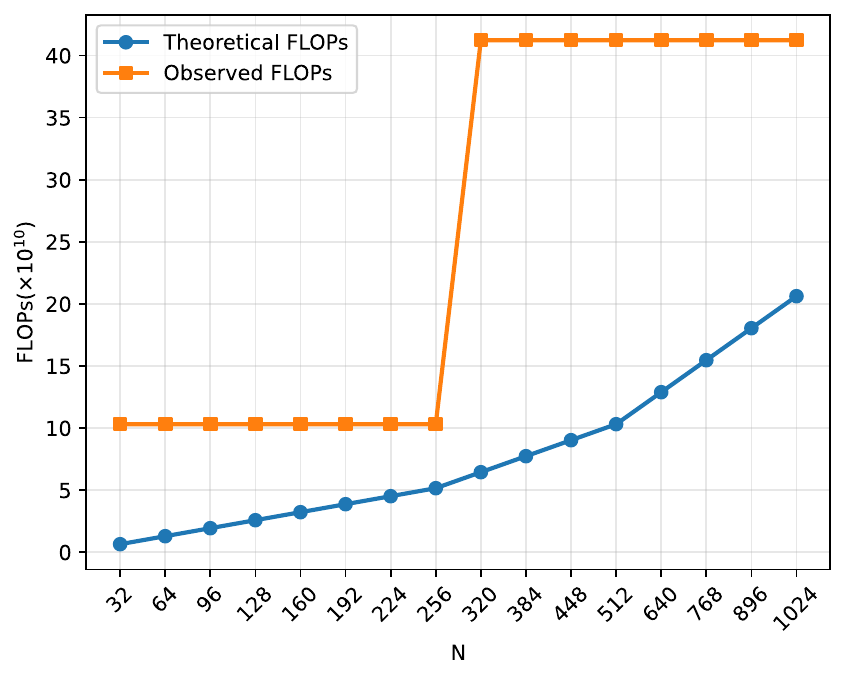}
            \caption{FLOPs, $k=8$}
            \label{fig:moe-upper-sglang-a800-flops-k8}
        \end{subfigure}

        \vspace{0.8em}
        \begin{subfigure}[t]{0.24\textwidth}
            \centering
            \includegraphics[width=\linewidth,height=0.75\linewidth,keepaspectratio]{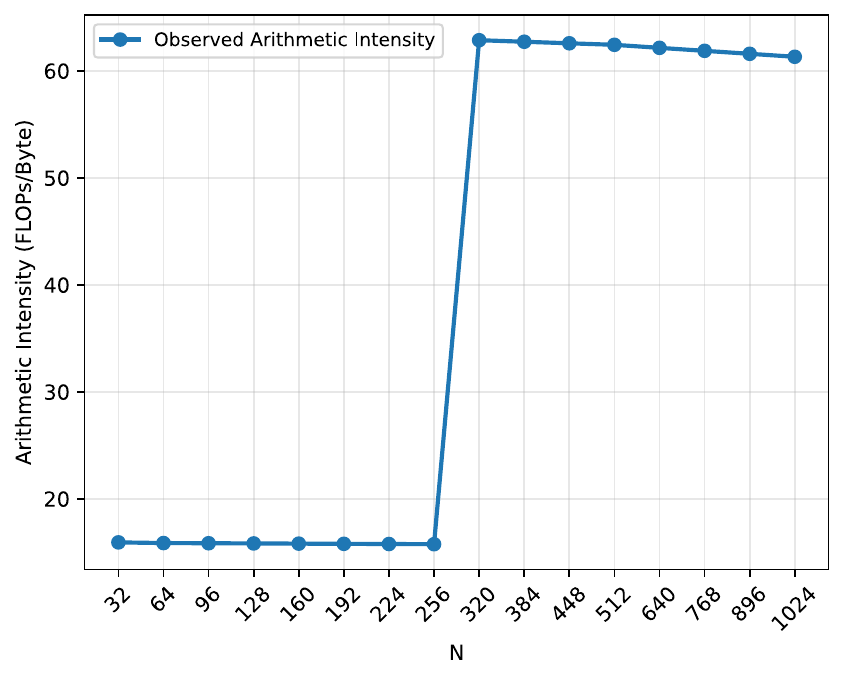}
            \caption{AI, $k=16$}
            \label{fig:moe-upper-sglang-a800-ai-k16}
        \end{subfigure}
        \hfill
        \begin{subfigure}[t]{0.24\textwidth}
            \centering
            \includegraphics[width=\linewidth,height=0.75\linewidth,keepaspectratio]{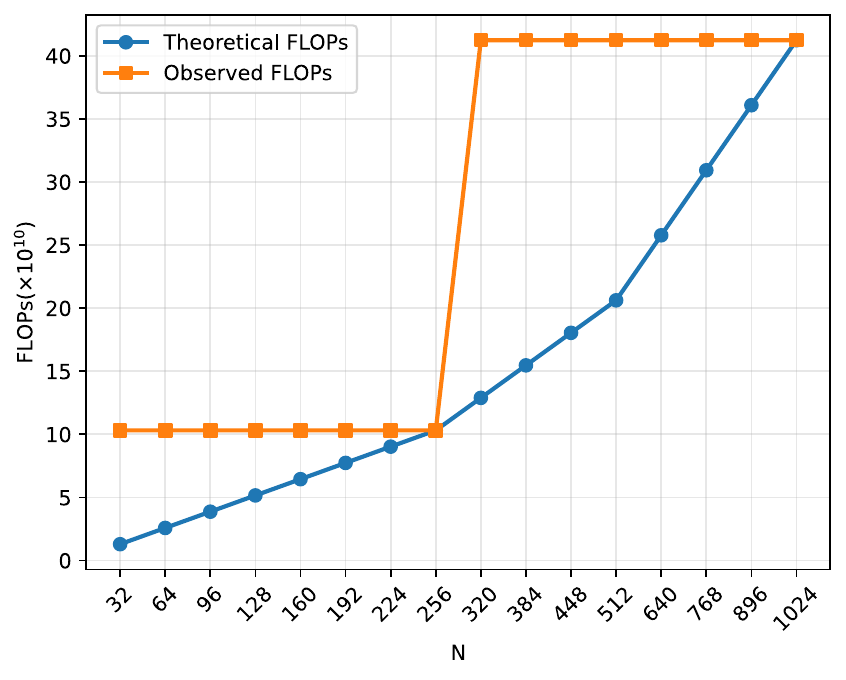}
            \caption{FLOPs, $k=16$}
            \label{fig:moe-upper-sglang-a800-flops-k16}
        \end{subfigure}
        \hfill
        \begin{subfigure}[t]{0.24\textwidth}
            \centering
            \includegraphics[width=\linewidth,height=0.75\linewidth,keepaspectratio]{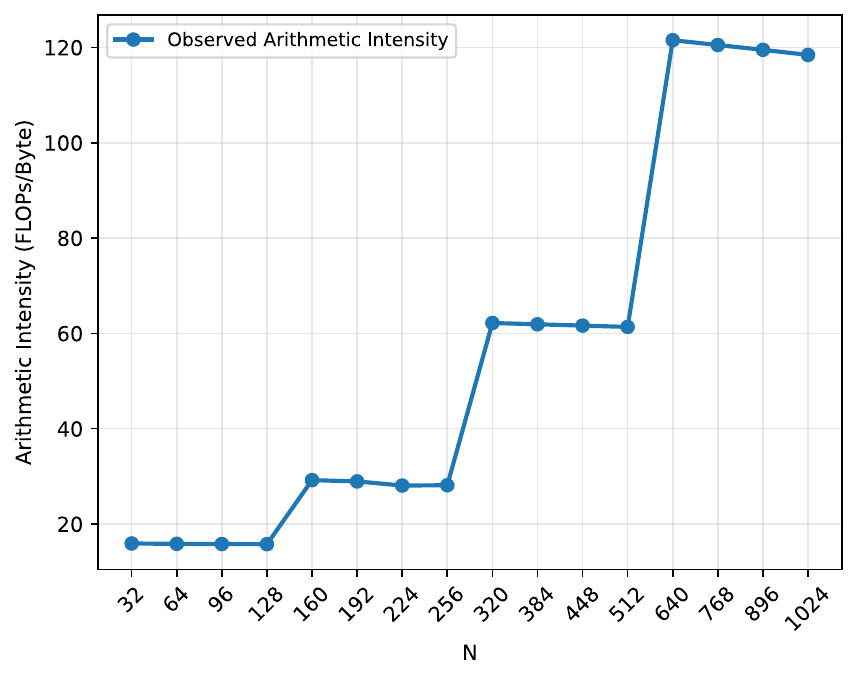}
            \caption{AI, $k=32$}
            \label{fig:moe-upper-sglang-a800-ai-k32}
        \end{subfigure}
        \hfill
        \begin{subfigure}[t]{0.24\textwidth}
            \centering
            \includegraphics[width=\linewidth,height=0.75\linewidth,keepaspectratio]{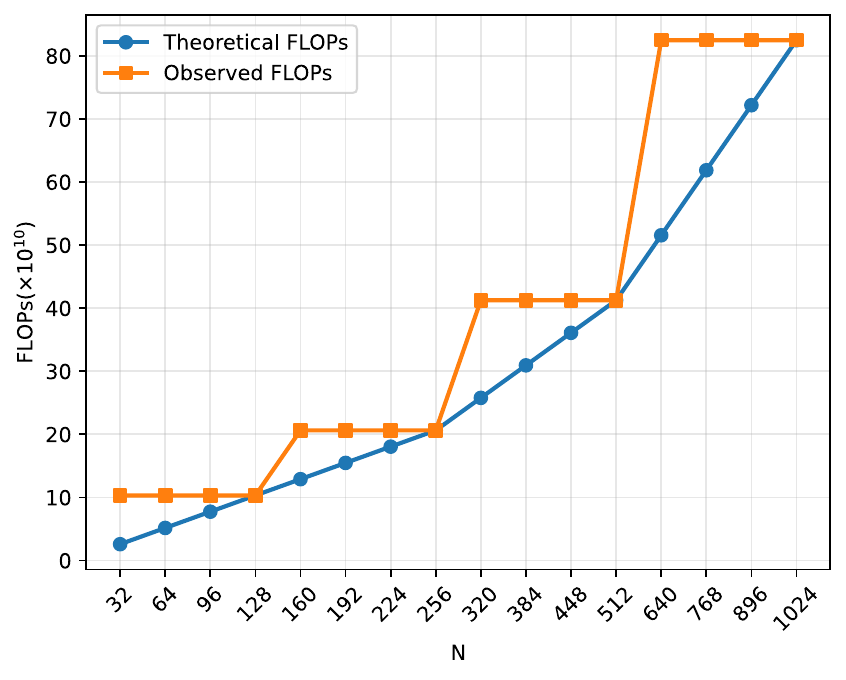}
            \caption{FLOPs, $k=32$}
            \label{fig:moe-upper-sglang-a800-flops-k32}
        \end{subfigure}
    \end{minipage}%
    }

    \caption{
    MoE FFN evaluation for load-balanced routing, the upper-bound case, with \textbf{SGLang} on \textbf{NVIDIA A800}.
    }
    \label{fig:moe-upper-sglang-a800}
\end{figure*}

\begin{figure*}[!p]
    \centering
    \scalebox{1}[1.0]{%
    \begin{minipage}{\textwidth}
        \centering

        \begin{subfigure}[t]{0.24\textwidth}
            \centering
            \includegraphics[width=\linewidth]{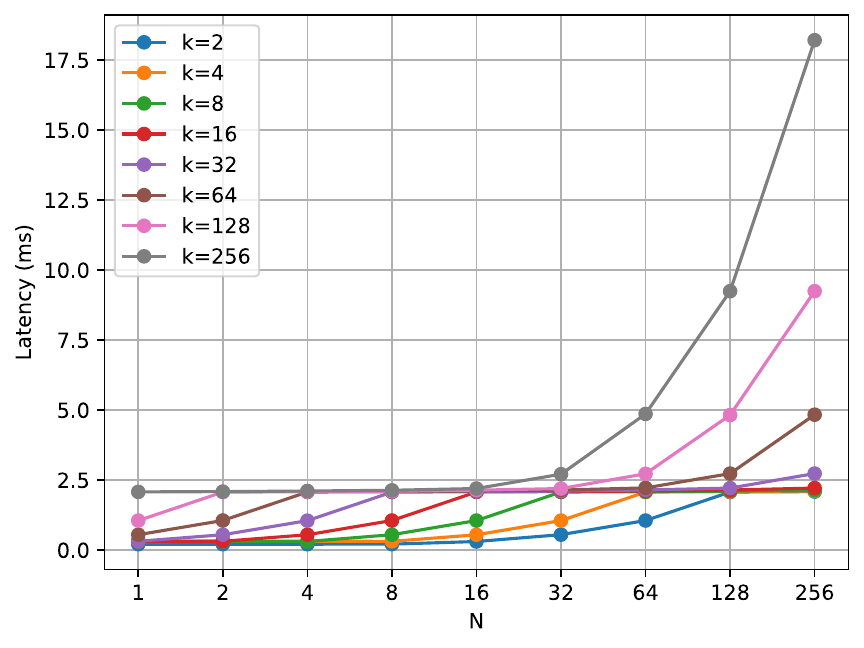}
            \caption{Latency vs. $N$}
            \label{fig:moe-upper-sglang-h800-latency}
        \end{subfigure}
        \hfill
        \begin{subfigure}[t]{0.24\textwidth}
            \centering
            \includegraphics[width=\linewidth]{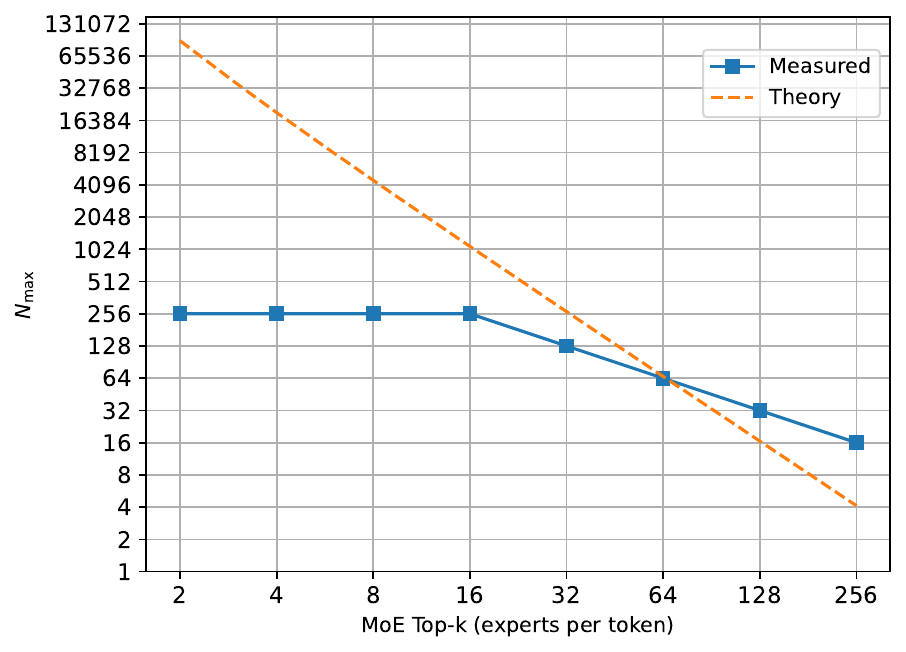}
            \caption{$N_{\max}$ vs. $k$}
            \label{fig:moe-upper-sglang-h800-nmax}
        \end{subfigure}
        \hfill
        \begin{subfigure}[t]{0.24\textwidth}
            \centering
            \includegraphics[width=\linewidth,height=0.75\linewidth,keepaspectratio]{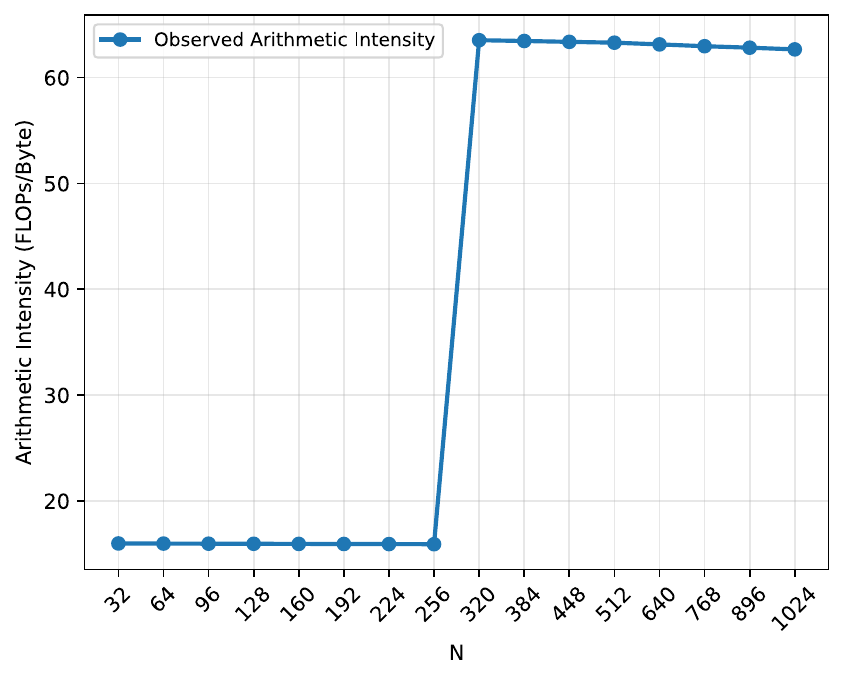}
            \caption{AI, $k=8$}
            \label{fig:moe-upper-sglang-h800-ai-k8}
        \end{subfigure}
        \hfill
        \begin{subfigure}[t]{0.24\textwidth}
            \centering
            \includegraphics[width=\linewidth,height=0.75\linewidth,keepaspectratio]{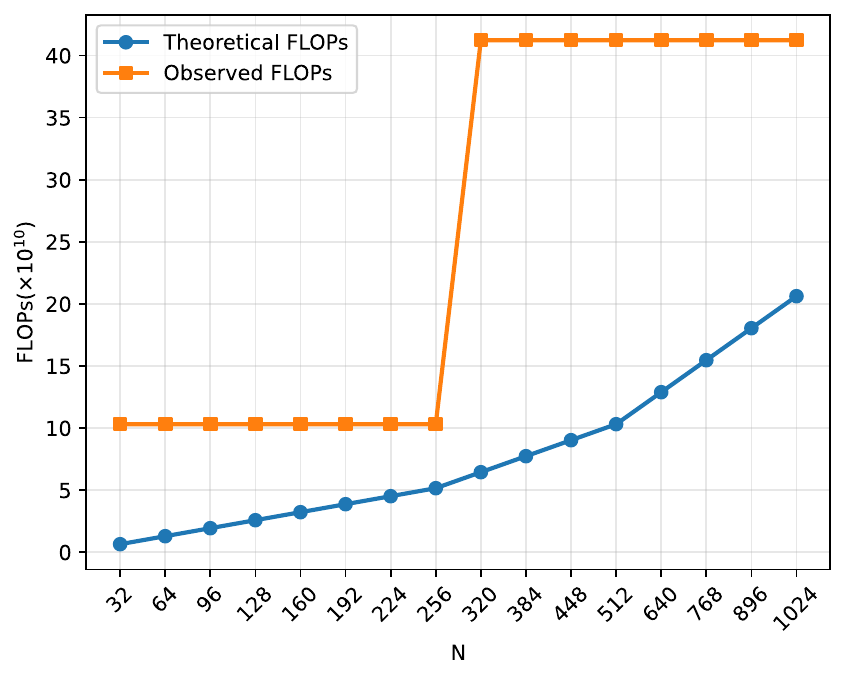}
            \caption{FLOPs, $k=8$}
            \label{fig:moe-upper-sglang-h800-flops-k8}
        \end{subfigure}

        \vspace{0.8em}
        \begin{subfigure}[t]{0.24\textwidth}
            \centering
            \includegraphics[width=\linewidth,height=0.75\linewidth,keepaspectratio]{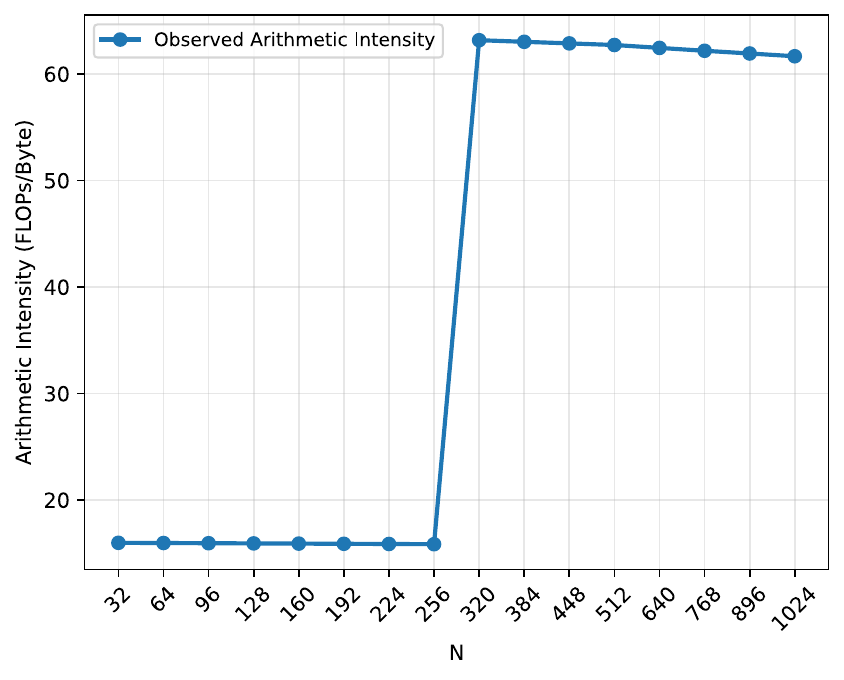}
            \caption{AI, $k=16$}
            \label{fig:moe-upper-sglang-h800-ai-k16}
        \end{subfigure}
        \hfill
        \begin{subfigure}[t]{0.24\textwidth}
            \centering
            \includegraphics[width=\linewidth,height=0.75\linewidth,keepaspectratio]{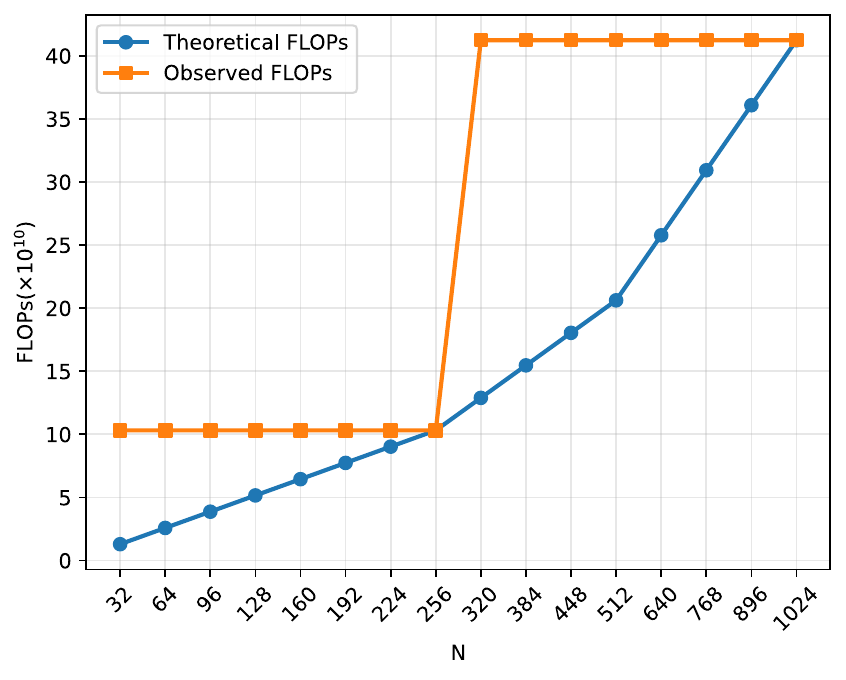}
            \caption{FLOPs, $k=16$}
            \label{fig:moe-upper-sglang-h800-flops-k16}
        \end{subfigure}
        \hfill
        \begin{subfigure}[t]{0.24\textwidth}
            \centering
            \includegraphics[width=\linewidth,height=0.75\linewidth,keepaspectratio]{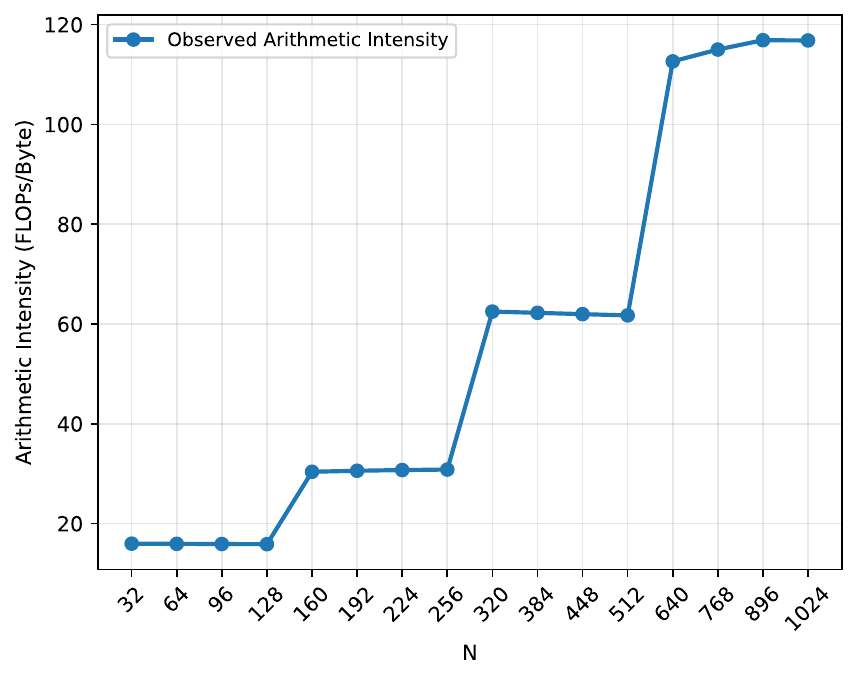}
            \caption{AI, $k=32$}
            \label{fig:moe-upper-sglang-h800-ai-k32}
        \end{subfigure}
        \hfill
        \begin{subfigure}[t]{0.24\textwidth}
            \centering
            \includegraphics[width=\linewidth,height=0.75\linewidth,keepaspectratio]{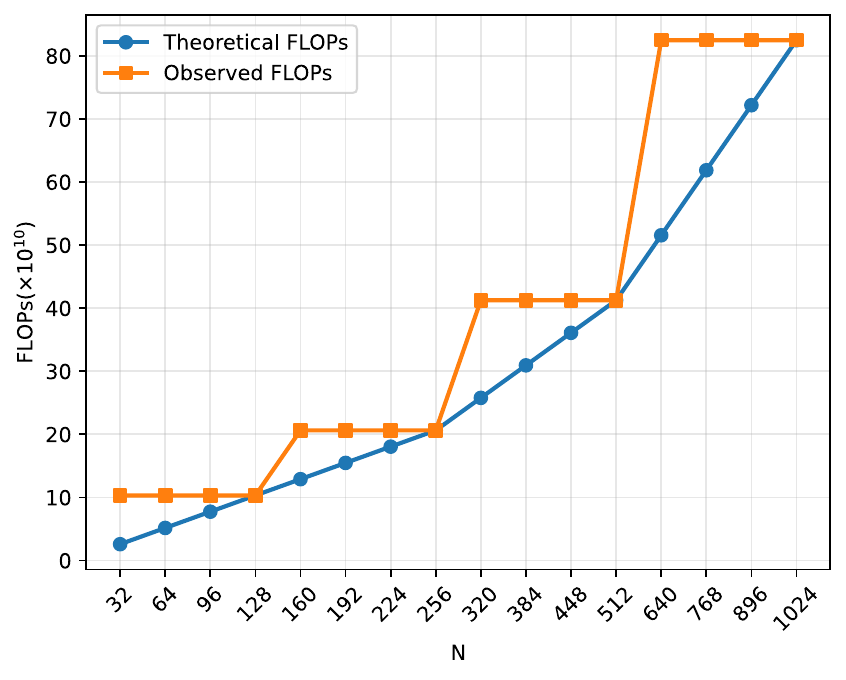}
            \caption{FLOPs, $k=32$}
            \label{fig:moe-upper-sglang-h800-flops-k32}
        \end{subfigure}
    \end{minipage}%
    }

    \caption{
    MoE FFN evaluation for load-balanced routing, the upper-bound case, with \textbf{SGLang} on \textbf{NVIDIA H800}.
    }
    \label{fig:moe-upper-sglang-h800}
\end{figure*}

Figures~\ref{fig:moe-upper-sglang-h20}, \ref{fig:moe-upper-sglang-a800}, and \ref{fig:moe-upper-sglang-h800} report the corresponding load-balanced results using the SGLang fused MoE backend. The same qualitative behavior appears under SGLang. The latency curves show near-free plateaus, the extracted $N_{\max}$ values follow the sparsity-dependent upper-bound trend, and the arithmetic-intensity and FLOPs profiles change in discrete steps as $N$ crosses backend granularity boundaries. As shown in Figures~\ref{fig:moe-upper-sglang-h20}\subref{fig:moe-upper-sglang-h20-nmax}, \ref{fig:moe-upper-sglang-a800}\subref{fig:moe-upper-sglang-a800-nmax}, and \ref{fig:moe-upper-sglang-h800}\subref{fig:moe-upper-sglang-h800-nmax}, this trend is consistent across all evaluated GPU platforms.

Overall, the load-balanced results show that MoE FFN NFP in the upper-bound case is primarily determined by aggregate expert-token padding slack. The agreement between vLLM and SGLang indicates that this behavior is not specific to one backend, but arises from the common granularity structure of fused MoE execution.

\clearpage

\vspace{-0.2cm}

\vspace{-0.2cm}

\clearpage

\vspace{-0.2cm}

\vspace{-0.2cm}

\vspace{-0.2cm}
\clearpage

\subsubsection{Load-skewed Routing as Lower Bound}

\begin{figure*}[!p]
    \centering
    \scalebox{1}[1.0]{%
    \begin{minipage}{\textwidth}
        \centering

        \begin{subfigure}[t]{0.24\textwidth}
            \centering
            \includegraphics[width=\linewidth]{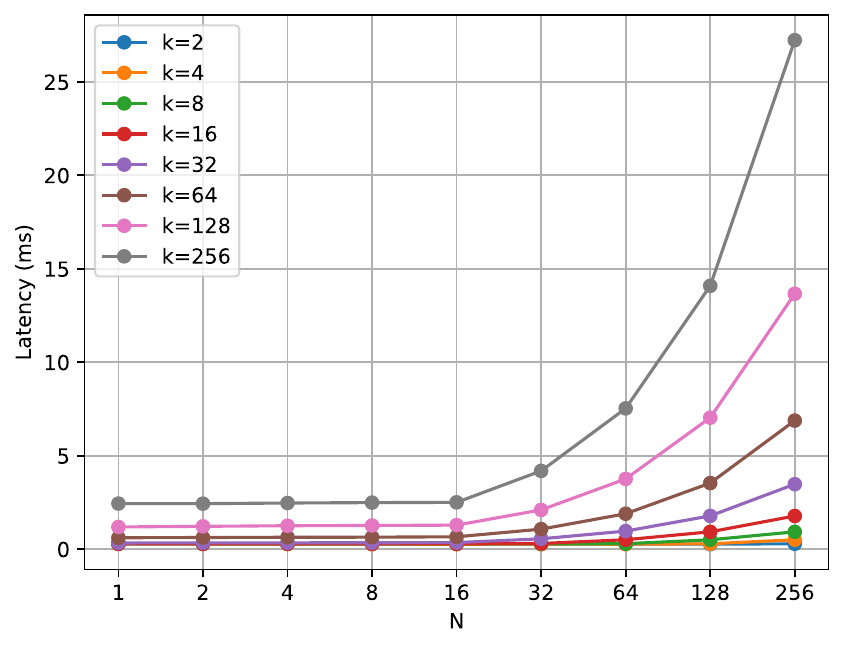}
            \caption{Latency vs. $N$}
            \label{fig:moe-lower-vllm-h20-latency}
        \end{subfigure}
        \hfill
        \begin{subfigure}[t]{0.24\textwidth}
            \centering
            \includegraphics[width=\linewidth]{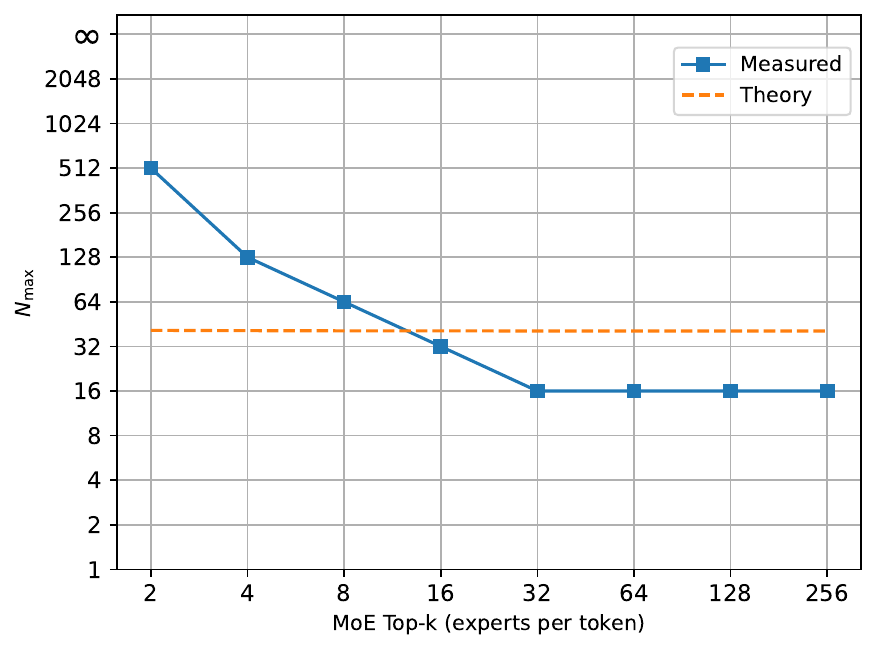}
            \caption{$N_{\max}$ vs. $k$}
            \label{fig:moe-lower-vllm-h20-nmax}
        \end{subfigure}
        \hfill
        \begin{subfigure}[t]{0.24\textwidth}
            \centering
            \includegraphics[width=\linewidth]{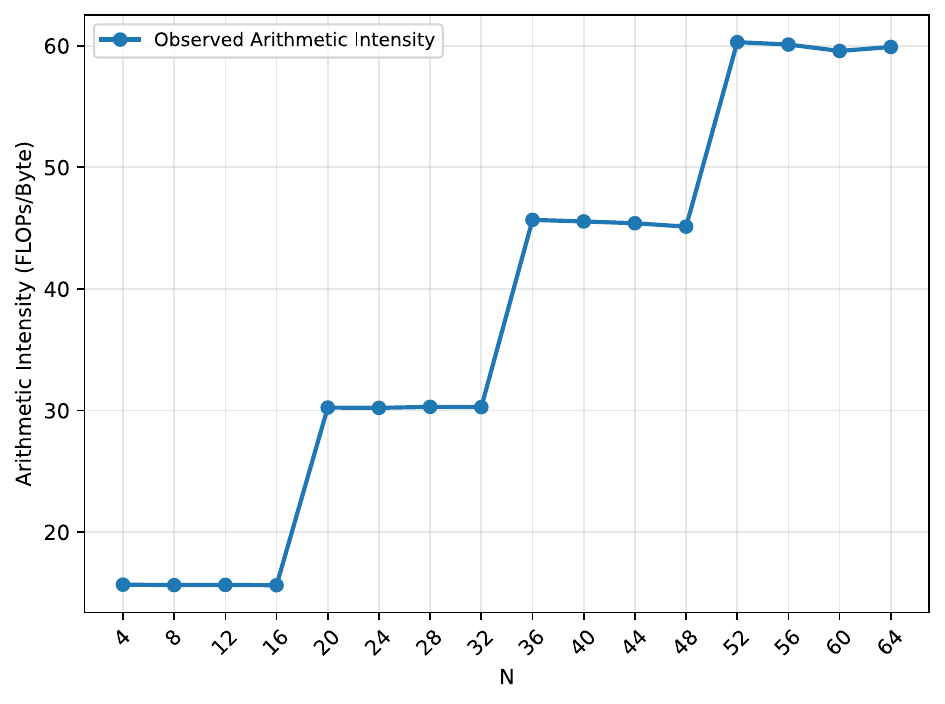}
            \caption{AI, $k=8$}
            \label{fig:moe-lower-vllm-h20-ai-k8}
        \end{subfigure}
        \hfill
        \begin{subfigure}[t]{0.24\textwidth}
            \centering
            \includegraphics[width=\linewidth]{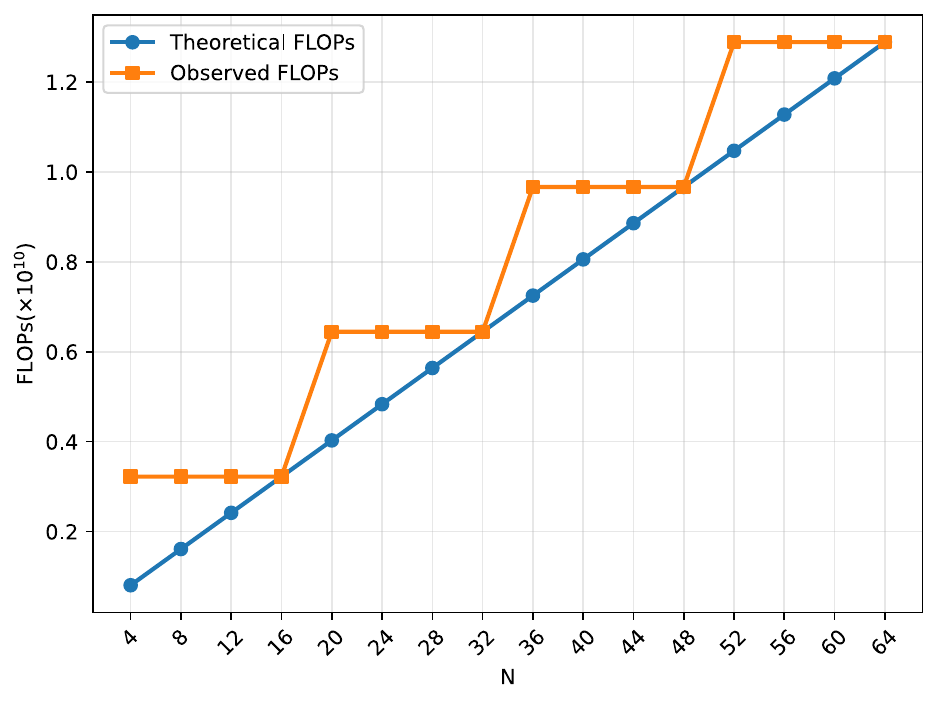}
            \caption{FLOPs, $k=8$}
            \label{fig:moe-lower-vllm-h20-flops-k8}
        \end{subfigure}

        \vspace{0.8em}

        \begin{subfigure}[t]{0.24\textwidth}
            \centering
            \includegraphics[width=\linewidth]{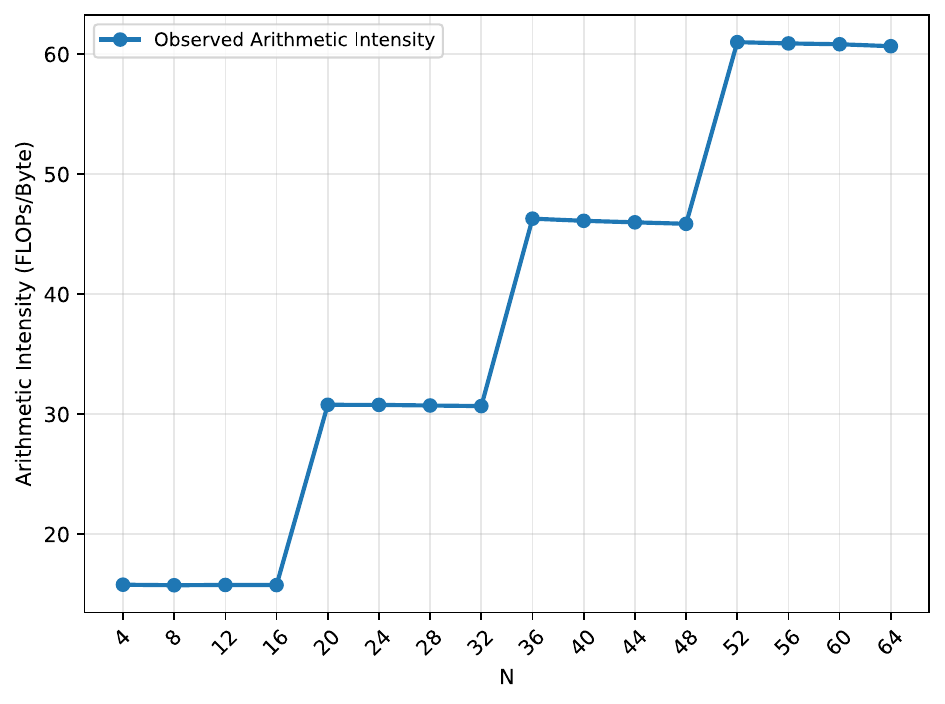}
            \caption{AI, $k=16$}
            \label{fig:moe-lower-vllm-h20-ai-k16}
        \end{subfigure}
        \hfill
        \begin{subfigure}[t]{0.24\textwidth}
            \centering
            \includegraphics[width=\linewidth]{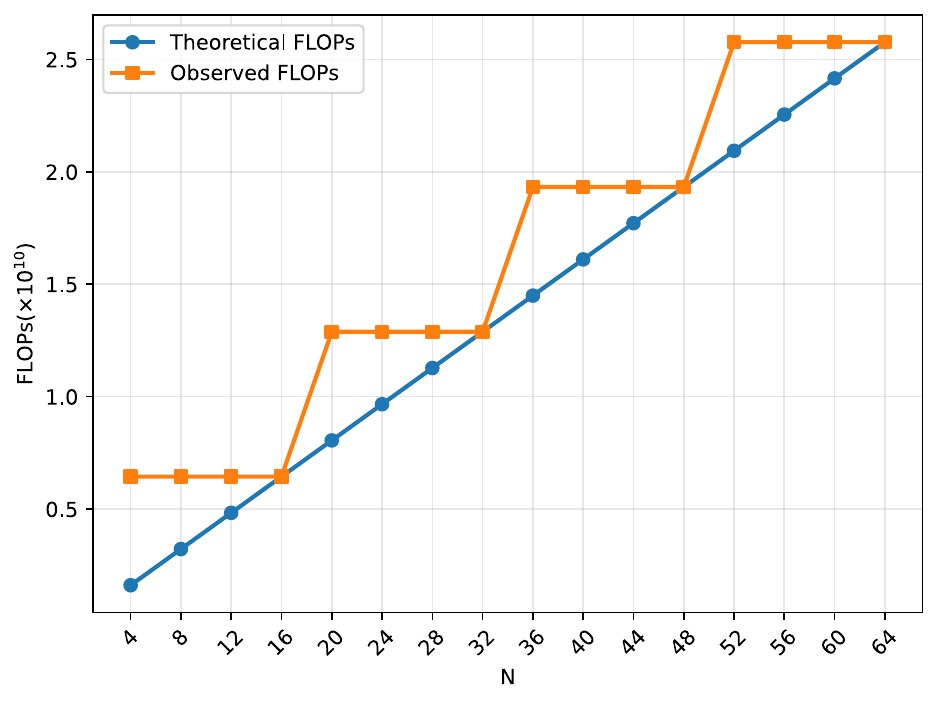}
            \caption{FLOPs, $k=16$}
            \label{fig:moe-lower-vllm-h20-flops-k16}
        \end{subfigure}
        \hfill
        \begin{subfigure}[t]{0.24\textwidth}
            \centering
            \includegraphics[width=\linewidth]{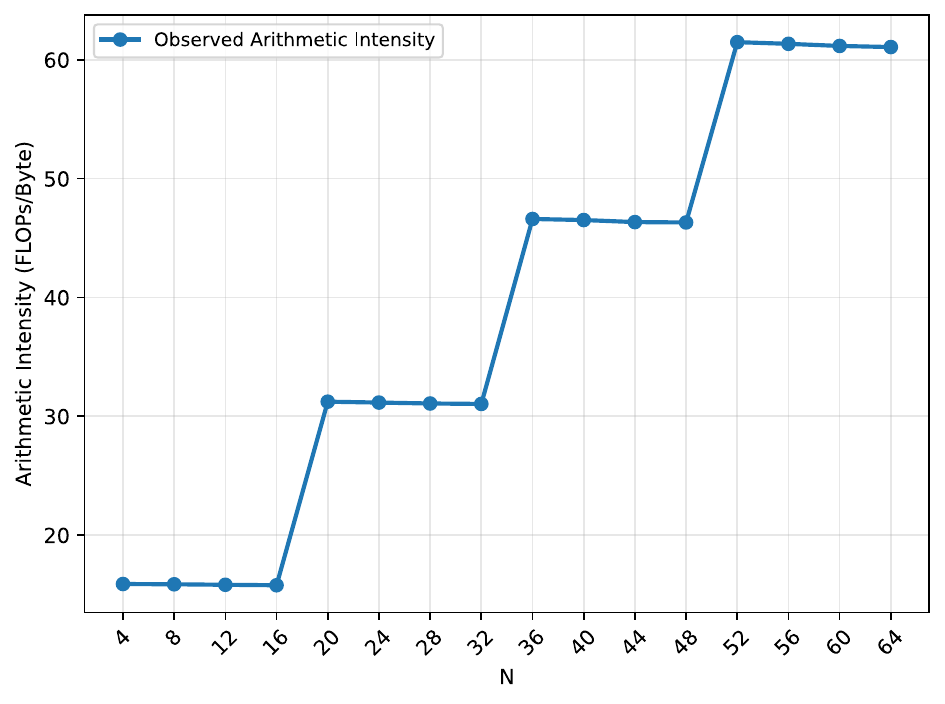}
            \caption{AI, $k=32$}
            \label{fig:moe-lower-vllm-h20-ai-k32}
        \end{subfigure}
        \hfill
        \begin{subfigure}[t]{0.24\textwidth}
            \centering
            \includegraphics[width=\linewidth]{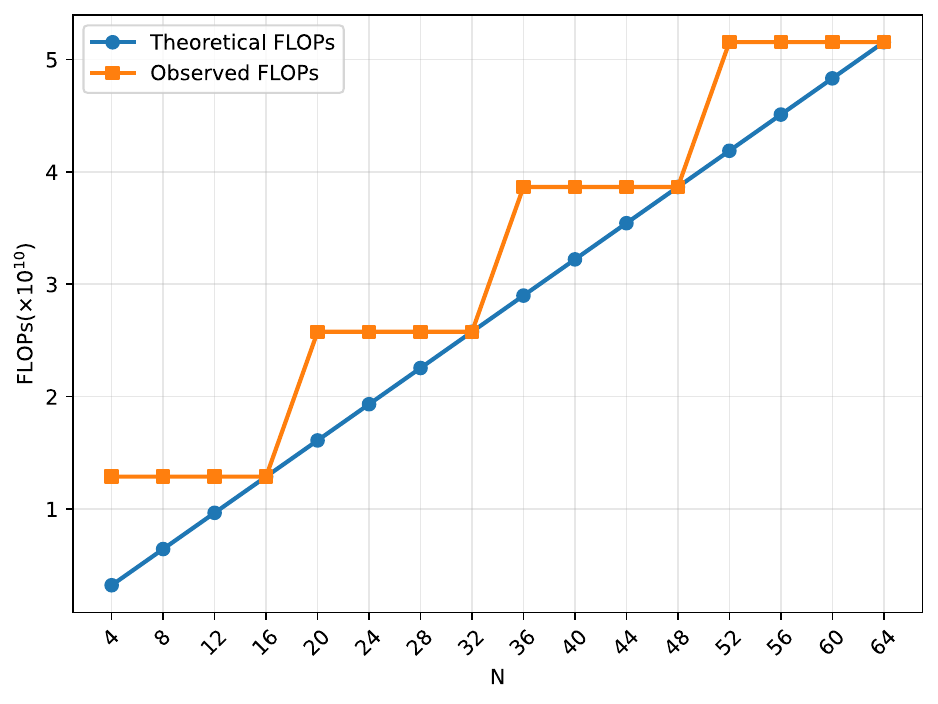}
            \caption{FLOPs, $k=32$}
            \label{fig:moe-lower-vllm-h20-flops-k32}
        \end{subfigure}
    \end{minipage}%
    }

    \caption{
    MoE FFN evaluation for load-skewed routing, the lower-bound case, with \textbf{vLLM} on \textbf{NVIDIA H20}.
    }
    \label{fig:moe-lower-vllm-h20}
\end{figure*}

\begin{figure*}[!p]
    \centering
    \scalebox{1}[1.0]{%
    \begin{minipage}{\textwidth}
        \centering

        \begin{subfigure}[t]{0.24\textwidth}
            \centering
            \includegraphics[width=\linewidth]{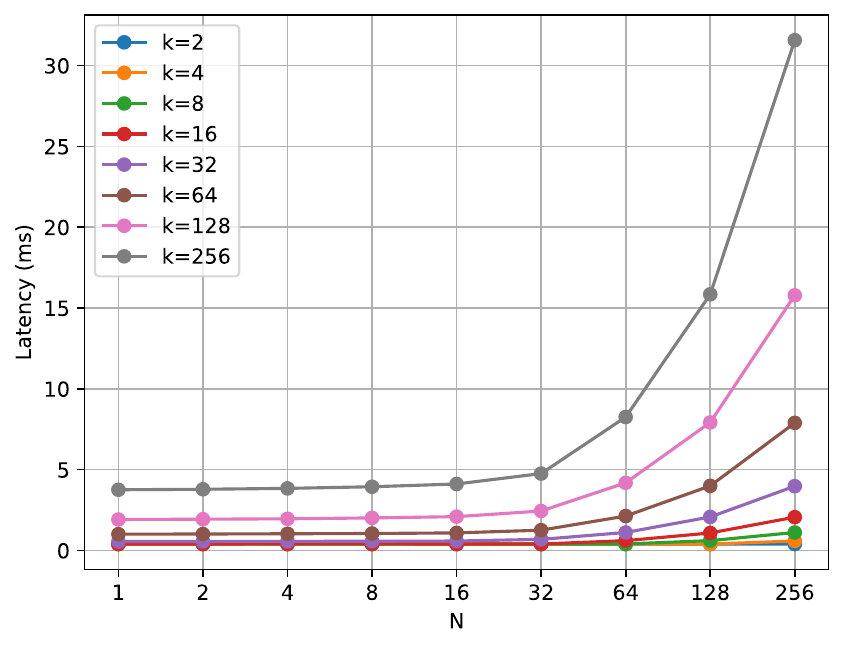}
            \caption{Latency vs. $N$}
            \label{fig:moe-lower-vllm-a800-latency}
        \end{subfigure}
        \hfill
        \begin{subfigure}[t]{0.24\textwidth}
            \centering
            \includegraphics[width=\linewidth]{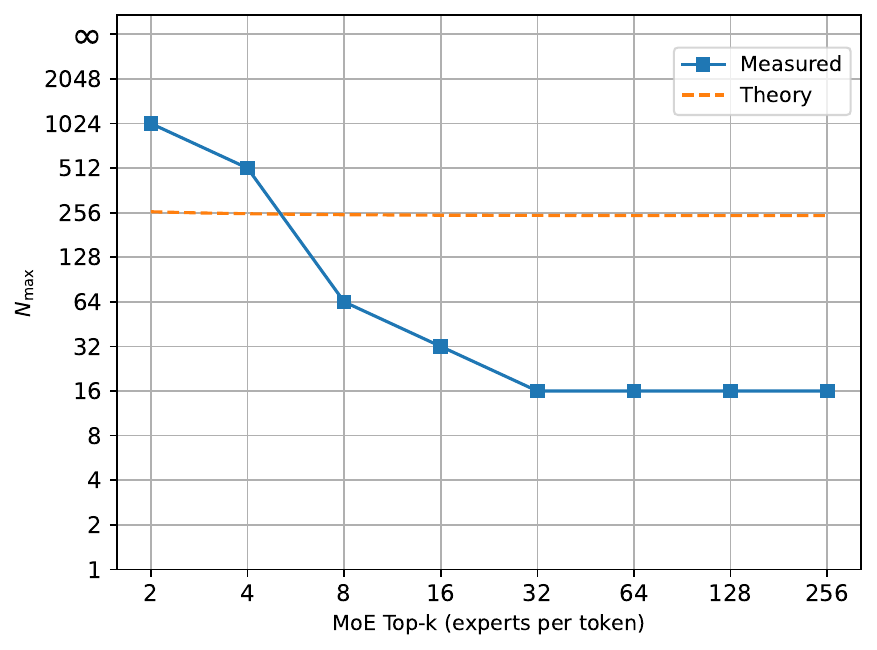}
            \caption{$N_{\max}$ vs. $k$}
            \label{fig:moe-lower-vllm-a800-nmax}
        \end{subfigure}
        \hfill
        \begin{subfigure}[t]{0.24\textwidth}
            \centering
            \includegraphics[width=\linewidth]{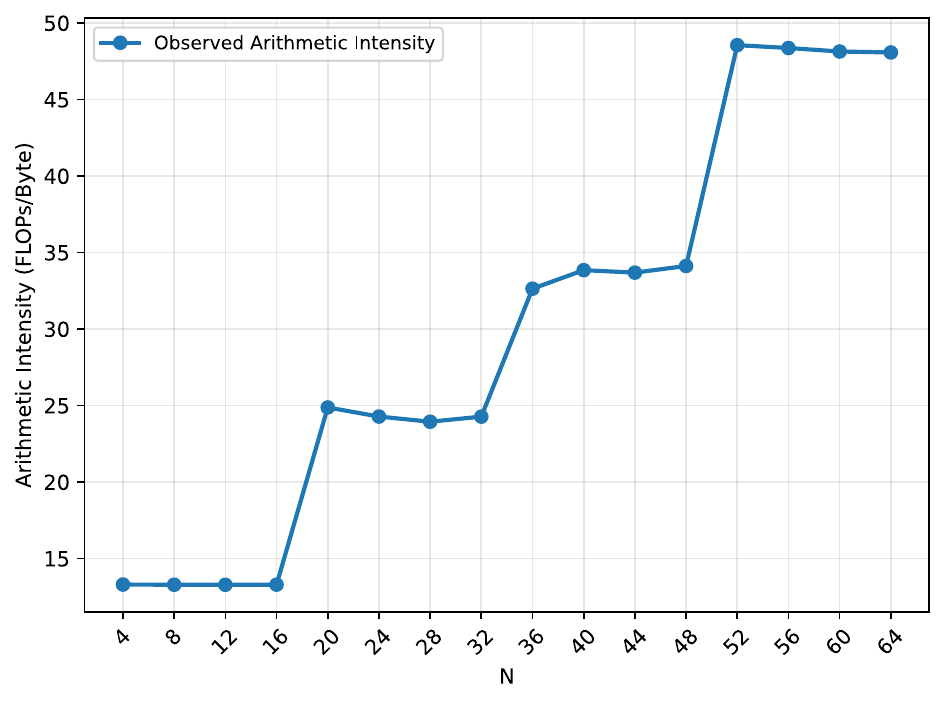}
            \caption{AI, $k=8$}
            \label{fig:moe-lower-vllm-a800-ai-k8}
        \end{subfigure}
        \hfill
        \begin{subfigure}[t]{0.24\textwidth}
            \centering
            \includegraphics[width=\linewidth]{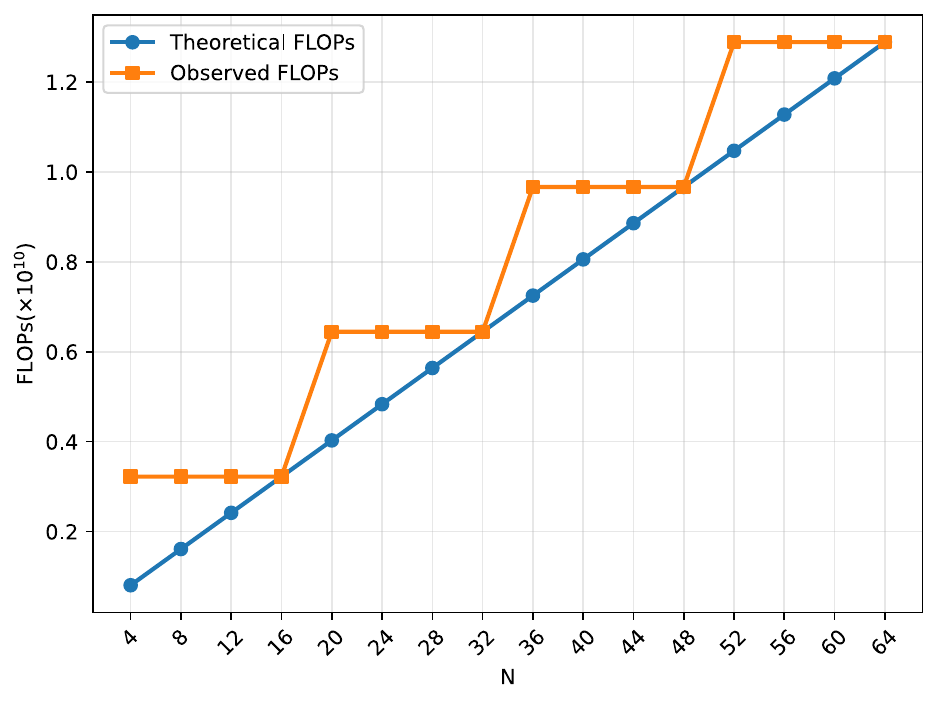}
            \caption{FLOPs, $k=8$}
            \label{fig:moe-lower-vllm-a800-flops-k8}
        \end{subfigure}

        \vspace{0.8em}

        \begin{subfigure}[t]{0.24\textwidth}
            \centering
            \includegraphics[width=\linewidth]{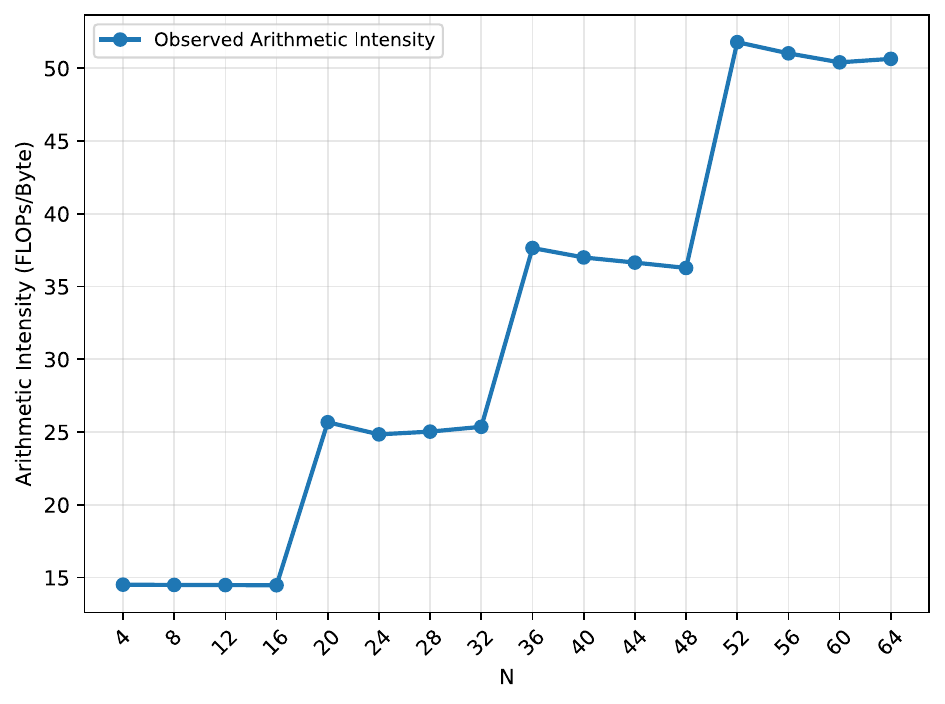}
            \caption{AI, $k=16$}
            \label{fig:moe-lower-vllm-a800-ai-k16}
        \end{subfigure}
        \hfill
        \begin{subfigure}[t]{0.24\textwidth}
            \centering
            \includegraphics[width=\linewidth]{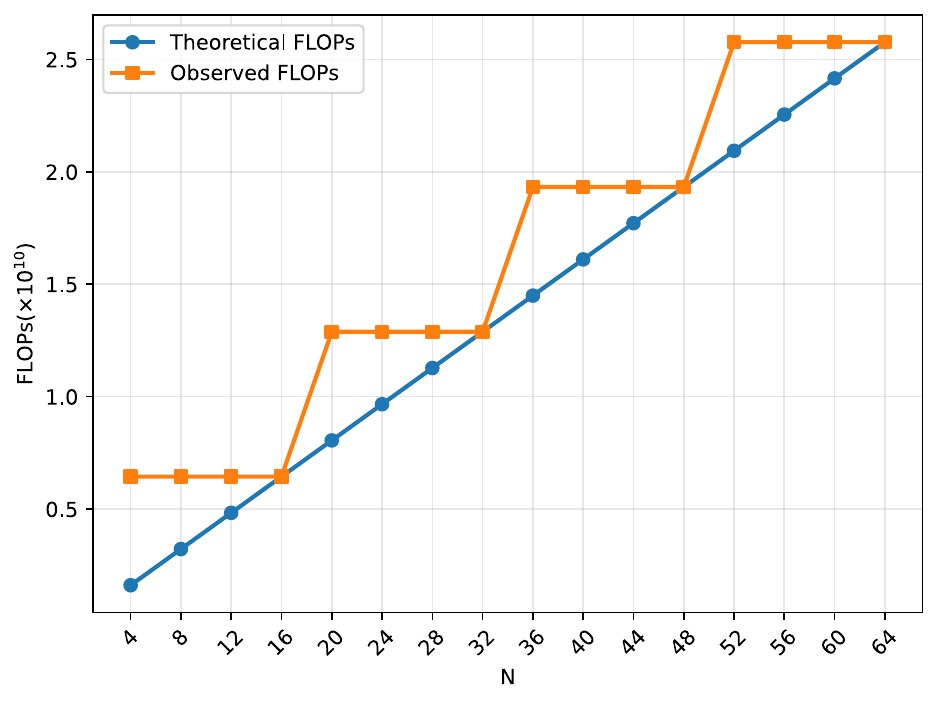}
            \caption{FLOPs, $k=16$}
            \label{fig:moe-lower-vllm-a800-flops-k16}
        \end{subfigure}
        \hfill
        \begin{subfigure}[t]{0.24\textwidth}
            \centering
            \includegraphics[width=\linewidth]{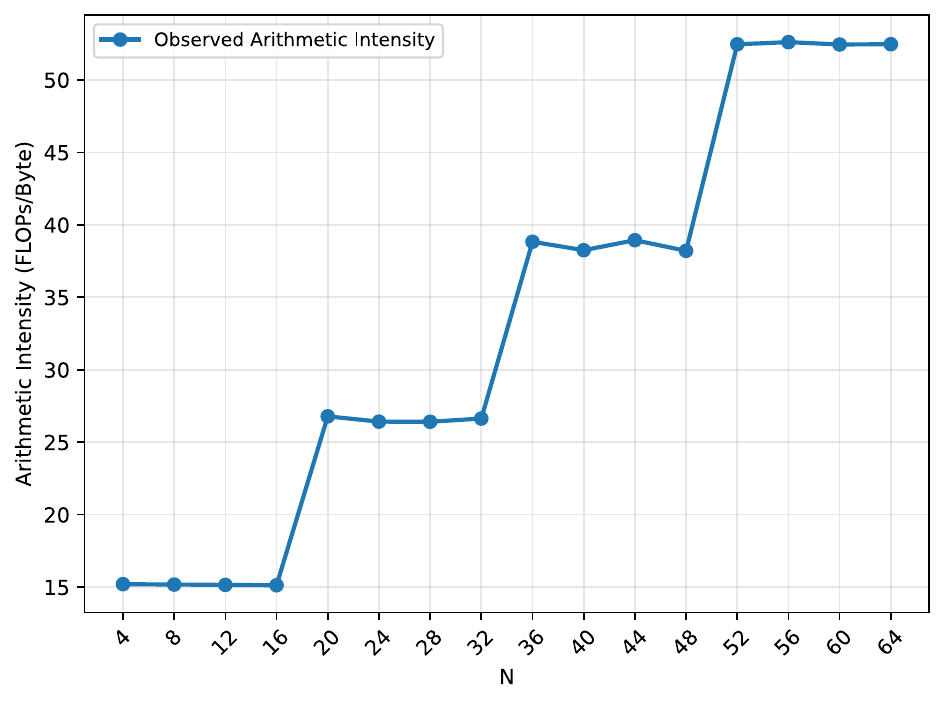}
            \caption{AI, $k=32$}
            \label{fig:moe-lower-vllm-a800-ai-k32}
        \end{subfigure}
        \hfill
        \begin{subfigure}[t]{0.24\textwidth}
            \centering
            \includegraphics[width=\linewidth]{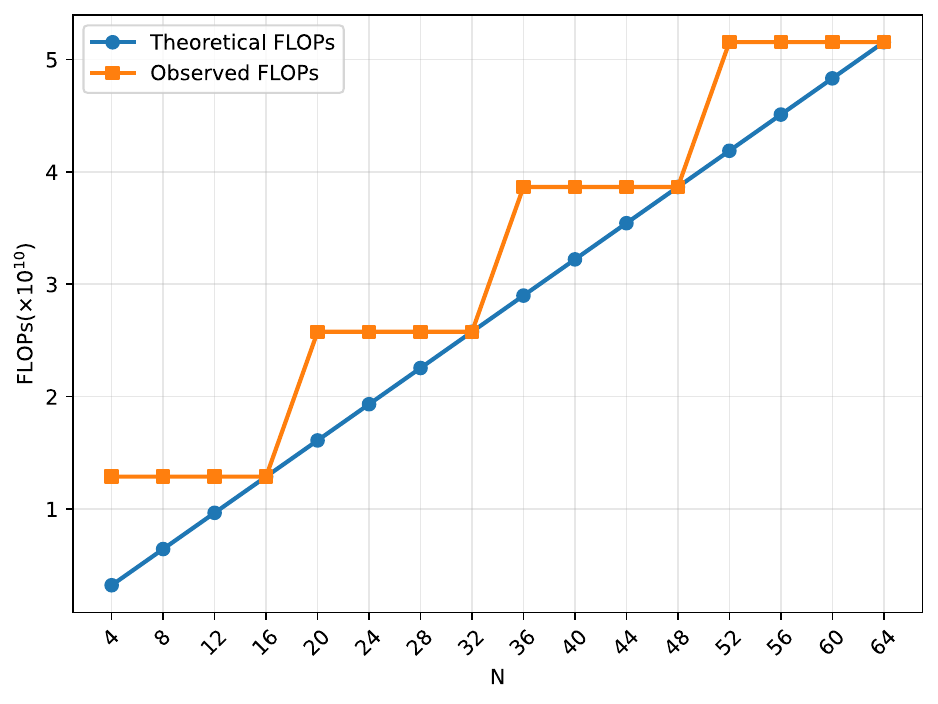}
            \caption{FLOPs, $k=32$}
            \label{fig:moe-lower-vllm-a800-flops-k32}
        \end{subfigure}
    \end{minipage}%
    }

    \caption{
    MoE FFN evaluation for load-skewed routing, the lower-bound case, with \textbf{vLLM} on \textbf{NVIDIA A800}.
    }
    \label{fig:moe-lower-vllm-a800}
\end{figure*}

\begin{figure*}[!p]
    \centering
    \scalebox{1}[1.0]{%
    \begin{minipage}{\textwidth}
        \centering

        \begin{subfigure}[t]{0.24\textwidth}
            \centering
            \includegraphics[width=\linewidth]{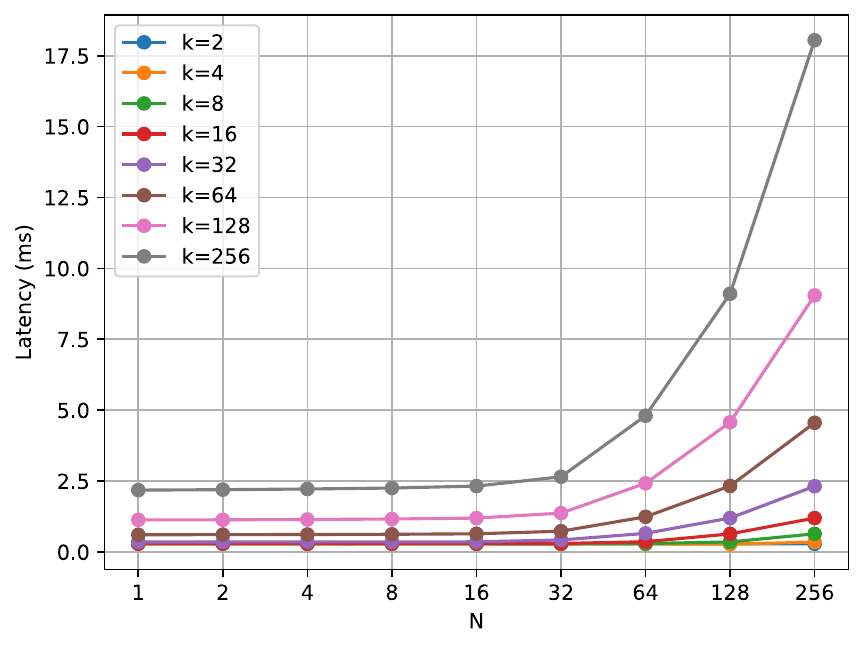}
            \caption{Latency vs. $N$}
            \label{fig:moe-lower-vllm-h800-latency}
        \end{subfigure}
        \hfill
        \begin{subfigure}[t]{0.24\textwidth}
            \centering
            \includegraphics[width=\linewidth]{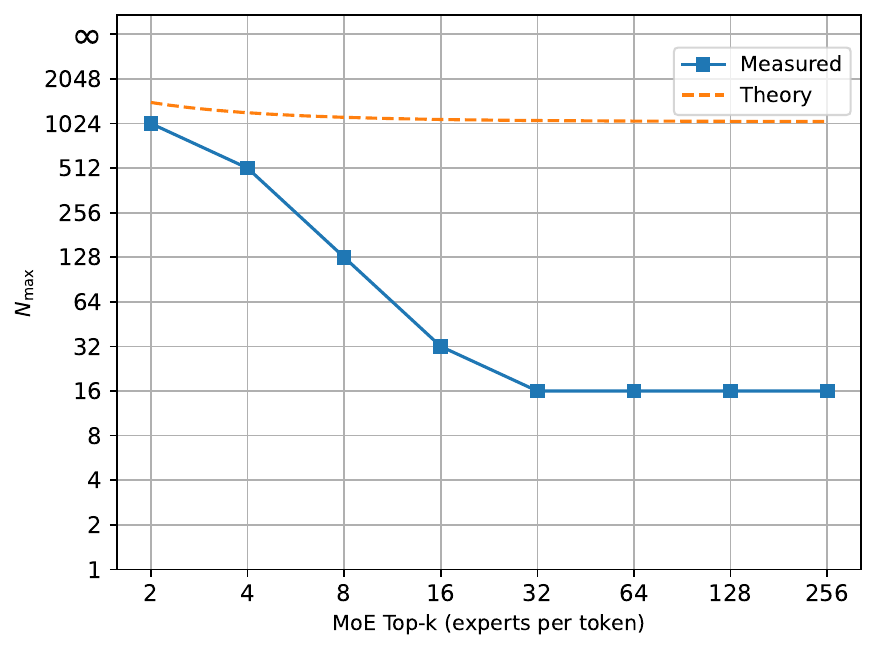}
            \caption{$N_{\max}$ vs. $k$}
            \label{fig:moe-lower-vllm-h800-nmax}
        \end{subfigure}
        \hfill
        \begin{subfigure}[t]{0.24\textwidth}
            \centering
            \includegraphics[width=\linewidth]{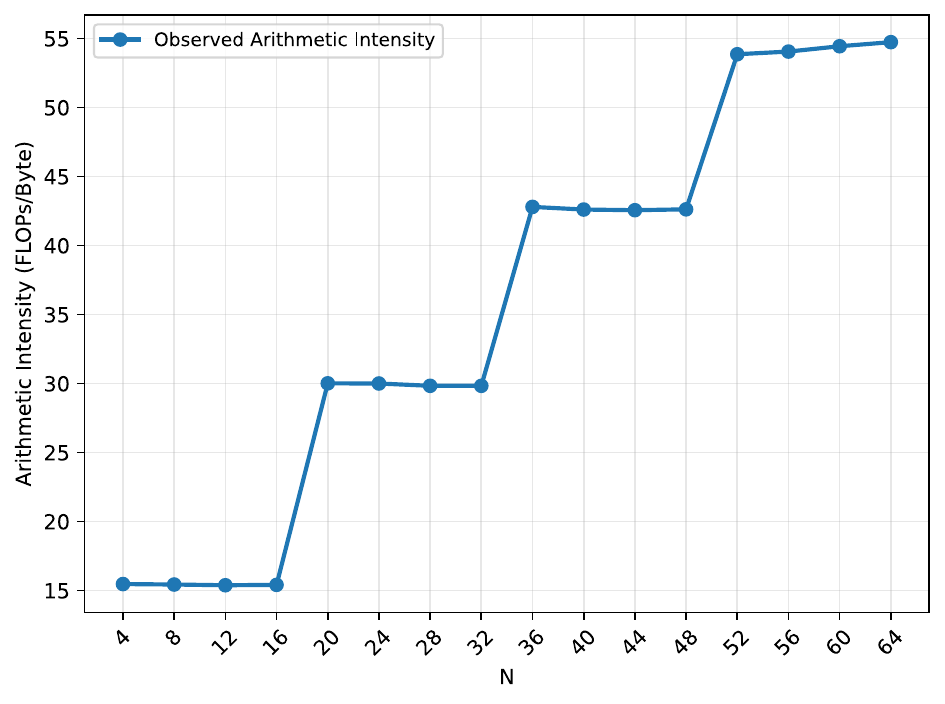}
            \caption{AI, $k=8$}
            \label{fig:moe-lower-vllm-h800-ai-k8}
        \end{subfigure}
        \hfill
        \begin{subfigure}[t]{0.24\textwidth}
            \centering
            \includegraphics[width=\linewidth]{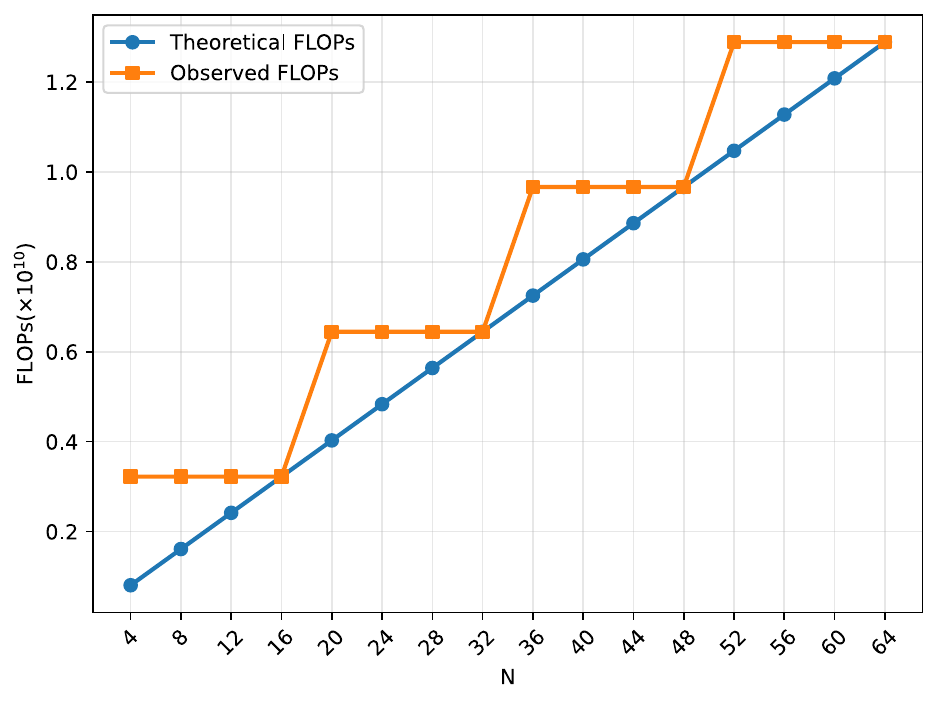}
            \caption{FLOPs, $k=8$}
            \label{fig:moe-lower-vllm-h800-flops-k8}
        \end{subfigure}

        \vspace{0.8em}

        \begin{subfigure}[t]{0.24\textwidth}
            \centering
            \includegraphics[width=\linewidth]{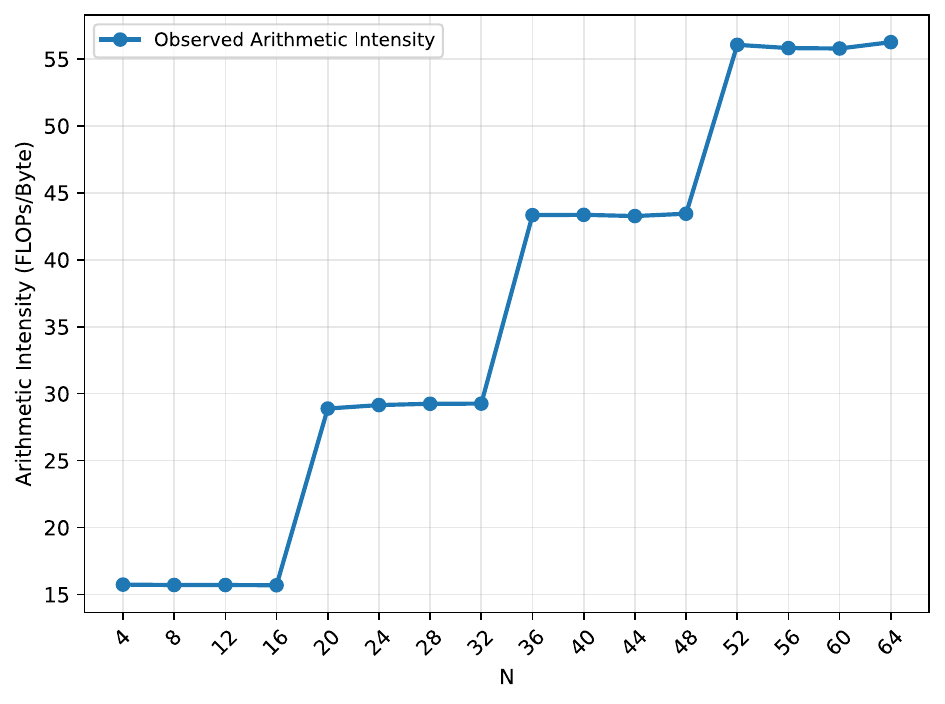}
            \caption{AI, $k=16$}
            \label{fig:moe-lower-vllm-h800-ai-k16}
        \end{subfigure}
        \hfill
        \begin{subfigure}[t]{0.24\textwidth}
            \centering
            \includegraphics[width=\linewidth]{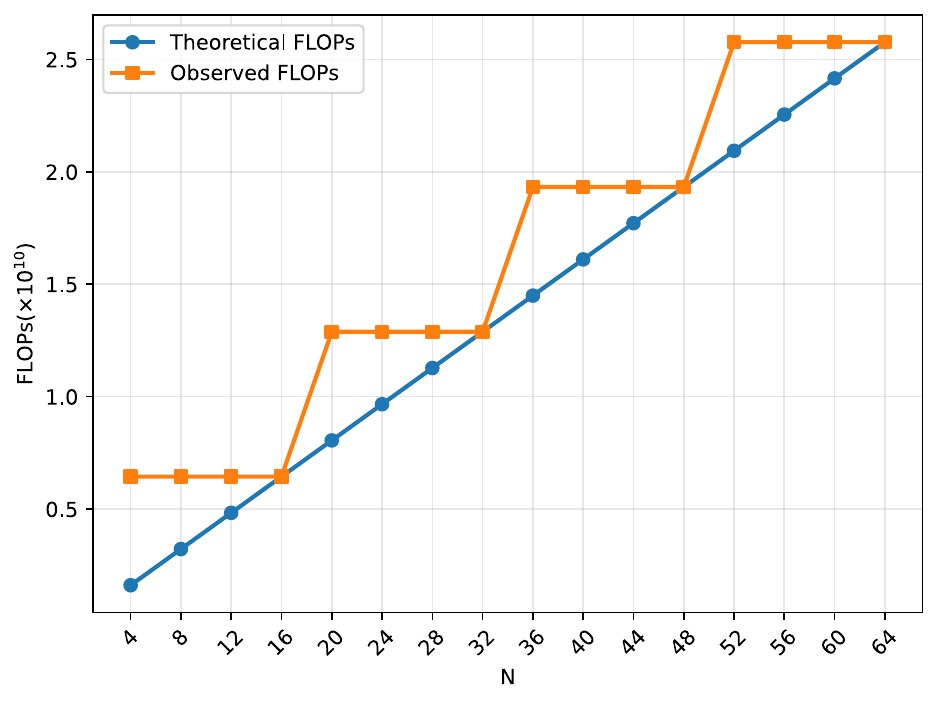}
            \caption{FLOPs, $k=16$}
            \label{fig:moe-lower-vllm-h800-flops-k16}
        \end{subfigure}
        \hfill
        \begin{subfigure}[t]{0.24\textwidth}
            \centering
            \includegraphics[width=\linewidth]{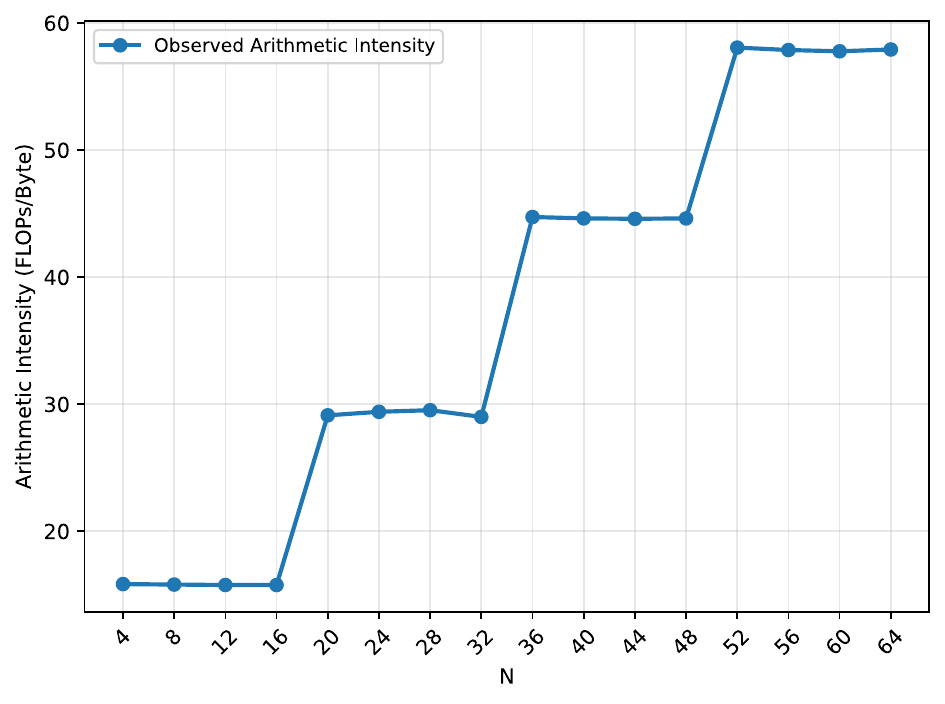}
            \caption{AI, $k=32$}
            \label{fig:moe-lower-vllm-h800-ai-k32}
        \end{subfigure}
        \hfill
        \begin{subfigure}[t]{0.24\textwidth}
            \centering
            \includegraphics[width=\linewidth]{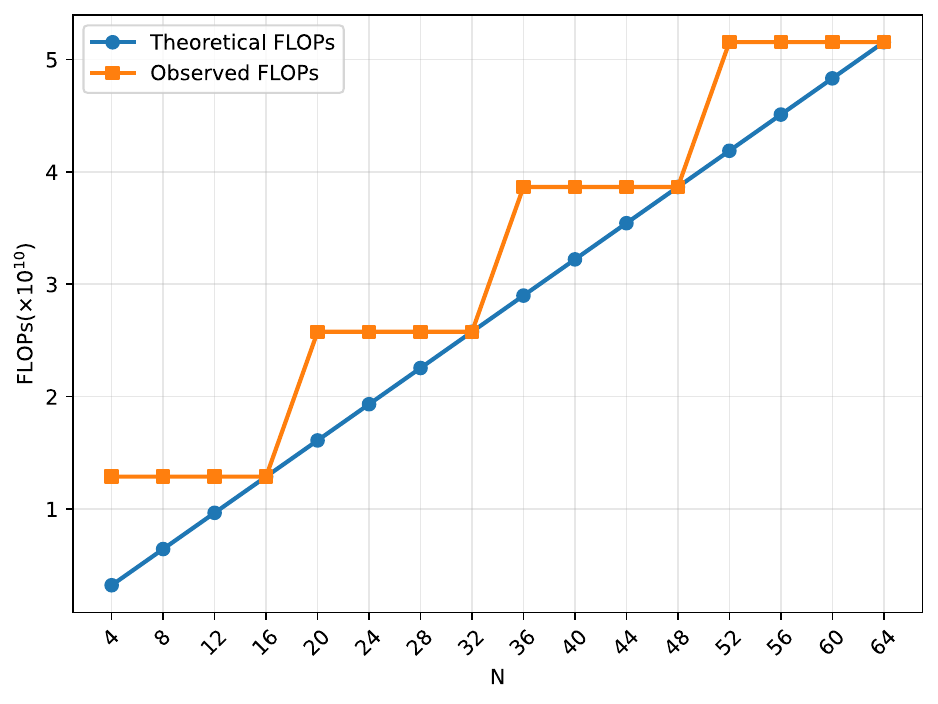}
            \caption{FLOPs, $k=32$}
            \label{fig:moe-lower-vllm-h800-flops-k32}
        \end{subfigure}
    \end{minipage}%
    }

    \caption{
    MoE FFN evaluation for load-skewed routing, the lower-bound case, with \textbf{vLLM} on \textbf{NVIDIA H800}.
    }
    \label{fig:moe-lower-vllm-h800}
\end{figure*}

Figures~\ref{fig:moe-lower-vllm-h20}, \ref{fig:moe-lower-vllm-a800}, and \ref{fig:moe-lower-vllm-h800} report the load-skewed MoE FFN results using the vLLM fused MoE backend. In this setting, all tokens are routed to the same selected experts, so larger N can only exploit padding slack within a fixed set of expert-token blocks. The latency curves still exhibit near-free regions with staircase-like transitions, but the boundary is much smaller than in the load-balanced case. This is because load-skewed routing does not create aggregate slack across many activated experts.

The extracted boundaries confirm the lower-bound behavior. Unlike the load-balanced case, increasing sparsity does not expose additional expert-distributed padding capacity. Instead, $N_{\max}$ is largely fixed by the backend's expert-token padding granularity. The arithmetic-intensity and runtime-FLOPs profiles show discrete changes as $N$ crosses granularity boundaries, again indicating that the observed near-free region is produced by padded fused-kernel execution rather than by the idle-compute baseline.

\begin{figure*}[!p]
    \centering
    \scalebox{1}[1.0]{%
    \begin{minipage}{\textwidth}
        \centering

        \begin{subfigure}[t]{0.24\textwidth}
            \centering
            \includegraphics[width=\linewidth]{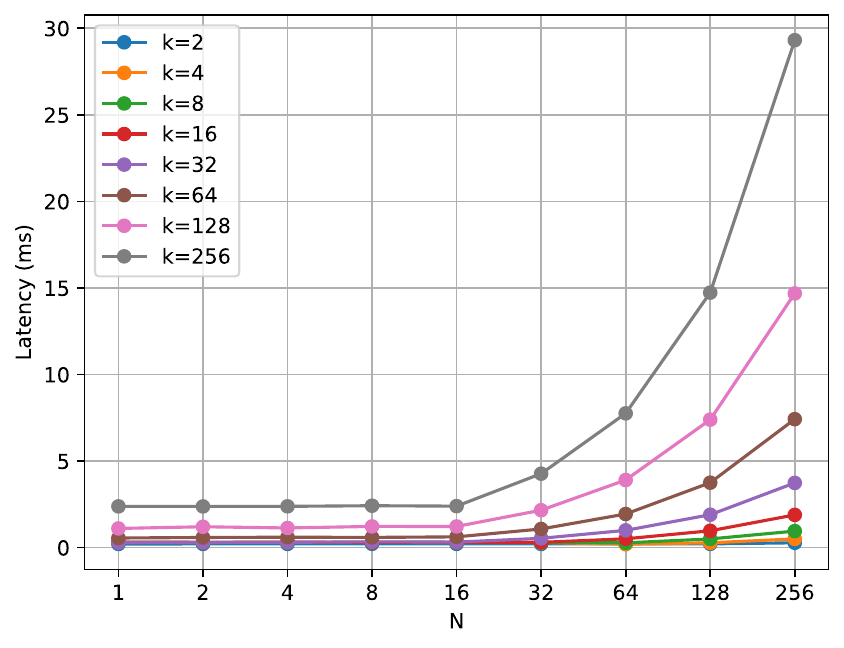}
            \caption{Latency vs. $N$}
            \label{fig:moe-lower-sglang-h20-latency}
        \end{subfigure}
        \hfill
        \begin{subfigure}[t]{0.24\textwidth}
            \centering
            \includegraphics[width=\linewidth]{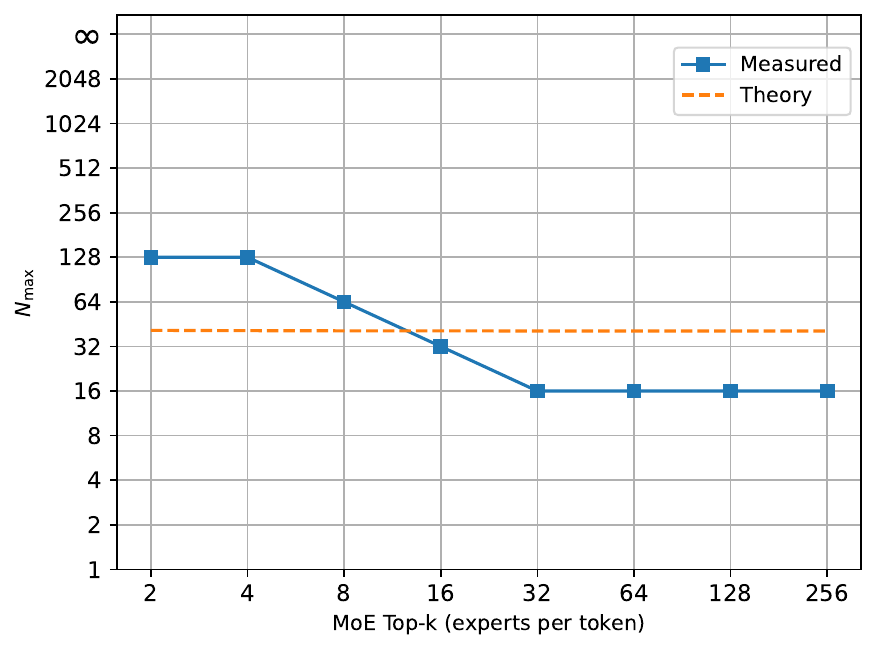}
            \caption{$N_{\max}$ vs. $k$}
            \label{fig:moe-lower-sglang-h20-nmax}
        \end{subfigure}
        \hfill
        \begin{subfigure}[t]{0.24\textwidth}
            \centering
            \includegraphics[width=\linewidth]{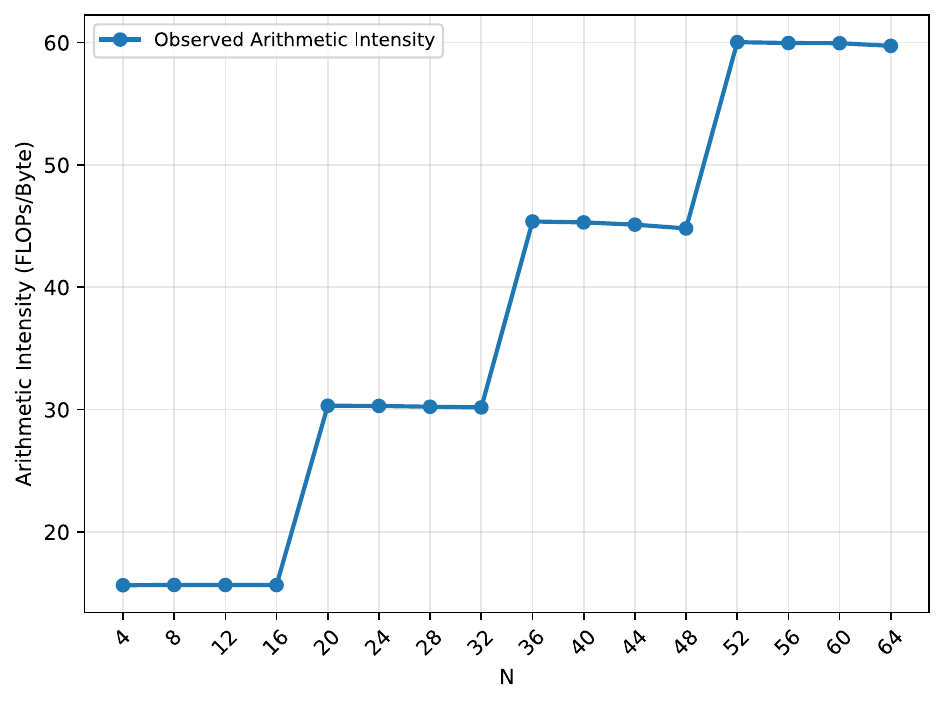}
            \caption{AI, $k=8$}
            \label{fig:moe-lower-sglang-h20-ai-k8}
        \end{subfigure}
        \hfill
        \begin{subfigure}[t]{0.24\textwidth}
            \centering
            \includegraphics[width=\linewidth]{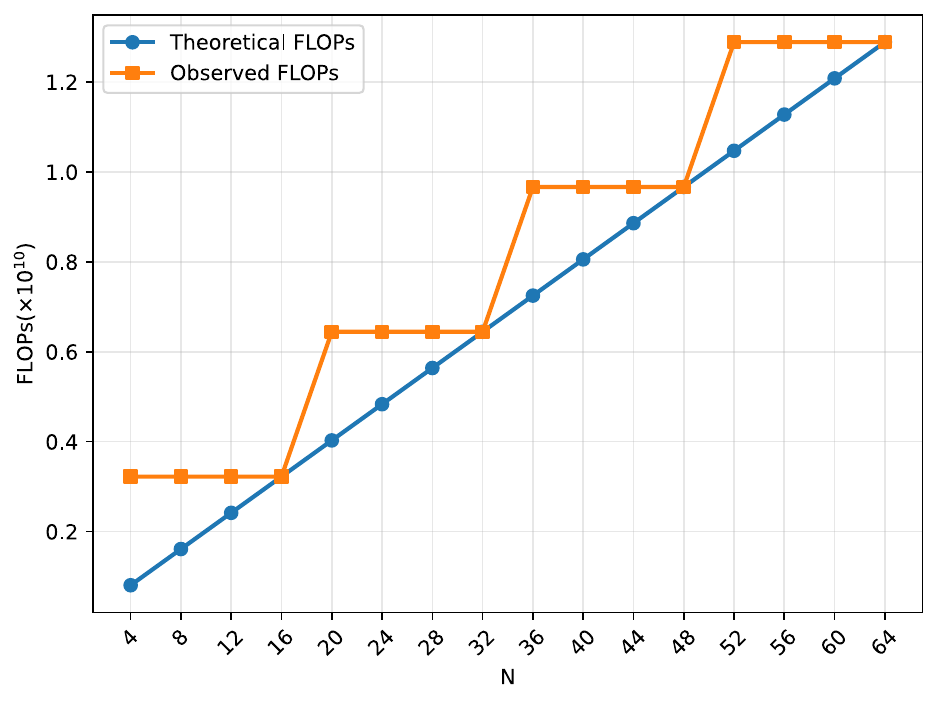}
            \caption{FLOPs, $k=8$}
            \label{fig:moe-lower-sglang-h20-flops-k8}
        \end{subfigure}

        \vspace{0.8em}

        \begin{subfigure}[t]{0.24\textwidth}
            \centering
            \includegraphics[width=\linewidth]{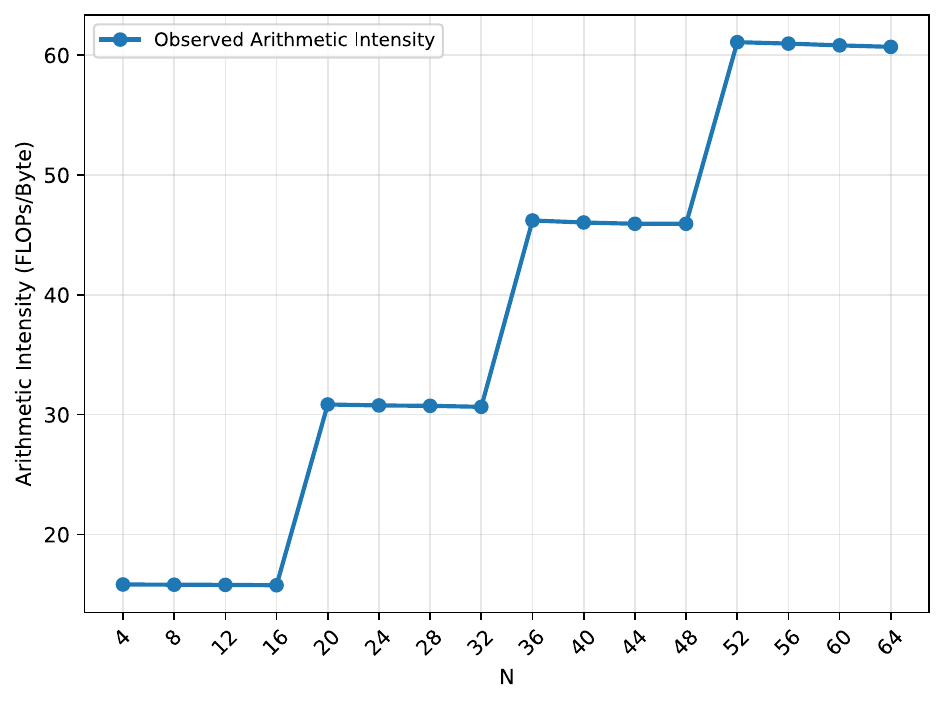}
            \caption{AI, $k=16$}
            \label{fig:moe-lower-sglang-h20-ai-k16}
        \end{subfigure}
        \hfill
        \begin{subfigure}[t]{0.24\textwidth}
            \centering
            \includegraphics[width=\linewidth]{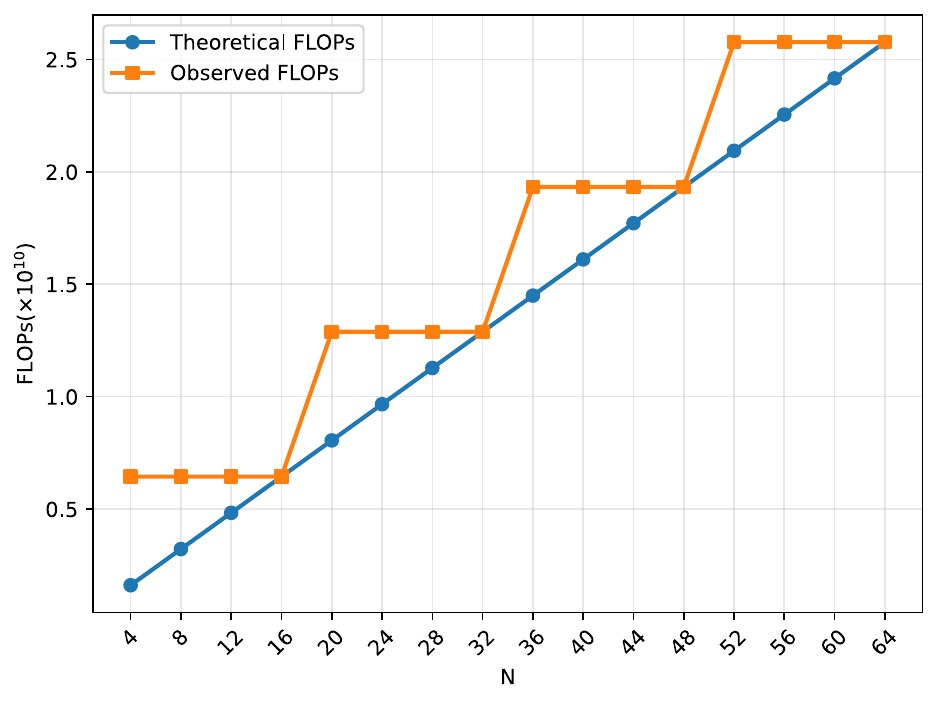}
            \caption{FLOPs, $k=16$}
            \label{fig:moe-lower-sglang-h20-flops-k16}
        \end{subfigure}
        \hfill
        \begin{subfigure}[t]{0.24\textwidth}
            \centering
            \includegraphics[width=\linewidth]{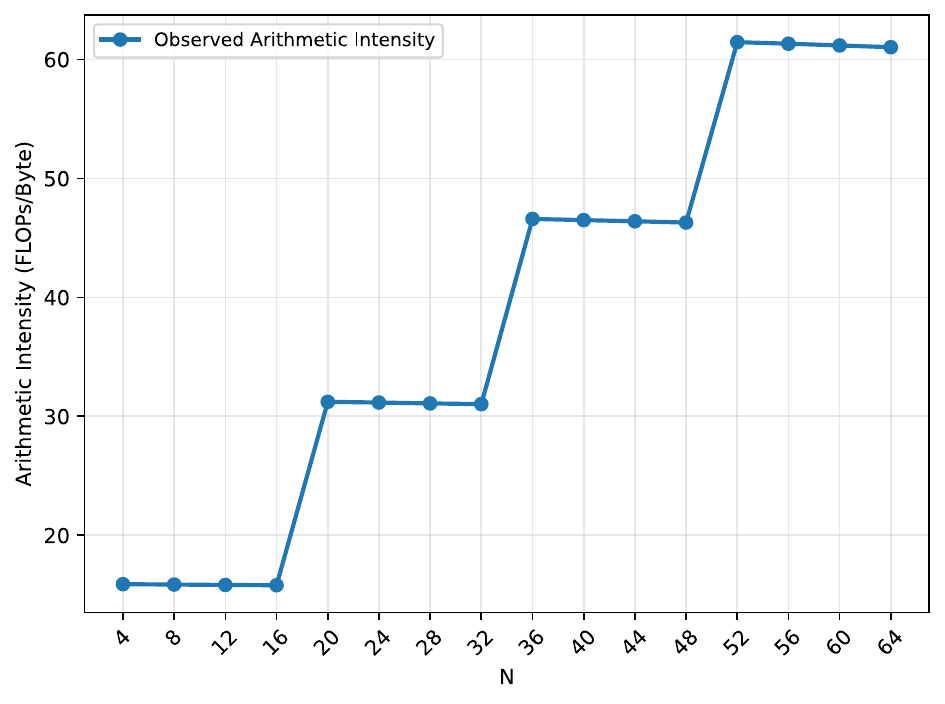}
            \caption{AI, $k=32$}
            \label{fig:moe-lower-sglang-h20-ai-k32}
        \end{subfigure}
        \hfill
        \begin{subfigure}[t]{0.24\textwidth}
            \centering
            \includegraphics[width=\linewidth]{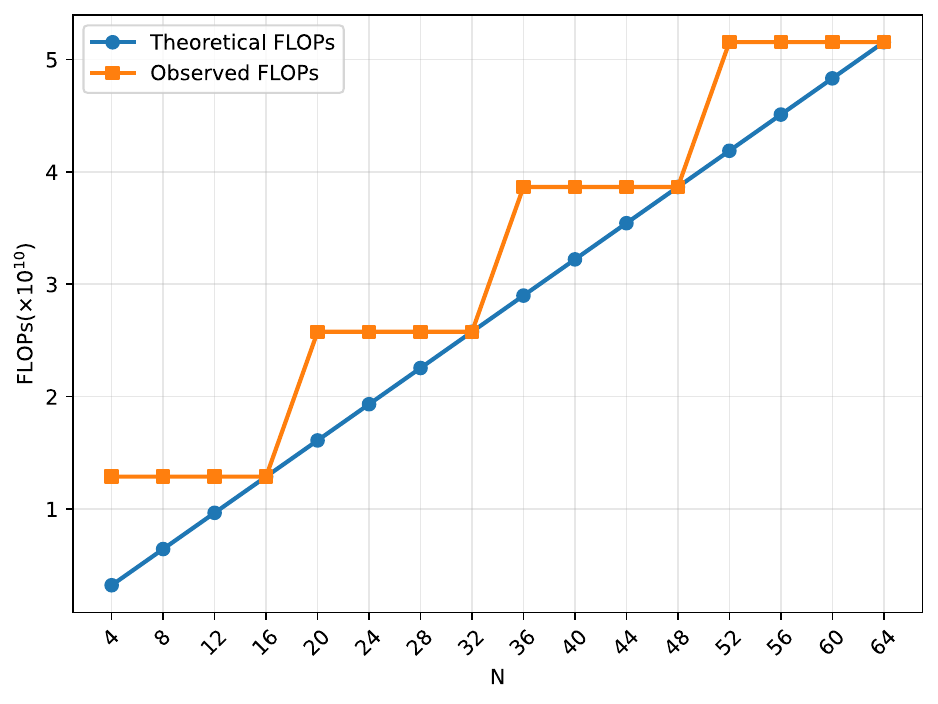}
            \caption{FLOPs, $k=32$}
            \label{fig:moe-lower-sglang-h20-flops-k32}
        \end{subfigure}
    \end{minipage}%
    }

    \caption{
    MoE FFN evaluation for load-skewed routing, the lower-bound case, with \textbf{SGLang} on \textbf{NVIDIA H20}.
    }
    \label{fig:moe-lower-sglang-h20}
\end{figure*}

\begin{figure*}[!p]
    \centering
    \scalebox{1}[1.0]{%
    \begin{minipage}{\textwidth}
        \centering

        \begin{subfigure}[t]{0.24\textwidth}
            \centering
            \includegraphics[width=\linewidth]{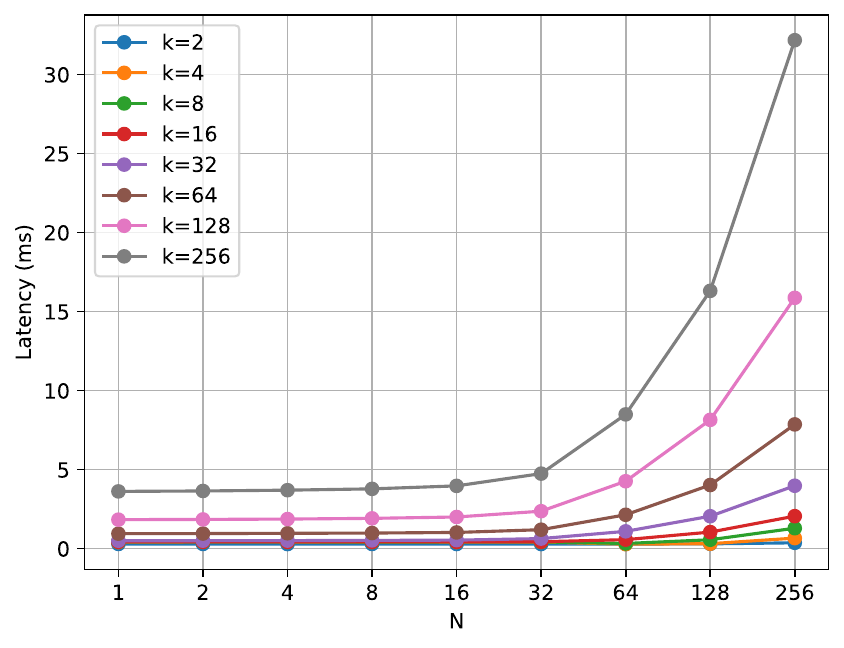}
            \caption{Latency vs. $N$}
            \label{fig:moe-lower-sglang-a800-latency}
        \end{subfigure}
        \hfill
        \begin{subfigure}[t]{0.24\textwidth}
            \centering
            \includegraphics[width=\linewidth]{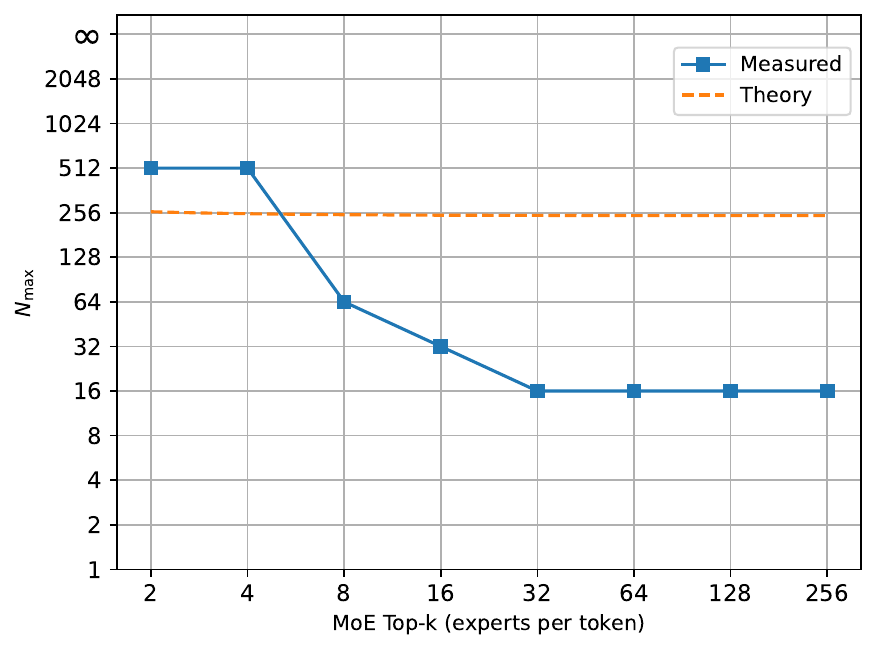}
            \caption{$N_{\max}$ vs. $k$}
            \label{fig:moe-lower-sglang-a800-nmax}
        \end{subfigure}
        \hfill
        \begin{subfigure}[t]{0.24\textwidth}
            \centering
            \includegraphics[width=\linewidth]{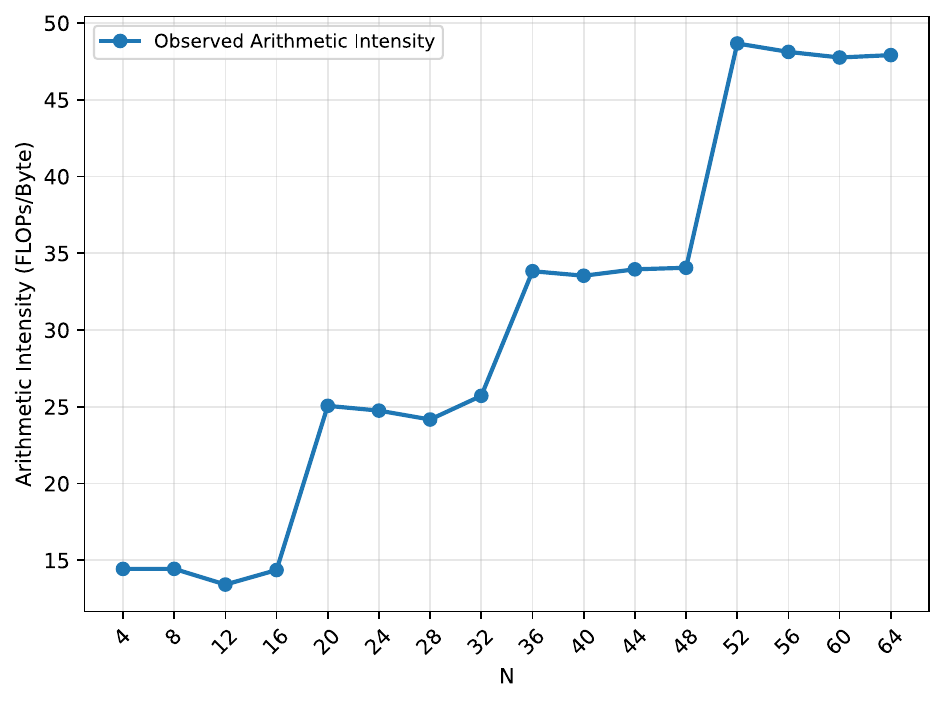}
            \caption{AI, $k=8$}
            \label{fig:moe-lower-sglang-a800-ai-k8}
        \end{subfigure}
        \hfill
        \begin{subfigure}[t]{0.24\textwidth}
            \centering
            \includegraphics[width=\linewidth]{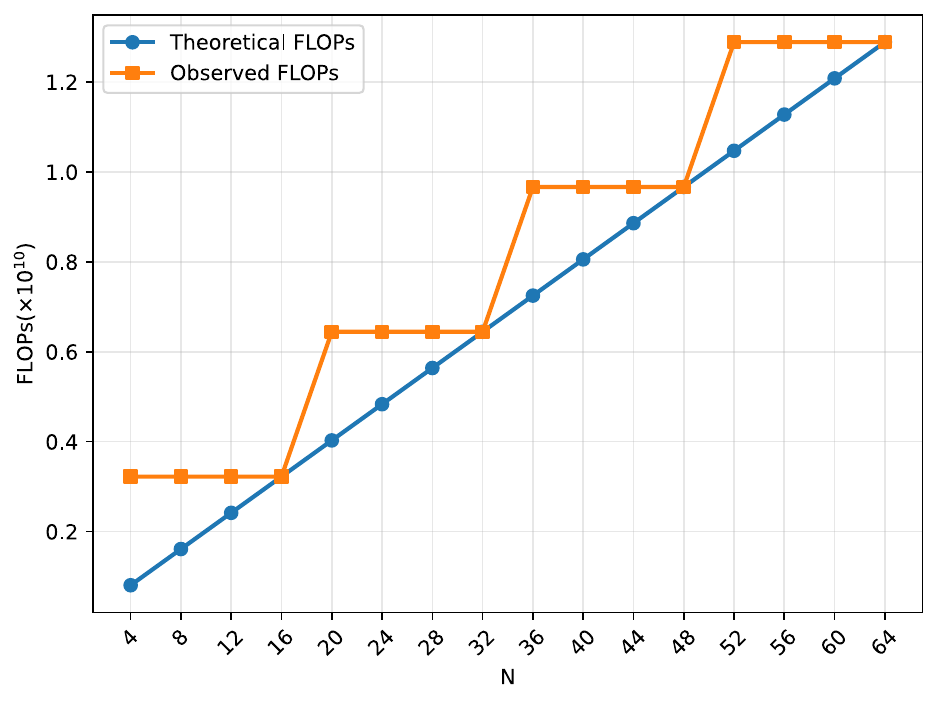}
            \caption{FLOPs, $k=8$}
            \label{fig:moe-lower-sglang-a800-flops-k8}
        \end{subfigure}

        \vspace{0.8em}

        \begin{subfigure}[t]{0.24\textwidth}
            \centering
            \includegraphics[width=\linewidth]{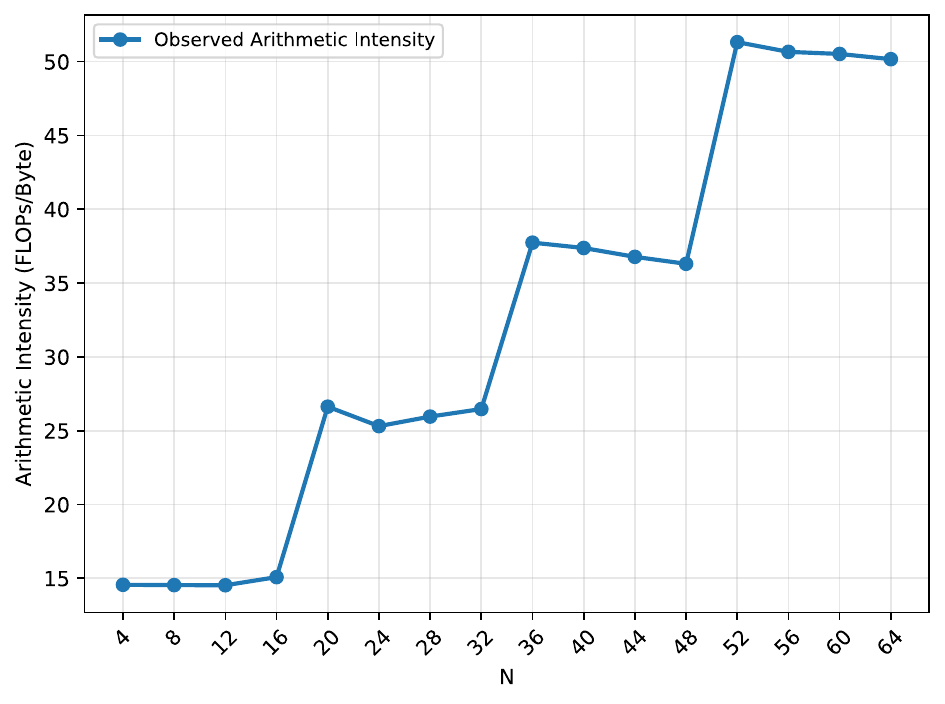}
            \caption{AI, $k=16$}
            \label{fig:moe-lower-sglang-a800-ai-k16}
        \end{subfigure}
        \hfill
        \begin{subfigure}[t]{0.24\textwidth}
            \centering
            \includegraphics[width=\linewidth]{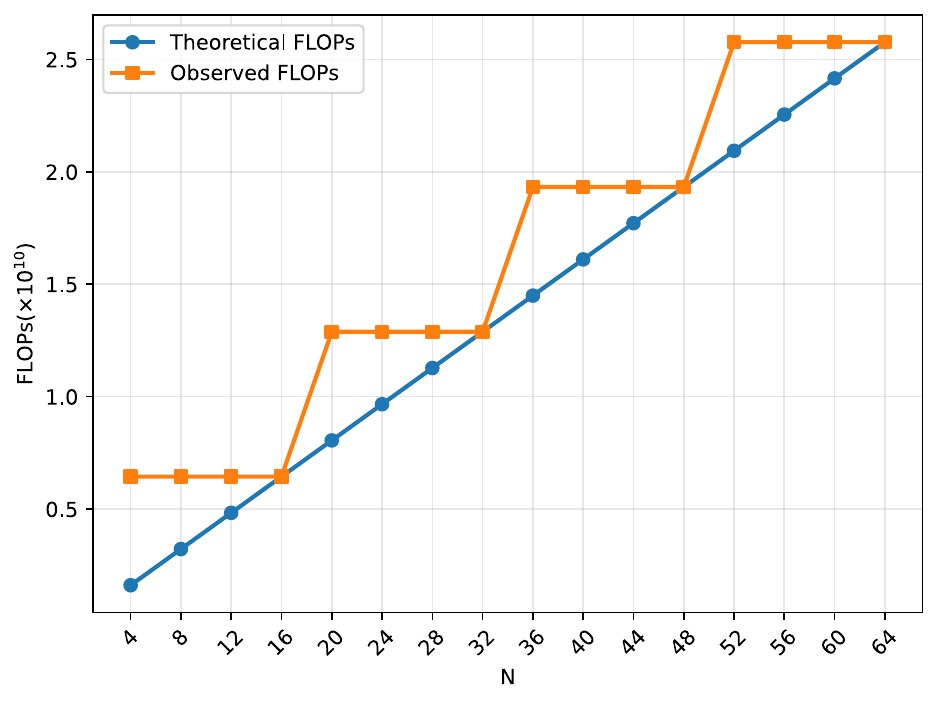}
            \caption{FLOPs, $k=16$}
            \label{fig:moe-lower-sglang-a800-flops-k16}
        \end{subfigure}
        \hfill
        \begin{subfigure}[t]{0.24\textwidth}
            \centering
            \includegraphics[width=\linewidth]{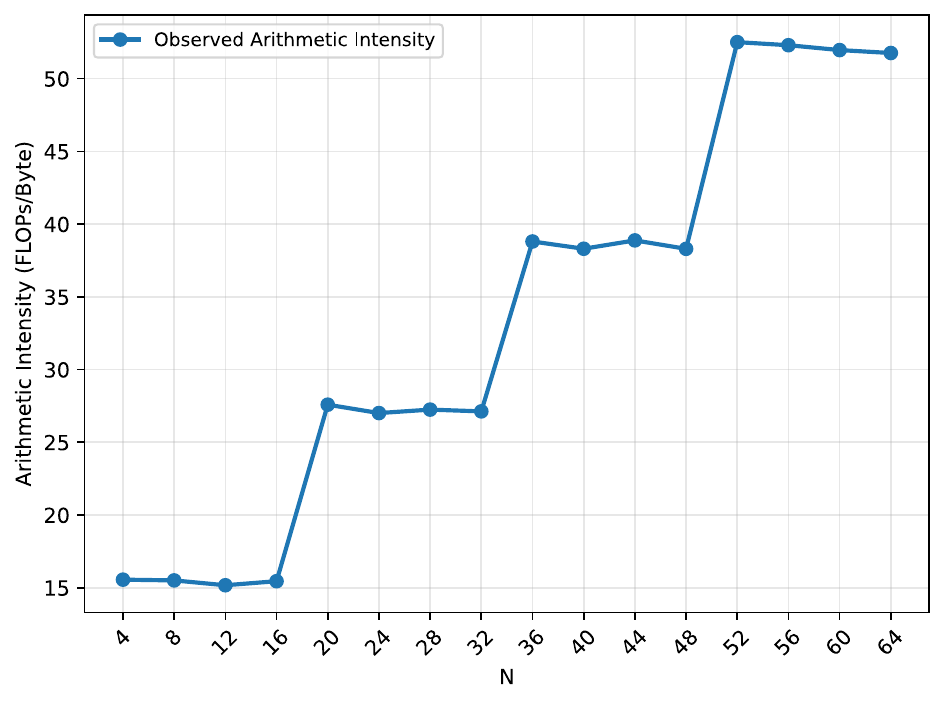}
            \caption{AI, $k=32$}
            \label{fig:moe-lower-sglang-a800-ai-k32}
        \end{subfigure}
        \hfill
        \begin{subfigure}[t]{0.24\textwidth}
            \centering
            \includegraphics[width=\linewidth]{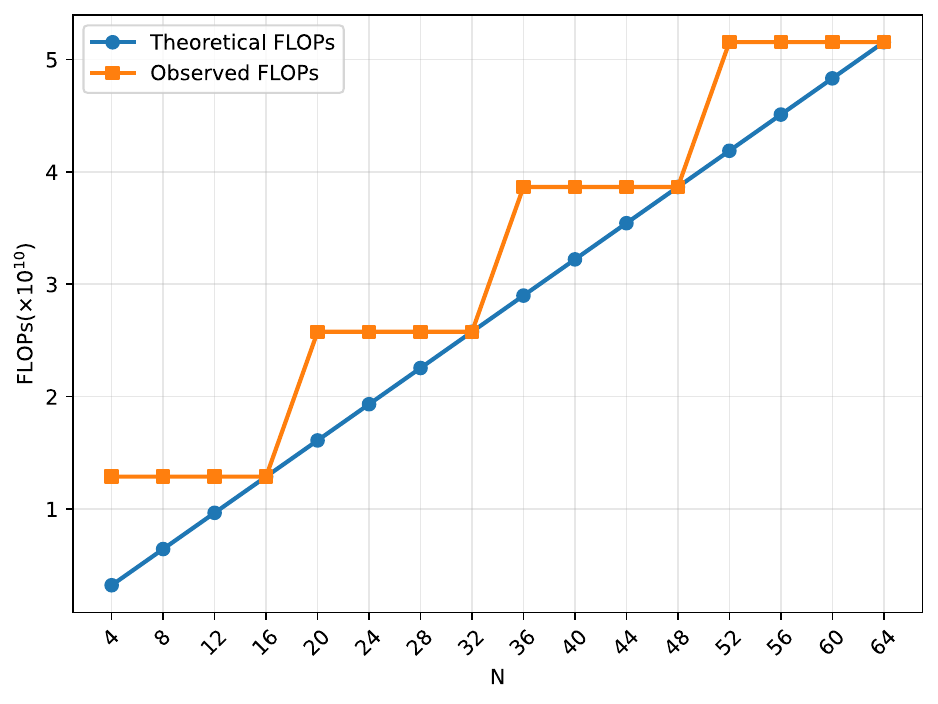}
            \caption{FLOPs, $k=32$}
            \label{fig:moe-lower-sglang-a800-flops-k32}
        \end{subfigure}
    \end{minipage}%
    }

    \caption{
    MoE FFN evaluation for load-skewed routing, the lower-bound case, with \textbf{SGLang} on \textbf{NVIDIA A800}.
    }
    \label{fig:moe-lower-sglang-a800}
\end{figure*}

\begin{figure*}[!p]
    \centering
    \scalebox{1}[1.0]{%
    \begin{minipage}{\textwidth}
        \centering

        \begin{subfigure}[t]{0.24\textwidth}
            \centering
            \includegraphics[width=\linewidth]{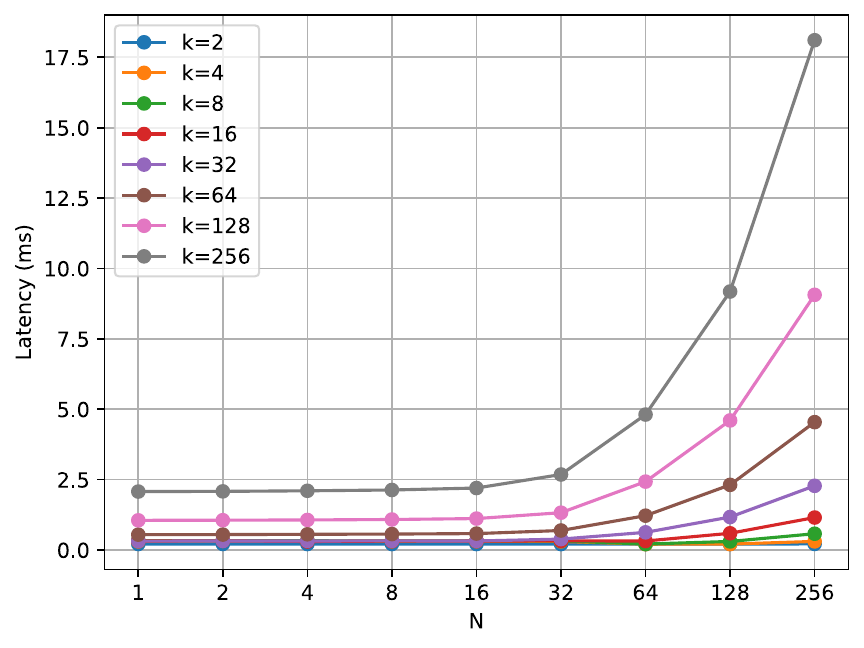}
            \caption{Latency vs. $N$}
            \label{fig:moe-lower-sglang-h800-latency}
        \end{subfigure}
        \hfill
        \begin{subfigure}[t]{0.24\textwidth}
            \centering
            \includegraphics[width=\linewidth]{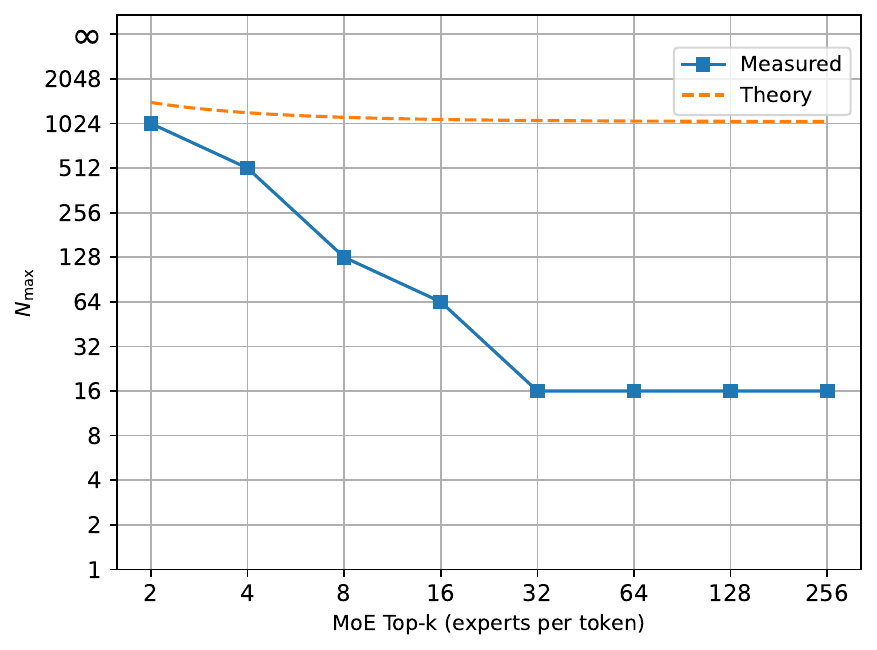}
            \caption{$N_{\max}$ vs. $k$}
            \label{fig:moe-lower-sglang-h800-nmax}
        \end{subfigure}
        \hfill
        \begin{subfigure}[t]{0.24\textwidth}
            \centering
            \includegraphics[width=\linewidth]{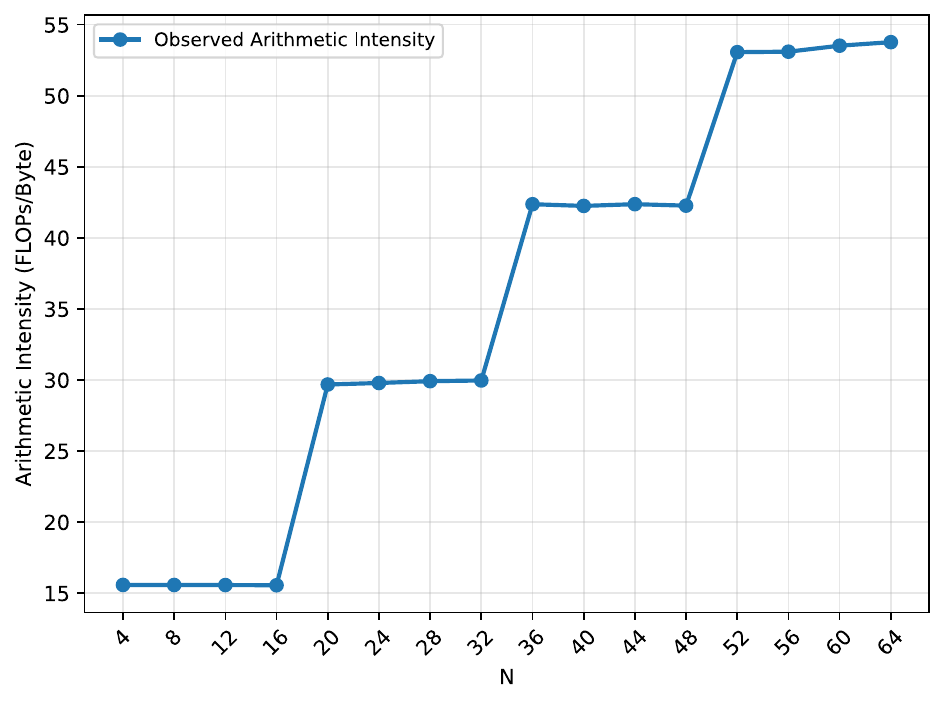}
            \caption{AI, $k=8$}
            \label{fig:moe-lower-sglang-h800-ai-k8}
        \end{subfigure}
        \hfill
        \begin{subfigure}[t]{0.24\textwidth}
            \centering
            \includegraphics[width=\linewidth]{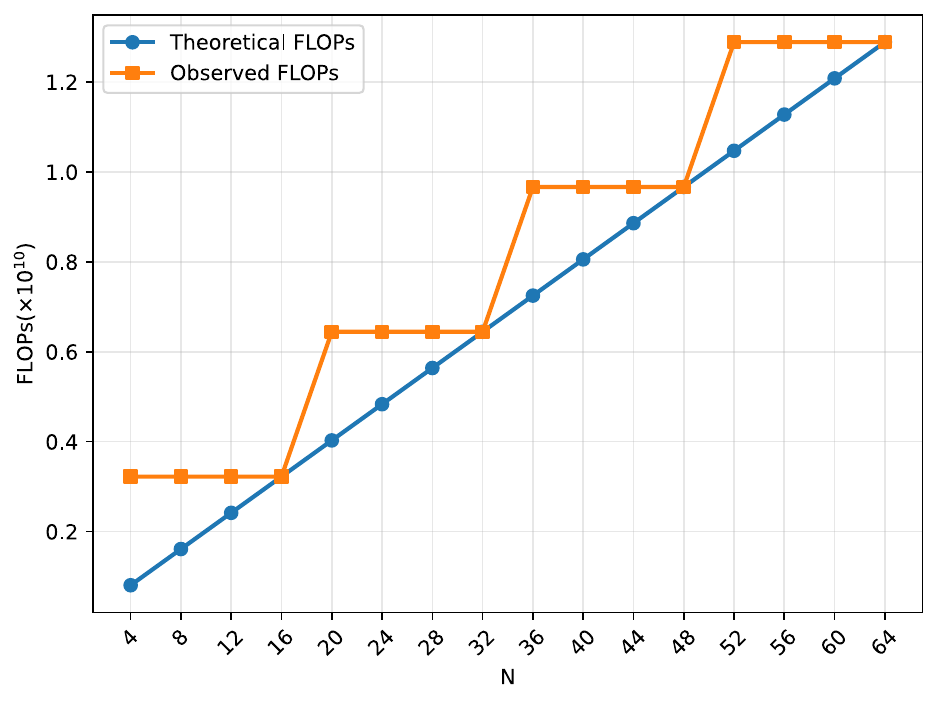}
            \caption{FLOPs, $k=8$}
            \label{fig:moe-lower-sglang-h800-flops-k8}
        \end{subfigure}

        \vspace{0.8em}

        \begin{subfigure}[t]{0.24\textwidth}
            \centering
            \includegraphics[width=\linewidth]{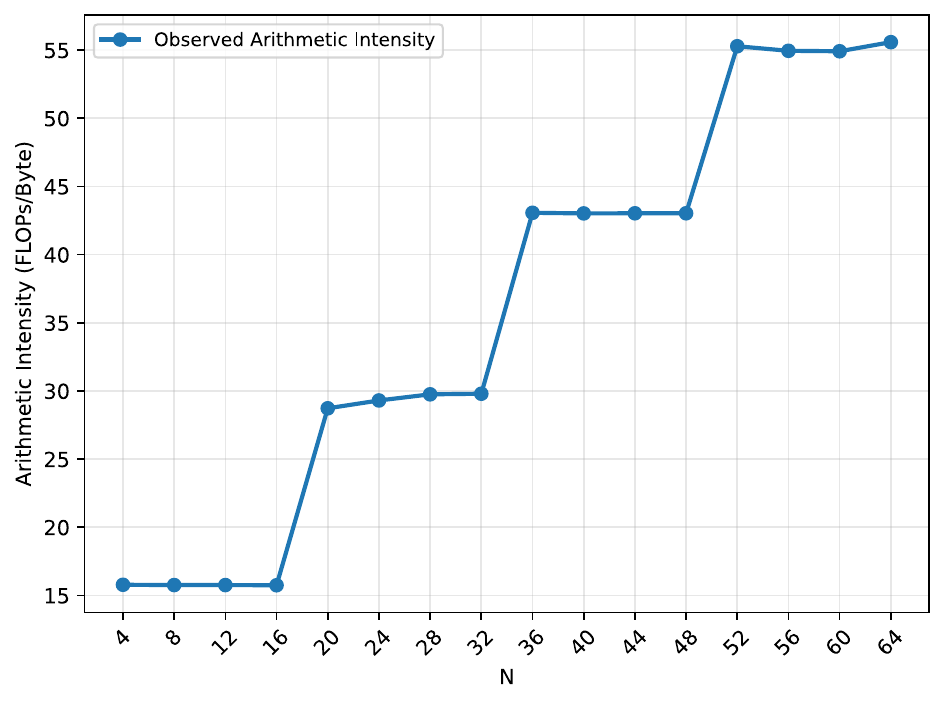}
            \caption{AI, $k=16$}
            \label{fig:moe-lower-sglang-h800-ai-k16}
        \end{subfigure}
        \hfill
        \begin{subfigure}[t]{0.24\textwidth}
            \centering
            \includegraphics[width=\linewidth]{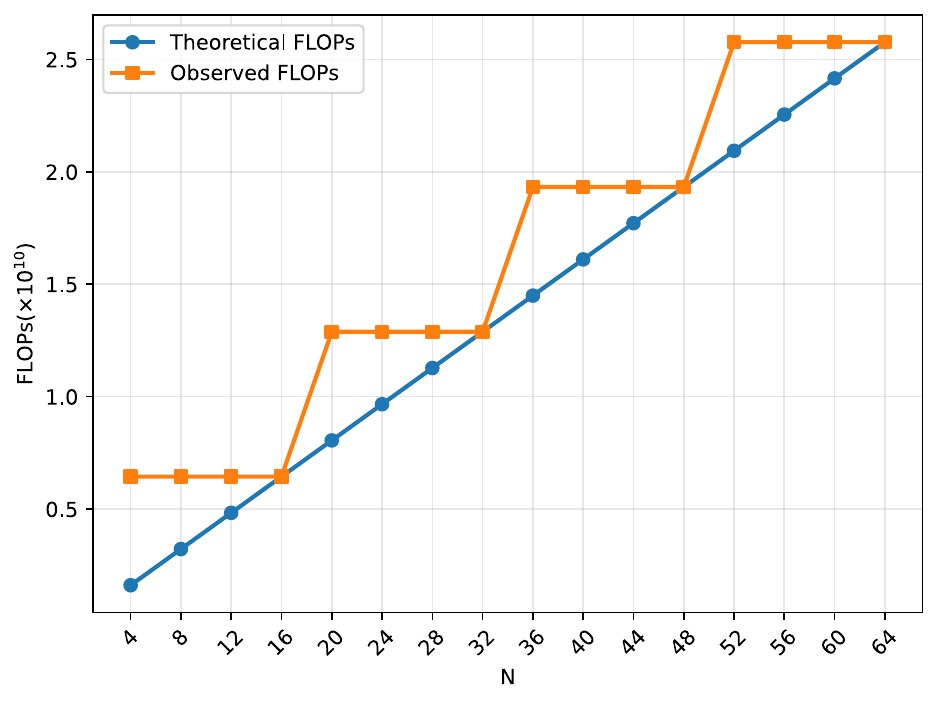}
            \caption{FLOPs, $k=16$}
            \label{fig:moe-lower-sglang-h800-flops-k16}
        \end{subfigure}
        \hfill
        \begin{subfigure}[t]{0.24\textwidth}
            \centering
            \includegraphics[width=\linewidth]{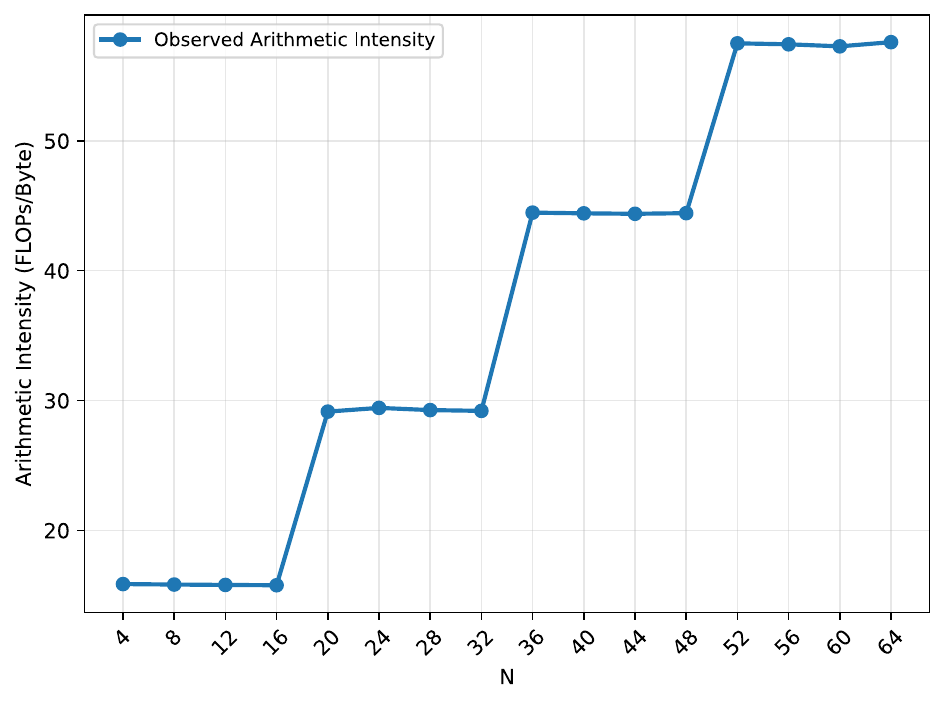}
            \caption{AI, $k=32$}
            \label{fig:moe-lower-sglang-h800-ai-k32}
        \end{subfigure}
        \hfill
        \begin{subfigure}[t]{0.24\textwidth}
            \centering
            \includegraphics[width=\linewidth]{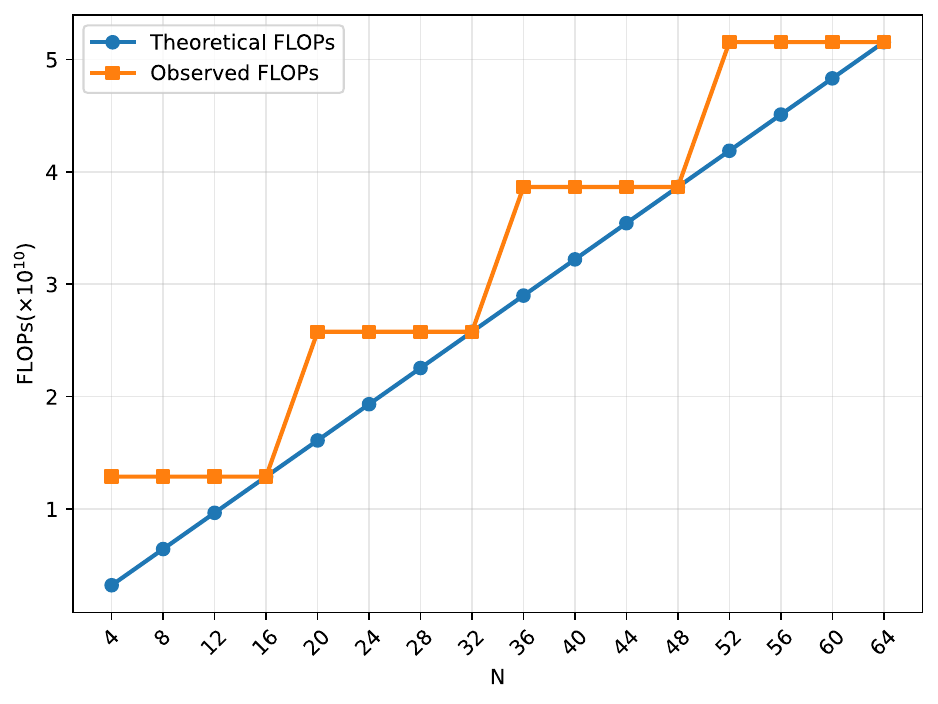}
            \caption{FLOPs, $k=32$}
            \label{fig:moe-lower-sglang-h800-flops-k32}
        \end{subfigure}
    \end{minipage}%
    }

    \caption{
    MoE FFN evaluation for load-skewed routing, the lower-bound case, with \textbf{SGLang} on \textbf{NVIDIA H800}.
    }
    \label{fig:moe-lower-sglang-h800}
\end{figure*}

Figures~\ref{fig:moe-lower-sglang-h20}, \ref{fig:moe-lower-sglang-a800}, and \ref{fig:moe-lower-sglang-h800} report the corresponding load-skewed results using the SGLang fused MoE backend. The same lower-bound behavior appears under SGLang: latency remains nearly flat within a granularity block and increases after the block is exceeded, while the extracted boundary remains primarily determined by the single expert-token padding granularity. This cross-backend consistency confirms that load-skewed routing exposes the granularity-fixed lower bound of MoE FFN NFP.

Taken together, the MoE FFN results support the mechanism identified in the main text. Across routing patterns, fused MoE backends, and evaluated single-GPU platforms, the MoE NFP boundary is governed primarily by implementation-induced kernel-granularity slack. Load-balanced routing exposes an aggregate upper-bound boundary that grows with sparsity, whereas load-skewed routing exposes a backend-granularity-fixed lower-bound boundary.

\clearpage

\clearpage

\clearpage

\subsection{Attention Layers Results}
\label{app:attention-results}

This section reports the complete Attention module-level results. These experiments complement the main-text analysis by validating the backend-specific query-granularity mechanism across attention backends and single-GPU platforms. We evaluate FlashAttention and FlashInfer on NVIDIA H20, A800, and H800. For each backend and GPU platform, we sweep the number of decode positions $N$ under multiple cached sequence lengths $L$ and extract $N_{\max}$ using the same 20\% latency tolerance. In each figure, the latency subfigure shows how latency changes as $N$ increases, the $N_{\max}$ subfigure compares the extracted boundary with the idle-compute prediction across sequence lengths, and the remaining subfigures report the arithmetic-intensity and runtime-FLOPs profiles for representative values of $L$.

\subsubsection{FlashAttention Backend}

\begin{figure*}[htbp]
    \centering

    \begin{subfigure}[t]{0.24\textwidth}
        \centering
        \includegraphics[width=\linewidth]{figs/Attention_FlashAtten/H20/fig1_latency_vs_N_multi_seq_len.pdf}
        \caption{Latency vs. $N$}
        \label{fig:attention-flashatten-h20-latency}
    \end{subfigure}
    \hfill
    \begin{subfigure}[t]{0.24\textwidth}
        \centering
        \includegraphics[width=\linewidth]{figs/Attention_FlashAtten/H20/fig2_nmax_vs_seq_len_with_theory.pdf}
        \caption{$N_{\max}$ vs. $L$}
        \label{fig:attention-flashatten-h20-nmax}
    \end{subfigure}
    \hfill
    \begin{subfigure}[t]{0.24\textwidth}
        \centering
        \includegraphics[width=\linewidth]{figs/Attention_FlashAtten/H20/fig3_arithmetic_intensity_vs_N_seq_len_1024_roofline.pdf}
        \caption{AI, $L=1024$}
        \label{fig:attention-flashatten-h20-ai-l1024}
    \end{subfigure}
    \hfill
    \begin{subfigure}[t]{0.24\textwidth}
        \centering
        \includegraphics[width=\linewidth]{figs/Attention_FlashAtten/H20/fig4_theoretical_vs_observed_flops_vs_N_seq_len_1024_roofline.pdf}
        \caption{FLOPs, $L=1024$}
        \label{fig:attention-flashatten-h20-flops-l1024}
    \end{subfigure}

    \vspace{0.6em}

    \begin{subfigure}[t]{0.24\textwidth}
        \centering
        \includegraphics[width=\linewidth]{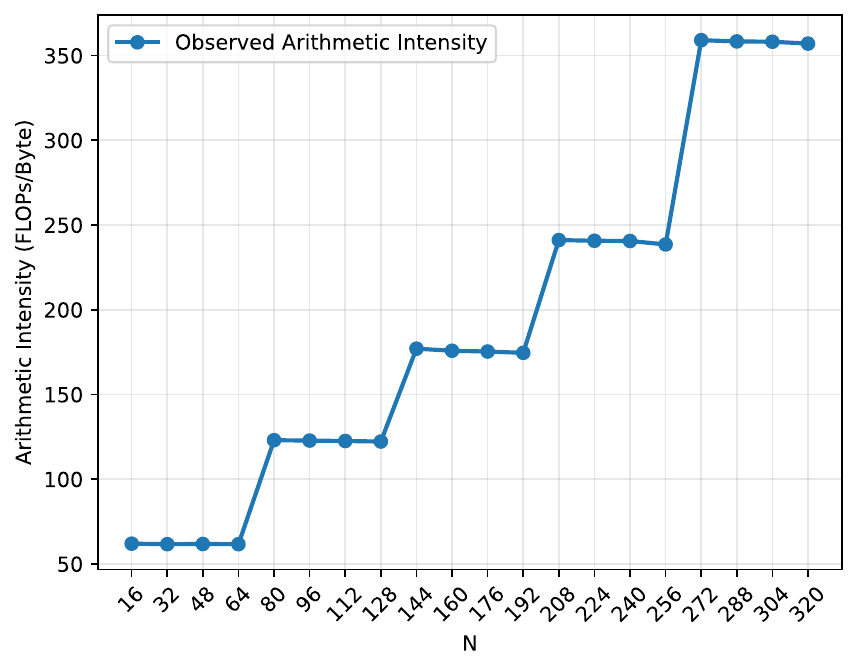}
        \caption{AI, $L=4096$}
        \label{fig:attention-flashatten-h20-ai-l4096}
    \end{subfigure}
    \hfill
    \begin{subfigure}[t]{0.24\textwidth}
        \centering
        \includegraphics[width=\linewidth]{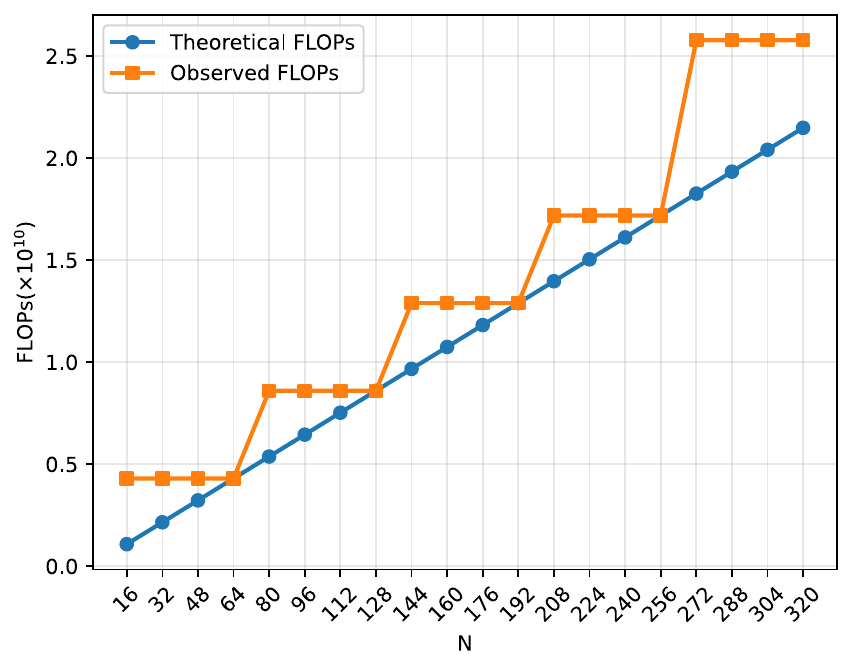}
        \caption{FLOPs, $L=4096$}
        \label{fig:attention-flashatten-h20-flops-l4096}
    \end{subfigure}
    \hfill
    \begin{subfigure}[t]{0.24\textwidth}
        \centering
        \includegraphics[width=\linewidth]{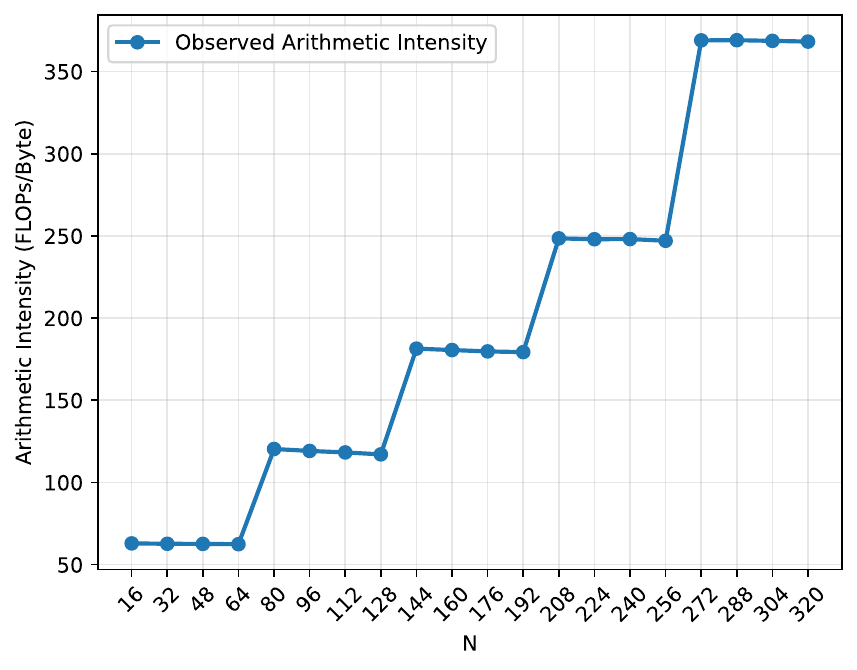}
        \caption{AI, $L=8192$}
        \label{fig:attention-flashatten-h20-ai-l8192}
    \end{subfigure}
    \hfill
    \begin{subfigure}[t]{0.24\textwidth}
        \centering
        \includegraphics[width=\linewidth]{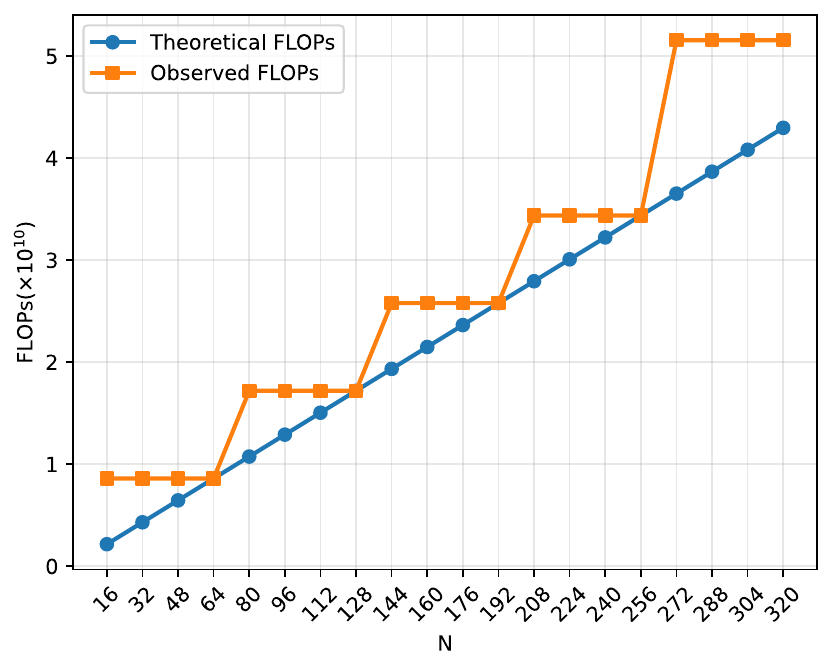}
        \caption{FLOPs, $L=8192$}
        \label{fig:attention-flashatten-h20-flops-l8192}
    \end{subfigure}

    \vspace{0.6em}

    \begin{subfigure}[t]{0.24\textwidth}
        \centering
        \includegraphics[width=\linewidth]{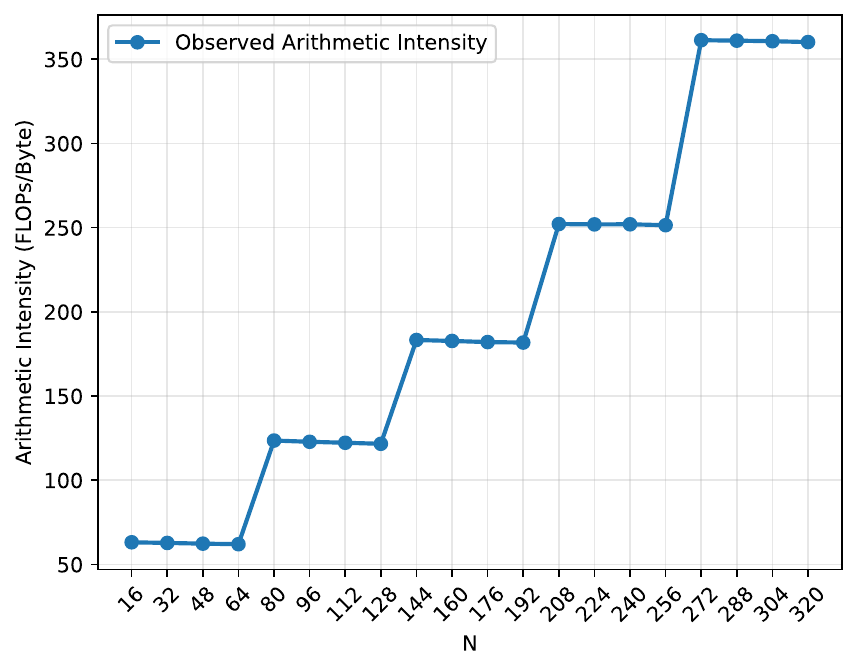}
        \caption{AI, $L=16384$}
        \label{fig:attention-flashatten-h20-ai-l16384}
    \end{subfigure}
    \hfill
    \begin{subfigure}[t]{0.24\textwidth}
        \centering
        \includegraphics[width=\linewidth]{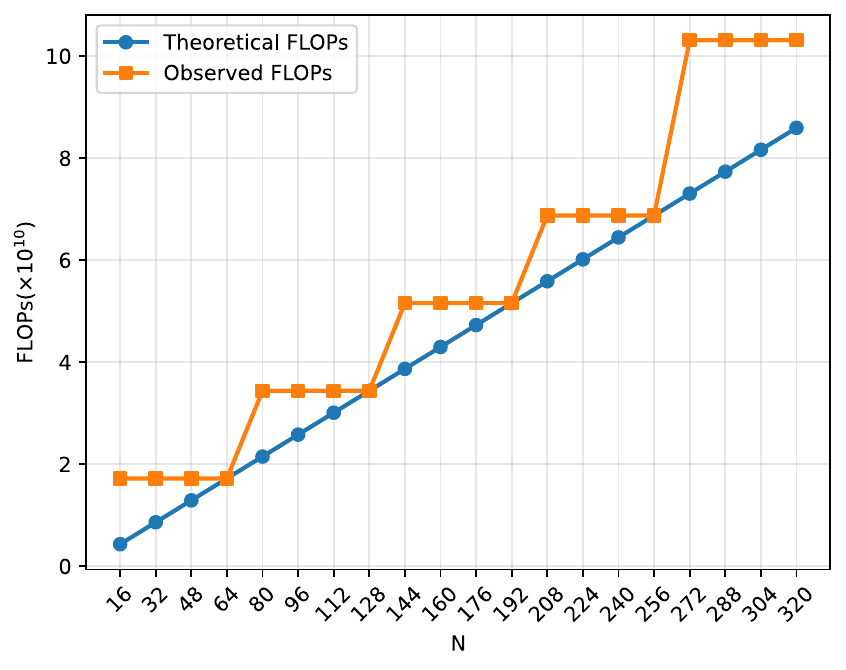}
        \caption{FLOPs, $L=16384$}
        \label{fig:attention-flashatten-h20-flops-l16384}
    \end{subfigure}
    \hfill
    \begin{subfigure}[t]{0.24\textwidth}
        \centering
        \includegraphics[width=\linewidth]{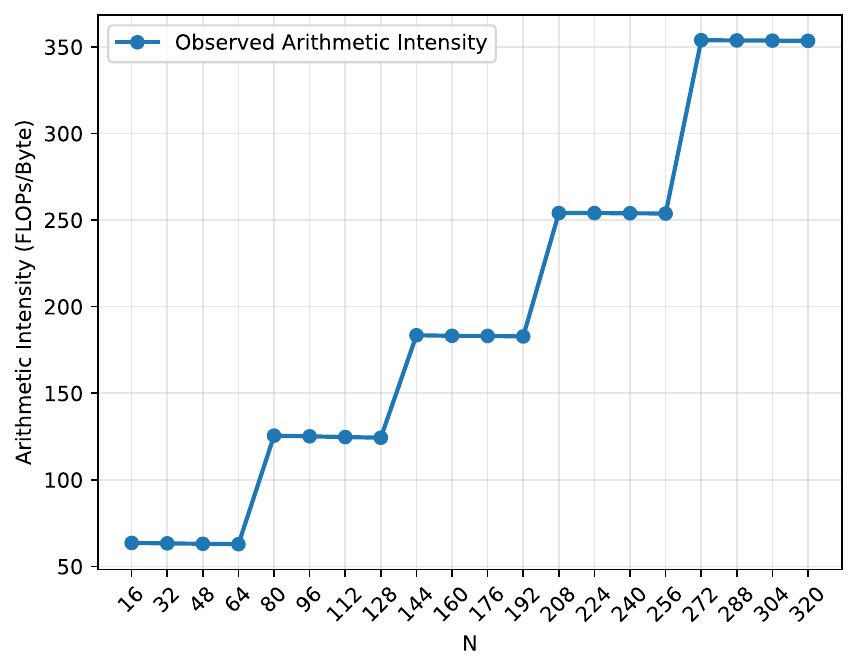}
        \caption{AI, $L=32768$}
        \label{fig:attention-flashatten-h20-ai-l32768}
    \end{subfigure}
    \hfill
    \begin{subfigure}[t]{0.24\textwidth}
        \centering
        \includegraphics[width=\linewidth]{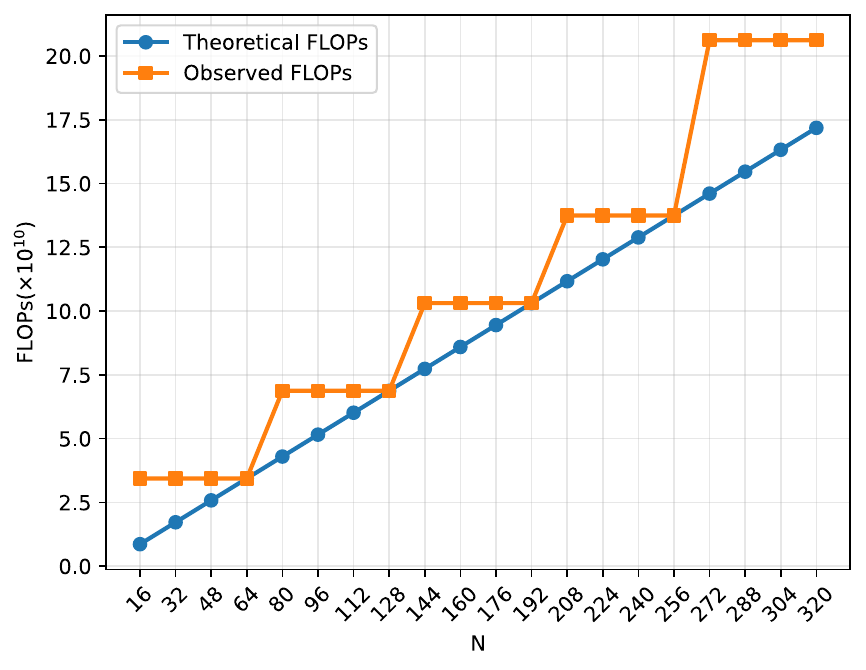}
        \caption{FLOPs, $L=32768$}
        \label{fig:attention-flashatten-h20-flops-l32768}
    \end{subfigure}

    \caption{
    Attention evaluation results with \textbf{FlashAttention} on \textbf{NVIDIA H20}.
    }
    \label{fig:attention-flashatten-h20}
\end{figure*}

\begin{figure*}[htbp]
    \centering

    \begin{subfigure}[t]{0.24\textwidth}
        \centering
        \includegraphics[width=\linewidth]{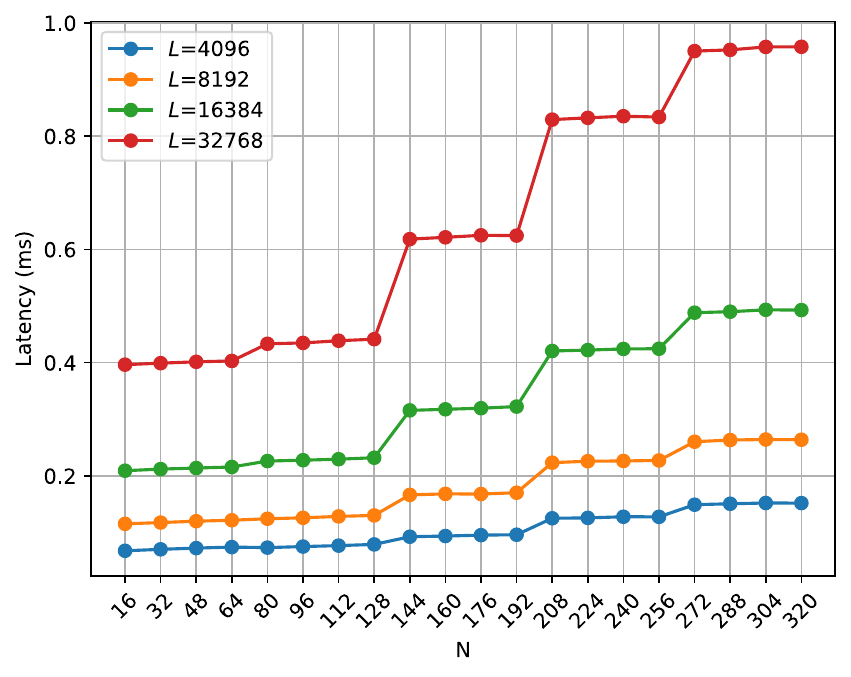}
        \caption{Latency vs. $N$}
        \label{fig:attention-flashatten-a800-latency}
    \end{subfigure}
    \hfill
    \begin{subfigure}[t]{0.24\textwidth}
        \centering
        \includegraphics[width=\linewidth]{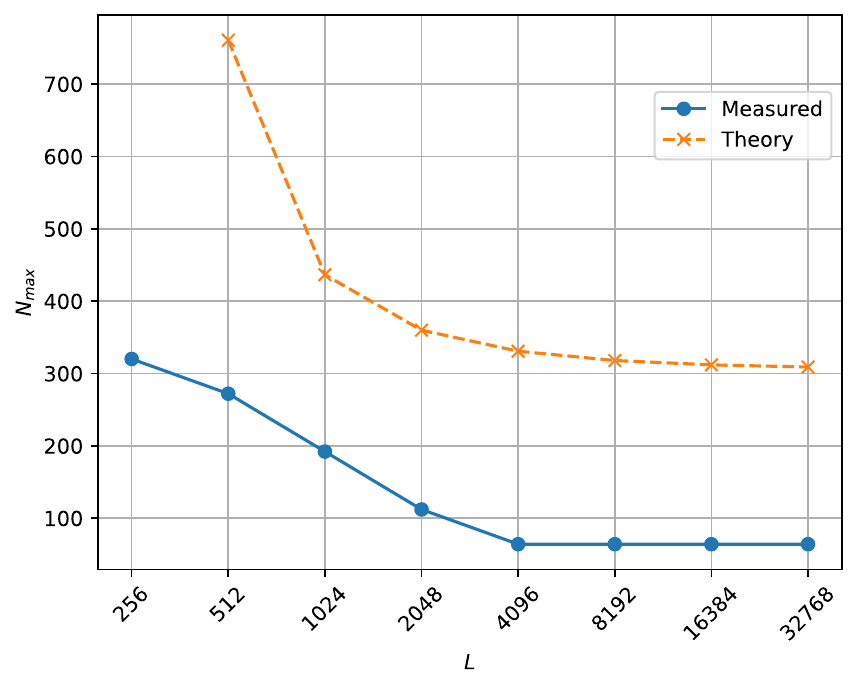}
        \caption{$N_{\max}$ vs. $L$}
        \label{fig:attention-flashatten-a800-nmax}
    \end{subfigure}
    \hfill
    \begin{subfigure}[t]{0.24\textwidth}
        \centering
        \includegraphics[width=\linewidth]{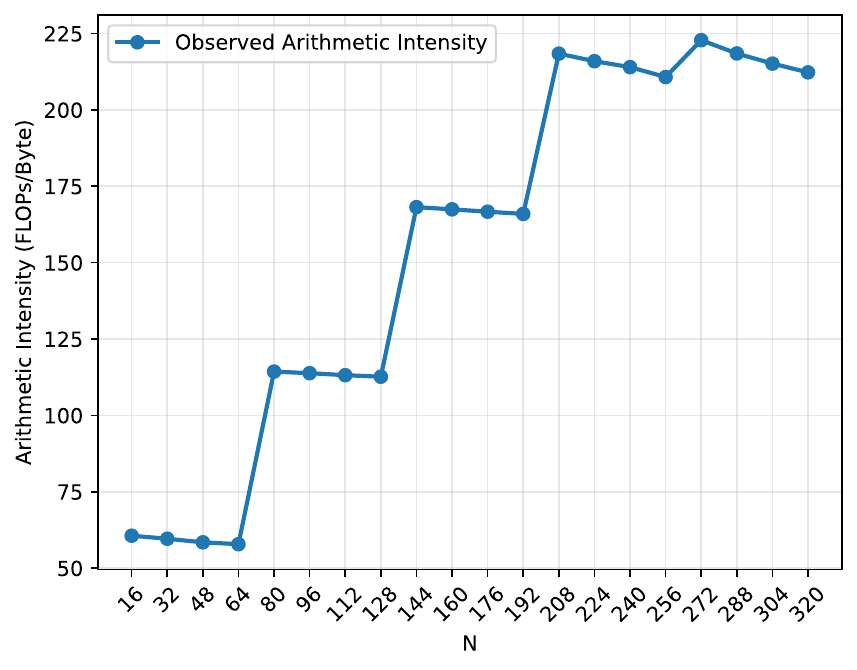}
        \caption{AI, $L=1024$}
        \label{fig:attention-flashatten-a800-ai-l1024}
    \end{subfigure}
    \hfill
    \begin{subfigure}[t]{0.24\textwidth}
        \centering
        \includegraphics[width=\linewidth]{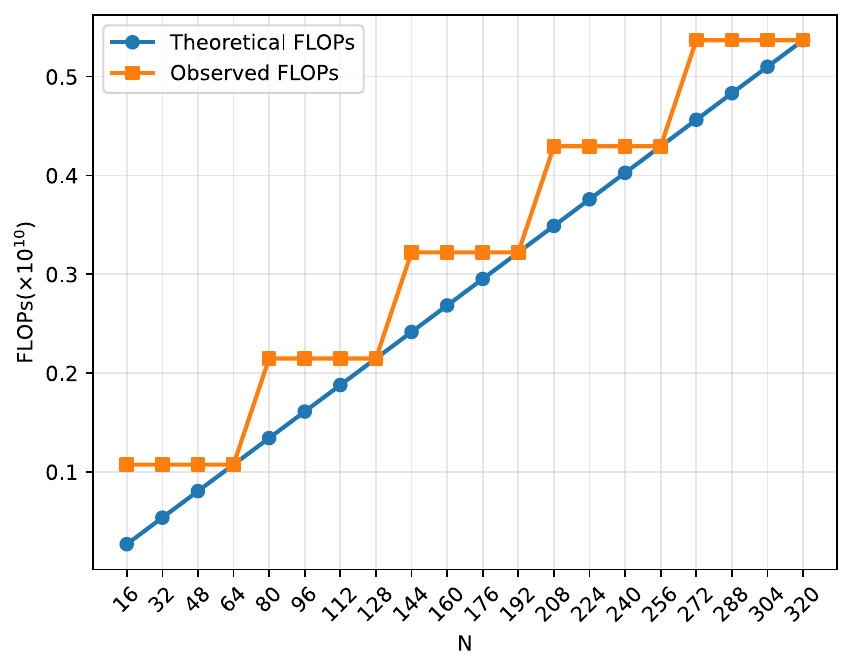}
        \caption{FLOPs, $L=1024$}
        \label{fig:attention-flashatten-a800-flops-l1024}
    \end{subfigure}

    \vspace{0.6em}

    \begin{subfigure}[t]{0.24\textwidth}
        \centering
        \includegraphics[width=\linewidth]{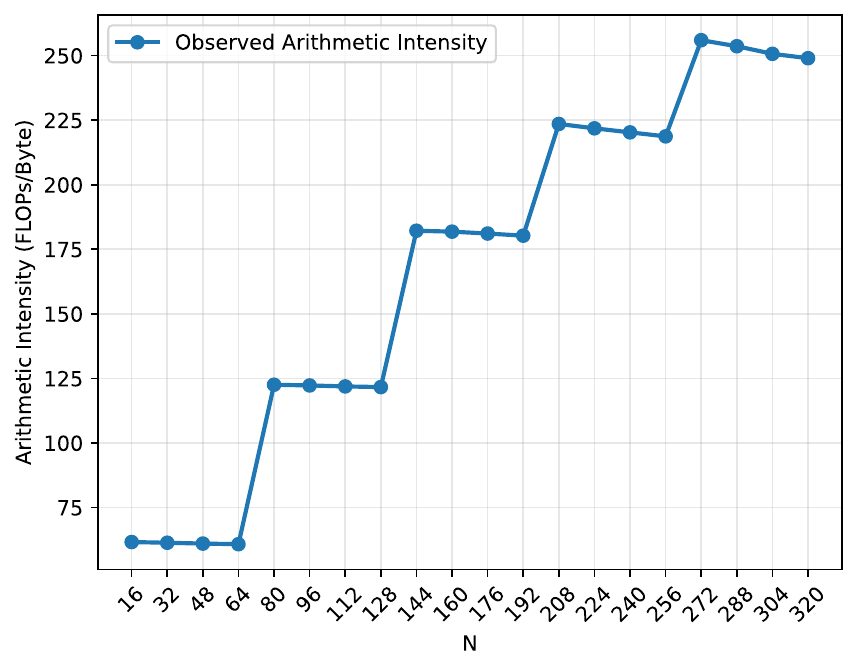}
        \caption{AI, $L=4096$}
        \label{fig:attention-flashatten-a800-ai-l4096}
    \end{subfigure}
    \hfill
    \begin{subfigure}[t]{0.24\textwidth}
        \centering
        \includegraphics[width=\linewidth]{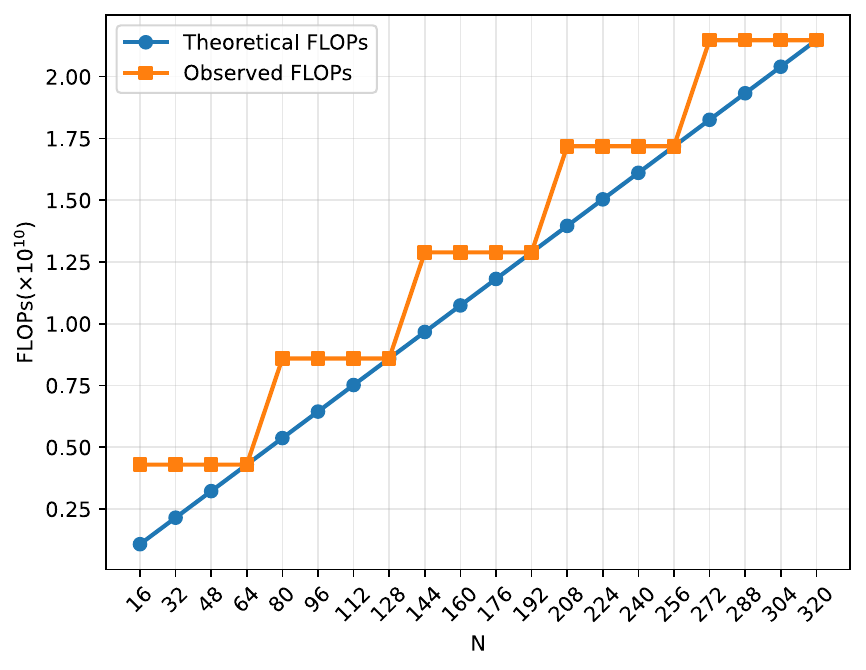}
        \caption{FLOPs, $L=4096$}
        \label{fig:attention-flashatten-a800-flops-l4096}
    \end{subfigure}
    \hfill
    \begin{subfigure}[t]{0.24\textwidth}
        \centering
        \includegraphics[width=\linewidth]{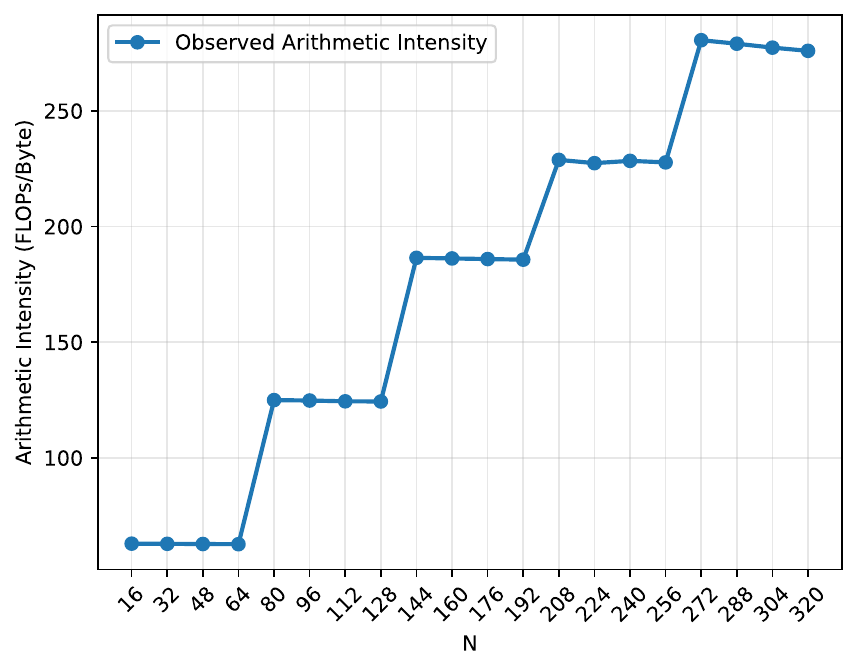}
        \caption{AI, $L=8192$}
        \label{fig:attention-flashatten-a800-ai-l8192}
    \end{subfigure}
    \hfill
    \begin{subfigure}[t]{0.24\textwidth}
        \centering
        \includegraphics[width=\linewidth]{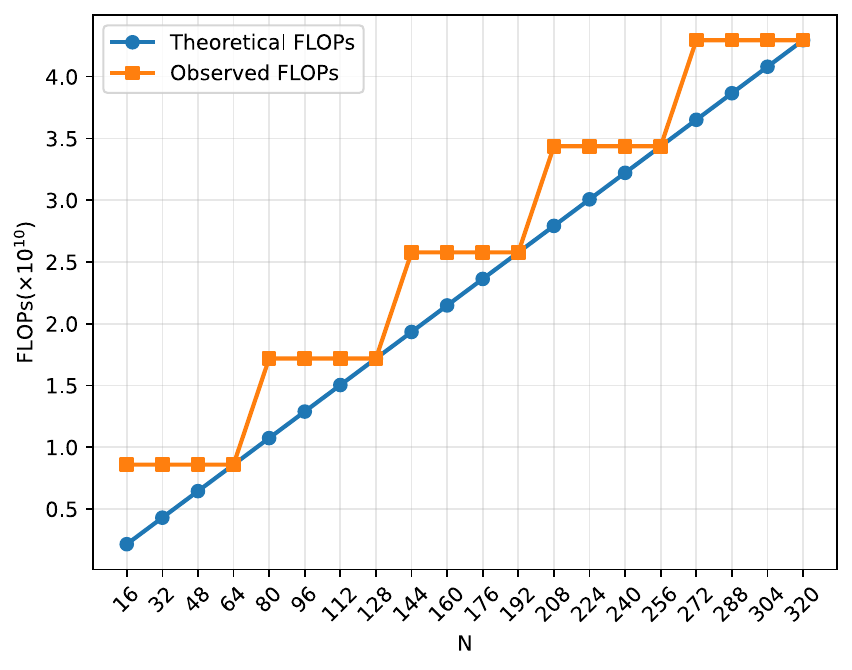}
        \caption{FLOPs, $L=8192$}
        \label{fig:attention-flashatten-a800-flops-l8192}
    \end{subfigure}

    \vspace{0.6em}

    \begin{subfigure}[t]{0.24\textwidth}
        \centering
        \includegraphics[width=\linewidth]{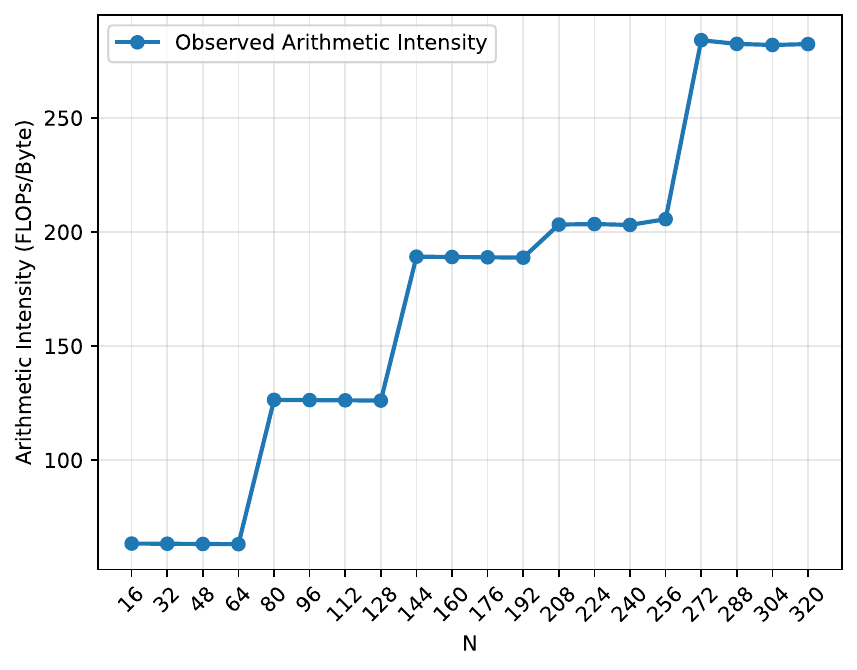}
        \caption{AI, $L=16384$}
        \label{fig:attention-flashatten-a800-ai-l16384}
    \end{subfigure}
    \hfill
    \begin{subfigure}[t]{0.24\textwidth}
        \centering
        \includegraphics[width=\linewidth]{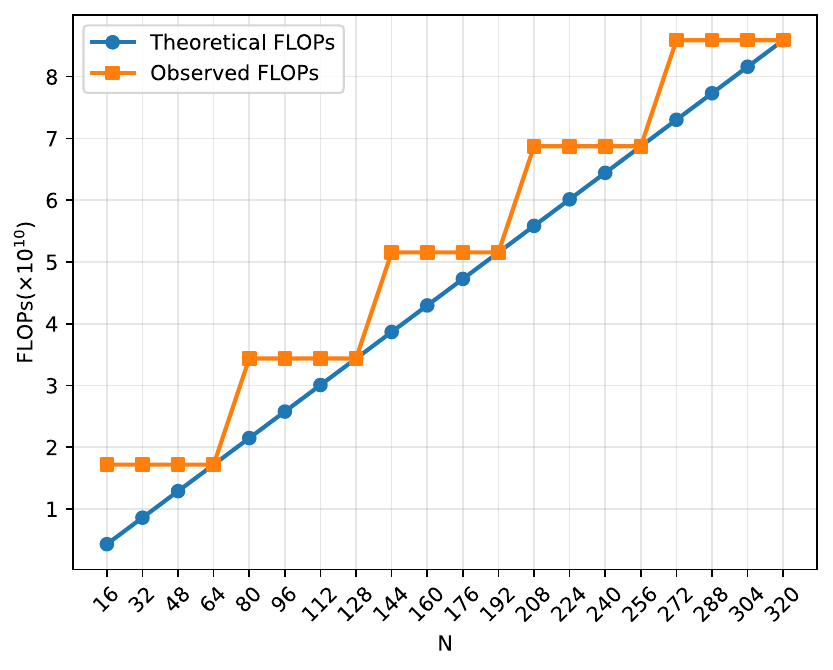}
        \caption{FLOPs, $L=16384$}
        \label{fig:attention-flashatten-a800-flops-l16384}
    \end{subfigure}
    \hfill
    \begin{subfigure}[t]{0.24\textwidth}
        \centering
        \includegraphics[width=\linewidth]{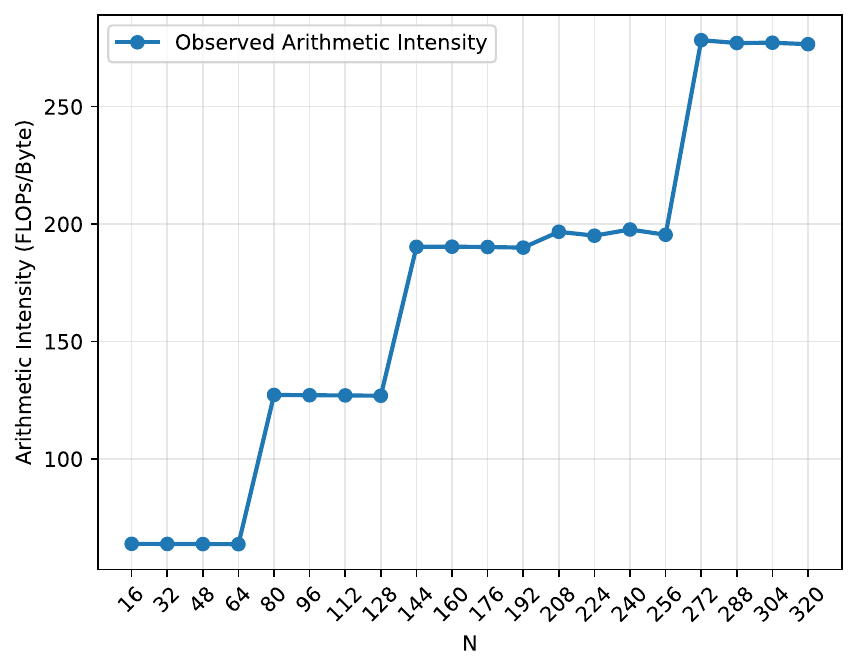}
        \caption{AI, $L=32768$}
        \label{fig:attention-flashatten-a800-ai-l32768}
    \end{subfigure}
    \hfill
    \begin{subfigure}[t]{0.24\textwidth}
        \centering
        \includegraphics[width=\linewidth]{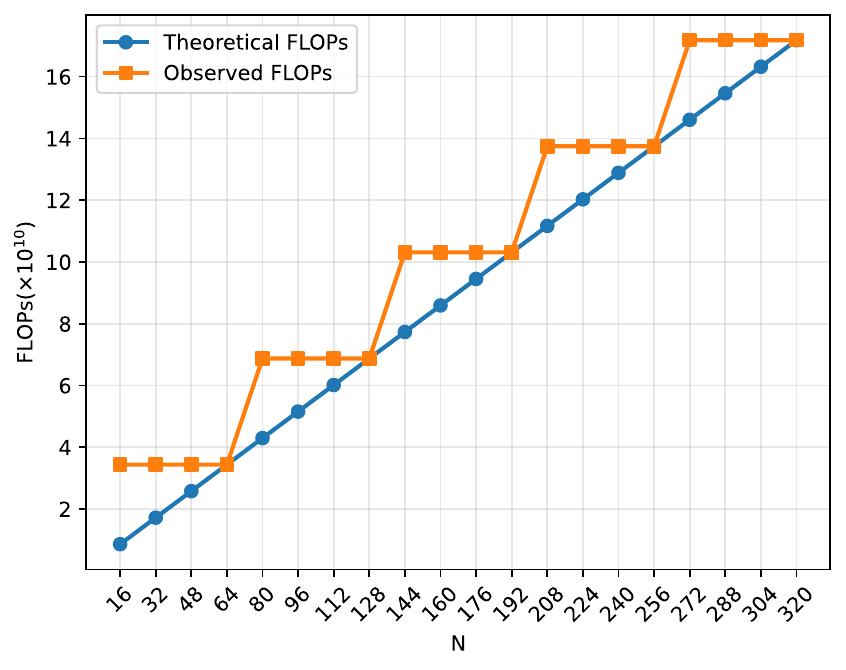}
        \caption{FLOPs, $L=32768$}
        \label{fig:attention-flashatten-a800-flops-l32768}
    \end{subfigure}

    \caption{
    Attention evaluation results with \textbf{FlashAttention} on \textbf{NVIDIA A800}.
    }
    \label{fig:attention-flashatten-a800}
\end{figure*}

\begin{figure*}[htbp]
    \centering

    \begin{subfigure}[t]{0.24\textwidth}
        \centering
        \includegraphics[width=\linewidth]{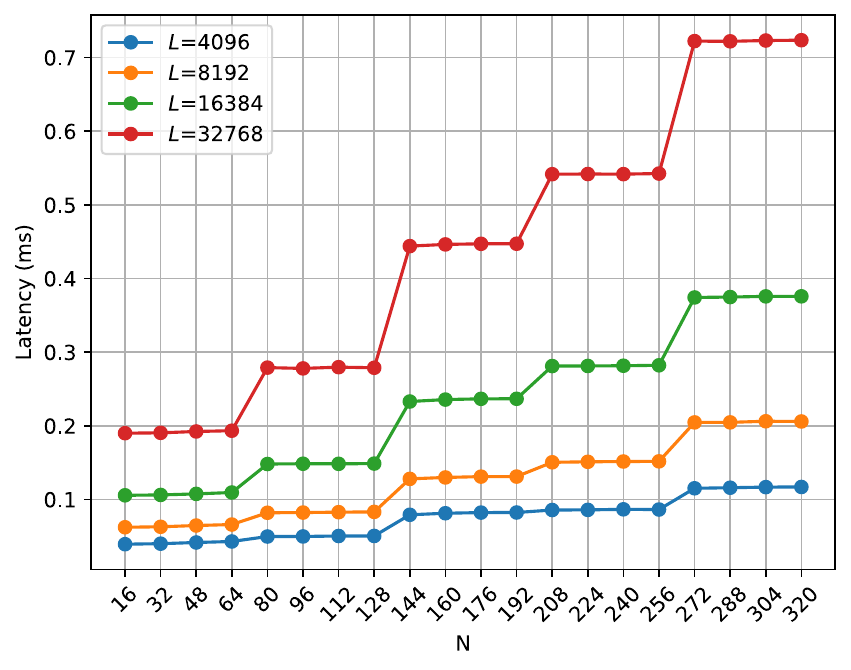}
        \caption{Latency vs. $N$}
        \label{fig:attention-flashatten-h800-latency}
    \end{subfigure}
    \hfill
    \begin{subfigure}[t]{0.24\textwidth}
        \centering
        \includegraphics[width=\linewidth]{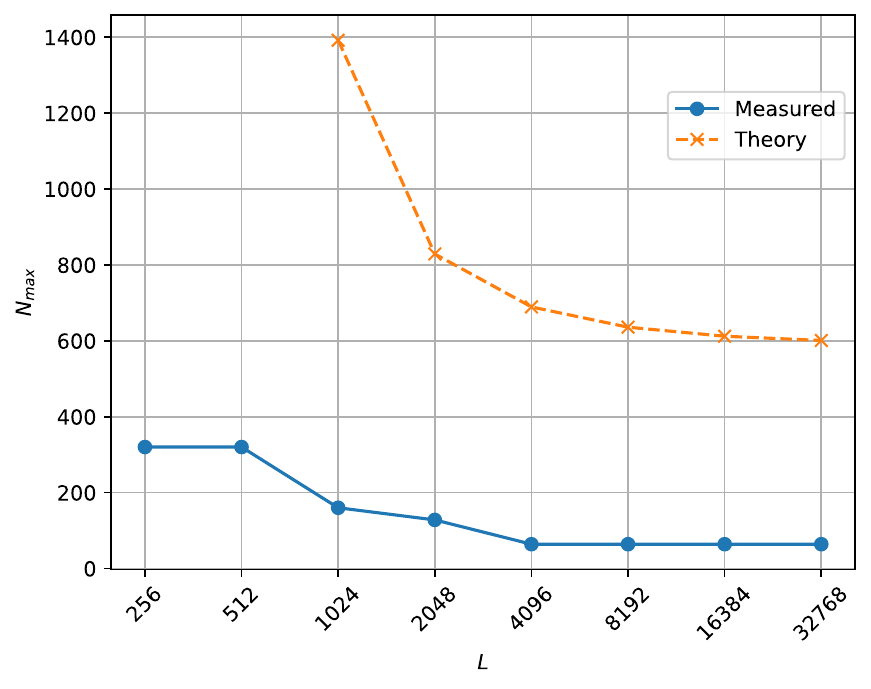}
        \caption{$N_{\max}$ vs. $L$}
        \label{fig:attention-flashatten-h800-nmax}
    \end{subfigure}
    \hfill
    \begin{subfigure}[t]{0.24\textwidth}
        \centering
        \includegraphics[width=\linewidth]{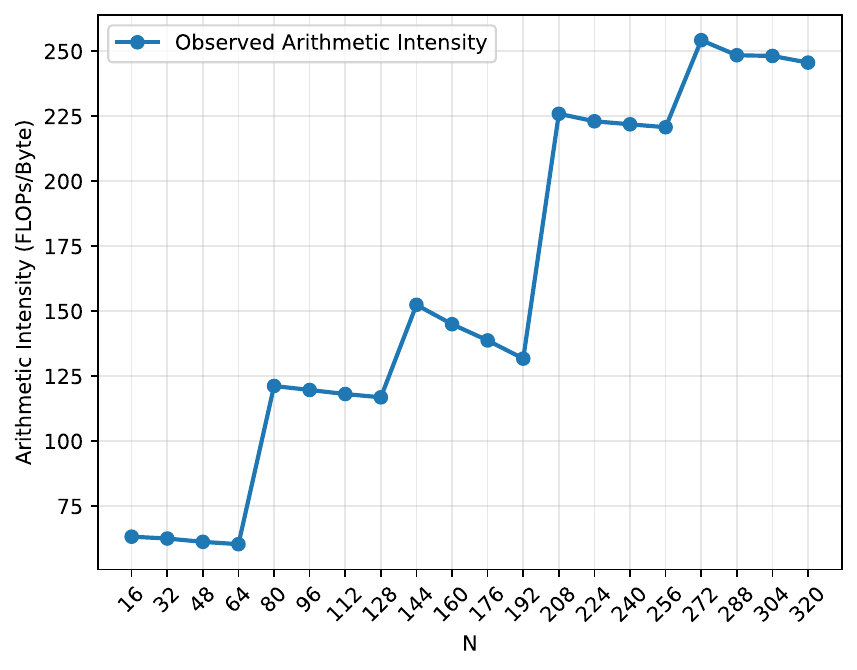}
        \caption{AI, $L=1024$}
        \label{fig:attention-flashatten-h800-ai-l1024}
    \end{subfigure}
    \hfill
    \begin{subfigure}[t]{0.24\textwidth}
        \centering
        \includegraphics[width=\linewidth]{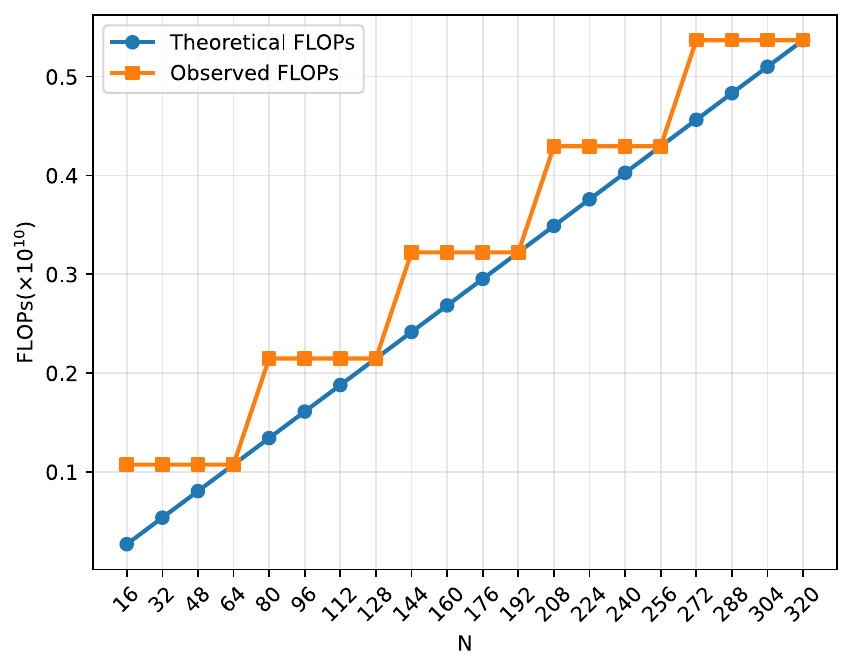}
        \caption{FLOPs, $L=1024$}
        \label{fig:attention-flashatten-h800-flops-l1024}
    \end{subfigure}

    \vspace{0.6em}

    \begin{subfigure}[t]{0.24\textwidth}
        \centering
        \includegraphics[width=\linewidth]{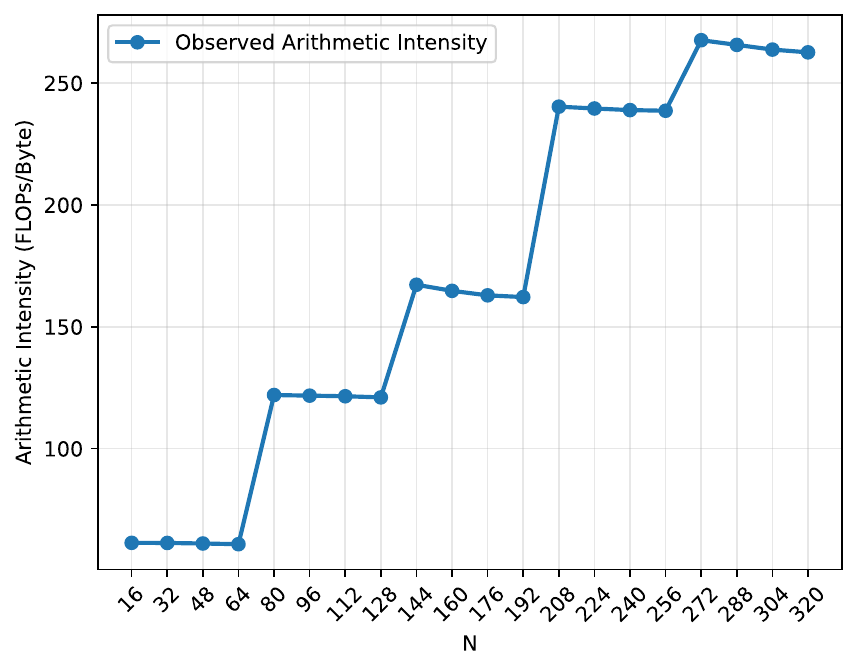}
        \caption{AI, $L=4096$}
        \label{fig:attention-flashatten-h800-ai-l4096}
    \end{subfigure}
    \hfill
    \begin{subfigure}[t]{0.24\textwidth}
        \centering
        \includegraphics[width=\linewidth]{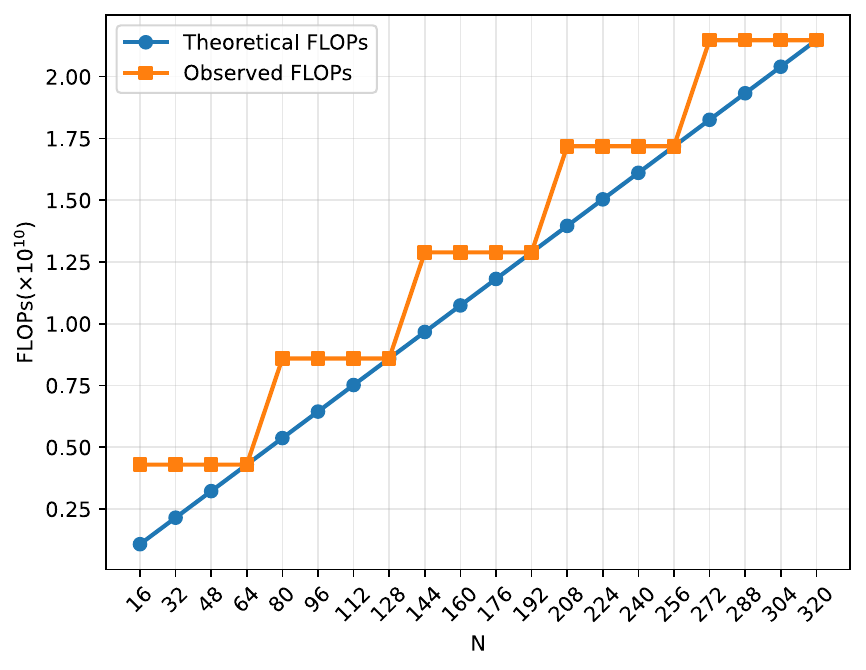}
        \caption{FLOPs, $L=4096$}
        \label{fig:attention-flashatten-h800-flops-l4096}
    \end{subfigure}
    \hfill
    \begin{subfigure}[t]{0.24\textwidth}
        \centering
        \includegraphics[width=\linewidth]{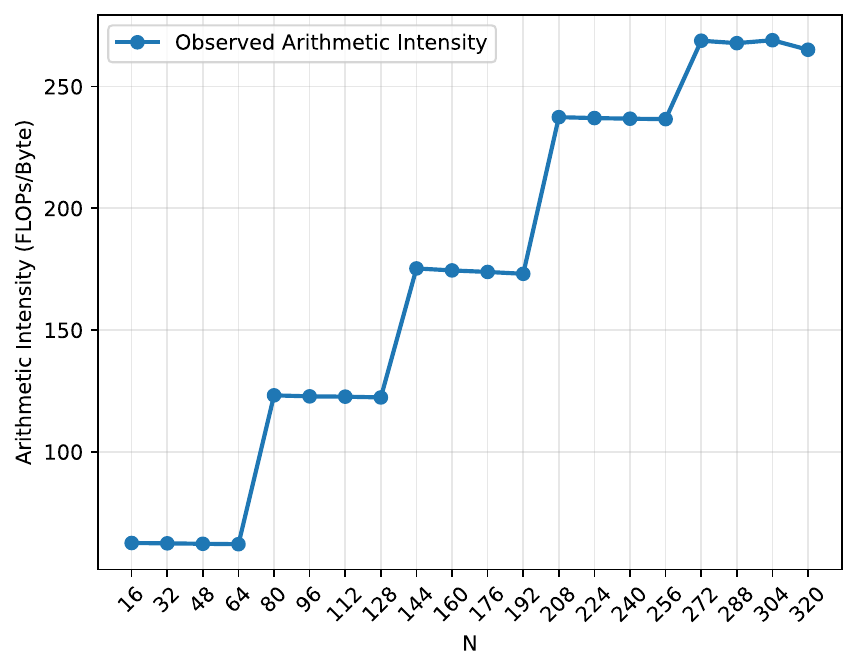}
        \caption{AI, $L=8192$}
        \label{fig:attention-flashatten-h800-ai-l8192}
    \end{subfigure}
    \hfill
    \begin{subfigure}[t]{0.24\textwidth}
        \centering
        \includegraphics[width=\linewidth]{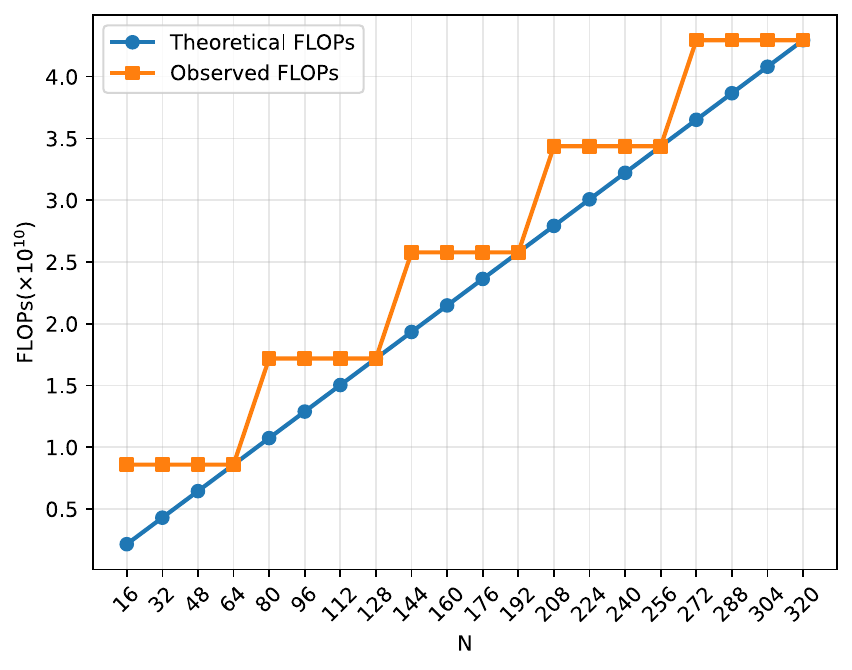}
        \caption{FLOPs, $L=8192$}
        \label{fig:attention-flashatten-h800-flops-l8192}
    \end{subfigure}

    \vspace{0.6em}

    \begin{subfigure}[t]{0.24\textwidth}
        \centering
        \includegraphics[width=\linewidth]{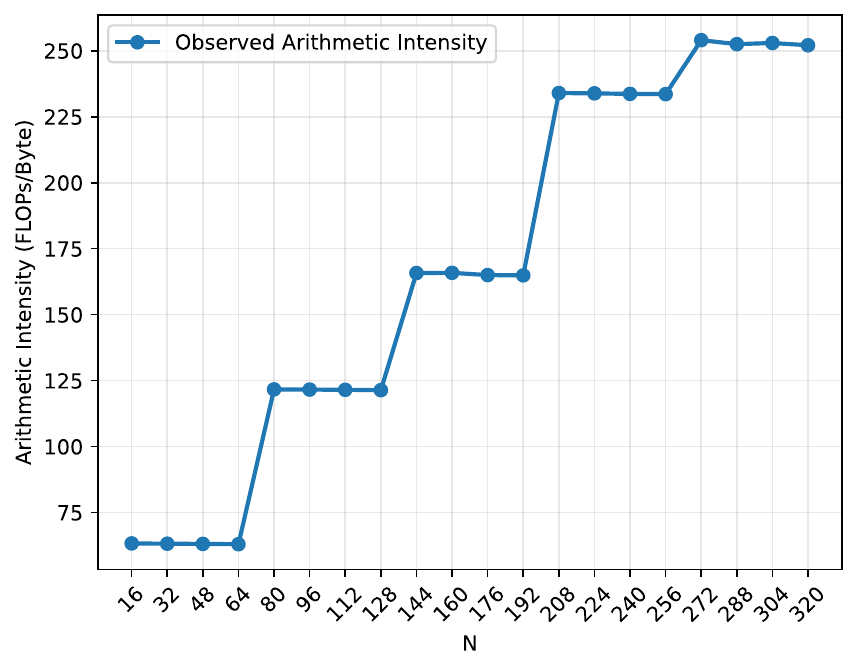}
        \caption{AI, $L=16384$}
        \label{fig:attention-flashatten-h800-ai-l16384}
    \end{subfigure}
    \hfill
    \begin{subfigure}[t]{0.24\textwidth}
        \centering
        \includegraphics[width=\linewidth]{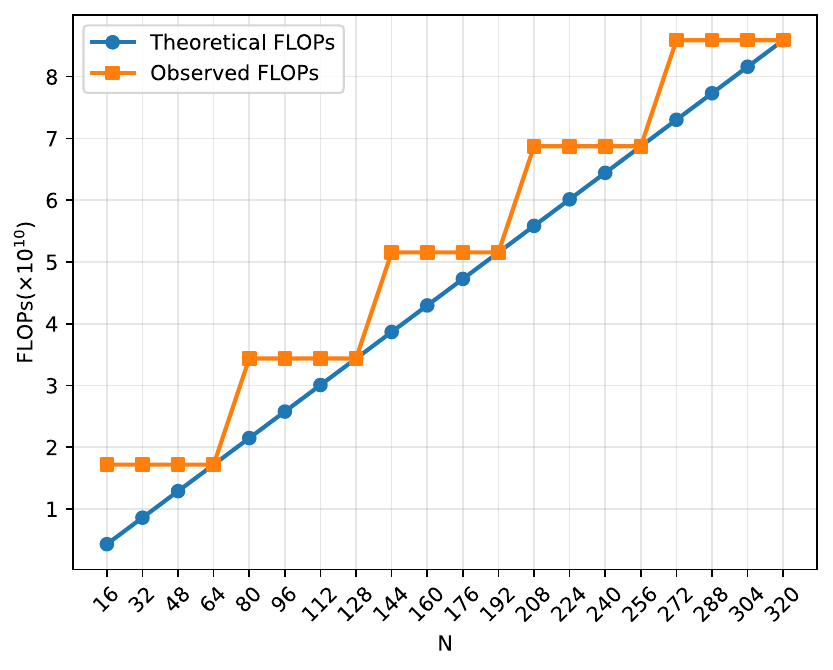}
        \caption{FLOPs, $L=16384$}
        \label{fig:attention-flashatten-h800-flops-l16384}
    \end{subfigure}
    \hfill
    \begin{subfigure}[t]{0.24\textwidth}
        \centering
        \includegraphics[width=\linewidth]{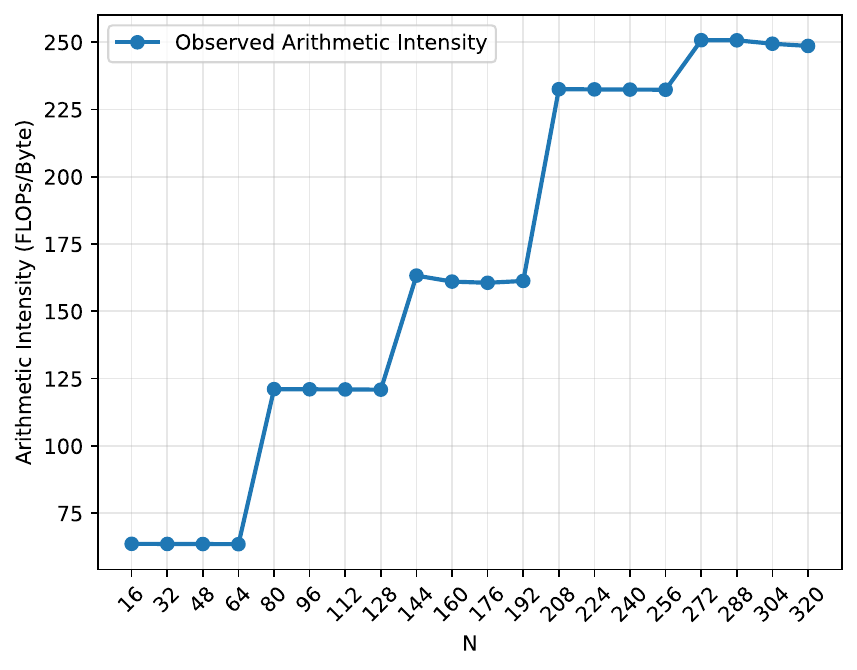}
        \caption{AI, $L=32768$}
        \label{fig:attention-flashatten-h800-ai-l32768}
    \end{subfigure}
    \hfill
    \begin{subfigure}[t]{0.24\textwidth}
        \centering
        \includegraphics[width=\linewidth]{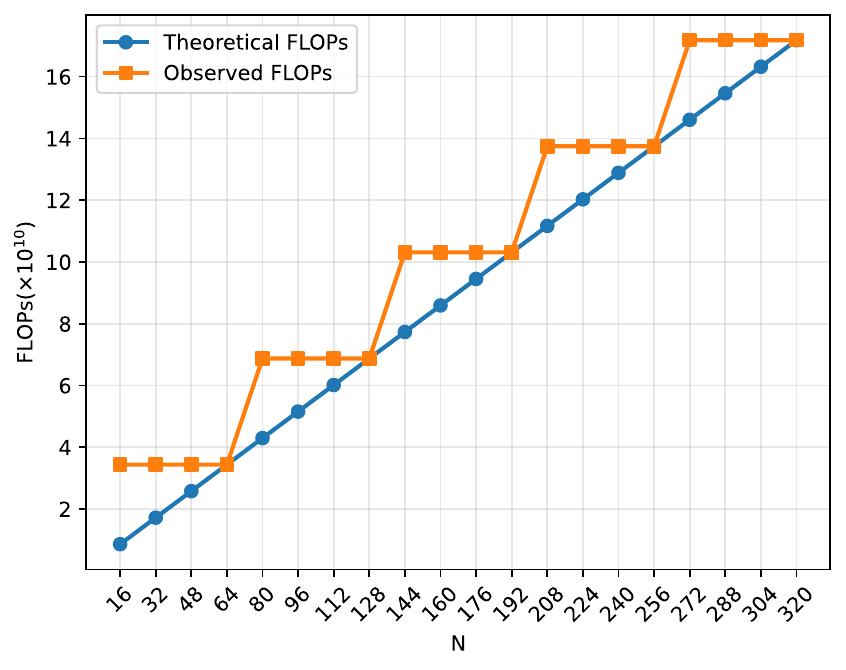}
        \caption{FLOPs, $L=32768$}
        \label{fig:attention-flashatten-h800-flops-l32768}
    \end{subfigure}

    \caption{
    Attention evaluation results with \textbf{FlashAttention} on \textbf{NVIDIA H800}.
    }
    \label{fig:attention-flashatten-h800}
\end{figure*}

Figures~\ref{fig:attention-flashatten-h20}, \ref{fig:attention-flashatten-a800}, and \ref{fig:attention-flashatten-h800} report the Attention results using the FlashAttention backend. Across all evaluated GPU platforms, the latency curves exhibit clear staircase behavior: latency remains nearly flat over several consecutive values of $N$ and increases only when $N$ crosses a backend granularity boundary. This confirms the existence of module-level NFP in Attention.

The extracted boundaries do not follow the idle-compute prediction. As shown in Figures~\ref{fig:attention-flashatten-h20}\subref{fig:attention-flashatten-h20-nmax}, \ref{fig:attention-flashatten-a800}\subref{fig:attention-flashatten-a800-nmax}, and \ref{fig:attention-flashatten-h800}\subref{fig:attention-flashatten-h800-nmax}, the measured $N_{\max}$ remains largely stable across cached sequence lengths $L$, whereas the KV-cache idle-compute baseline predicts an $L$-dependent boundary. This mismatch indicates that the Attention NFP boundary is not primarily governed by the compute-to-memory balance induced by KV-cache access.

The profiling results explain this mismatch. The achieved arithmetic intensity and runtime FLOPs change in discrete steps rather than smoothly with $N$. This indicates that FlashAttention executes padded or tiled query blocks, so larger N can remain within an existing backend granularity until the next boundary is crossed. Thus, the FlashAttention boundary is determined mainly by backend-specific query padding and tiling granularity.


\clearpage

\subsubsection{FlashInfer Backend}
\begin{figure*}[htbp]
    \centering

    \begin{subfigure}[t]{0.24\textwidth}
        \centering
        \includegraphics[width=\linewidth]{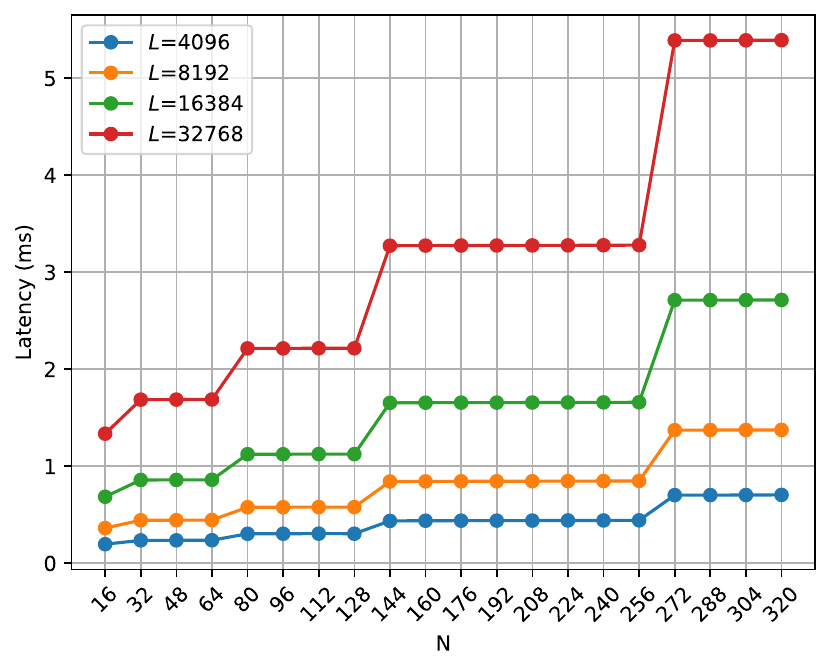}
        \caption{Latency vs. $N$}
        \label{fig:attention-flashinfer-h20-latency}
    \end{subfigure}
    \hfill
    \begin{subfigure}[t]{0.24\textwidth}
        \centering
        \includegraphics[width=\linewidth]{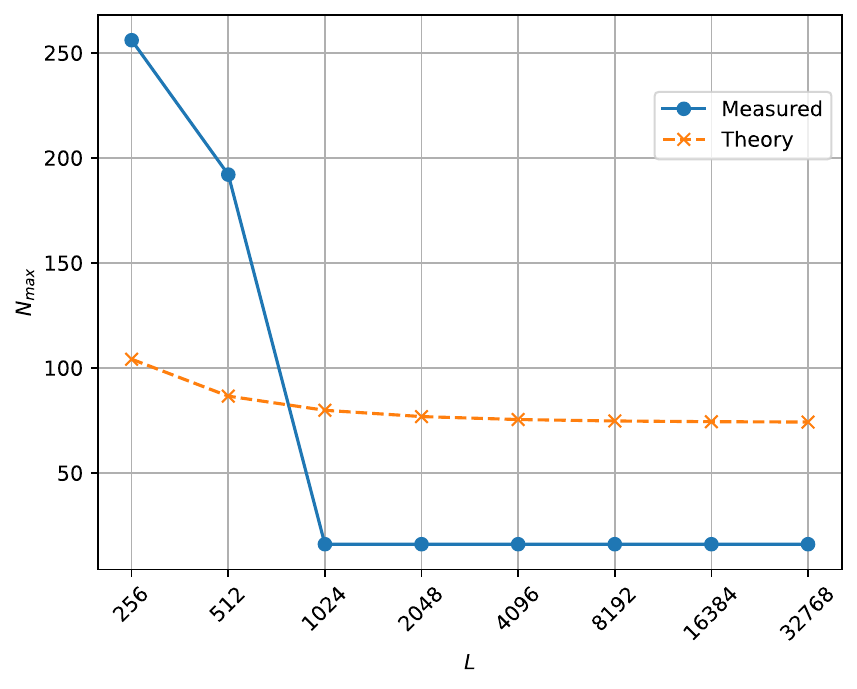}
        \caption{$N_{\max}$ vs. $L$}
        \label{fig:attention-flashinfer-h20-nmax}
    \end{subfigure}
    \hfill
    \begin{subfigure}[t]{0.24\textwidth}
        \centering
        \includegraphics[width=\linewidth]{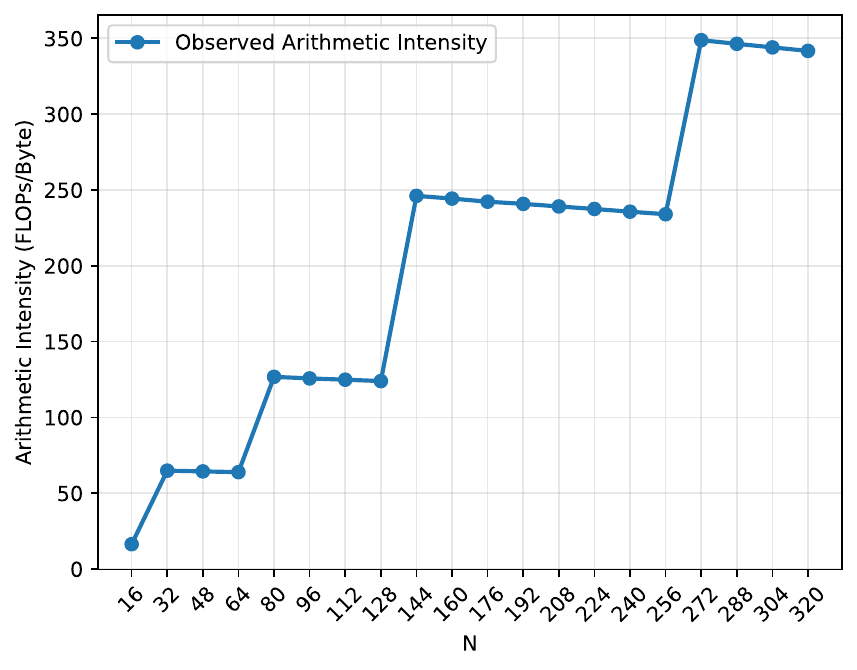}
        \caption{AI, $L=1024$}
        \label{fig:attention-flashinfer-h20-ai-l1024}
    \end{subfigure}
    \hfill
    \begin{subfigure}[t]{0.24\textwidth}
        \centering
        \includegraphics[width=\linewidth]{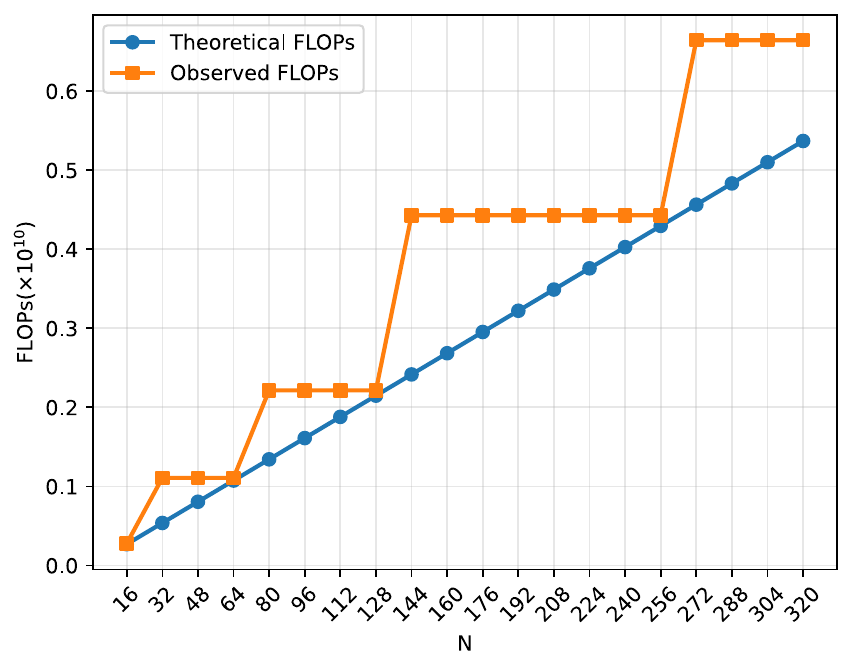}
        \caption{FLOPs, $L=1024$}
        \label{fig:attention-flashinfer-h20-flops-l1024}
    \end{subfigure}

    \vspace{0.6em}

    \begin{subfigure}[t]{0.24\textwidth}
        \centering
        \includegraphics[width=\linewidth]{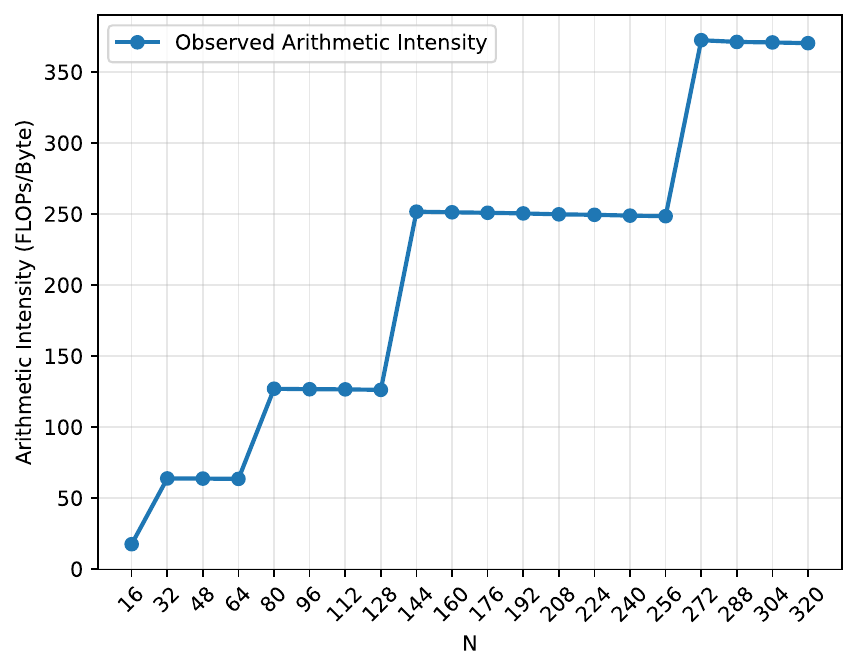}
        \caption{AI, $L=4096$}
        \label{fig:attention-flashinfer-h20-ai-l4096}
    \end{subfigure}
    \hfill
    \begin{subfigure}[t]{0.24\textwidth}
        \centering
        \includegraphics[width=\linewidth]{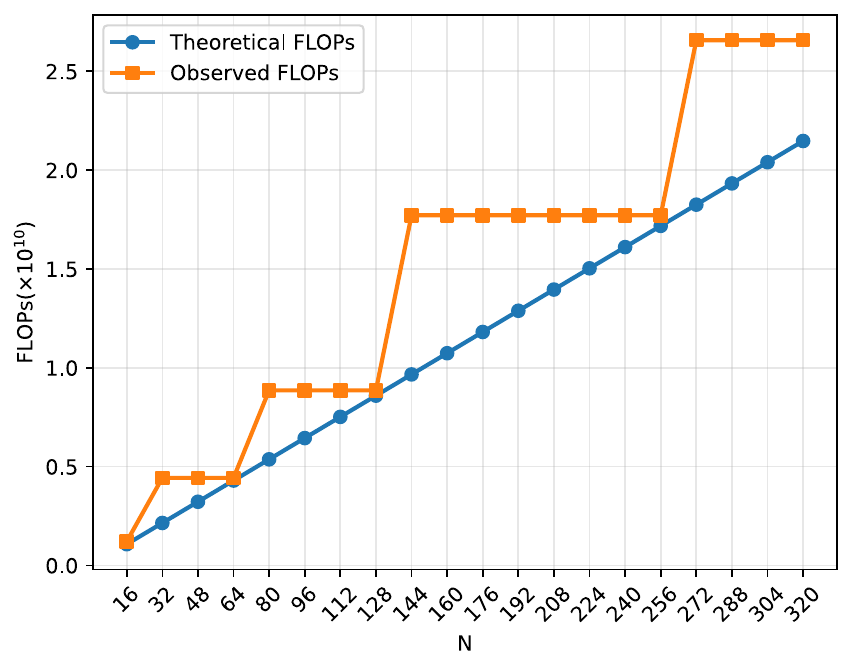}
        \caption{FLOPs, $L=4096$}
        \label{fig:attention-flashinfer-h20-flops-l4096}
    \end{subfigure}
    \hfill
    \begin{subfigure}[t]{0.24\textwidth}
        \centering
        \includegraphics[width=\linewidth]{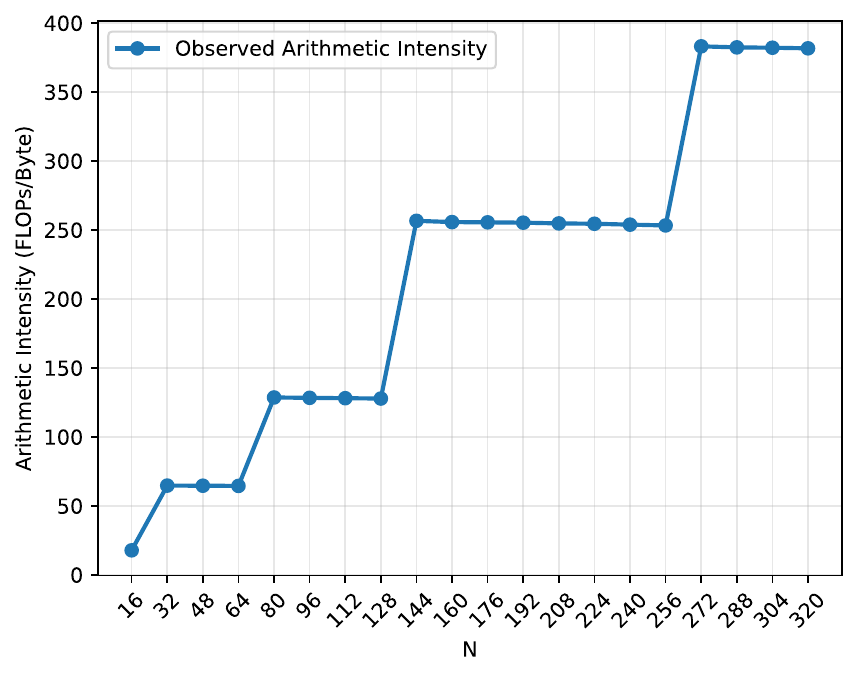}
        \caption{AI, $L=8192$}
        \label{fig:attention-flashinfer-h20-ai-l8192}
    \end{subfigure}
    \hfill
    \begin{subfigure}[t]{0.24\textwidth}
        \centering
        \includegraphics[width=\linewidth]{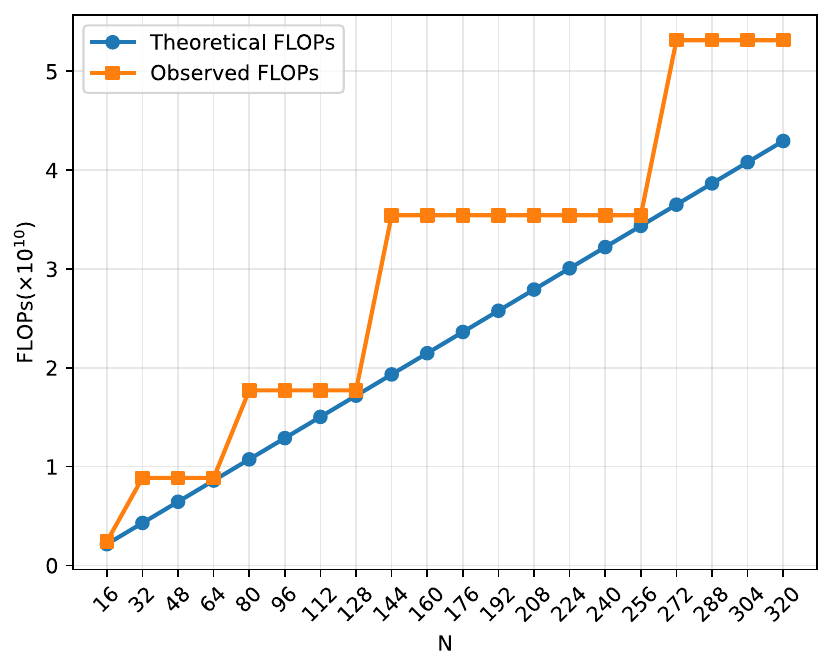}
        \caption{FLOPs, $L=8192$}
        \label{fig:attention-flashinfer-h20-flops-l8192}
    \end{subfigure}

    \vspace{0.6em}

    \begin{subfigure}[t]{0.24\textwidth}
        \centering
        \includegraphics[width=\linewidth]{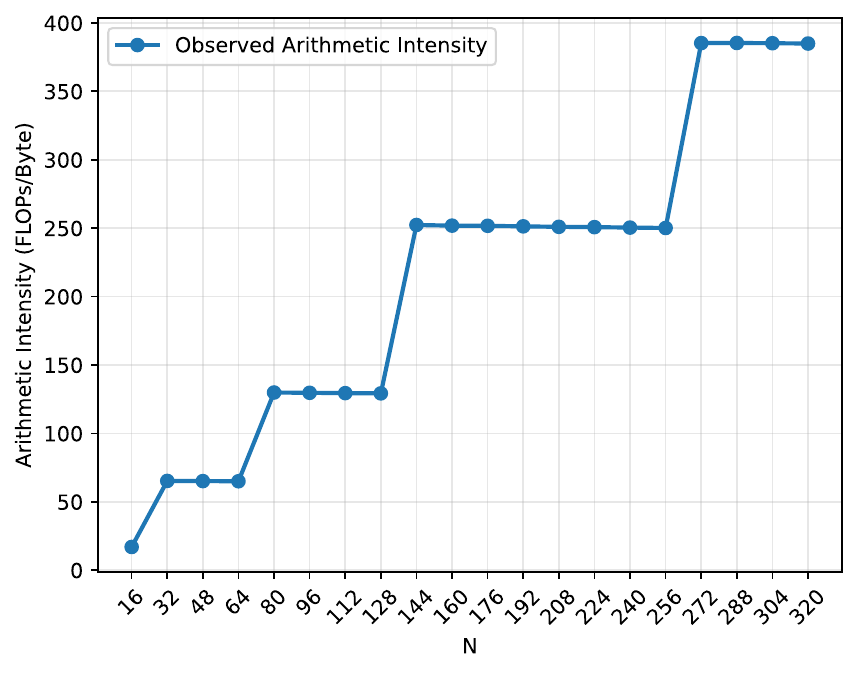}
        \caption{AI, $L=16384$}
        \label{fig:attention-flashinfer-h20-ai-l16384}
    \end{subfigure}
    \hfill
    \begin{subfigure}[t]{0.24\textwidth}
        \centering
        \includegraphics[width=\linewidth]{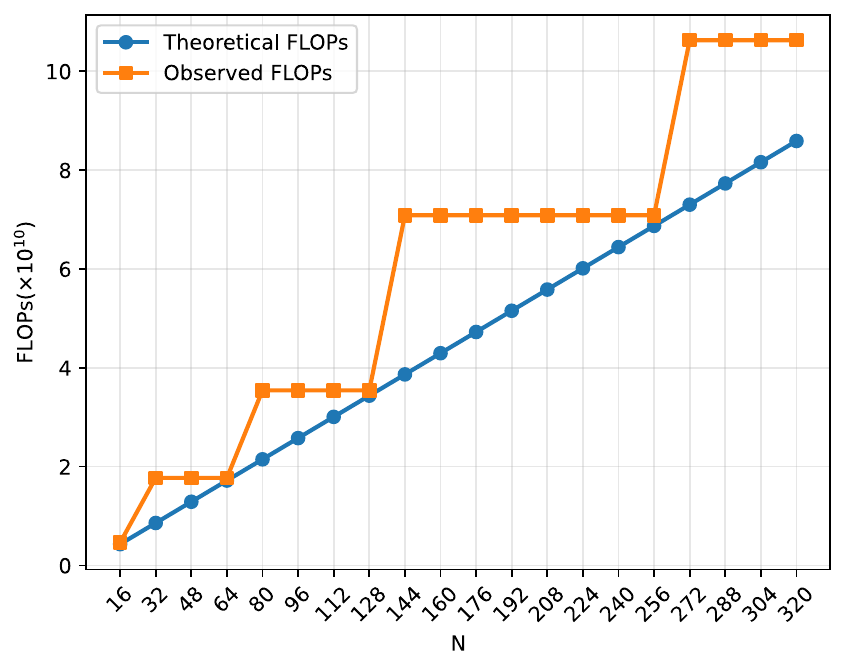}
        \caption{FLOPs, $L=16384$}
        \label{fig:attention-flashinfer-h20-flops-l16384}
    \end{subfigure}
    \hfill
    \begin{subfigure}[t]{0.24\textwidth}
        \centering
        \includegraphics[width=\linewidth]{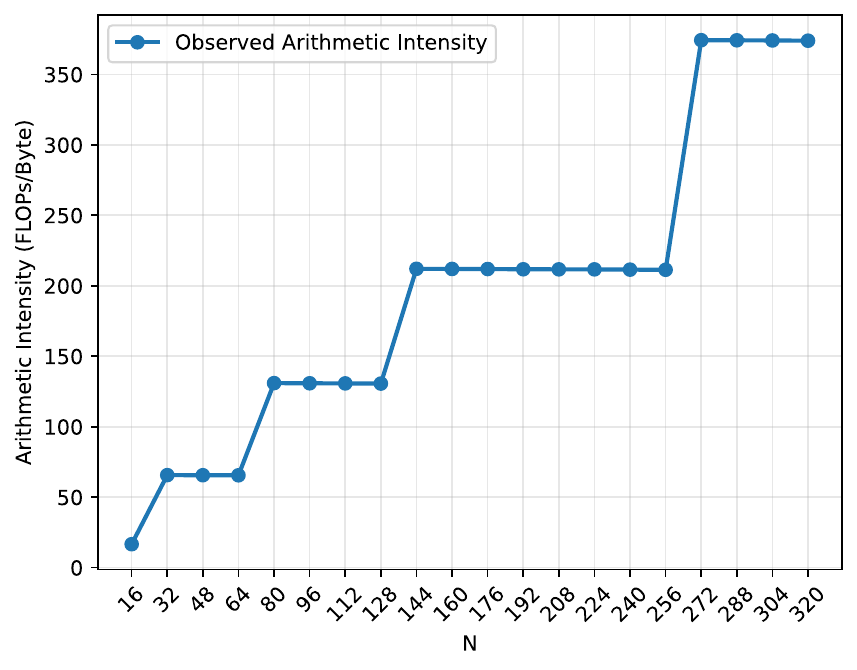}
        \caption{AI, $L=32768$}
        \label{fig:attention-flashinfer-h20-ai-l32768}
    \end{subfigure}
    \hfill
    \begin{subfigure}[t]{0.24\textwidth}
        \centering
        \includegraphics[width=\linewidth]{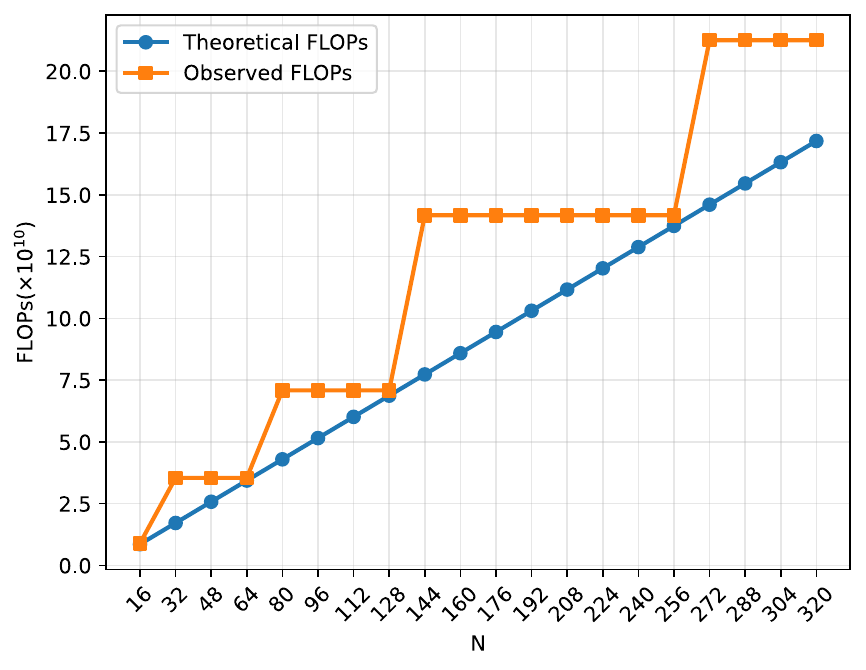}
        \caption{FLOPs, $L=32768$}
        \label{fig:attention-flashinfer-h20-flops-l32768}
    \end{subfigure}

    \caption{
    Attention evaluation results with \textbf{FlashInfer} on \textbf{NVIDIA H20}.
    }
    \label{fig:attention-flashinfer-h20}
\end{figure*}

\begin{figure*}[htbp]
    \centering

    \begin{subfigure}[t]{0.24\textwidth}
        \centering
        \includegraphics[width=\linewidth]{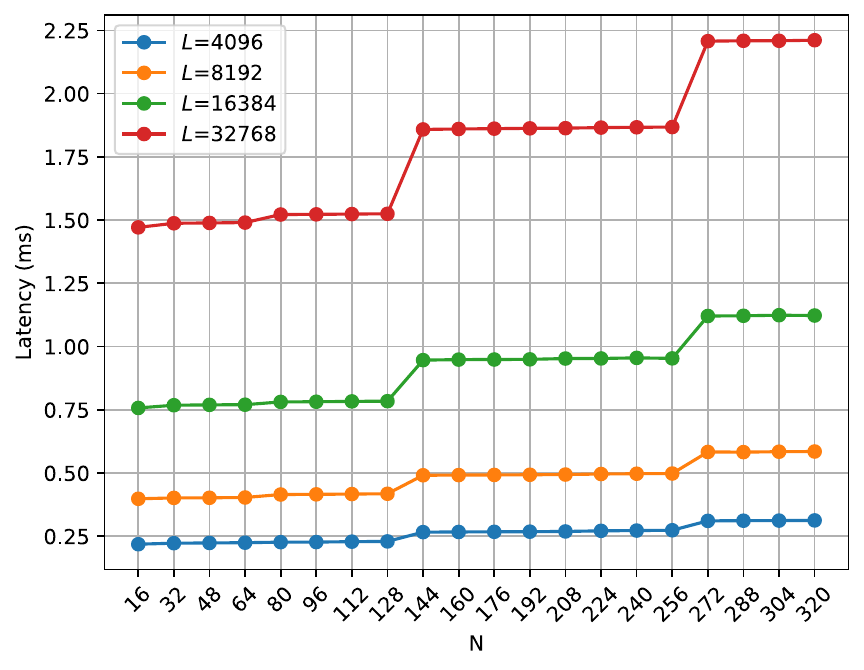}
        \caption{Latency vs. $N$}
        \label{fig:attention-flashinfer-a800-latency}
    \end{subfigure}
    \hfill
    \begin{subfigure}[t]{0.24\textwidth}
        \centering
        \includegraphics[width=\linewidth]{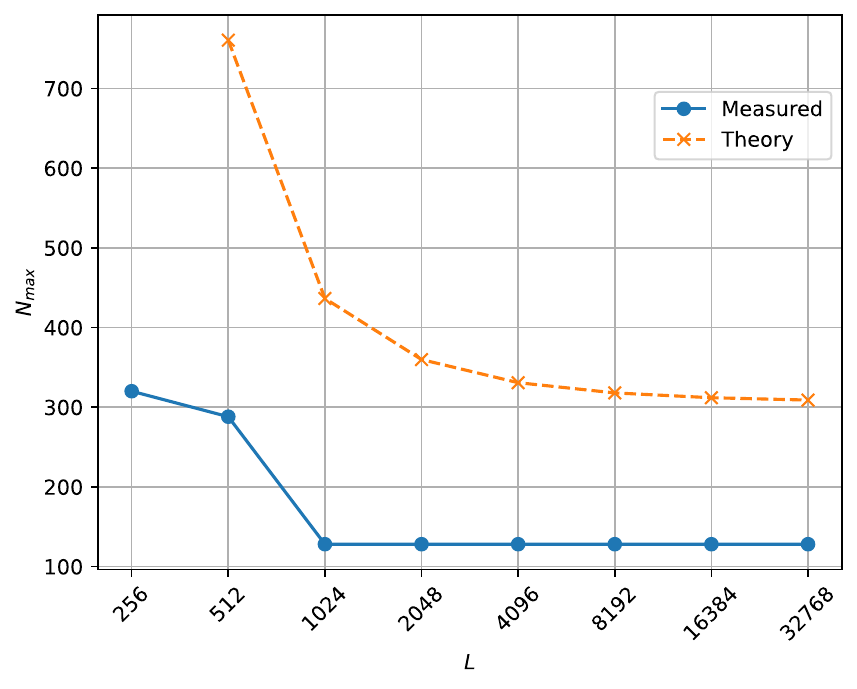}
        \caption{$N_{\max}$ vs. $L$}
        \label{fig:attention-flashinfer-a800-nmax}
    \end{subfigure}
    \hfill
    \begin{subfigure}[t]{0.24\textwidth}
        \centering
        \includegraphics[width=\linewidth]{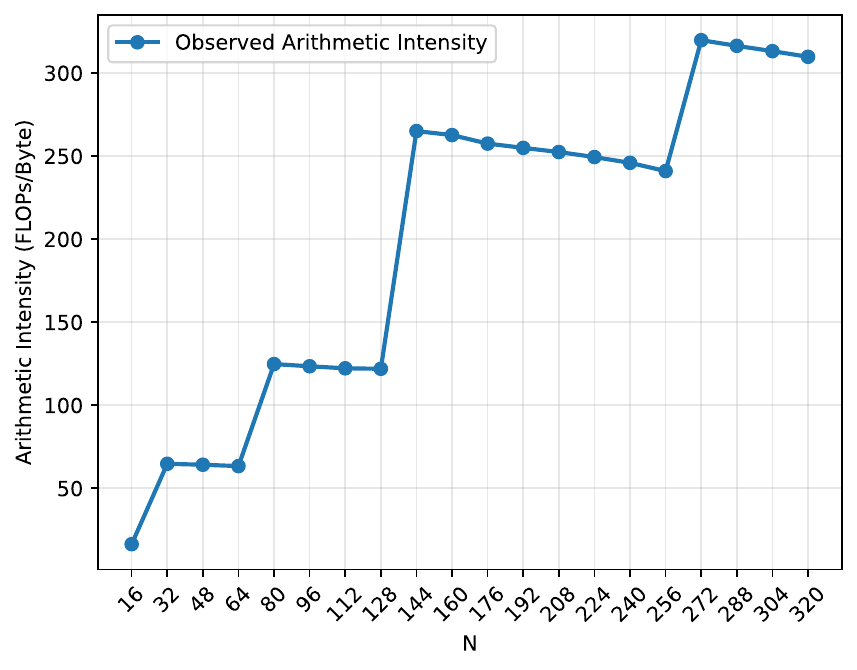}
        \caption{AI, $L=1024$}
        \label{fig:attention-flashinfer-a800-ai-l1024}
    \end{subfigure}
    \hfill
    \begin{subfigure}[t]{0.24\textwidth}
        \centering
        \includegraphics[width=\linewidth]{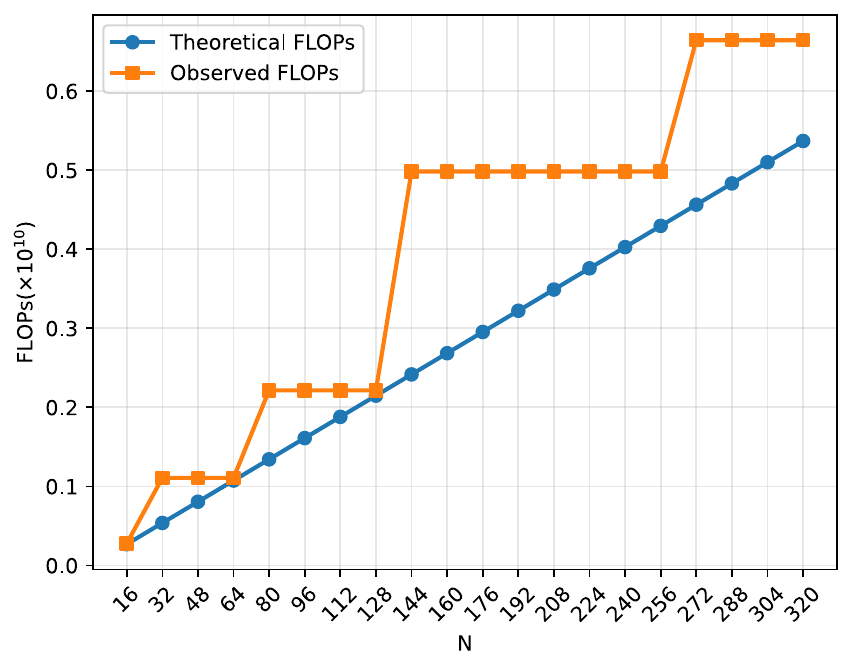}
        \caption{FLOPs, $L=1024$}
        \label{fig:attention-flashinfer-a800-flops-l1024}
    \end{subfigure}

    \vspace{0.6em}

    \begin{subfigure}[t]{0.24\textwidth}
        \centering
        \includegraphics[width=\linewidth]{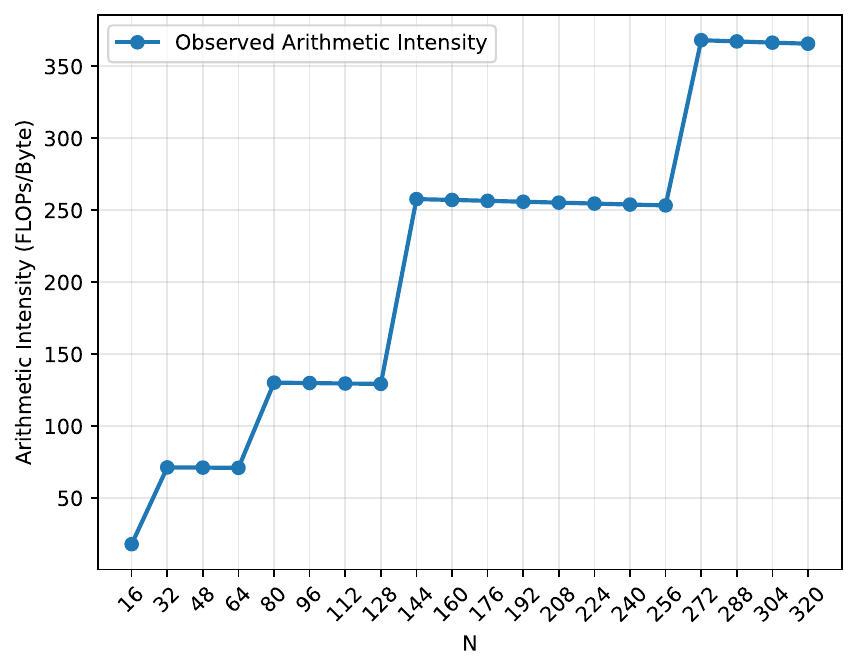}
        \caption{AI, $L=4096$}
        \label{fig:attention-flashinfer-a800-ai-l4096}
    \end{subfigure}
    \hfill
    \begin{subfigure}[t]{0.24\textwidth}
        \centering
        \includegraphics[width=\linewidth]{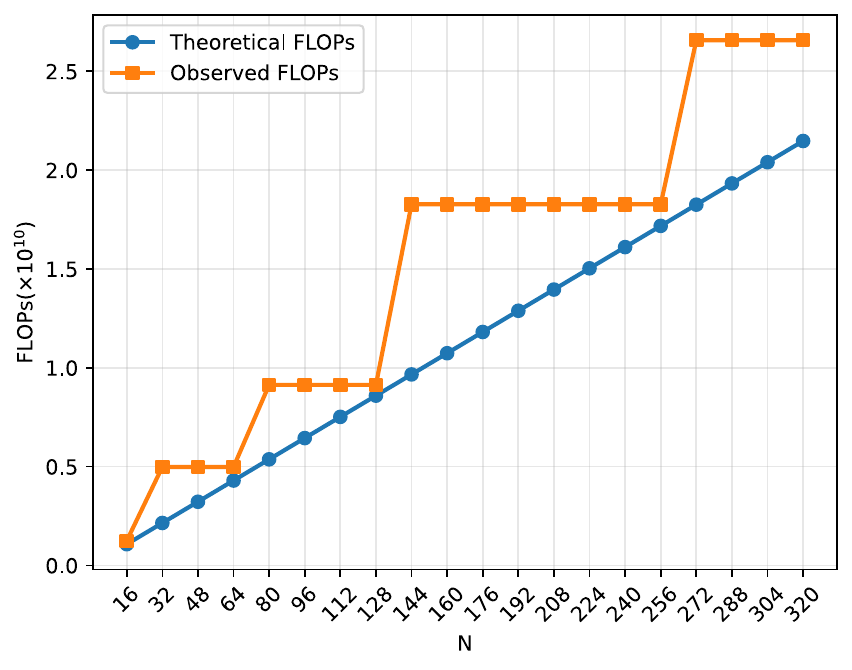}
        \caption{FLOPs, $L=4096$}
        \label{fig:attention-flashinfer-a800-flops-l4096}
    \end{subfigure}
    \hfill
    \begin{subfigure}[t]{0.24\textwidth}
        \centering
        \includegraphics[width=\linewidth]{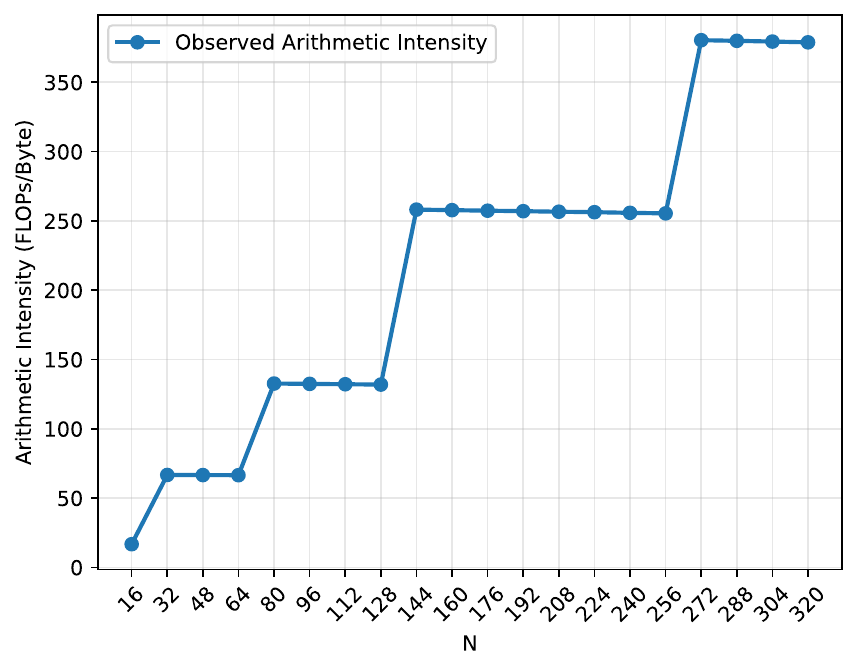}
        \caption{AI, $L=8192$}
        \label{fig:attention-flashinfer-a800-ai-l8192}
    \end{subfigure}
    \hfill
    \begin{subfigure}[t]{0.24\textwidth}
        \centering
        \includegraphics[width=\linewidth]{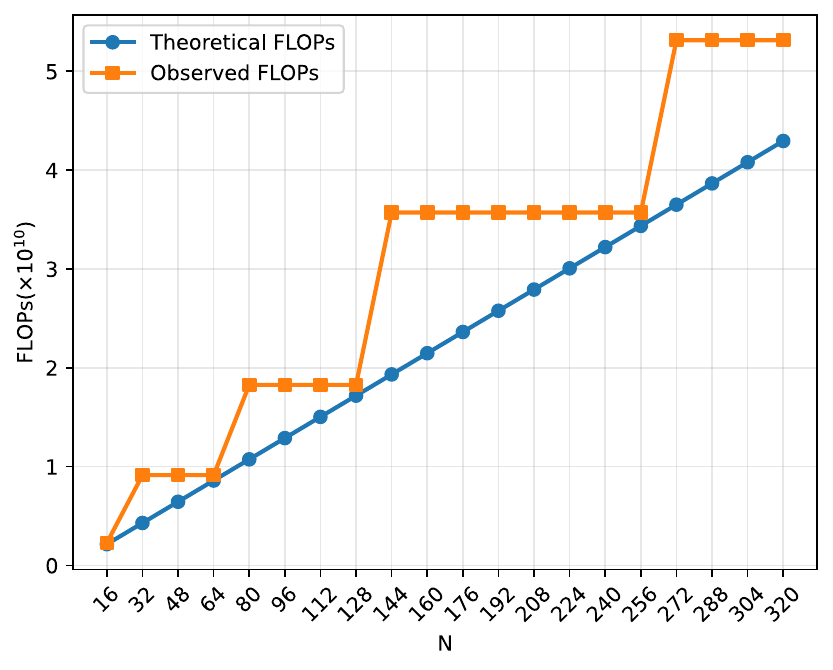}
        \caption{FLOPs, $L=8192$}
        \label{fig:attention-flashinfer-a800-flops-l8192}
    \end{subfigure}

    \vspace{0.6em}

    \begin{subfigure}[t]{0.24\textwidth}
        \centering
        \includegraphics[width=\linewidth]{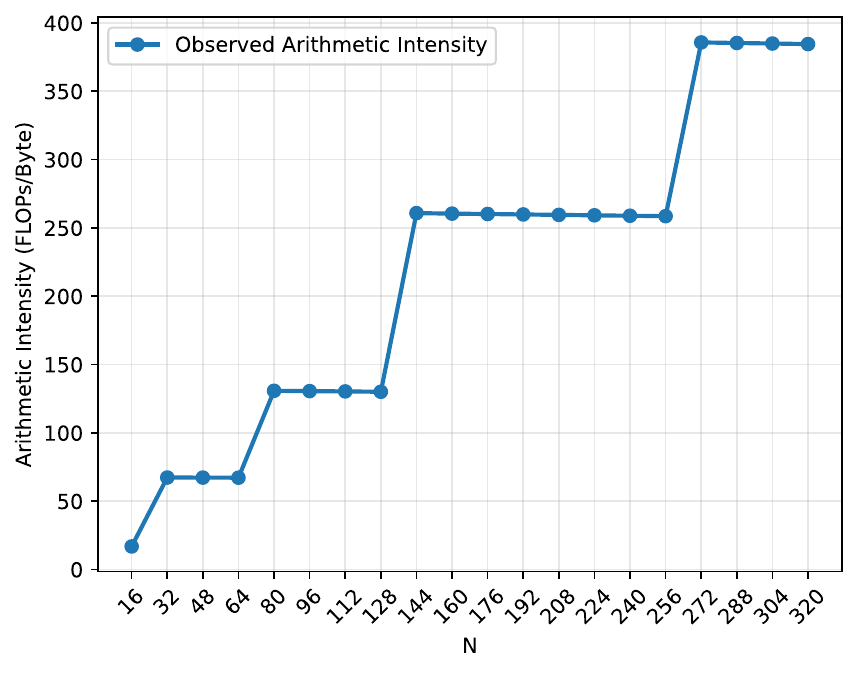}
        \caption{AI, $L=16384$}
        \label{fig:attention-flashinfer-a800-ai-l16384}
    \end{subfigure}
    \hfill
    \begin{subfigure}[t]{0.24\textwidth}
        \centering
        \includegraphics[width=\linewidth]{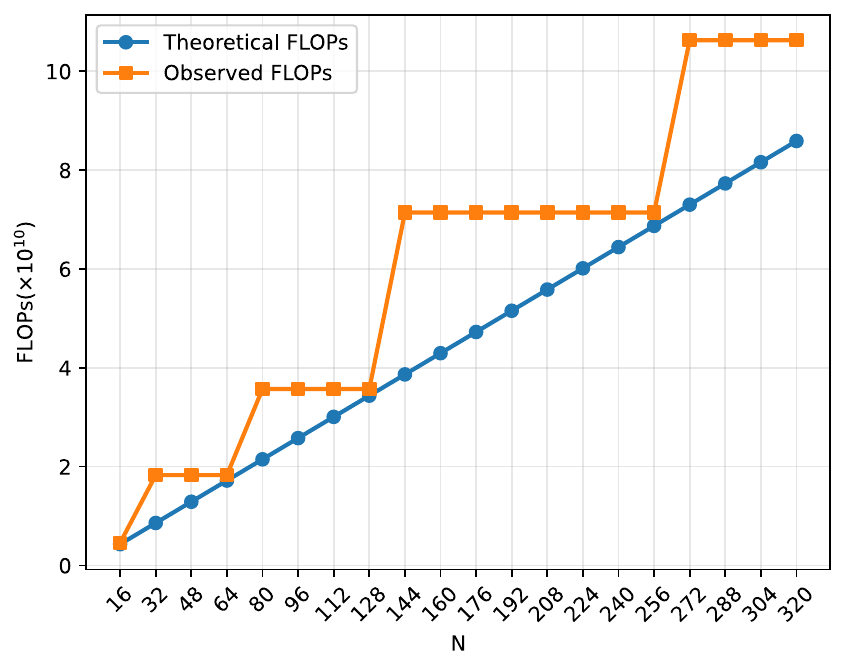}
        \caption{FLOPs, $L=16384$}
        \label{fig:attention-flashinfer-a800-flops-l16384}
    \end{subfigure}
    \hfill
    \begin{subfigure}[t]{0.24\textwidth}
        \centering
        \includegraphics[width=\linewidth]{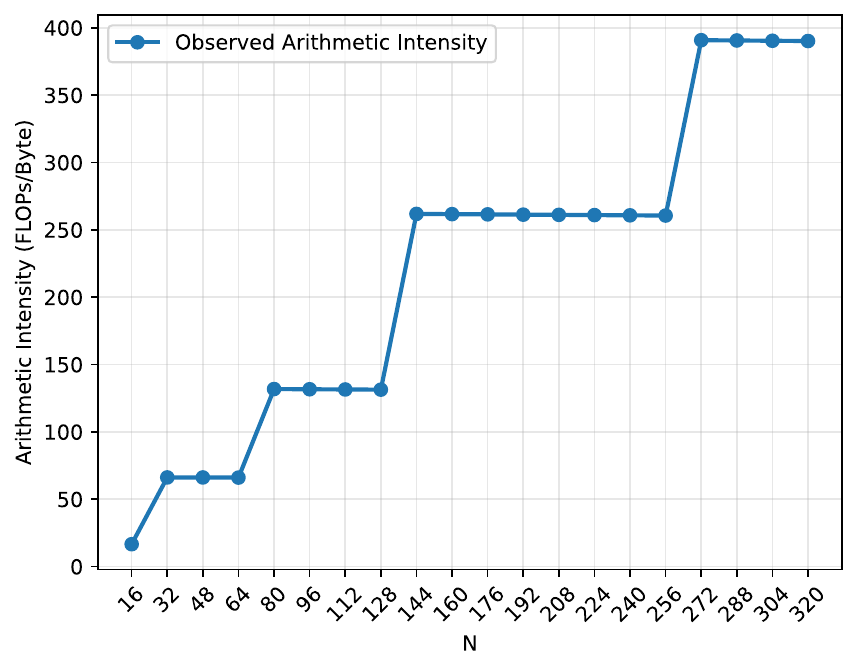}
        \caption{AI, $L=32768$}
        \label{fig:attention-flashinfer-a800-ai-l32768}
    \end{subfigure}
    \hfill
    \begin{subfigure}[t]{0.24\textwidth}
        \centering
        \includegraphics[width=\linewidth]{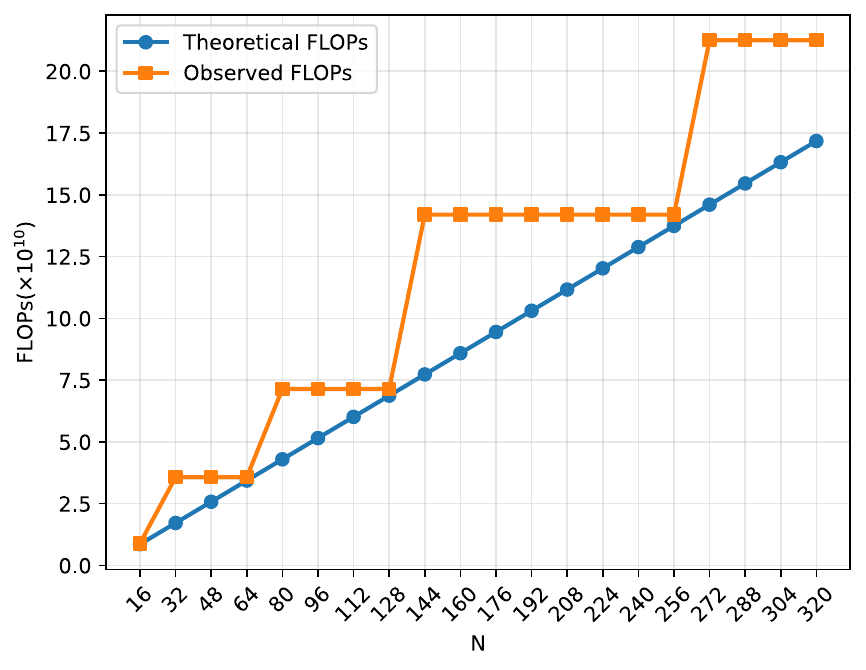}
        \caption{FLOPs, $L=32768$}
        \label{fig:attention-flashinfer-a800-flops-l32768}
    \end{subfigure}

    \caption{
    Attention evaluation results with \textbf{FlashInfer} on \textbf{NVIDIA A800}.
    }
    \label{fig:attention-flashinfer-a800}
\end{figure*}

\begin{figure*}[htbp]
    \centering

    \begin{subfigure}[t]{0.24\textwidth}
        \centering
        \includegraphics[width=\linewidth]{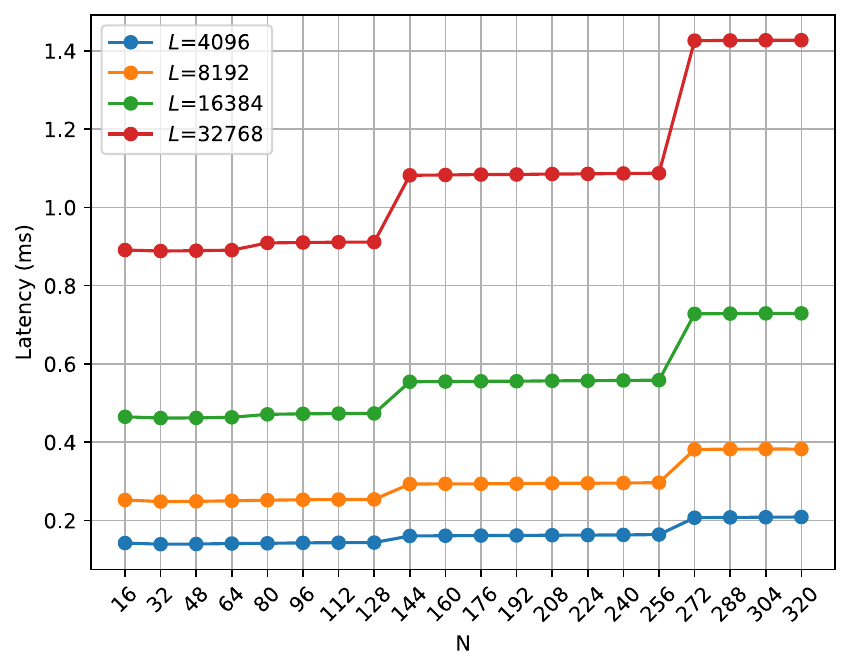}
        \caption{Latency vs. $N$}
        \label{fig:attention-flashinfer-h800-latency}
    \end{subfigure}
    \hfill
    \begin{subfigure}[t]{0.24\textwidth}
        \centering
        \includegraphics[width=\linewidth]{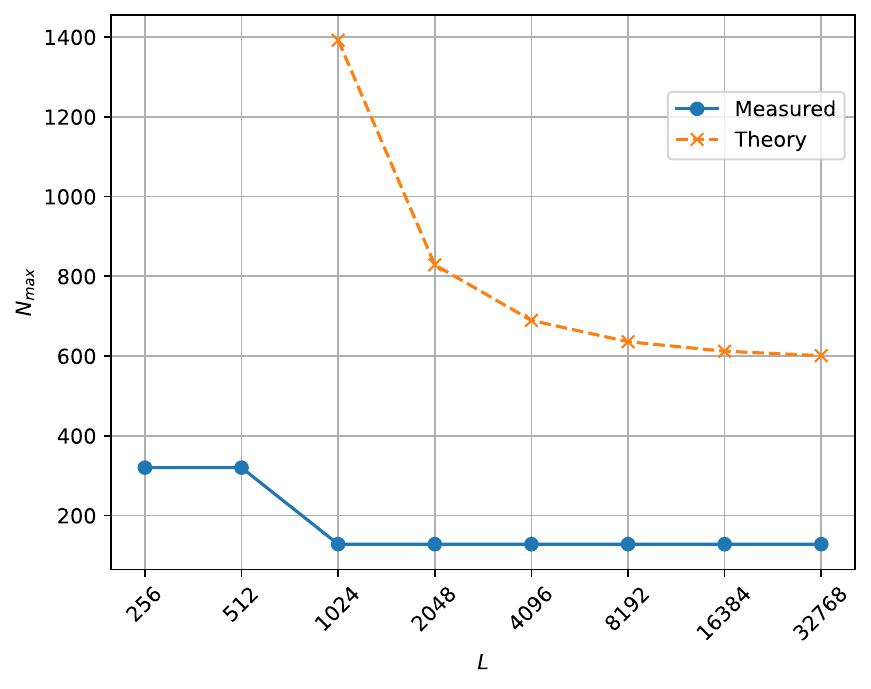}
        \caption{$N_{\max}$ vs. $L$}
        \label{fig:attention-flashinfer-h800-nmax}
    \end{subfigure}
    \hfill
    \begin{subfigure}[t]{0.24\textwidth}
        \centering
        \includegraphics[width=\linewidth]{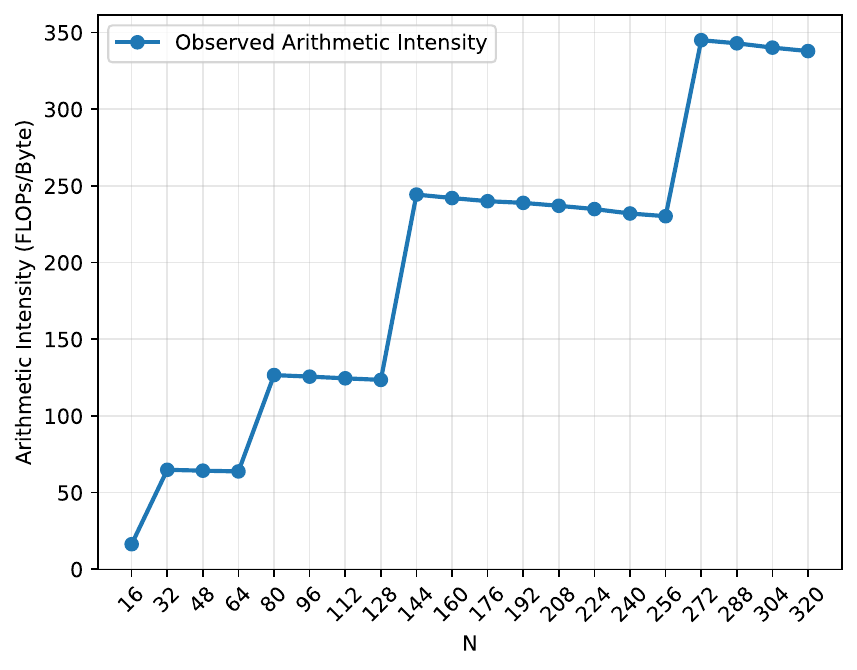}
        \caption{AI, $L=1024$}
        \label{fig:attention-flashinfer-h800-ai-l1024}
    \end{subfigure}
    \hfill
    \begin{subfigure}[t]{0.24\textwidth}
        \centering
        \includegraphics[width=\linewidth]{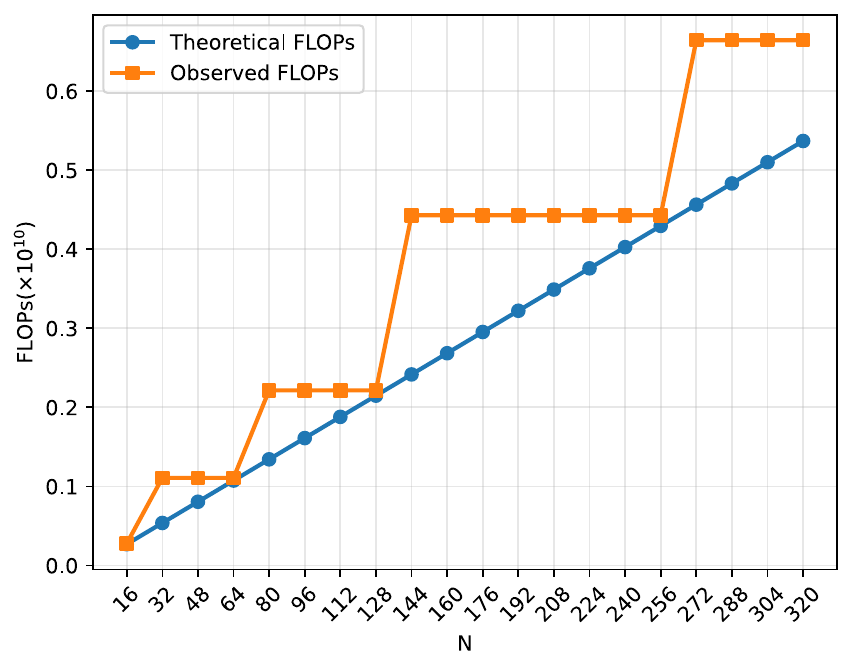}
        \caption{FLOPs, $L=1024$}
        \label{fig:attention-flashinfer-h800-flops-l1024}
    \end{subfigure}

    \vspace{0.6em}

    \begin{subfigure}[t]{0.24\textwidth}
        \centering
        \includegraphics[width=\linewidth]{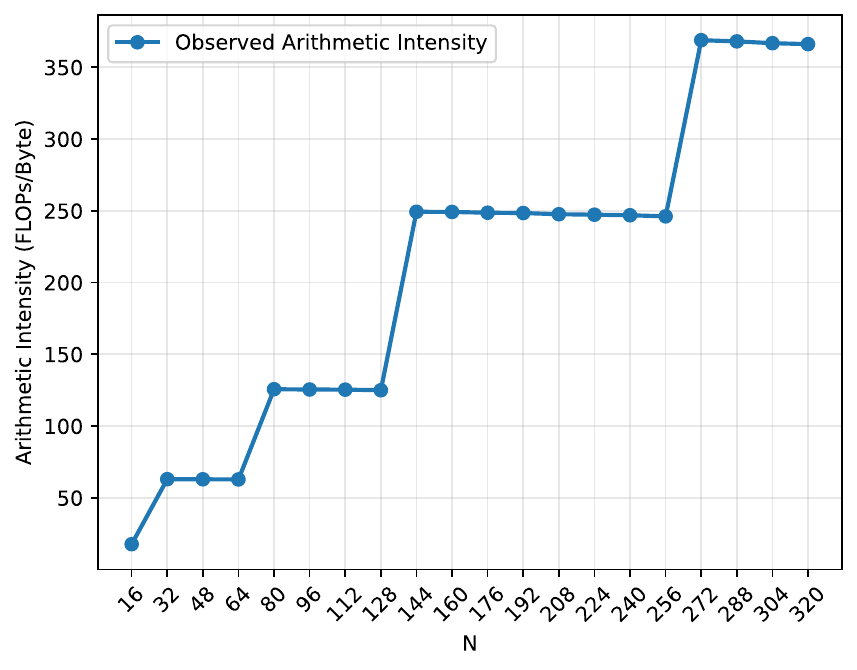}
        \caption{AI, $L=4096$}
        \label{fig:attention-flashinfer-h800-ai-l4096}
    \end{subfigure}
    \hfill
    \begin{subfigure}[t]{0.24\textwidth}
        \centering
        \includegraphics[width=\linewidth]{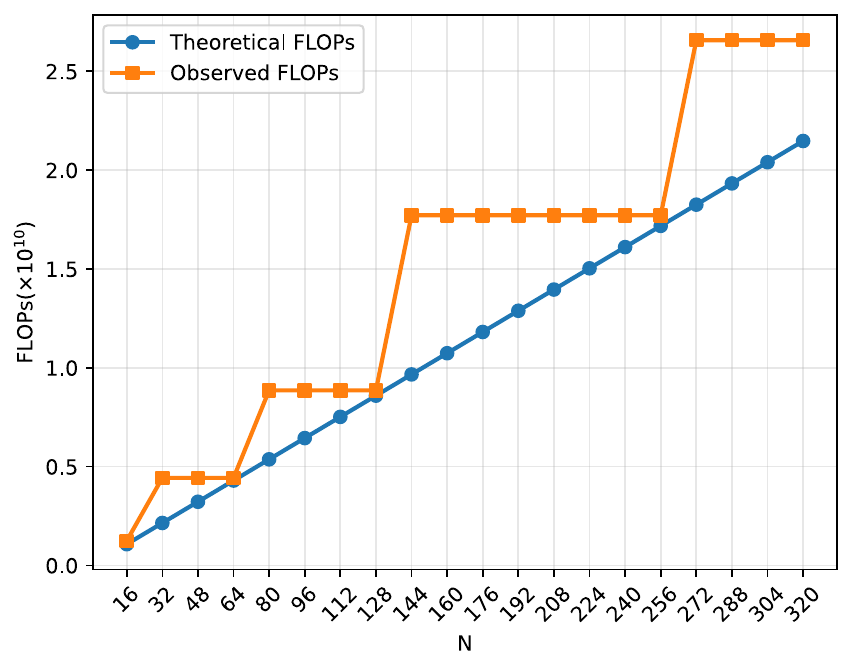}
        \caption{FLOPs, $L=4096$}
        \label{fig:attention-flashinfer-h800-flops-l4096}
    \end{subfigure}
    \hfill
    \begin{subfigure}[t]{0.24\textwidth}
        \centering
        \includegraphics[width=\linewidth]{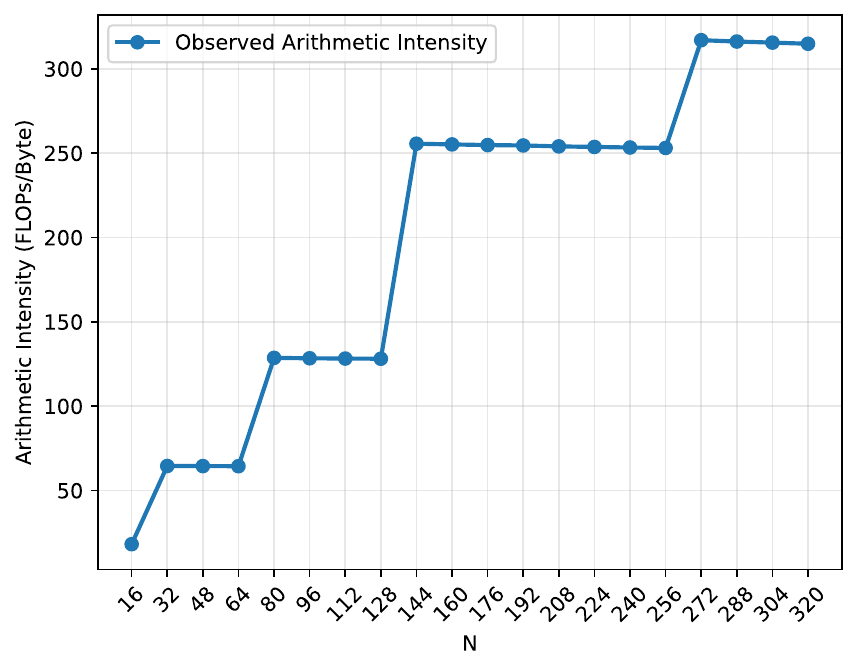}
        \caption{AI, $L=8192$}
        \label{fig:attention-flashinfer-h800-ai-l8192}
    \end{subfigure}
    \hfill
    \begin{subfigure}[t]{0.24\textwidth}
        \centering
        \includegraphics[width=\linewidth]{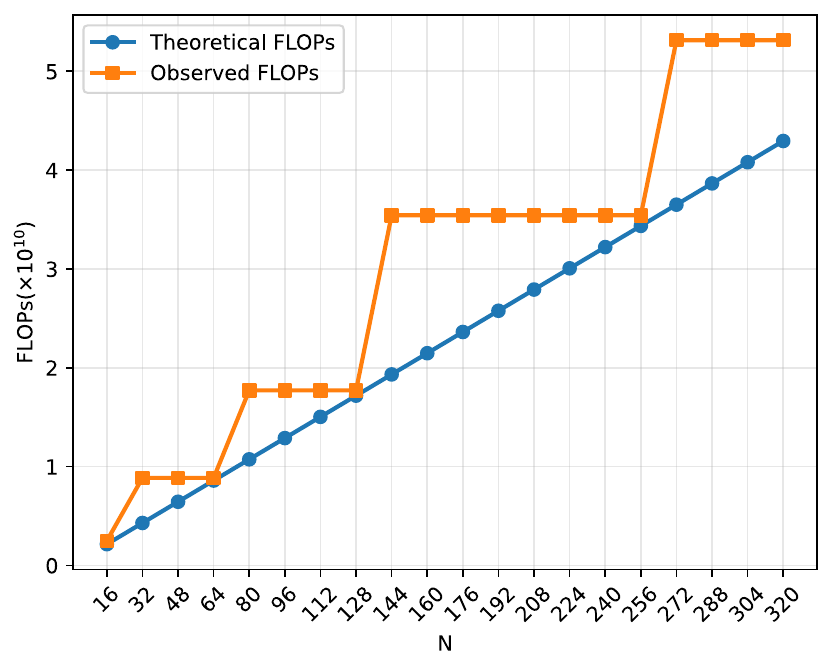}
        \caption{FLOPs, $L=8192$}
        \label{fig:attention-flashinfer-h800-flops-l8192}
    \end{subfigure}

    \vspace{0.6em}

    \begin{subfigure}[t]{0.24\textwidth}
        \centering
        \includegraphics[width=\linewidth]{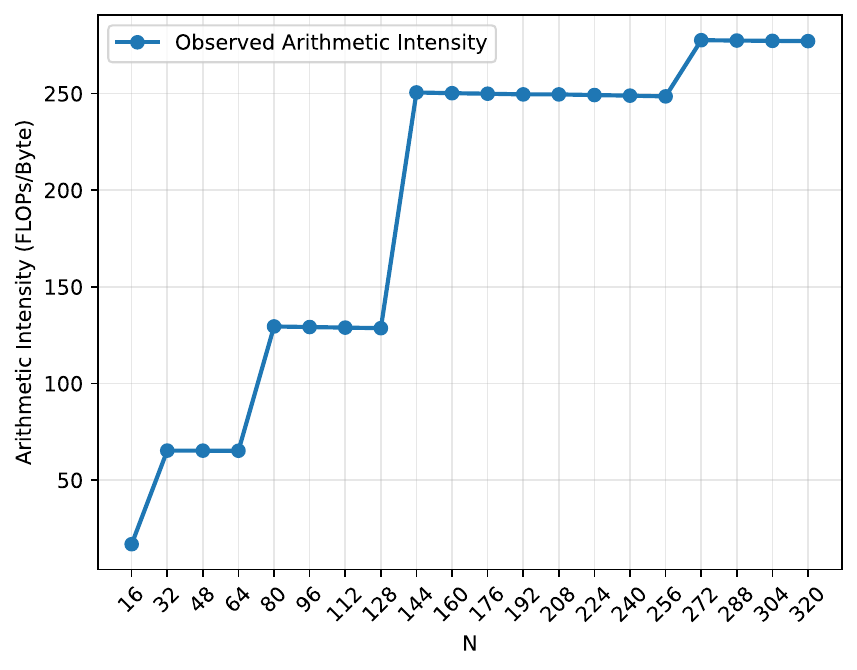}
        \caption{AI, $L=16384$}
        \label{fig:attention-flashinfer-h800-ai-l16384}
    \end{subfigure}
    \hfill
    \begin{subfigure}[t]{0.24\textwidth}
        \centering
        \includegraphics[width=\linewidth]{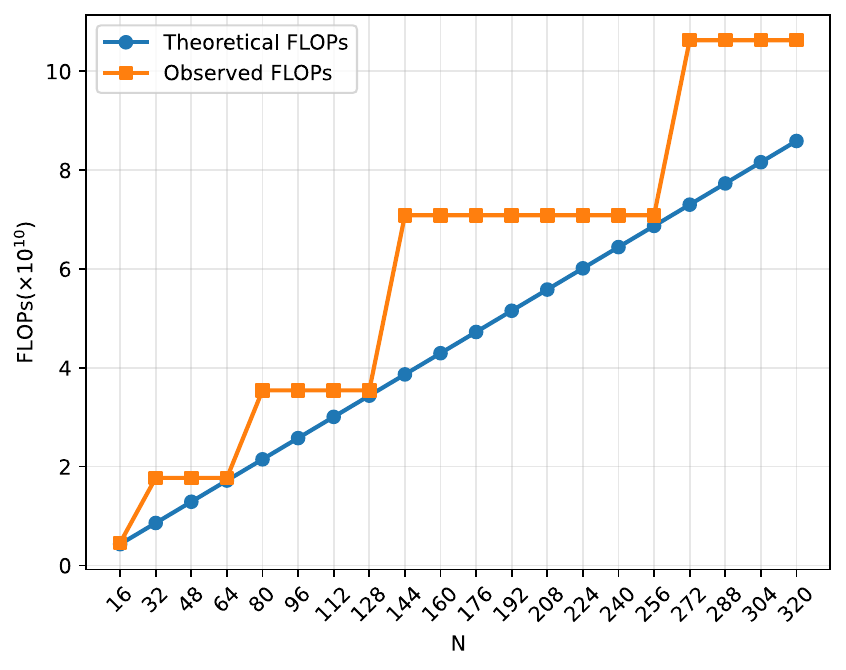}
        \caption{FLOPs, $L=16384$}
        \label{fig:attention-flashinfer-h800-flops-l16384}
    \end{subfigure}
    \hfill
    \begin{subfigure}[t]{0.24\textwidth}
        \centering
        \includegraphics[width=\linewidth]{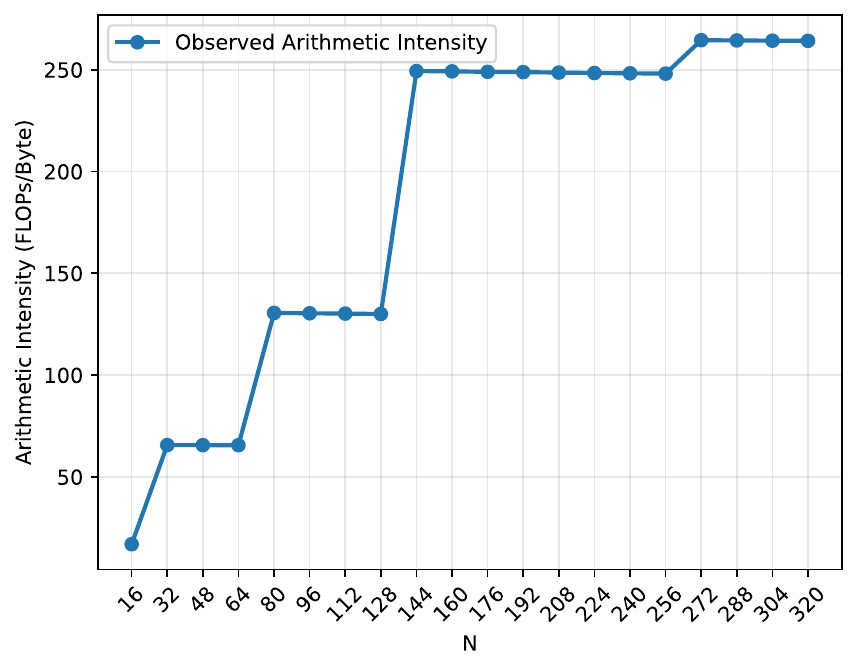}
        \caption{AI, $L=32768$}
        \label{fig:attention-flashinfer-h800-ai-l32768}
    \end{subfigure}
    \hfill
    \begin{subfigure}[t]{0.24\textwidth}
        \centering
        \includegraphics[width=\linewidth]{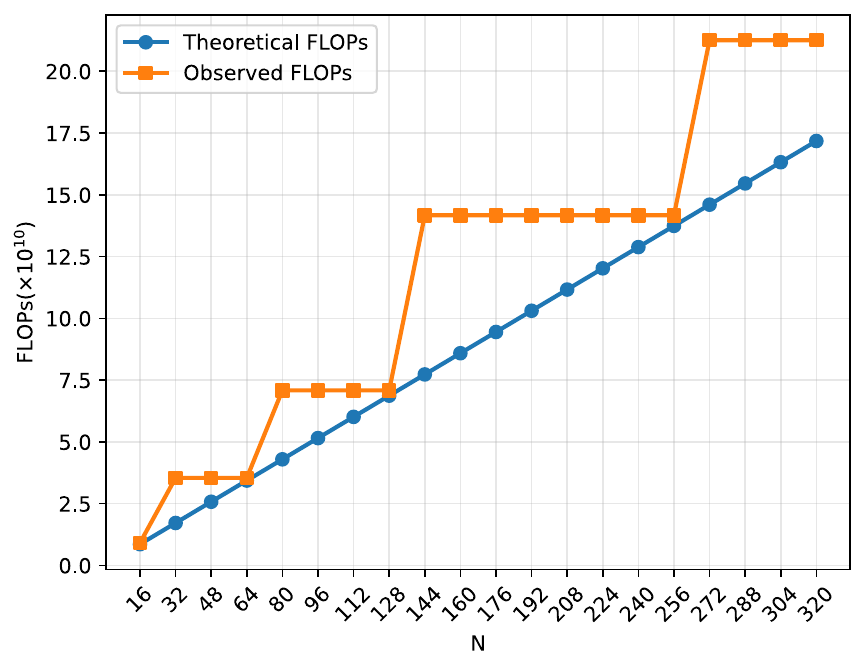}
        \caption{FLOPs, $L=32768$}
        \label{fig:attention-flashinfer-h800-flops-l32768}
    \end{subfigure}

    \caption{
    Attention evaluation results with \textbf{FlashInfer} on \textbf{NVIDIA H800}.
    }
    \label{fig:attention-flashinfer-h800}
\end{figure*}

Figures~\ref{fig:attention-flashinfer-h20}, \ref{fig:attention-flashinfer-a800}, and \ref{fig:attention-flashinfer-h800} report the corresponding Attention results using the FlashInfer backend. The same qualitative behavior appears under FlashInfer. The latency curves show near-free plateaus followed by discrete jumps, indicating that larger N remains near-free within a backend granularity block and becomes expensive only after crossing the next boundary.

The extracted boundaries again remain largely independent of cached sequence length. As shown in Figures~\ref{fig:attention-flashinfer-h20}\subref{fig:attention-flashinfer-h20-nmax}, \ref{fig:attention-flashinfer-a800}\subref{fig:attention-flashinfer-a800-nmax}, and \ref{fig:attention-flashinfer-h800}\subref{fig:attention-flashinfer-h800-nmax}, the measured $N_{\max}$ does not track the $L$-dependent idle-compute prediction. Instead, it is primarily determined by the query granularity of the FlashInfer backend. The arithmetic-intensity and FLOPs profiles further show discrete changes as $N$ increases, confirming that the observed near-free region is produced by padded backend execution rather than a smooth resource-balance transition.

The agreement between FlashAttention and FlashInfer confirms that the Attention NFP mechanism is not specific to one backend implementation. At the same time, the actual boundary value is backend-specific, since different attention kernels use different query tiling and padding rules. Therefore, Attention contributes an implementation-dependent granularity term $M_{\mathrm{attn}}$ to the model-level NFP principle.

Overall, the Attention results confirm the mechanism identified in the main text. Across attention backends and evaluated single-GPU platforms, the Attention NFP boundary is governed primarily by backend-specific query padding and tiling rather than by the KV-cache idle-compute baseline. This explains why the measured boundary is largely independent of cached sequence length $L$ and supports the use of $M_{\mathrm{attn}}$ in the model-level NFP principle.


\clearpage

\section{Implementation Evidence for MoE Expert-Token Padding}
\label{app:moe-padding}

The MoE FFN results in Appendix~\ref{app:moe-ffn-results} show staircase latency, arithmetic intensity, and runtime FLOPs, indicating that fused MoE execution performs padded computation beyond the logical routed-token workload. This appendix provides source-level evidence for this mechanism. We inspect the fused MoE implementations in vLLM and SGLang and show that both backends align expert-token workloads to implementation-defined block sizes before expert GEMM execution. This alignment creates kernel-granularity slack: larger N can remain within existing padded expert-token blocks until the next block boundary is crossed.

\subsection{vLLM Fused-MoE Expert-Token Alignment}
\label{app:moe-padding-vllm}

Tables~\ref{tab:vllm-chunked-fused-moe-padding-path} and~\ref{tab:vllm-fused-moe-padding-path} show the relevant vLLM fused MoE execution paths. In both the standard and chunked paths, vLLM first selects a kernel configuration from the MoE problem size and obtains a matrix-multiplication row-block parameter \texttt{BLOCK\_SIZE\_M}. Before launching the expert GEMMs, the backend passes this block size to \texttt{moe\_align\_block\_size}, which constructs the expert-token layout used by the fused MoE kernel.

Table~\ref{tab:vllm-moe-align-block-size} shows the implementation of this alignment helper. The function explicitly aligns the routed token distribution across experts to the matrix-multiplication block size, allocates padded token and expert-block buffers, invokes the backend alignment operator, and returns \texttt{num\_tokens\_post\_pad}. Thus, the expert GEMMs operate on padded expert-token blocks rather than exactly the logical routed-token counts.

This implementation explains the staircase behavior observed in the MoE profiling results. When increasing N increases the logical routed-token count but remain within the same \texttt{BLOCK\_SIZE\_M}-aligned expert-token block, the executed GEMM shape does not change. Once the logical workload crosses a block boundary, the padded shape increases discretely, producing the observed jumps in latency and runtime FLOPs.

\begin{table*}[htbp]
\centering

\begin{minipage}{0.82\textwidth}

\noindent{\color{coderule}\rule{\linewidth}{0.45pt}}
\vspace{-0.4em}

\begin{lstlisting}[
    style=prettyPythonKernel,
    basicstyle=\ttfamily\footnotesize
]
for chunk in range((num_tokens // CHUNK_SIZE) + 1):
    begin_chunk_idx, end_chunk_idx = (
        chunk * CHUNK_SIZE,
        min((chunk + 1) * CHUNK_SIZE, num_tokens),
    )

    curr_hidden_states = hidden_states[begin_chunk_idx:end_chunk_idx]
    tokens_in_chunk, _ = curr_hidden_states.size()

    if tokens_in_chunk == 0:
        break

    if tokens_in_chunk < CHUNK_SIZE and chunk > 0:
        config = get_config_func(tokens_in_chunk)

    curr_topk_ids = topk_ids[begin_chunk_idx:end_chunk_idx]
    curr_topk_weights = topk_weights[begin_chunk_idx:end_chunk_idx]

    # ...

    sorted_token_ids, expert_ids, num_tokens_post_padded = (
        moe_align_block_size(curr_topk_ids, config['BLOCK_SIZE_M'],
                             global_num_experts, expert_map))
\end{lstlisting}

\vspace{-0.9em}
\noindent{\color{coderule}\rule{\linewidth}{0.45pt}}

\end{minipage}

\vspace{0.3em}

\caption{
Source-code evidence for expert-token padding in the \textbf{vLLM chunked fused MoE} path. The backend slices tokens and routing metadata into chunks, optionally reselects the kernel configuration for a tail chunk, and applies \texttt{moe\_align\_block\_size} with \texttt{BLOCK\_SIZE\_M} to each chunk before expert GEMM execution.
}
\label{tab:vllm-chunked-fused-moe-padding-path}

\end{table*}

\begin{table*}[htbp]
\centering

\begin{minipage}{0.82\textwidth}

\noindent{\color{coderule}\rule{\linewidth}{0.45pt}}
\vspace{-0.4em}

\begin{lstlisting}[style=prettyPythonKernel]
E, num_tokens, N, K, top_k_num = mk._moe_problem_size(
    hidden_states, w1, w2, topk_ids)

if global_num_experts == -1:
    global_num_experts = E

config = try_get_optimal_moe_config(
    w1.size(),
    w2.size(),
    top_k_num,
    self.quant_config.config_name(hidden_states.dtype),
    num_tokens,
    block_shape=self.block_shape,
)

# ...

sorted_token_ids, expert_ids, num_tokens_post_padded = (
    moe_align_block_size(topk_ids, config['BLOCK_SIZE_M'],
                         global_num_experts, expert_map))
\end{lstlisting}

\vspace{-0.9em}
\noindent{\color{coderule}\rule{\linewidth}{0.45pt}}

\end{minipage}

\vspace{0.3em}

\caption{
Source-code evidence for expert-token padding in the standard \textbf{vLLM fused MoE} path. The backend selects a kernel configuration from the MoE problem size and passes \texttt{BLOCK\_SIZE\_M} to \texttt{moe\_align\_block\_size}, which produces padded expert-token assignments before expert GEMM execution.
}
\label{tab:vllm-fused-moe-padding-path}

\end{table*}

\begin{table*}[htbp]
\centering

\begin{minipage}{0.82\textwidth}

\noindent{\color{coderule}\rule{\linewidth}{0.45pt}}
\vspace{-0.4em}

\begin{lstlisting}[style=prettyPythonKernel]
def moe_align_block_size(
    topk_ids: torch.Tensor,
    block_size: int,
    num_experts: int,
    expert_map: Optional[torch.Tensor] = None,
    pad_sorted_ids: bool = False
) -> tuple[torch.Tensor, torch.Tensor, torch.Tensor]:
    """
    Aligns the token distribution across experts to be compatible with block
    size for matrix multiplication.
    """
    max_num_tokens_padded = topk_ids.numel() + num_experts * (block_size - 1)
    if pad_sorted_ids:
        max_num_tokens_padded = round_up(max_num_tokens_padded, block_size)
    sorted_ids = torch.empty((max_num_tokens_padded, ),
                             dtype=torch.int32,
                             device=topk_ids.device)
    max_num_m_blocks = triton.cdiv(max_num_tokens_padded, block_size)
    expert_ids = torch.empty((max_num_m_blocks, ),
                             dtype=torch.int32,
                             device=topk_ids.device)
    num_tokens_post_pad = torch.empty((1),
                                      dtype=torch.int32,
                                      device=topk_ids.device)
    ops.moe_align_block_size(topk_ids, num_experts, block_size, sorted_ids,
                             expert_ids, num_tokens_post_pad)
    if expert_map is not None:
        expert_ids = expert_map[expert_ids]

    return sorted_ids, expert_ids, num_tokens_post_pad
\end{lstlisting}

\vspace{-0.9em}
\noindent{\color{coderule}\rule{\linewidth}{0.45pt}}

\end{minipage}

\vspace{0.3em}

\caption{
Implementation of expert-token block alignment in \textbf{vLLM fused MoE}. The helper function \texttt{moe\_align\_block\_size} explicitly aligns routed tokens to the matrix-multiplication block size, allocates padded token buffers, invokes the backend alignment operator, and returns \texttt{num\_tokens\_post\_pad}, providing direct evidence that fused MoE execution pads expert-token workloads to a block granularity.
}
\label{tab:vllm-moe-align-block-size}

\end{table*}

\clearpage

\clearpage

\subsection{SGLang Fused-MoE Expert-Token Alignment}
\label{app:moe-padding-sglang}

SGLang follows the same expert-token alignment pattern. As shown in Table~\ref{tab:sglang-fused-moe-padding-path}, the fused MoE backend derives the top-$k$ routing tensors, selects a Triton MoE kernel configuration containing \texttt{BLOCK\_SIZE\_M}, slices the routing metadata for the current token chunk, and invokes \texttt{moe\_align\_block\_size} before expert GEMM execution. The returned \texttt{num\_tokens\_post\_padded} indicates that the routed expert-token workload has been expanded to a padded block-granular representation.

Table~\ref{tab:sglang-moe-align-block-size} provides the corresponding helper implementation. Similar to vLLM, the function aligns routed tokens to the matrix-multiplication block size, allocates padded token and expert buffers according to \texttt{max\_num\_tokens\_padded}, invokes the backend alignment operator, and returns \texttt{num\_tokens\_post\_pad}. Therefore, SGLang also executes fused MoE computation at expert-token block granularity.

The agreement between vLLM and SGLang shows that the MoE NFP mechanism is not an artifact of a single backend. Instead, both practical fused MoE implementations use the same structural pattern: routed tokens are grouped by expert and padded to block-size multiples before expert GEMM execution.

\begin{table*}[htbp]
\centering

\begin{minipage}{0.82\textwidth}

\noindent{\color{coderule}\rule{\linewidth}{0.45pt}}
\vspace{-0.4em}

\begin{lstlisting}[style=prettyPythonKernel]
topk_weights, topk_ids, _ = topk_output

...

get_config_func = functools.partial(
    try_get_optimal_moe_config,
    w1.shape,
    (w2.shape[0], w2.shape[1], w2.shape[2] - padded_size),
    topk_ids.shape[1],
    config_dtype,
    block_shape=block_shape,
    per_channel_quant=per_channel_quant,
    return_down_config=True,
)

config, (down_config, max_block_m) = get_config_func(M)

...

curr_topk_ids = topk_ids[begin_chunk_idx:end_chunk_idx]

curr_topk_weights = topk_weights[begin_chunk_idx:end_chunk_idx]

...

sorted_token_ids, expert_ids, num_tokens_post_padded = moe_align_block_size(
    curr_topk_ids, config["BLOCK_SIZE_M"], E
)
\end{lstlisting}

\vspace{-0.9em}
\noindent{\color{coderule}\rule{\linewidth}{0.45pt}}

\end{minipage}

\vspace{0.3em}

\caption{
Source-code evidence for expert-token padding in \textbf{SGLang fused MoE}. The backend derives the top-$k$ routing tensors, selects a Triton MoE kernel configuration that includes \texttt{BLOCK\_SIZE\_M}, slices the routing metadata for the current token chunk, and applies \texttt{moe\_align\_block\_size} before expert GEMM execution. The returned \texttt{num\_tokens\_post\_padded} shows that SGLang also pads routed expert-token workloads to a block granularity.
}
\label{tab:sglang-fused-moe-padding-path}

\end{table*}

\begin{table*}[htbp]
\centering

\begin{minipage}{0.82\textwidth}

\noindent{\color{coderule}\rule{\linewidth}{0.45pt}}
\vspace{-0.4em}

\begin{lstlisting}[style=prettyPythonKernel]
def moe_align_block_size(
    topk_ids: torch.Tensor, block_size: int, num_experts: int
) -> Tuple[torch.Tensor, torch.Tensor, torch.Tensor]:
    """
    Aligns the token distribution across experts to be compatible with block
    size for matrix multiplication.
    """

    if topk_ids.numel() < num_experts + 1:
        max_num_tokens_padded = topk_ids.numel() * block_size
    else:
        max_num_tokens_padded = topk_ids.numel() + (num_experts + 1) * (block_size - 1)

    sorted_ids = torch.empty(
        (max_num_tokens_padded,), dtype=torch.int32, device=topk_ids.device
    )

    max_num_m_blocks = triton.cdiv(max_num_tokens_padded, block_size)

    expert_ids = torch.empty(
        (max_num_m_blocks,), dtype=torch.int32, device=topk_ids.device
    )

    num_tokens_post_pad = torch.empty((1), dtype=torch.int32, device=topk_ids.device)

    # In EP, expert_ids for filtered experts are -1. We have num_experts + 1 ids in total.
    cumsum_buffer = torch.empty(
        (num_experts + 2,), dtype=torch.int32, device=topk_ids.device
    )

    sgl_moe_align_block_size(
        topk_ids,
        num_experts + 1,
        block_size,
        sorted_ids,
        expert_ids,
        num_tokens_post_pad,
        cumsum_buffer,
        True,
    )

    return sorted_ids, expert_ids, num_tokens_post_pad
\end{lstlisting}

\vspace{-0.9em}
\noindent{\color{coderule}\rule{\linewidth}{0.45pt}}

\end{minipage}

\vspace{0.3em}

\caption{
Implementation of expert-token block alignment in \textbf{SGLang fused MoE}. The helper function \texttt{moe\_align\_block\_size} explicitly aligns routed tokens to the matrix-multiplication block size, allocates padded token and expert buffers based on \texttt{max\_num\_tokens\_padded}, invokes the backend alignment operator, and returns \texttt{num\_tokens\_post\_pad}. This provides direct evidence that SGLang pads expert-token workloads to a block granularity before fused MoE execution.
}
\label{tab:sglang-moe-align-block-size}

\end{table*}

\clearpage

\subsection{From MoE Expert-Token Alignment to MoE NFP Boundaries}
\label{app:moe-padding-rules}

This section maps the backend kernel-selection rules to the concrete parameter values used in the NFP principle: the padding granularity $M_{\mathrm{moe}}$ and the branch-validity bound $\tau$.
The source-code evidence above shows that fused MoE kernels pad routed expert-token workloads to a row-block granularity before expert GEMM execution. We identify the backend-selected row-block size with the MoE padding granularity used in the NFP principle:
\begin{equation}
M_{\mathrm{moe}} := \texttt{BLOCK\_SIZE\_M}.
\end{equation}
For an expert $e$ with $m_e(N)$ logical routed tokens, the fused MoE kernel executes the padded count
\begin{equation}
\widetilde{m}_e(N)
=
\left\lceil
\frac{m_e(N)}{M_{\mathrm{moe}}}
\right\rceil
M_{\mathrm{moe}} .
\end{equation}
Thus, increasing N remains near-free as long as it does not increase the padded expert-token counts $\widetilde{m}_e(N)$ or trigger a new kernel configuration.

To instantiate $M_{\mathrm{moe}}$ in practice, we inspect the backend rules that select \texttt{BLOCK\_SIZE\_M}. Tables~\ref{tab:vllm-moe-block-size-m} and~\ref{tab:sglang-moe-block-size-m} summarize these source-derived rules for vLLM and SGLang fused MoE fallback kernels. In these tables, $M$ denotes the token dimension seen by the kernel configuration, while $M_{\mathrm{moe}}$ denotes the resulting expert-token padding granularity. For a given backend, version, numerical format or quantization mode, and baseline token shape, the corresponding table determines both the current $M_{\mathrm{moe}}$ and the range of $M$ over which this value remains valid.

This distinction is important for deriving the load-balanced upper bound. Suppose the baseline workload selects $M_{\mathrm{moe}}$ and remains in the same kernel-configuration branch up to token dimension $\tau$. Under load-balanced routing, padding slack is aggregated across activated experts. With $E$ experts and top-$k$ routing, the padding-capacity bound is approximately $M_{\mathrm{moe}}E/k$. However, this bound is valid only while the same \texttt{BLOCK\_SIZE\_M} branch is used. Therefore, the backend-aware MoE upper bound is
\begin{equation}
N_{\max}^{\mathrm{moe,bal}}
\approx
\min\left(
\frac{M_{\mathrm{moe}}E}{k},
\tau,
M_{\mathrm{attn}}
\right).
\end{equation}
For the evaluated vLLM v0.9.1 and SGLang BF16/FP16 fallback configurations, the small-token branch uses $M_{\mathrm{moe}}=16$ and remains valid while $M\le E$. Hence $\tau=E$, giving the specialized form used in our experiments:
\begin{equation}
N_{\max}^{\mathrm{moe,bal}}
\approx
\min\left(
\frac{M_{\mathrm{moe}}E}{k},
E,
M_{\mathrm{attn}}
\right).
\end{equation}

For load-skewed routing, the activated expert set is fixed from the single-position baseline. Larger N can only use the slack within the same selected expert-token blocks, so the MoE-side boundary is fixed by the local padding granularity:
\begin{equation}
N_{\max}^{\mathrm{moe,skew}}
\approx
\min\left(
M_{\mathrm{moe}},
M_{\mathrm{attn}}
\right).
\end{equation}
In this case, the kernel-configuration branch limit is typically non-binding in the small-$N$ regime.

Overall, Tables~\ref{tab:vllm-moe-block-size-m} and~\ref{tab:sglang-moe-block-size-m} provide the practical rule for instantiating $M_{\mathrm{moe}}$. The padding equation determines the available expert-token slack, while the branch-validity bound $\tau$ ensures that the principle is applied within the backend branch that selected the current $M_{\mathrm{moe}}$.

\begin{table}[htbp]
\centering
\renewcommand{\arraystretch}{1.15}
\begin{tabularx}{\linewidth}{>{\raggedright\arraybackslash}p{0.26\linewidth}
                            >{\raggedright\arraybackslash}X
                            >{\centering\arraybackslash}p{0.18\linewidth}}
\toprule
\textbf{vLLM version} & \textbf{Condition} & \textbf{\texttt{BLOCK\_SIZE\_M}} \\
\midrule
v0.9.0--v0.16.0
    & \texttt{M <= E} & 16 \\
    & otherwise & 64 \\
\midrule
v0.17.0--v0.20.2
    & \texttt{M <= 32} & 16 \\
    & \texttt{32 < M <= 96} & 32 \\
    & \texttt{96 < M <= 512} & 64 \\
    & \texttt{M > 512} & 128 \\
\bottomrule
\end{tabularx}

\caption{Source-derived \texttt{BLOCK\_SIZE\_M} selection rules for vLLM fused MoE fallback kernels. Here $M$ denotes the token dimension used by the kernel configuration, and $E$ denotes the number of experts. For a branch selected at the baseline workload, the corresponding \texttt{BLOCK\_SIZE\_M} instantiates the MoE padding granularity $M_{\mathrm{moe}}$, while the branch condition determines the validity range $\tau$ over which this $M_{\mathrm{moe}}$ remains applicable. In the evaluated vLLM v0.9.1 configuration, the small-token branch uses $M_{\mathrm{moe}}=16$ for $M\le E$.
}
\label{tab:vllm-moe-block-size-m}
\end{table}

\begin{table}[htbp]
\centering
\renewcommand{\arraystretch}{1.15}
\begin{tabularx}{\linewidth}{>{\raggedright\arraybackslash}p{0.32\linewidth}
                            >{\raggedright\arraybackslash}X
                            >{\centering\arraybackslash}p{0.18\linewidth}}
\toprule
\textbf{Quantization} &
\textbf{Condition} &
\textbf{\texttt{BLOCK\_SIZE\_M}} \\
\midrule
bf16/fp16
& \texttt{M <= E}
& 16 \\
& \texttt{M > E}
& 64 \\
\midrule
per-tensor FP8
& \texttt{M <= E}
& 64 \\
& \texttt{M > E}
& 128 \\
\midrule
block-wise FP8
& any \texttt{M}
& 64 \\
\bottomrule
\end{tabularx}
\caption{
Source-derived \texttt{BLOCK\_SIZE\_M} selection rules for SGLang Triton fused MoE fallback kernels. Here $M$ denotes the token dimension used by the kernel configuration, and $E$ denotes the number of experts. For a branch selected at the baseline workload, the corresponding \texttt{BLOCK\_SIZE\_M} instantiates the MoE padding granularity $M_{\mathrm{moe}}$, while the branch condition determines the validity range $\tau$ over which this $M_{\mathrm{moe}}$ remains applicable. The selected branch depends on the quantization mode; in the evaluated SGLang BF16/FP16 configuration, the small-token branch uses $M_{\mathrm{moe}}=16$ for $M\le E$.
}
\label{tab:sglang-moe-block-size-m}
\end{table}

\clearpage

\section{Implementation Evidence for Attention Query-Tile Granularity}
\label{app:attention-query-tile}

The Attention results in Appendix~\ref{app:attention-results} show staircase latency, arithmetic intensity, and runtime FLOPs, while the measured NFP boundary remains largely independent of the cached sequence length $L$. This appendix provides source-level evidence for this behavior. We show that practical attention backends execute query positions in backend-defined query tiles. This query-tile granularity allows larger N to remain within an existing tile until the next tile boundary is crossed.

\subsection{FlashAttention Query-Tile Scheduling}
\label{app:attention-query-tile-flashattention}

Tables~\ref{tab:flashattn-query-tile-kernel} and~\ref{tab:flashattn-query-tile-launch} show how FlashAttention implements query-tile execution. On the kernel side, Table~\ref{tab:flashattn-query-tile-kernel} shows that the device function receives a query-tile index \texttt{m\_block}, obtains \texttt{kBlockM} from \texttt{Kernel\_traits}, and uses it to tile the query tensor with shape \texttt{(kBlockM, kHeadDim)}. This indicates that each FlashAttention query tile covers up to \texttt{kBlockM} query positions.

On the launch side, Table~\ref{tab:flashattn-query-tile-launch} shows that the backend computes the number of query tiles by applying ceiling division to \texttt{seqlen\_q} with respect to \texttt{kBlockM}. The resulting tile count is used as a kernel grid dimension, and the launch path also checks whether \texttt{seqlen\_q} is aligned to \texttt{kBlockM}. Thus, the amount of launched attention work changes at query-tile boundaries.

Together, these two code paths show that increasing the number of decode positions does not continuously change the executed FlashAttention workload. Larger N remains within the same query tile until the query length crosses the next \texttt{kBlockM}-aligned boundary, at which point an additional query tile is launched. Therefore, \texttt{kBlockM} is the FlashAttention instantiation of the backend query-tile granularity $M_{\mathrm{attn}}$.

\clearpage
\begin{table*}[htbp]
\centering

\begin{minipage}{0.82\textwidth}

\noindent{\color{coderule}\rule{\linewidth}{0.45pt}}
\vspace{-0.4em}

\begin{lstlisting}[style=prettyCppKernel]
template<typename Kernel_traits, bool Is_dropout, bool Is_causal, bool Is_local, bool Has_alibi, bool Is_even_MN, bool Is_even_K, bool Is_softcap, bool Return_softmax, typename Params>
inline __device__ void compute_attn_1rowblock(const Params &params, const int bidb, const int bidh, const int m_block) {
    using Element = typename Kernel_traits::Element;
    using ElementAccum = typename Kernel_traits::ElementAccum;
    using index_t = typename Kernel_traits::index_t;

    // Shared memory.
    extern __shared__ char smem_[];

    // The thread index.
    const int tidx = threadIdx.x;

    constexpr int kBlockM = Kernel_traits::kBlockM;
    constexpr int kBlockN = Kernel_traits::kBlockN;
    constexpr int kHeadDim = Kernel_traits::kHeadDim;

    ...

    Tensor mQ = make_tensor(make_gmem_ptr(reinterpret_cast<Element*>(params.q_ptr)
                         + binfo.q_offset(params.q_batch_stride, params.q_row_stride, bidb)),
                            make_shape(binfo.actual_seqlen_q, params.h, params.d),
                            make_stride(params.q_row_stride, params.q_head_stride, _1{}));

    Tensor gQ = local_tile(mQ(_, bidh, _), Shape<Int<kBlockM>, Int<kHeadDim>>{},
                           make_coord(m_block, 0));  // (kBlockM, kHeadDim)
\end{lstlisting}

\vspace{-0.9em}
\noindent{\color{coderule}\rule{\linewidth}{0.45pt}}

\end{minipage}

\vspace{0.3em}

\caption{
FlashAttention kernel-side query-tile execution. The device function receives the query-tile index \texttt{m\_block}, loads the query-tile height \texttt{kBlockM} from \texttt{Kernel\_traits}, and tiles the query tensor into \texttt{(kBlockM, kHeadDim)} regions. This shows that each query tile covers up to \texttt{kBlockM} query positions.
}
\label{tab:flashattn-query-tile-kernel}

\end{table*}
\begin{table*}[htbp]
\centering

\begin{minipage}{0.82\textwidth}

\noindent{\color{coderule}\rule{\linewidth}{0.45pt}}
\vspace{-0.4em}

\begin{lstlisting}[style=prettyCppKernel]
const int num_m_block = (params.seqlen_q + Kernel_traits::kBlockM - 1) / Kernel_traits::kBlockM;

dim3 grid(num_m_block, params.b, params.h);

const bool is_even_MN = params.cu_seqlens_q == nullptr && params.cu_seqlens_k == nullptr && params.seqlen_k % Kernel_traits::kBlockN == 0 && params.seqlen_q % Kernel_traits::kBlockM == 0;
\end{lstlisting}

\vspace{-0.9em}
\noindent{\color{coderule}\rule{\linewidth}{0.45pt}}

\end{minipage}

\vspace{0.3em}
\caption{
FlashAttention launch-side query-tile scheduling. The launch template computes the number of query tiles by applying ceiling division to \texttt{seqlen\_q} with respect to \texttt{kBlockM}, uses the resulting \texttt{num\_m\_block} as a grid dimension, and checks whether the query length is aligned to \texttt{kBlockM}.
}
\label{tab:flashattn-query-tile-launch}

\end{table*}

\clearpage

\subsection{FlashInfer Query-Tile Scheduling}
\label{app:attention-query-tile-flashinfer}

Tables~\ref{tab:flashinfer-query-tile-kernel} and~\ref{tab:flashinfer-query-tile-scheduler} show how FlashInfer implements query-tile execution. On the kernel side, Table~\ref{tab:flashinfer-query-tile-kernel} shows that \texttt{CTA\_TILE\_Q} is part of the kernel traits and is used to size the shared-memory buffers for query, output, and intermediate attention states. This indicates that \texttt{CTA\_TILE\_Q} defines the number of query rows covered by one FlashInfer query tile.

On the scheduler side, Table~\ref{tab:flashinfer-query-tile-scheduler} shows that FlashInfer selects a concrete \texttt{cta\_tile\_q}, computes the number of query tiles by applying ceiling division to \texttt{packed\_qo\_len}, and then enumerates \texttt{q\_tile\_idx} for execution. Thus, the scheduled attention work changes at query-tile boundaries rather than continuously with the logical query length.

Together, these two code paths show that larger N remains within the same FlashInfer query tile until the query workload crosses the next \texttt{CTA\_TILE\_Q}-aligned boundary. At that point, an additional query tile is scheduled. Therefore, \texttt{CTA\_TILE\_Q} is the FlashInfer instantiation of the backend query-tile granularity $M_{\mathrm{attn}}$.

\begin{table*}[htbp]
\centering

\begin{minipage}{0.9\textwidth}

\noindent{\color{coderule}\rule{\linewidth}{0.45pt}}
\vspace{-0.4em}

\begin{lstlisting}[style=prettyCppKernel]
template <uint32_t NUM_WARPS_KV, uint32_t CTA_TILE_Q, uint32_t CTA_TILE_KV, uint32_t HEAD_DIM_QK,
          uint32_t HEAD_DIM_VO, typename DTypeQ, typename DTypeKV, typename DTypeO>
struct SharedStorageQKVO {
  union {
    struct {
      alignas(16) DTypeQ q_smem[CTA_TILE_Q * HEAD_DIM_QK];
      ...
    };
    struct {  // NOTE(Zihao): synchronize attention states across warps
      alignas(
          16) std::conditional_t<NUM_WARPS_KV == 1, float[1],
                                 float[NUM_WARPS_KV * CTA_TILE_Q * HEAD_DIM_VO]> cta_sync_o_smem;
      alignas(16) std::conditional_t<NUM_WARPS_KV == 1, float2[1],
                                     float2[NUM_WARPS_KV * CTA_TILE_Q]> cta_sync_md_smem;
    };
    alignas(16) DTypeO smem_o[CTA_TILE_Q * HEAD_DIM_VO];
  };

  ...
};

template <MaskMode MASK_MODE_, uint32_t CTA_TILE_Q_, uint32_t NUM_MMA_Q_, uint32_t NUM_MMA_KV_,
          uint32_t NUM_MMA_D_QK_, uint32_t NUM_MMA_D_VO_, uint32_t NUM_WARPS_Q_,
          uint32_t NUM_WARPS_KV_, PosEncodingMode POS_ENCODING_MODE_, typename DTypeQ_,
          typename DTypeKV_, typename DTypeO_, typename DTypeQKAccum_, typename IdType_,
          typename AttentionVariant_>
struct KernelTraits {
  ...

  static constexpr uint32_t CTA_TILE_Q = CTA_TILE_Q_;
  ...

  using SharedStorage = SharedStorageQKVO<NUM_WARPS_KV, CTA_TILE_Q, CTA_TILE_KV, HEAD_DIM_QK,
                                          HEAD_DIM_VO, DTypeQ, DTypeKV, DTypeO>;

  ...
};
\end{lstlisting}

\vspace{-0.9em}
\noindent{\color{coderule}\rule{\linewidth}{0.45pt}}

\end{minipage}

\vspace{0.3em}

\caption{
FlashInfer kernel-side query-tile definition. The kernel traits expose \texttt{CTA\_TILE\_Q} as the query-tile size, and the shared-memory layout allocates query, output, and synchronization buffers with dimensions proportional to \texttt{CTA\_TILE\_Q}. This shows that each FlashInfer query tile covers up to \texttt{CTA\_TILE\_Q} query rows.
}
\label{tab:flashinfer-query-tile-kernel}

\end{table*}
\begin{table*}[htbp]
\centering
\begin{minipage}{0.82\textwidth}

\noindent{\color{coderule}\rule{\linewidth}{0.45pt}}
\vspace{-0.4em}

\begin{lstlisting}[style=prettyCppKernel]
template <typename IdType>
inline auto PrefillSplitQOKVIndptr(IdType* qo_indptr_h, IdType* kv_indptr_h,
                                   uint32_t total_num_rows, uint32_t batch_size,
                                   uint32_t num_qo_heads, uint32_t num_kv_heads, uint32_t head_dim,
                                   uint32_t page_size, uint32_t max_batch_size_if_split,
                                   bool enable_cuda_graph, int32_t window_left,
                                   int32_t fixed_split_size, bool disable_split_kv) {
  ...
  const uint32_t gqa_group_size = num_qo_heads / num_kv_heads;

  std::vector<int64_t> packed_qo_len_arr(batch_size), kv_len_arr(batch_size);
  for (uint32_t i = 0; i < batch_size; ++i) {
    packed_qo_len_arr[i] = int64_t(qo_indptr_h[i + 1] - qo_indptr_h[i]) * int64_t(gqa_group_size);

    ...
  }

  ...

  uint32_t cta_tile_q;
  ...
  cta_tile_q = FA2DetermineCtaTileQ(avg_packed_qo_len, head_dim);

  ...

  for (uint32_t request_idx = 0; request_idx < batch_size; ++request_idx) {
    const int64_t packed_qo_len = packed_qo_len_arr[request_idx];
    const int64_t num_tiles_q = ceil_div(packed_qo_len, cta_tile_q);

    ...

    for (uint32_t q_tile_idx = 0; q_tile_idx < num_tiles_q; ++q_tile_idx) {
      ...
      qo_tile_indices.push_back(q_tile_idx);
      ...
    }

    ...
  }

  ...
}
\end{lstlisting}

\vspace{-0.9em}
\noindent{\color{coderule}\rule{\linewidth}{0.45pt}}

\end{minipage}

\vspace{0.3em}

\caption{
FlashInfer scheduler-side query-tile scheduling. The scheduler selects a concrete \texttt{cta\_tile\_q}, computes the number of query tiles by applying ceiling division to \texttt{packed\_qo\_len}, and enumerates query-tile indices \texttt{q\_tile\_idx} for execution.
}
\label{tab:flashinfer-query-tile-scheduler}

\end{table*}

\clearpage

\subsection{From Query-Tile Granularity to Attention NFP Boundaries}
\label{app:attention-query-tile-rules}

The source-code evidence in Appendix~\ref{app:attention-query-tile-flashattention} and Appendix~\ref{app:attention-query-tile-flashinfer} shows that practical attention backends execute query positions in backend-defined query tiles. We use the query-tile size selected by the backend as the concrete implementation of the Attention granularity in the NFP principle:
\begin{equation}
M_{\mathrm{attn}}
:=
\begin{cases}
\texttt{kBlockM}, & \text{FlashAttention}, \\
\texttt{CTA\_TILE\_Q}, & \text{FlashInfer}.
\end{cases}
\end{equation}
Thus, $M_{\mathrm{attn}}$ is not a model-intrinsic constant. It is determined by the attention backend, kernel path, architecture, head dimension, and other kernel-configuration conditions.

For a decode forward with $N$ query positions, the backend executes query tiles rather than individual query positions. The number of executed query tiles is
\begin{equation}
n_{\mathrm{tile}}(N)
=
\left\lceil
\frac{N}{M_{\mathrm{attn}}}
\right\rceil ,
\end{equation}
and the corresponding tiled query capacity is
\begin{equation}
\widetilde{N}
=
n_{\mathrm{tile}}(N)M_{\mathrm{attn}}
=
\left\lceil
\frac{N}{M_{\mathrm{attn}}}
\right\rceil
M_{\mathrm{attn}} .
\end{equation}
Therefore, runtime work changes at query-tile boundaries rather than continuously with the logical query count. Increasing N remains near-free as long as it does not increase $n_{\mathrm{tile}}(N)$; once $N$ crosses the next $M_{\mathrm{attn}}$-aligned boundary, an additional query tile is executed.

To instantiate $M_{\mathrm{attn}}$ in practice, we inspect the backend rules that select the query-tile size. Tables~\ref{tab:fa2-kblockm}, \ref{tab:fa3-kblockm}, and~\ref{tab:flashinfer-cta-tile-q} summarize the source-derived rules for FlashAttention-2, FlashAttention-3, and FlashInfer, respectively. These rules specify which \texttt{kBlockM} or \texttt{CTA\_TILE\_Q} is selected under a given backend configuration, and therefore determine the $M_{\mathrm{attn}}$ used by the NFP principle.

These rules provide a practical procedure for determining the Attention-side NFP boundary. Given a backend and kernel configuration, we first identify the selected query-tile size and instantiate $M_{\mathrm{attn}}$. We then determine the largest N that can be used before the query workload crosses the next tile boundary. Starting from the single-position baseline, the Attention-side boundary is therefore approximated by
\begin{equation}
N_{\max}^{\mathrm{attn}}
\approx
M_{\mathrm{attn}} .
\end{equation}

This rule explains why the measured Attention boundary is largely independent of the cached sequence length $L$. Unlike the KV-cache idle-compute baseline, which predicts an $L$-dependent boundary, the realized boundary is controlled by backend query-tile granularity. Consequently, Attention contributes the implementation-dependent term $M_{\mathrm{attn}}$ to the model-level NFP principle, constraining both Dense and MoE models whenever the attention module is the first component to leave the near-free regime.

\clearpage

\subsubsection{FlashAttention-2 Query-Tile Rules}
\begin{table}[htbp]
\centering
\renewcommand{\arraystretch}{1.15}
\begin{tabularx}{\linewidth}{>{\raggedright\arraybackslash}p{0.24\linewidth}
                            >{\raggedright\arraybackslash}X
                            >{\centering\arraybackslash}p{0.16\linewidth}}
\toprule
\textbf{Kernel path} & \textbf{Condition} & \textbf{\texttt{kBlockM}} \\
\midrule
Non-split-KV forward & \texttt{headdim = 32} & 128 \\
                     & \texttt{headdim = 64} & 128 \\
                     & \texttt{headdim = 96}, SM8x and causal & 64 \\
                     & \texttt{headdim = 96}, otherwise & 128 \\
                     & \texttt{headdim = 128}, no dropout, SM8x and causal & 64 \\
                     & \texttt{headdim = 128}, otherwise & 128 \\
                     & \texttt{headdim = 192}, no dropout & 128 \\
                     & \texttt{headdim = 192}, with dropout & 64 \\
                     & \texttt{headdim = 256}, high shared-memory satisfied & 128 \\
                     & \texttt{headdim = 256}, otherwise & 64 \\
\midrule
Split-KV forward     & Main split-KV attention kernel & 64 \\
\bottomrule
\end{tabularx}

\caption{
Source-derived \texttt{kBlockM} selection rules for FlashAttention-2 forward kernels. Here \texttt{kBlockM} denotes the query-tile size used by the selected kernel path and therefore instantiates the Attention granularity $M_{\mathrm{attn}}$ in the NFP principle. The selected value depends on the forward kernel path, head dimension, architecture, causal mode, dropout, and shared-memory constraints.
}
\label{tab:fa2-kblockm}

\end{table}

\subsubsection{FlashAttention-3 Query-Tile Rules}
\begin{table}[htbp]
\centering
\renewcommand{\arraystretch}{1.15}
\begin{tabularx}{\linewidth}{>{\raggedright\arraybackslash}p{0.24\linewidth}
                            >{\raggedright\arraybackslash}X
                            >{\centering\arraybackslash}p{0.16\linewidth}}
\toprule
\textbf{Kernel path} & \textbf{Condition} & \textbf{\texttt{kBlockM}} \\
\midrule
SM90, BF16/FP16
    & \texttt{headdim $\leq$ 64} and \texttt{headdim\_v = 512} & 64 \\
    & \texttt{headdim $\leq$ 64} and \texttt{headdim\_v = 256} & 128 \\
    & \texttt{headdim $\leq$ 64} and other \texttt{headdim\_v} & 192 \\
    & \texttt{64 < headdim $\leq$ 96} & 192 \\
    & \texttt{96 < headdim $\leq$ 128} & 128 \\
    & \texttt{128 < headdim $\leq$ 192} & 128 \\
    & \texttt{headdim > 192} & 128 \\
\midrule
SM90, FP8 / non-16-bit
    & \texttt{headdim $\leq$ 64} & 192 \\
    & \texttt{64 < headdim $\leq$ 96} & 192 \\
    & \texttt{96 < headdim $\leq$ 128} & 128 \\
    & \texttt{128 < headdim $\leq$ 192} & 128 \\
    & \texttt{headdim > 192} & 128 \\
\midrule
SM8x, BF16/FP16
    & All listed \texttt{headdim} ranges & 128 \\
\midrule
SM8x, FP8 / non-16-bit
    & All listed \texttt{headdim} ranges & 128 \\
\bottomrule
\end{tabularx}
\caption{
Source-derived \texttt{kBlockM} selection rules for FlashAttention-3 forward kernels. The selected \texttt{kBlockM} defines the query-tile size and instantiates $M_{\mathrm{attn}}$ for FlashAttention-3. The rule depends on the GPU architecture, numerical format, head dimension, value-head dimension, and kernel path.
}
\label{tab:fa3-kblockm}

\end{table}

\clearpage

\subsubsection{FlashInfer Query-Tile Rules}
FlashInfer determines the query-tile size through a scheduler-side selector.
Let $x$ denote the packed query/output workload used by the scheduler to select
\texttt{CTA\_TILE\_Q}. Intuitively, $x$ summarizes the effective query-side length
seen by the scheduler after packing and grouped-query attention. The selected
\texttt{CTA\_TILE\_Q} then defines the FlashInfer query-tile size and instantiates
$M_{\mathrm{attn}}$ in the NFP principle. This connects the backend rule to NFP:
as the number of decode positions $N$ increases, the packed query/output workload
seen by the scheduler also increases, which may change the selected query-tile size
or the number of executed query tiles.

The array \texttt{qo\_indptr} is the index-pointer array for the packed
query/output tensor. The difference
\(\texttt{qo\_indptr}_{i+1}-\texttt{qo\_indptr}_{i}\) gives the number of
query/output rows contributed by request $i$. In the uniform multi-position
decode setting used in our analysis, each request contributes $N$ query positions,
so this difference is $N$.

FlashInfer uses different definitions of $x$ depending on whether CUDA Graph execution is enabled.

\begin{itemize}
    \item \textbf{CUDA Graph path.}
    The selector uses the maximum packed query/output length:
    \begin{equation}
    x = \texttt{max\_qo\_len}
      = (\texttt{total\_num\_rows} - b + 1)
        \cdot \texttt{gqa\_group\_size}.
    \end{equation}
    In our multi-position decode setting, \texttt{block\_size} corresponds to the
    number of query positions per request, i.e., $N$. Therefore, when
    \(\texttt{total\_num\_rows}=b\cdot N\), the selector input becomes
    \begin{equation}
    x =
    \bigl(b(N-1)+1\bigr)
    \cdot \texttt{gqa\_group\_size}.
    \end{equation}

    \item \textbf{Non-CUDA Graph path.}
    The selector uses the batch-average packed query/output length:
    \begin{equation}
    x = \texttt{avg\_packed\_qo\_len}
      = \frac{1}{b}
        \sum_{i=0}^{b-1}
        \bigl(\texttt{qo\_indptr}_{i+1}
        - \texttt{qo\_indptr}_{i}\bigr)
        \cdot \texttt{gqa\_group\_size}.
    \end{equation}
In our uniform multi-position decode setting, each request contributes $N$
query positions, so
\(\texttt{qo\_indptr}_{i+1}-\texttt{qo\_indptr}_{i}=N\). After multiplication
by \texttt{gqa\_group\_size}, the scheduler's packed query/output length scales
linearly with $N$.
\end{itemize}

\begin{table}[htbp]
\centering
\renewcommand{\arraystretch}{1.15}
\begin{tabularx}{0.82\linewidth}{>{\raggedright\arraybackslash}X
                            >{\centering\arraybackslash}p{0.24\linewidth}}
\toprule
\textbf{Condition} & \textbf{\texttt{CTA\_TILE\_Q}} \\
\midrule
\(x > 64\) and \(\texttt{head\_dim} < 256\) & 128 \\
Otherwise, Ampere or newer, and \(x > 16\) & 64 \\
Otherwise, Ampere or newer, and \(x \leq 16\) & 16 \\
Otherwise, Turing / pre-Ampere & 64 \\
\bottomrule
\end{tabularx}
\caption{
Source-derived \texttt{CTA\_TILE\_Q} selection rules for FlashInfer attention kernels. The scheduler computes a selector input $x$ from the packed query/output workload and uses it to select \texttt{CTA\_TILE\_Q}, which defines the FlashInfer query-tile size and instantiates $M_{\mathrm{attn}}$ in the NFP principle.
}
\label{tab:flashinfer-cta-tile-q}
\end{table}
\clearpage

\clearpage

\section{Model-Level Validation Implementation Details}
\label{sec:model_valid_setup}

\subsection{Overall Validation Protocol}

We validate the NFP principle at the full-model level by measuring the latency of a complete decode forward. For each model and serving configuration, a decode forward processes $N$ decode positions over a pre-allocated KV cache. The measured latency includes all transformer layers and the corresponding serving backend kernels, but excludes prefill, tokenization, sampling, data loading, and CPU-side preprocessing. This scope matches the system-side quantity modeled by the NFP principle while avoiding unrelated request-level overheads.

For each configuration, we sweep the number of decode positions $N$ and repeat the measurement 10 times. The measured boundary is extracted using the same criterion as in the module-level analysis:
\begin{equation}
    N_{\max}=N_{\max}(0.2).
\end{equation}
That is, $N_{\max}$ is the largest $N$ whose decode-forward latency remains within a 20\% increase over the baseline latency. Using the same tolerance keeps the full-model validation directly comparable with the module-level results and avoids introducing a separate boundary definition.

All full-model experiments are conducted on the same three single-GPU platforms used in the module-level analysis: NVIDIA H20, A800, and H800. These platforms cover different compute-to-memory balance points $\rho$, allowing us to test whether the full-model boundary follows the hardware-sensitive terms in the NFP principle while avoiding inter-GPU communication effects. Table~\ref{tab:hardware_platforms} summarizes the hardware platforms. 
For each setting, the predicted boundary is computed directly from the NFP principle using the model configuration, hardware balance point $\rho$, and backend granularity parameters. No fitted parameter is introduced in the model-level validation.

In addition to DLLMs, we validate the NFP principle on autoregressive (AR) models. At the abstraction level modeled by the NFP principle, the verification forward in speculative decoding or multi-token prediction shares the same multi-position decode paradigm as a DLLM forward: both process multiple decode positions through the transformer stack in a single pass, and the NFP boundary is determined by the same module-level granularity and resource-balance constraints. The principle is therefore expected to apply regardless of how the candidate decode positions are generated. AR validation is conducted on NVIDIA H800. Cross-platform generalization has already been established in the DLLM setting across H20, A800, and H800; the AR experiments target paradigm-independence rather than repeating hardware-generalization evidence.

\subsection{Dense Model Validation Implementation Details}
\label{sec:model_valid_dense}

For dense-model validation, we use WeDLM-8B, a representative dense DLLM. The model has 36 transformer layers, hidden size $d_{\mathrm{model}}=4096$, intermediate size $d_{\mathrm{ff}}=12288$, 32 attention heads, 8 key-value heads, and head dimension 128. The model is executed in BF16 precision.

We serve WeDLM-8B using a nano-vLLM-based serving stack with FlashAttention-2 as the attention backend. For each hardware platform and batch size, the dense-model prediction is instantiated as
\begin{equation}
    N^{\mathrm{dense}}_{\max}
    \approx
    \min\left(\frac{\rho s}{2b}, M_{\mathrm{attn}}\right),
\end{equation}
where $s=2$ for BF16, $b$ is the serving batch size, and $M_{\mathrm{attn}}$ is the query granularity of the FlashAttention-2 backend.

We evaluate sequence lengths $L\in\{128,256,384,512\}$. These lengths are chosen to test the prediction across different cached-context sizes while keeping all experiments within a single-GPU setting. For dense models, the primary hardware-sensitive term is the Dense FFN term $\rho s/2b$; therefore, we focus on varying batch size and hardware balance rather than pushing to very long contexts. Longer contexts at larger batch sizes exceed the single-GPU memory budget for this model and would require multi-GPU execution, introducing communication overheads that are outside the scope of this validation.

For AR verification, we use Qwen3-8B with the same architecture and serving configuration as WeDLM-8B. The evaluation parameters (batch sizes, sequence lengths) are kept identical, and the experiment is conducted on NVIDIA H800.

\clearpage

\subsection{MoE Model Validation Implementation Details}
\label{sec:model_valid_moe}

For MoE-model validation, we use LLaDA-2.1-mini, a representative MoE DLLM. The model has 20 transformer layers, hidden size $d_{\mathrm{model}}=2048$, intermediate size 5120, 16 attention heads, 4 key-value heads, and head dimension 128. The MoE layers contain $E=256$ experts, select $k=8$ experts per token, and use MoE intermediate size 512. The model is executed in BF16 precision.

We serve LLaDA-2.1-mini using SGLang with FlashInfer as the attention backend and the SGLang fused MoE backend for expert execution. This serving stack is used because the MoE NFP principle depends explicitly on both the fused-MoE expert-token padding granularity and the attention query granularity. For each setting, the predicted boundary is computed from the MoE NFP principle. For the load-balanced case, we use
\begin{equation}
    N^{\mathrm{moe,bal}}_{\max}
    \approx
    \min\left(\frac{M_{\mathrm{moe}}E}{k}, E, M_{\mathrm{attn}}\right).
\end{equation}
For the load-skewed case, we use
\begin{equation}
    N^{\mathrm{moe,skew}}_{\max}
    \approx
    \min\left(M_{\mathrm{moe}}, M_{\mathrm{attn}}\right).
\end{equation}
Here $M_{\mathrm{moe}}$ is the expert-token padding granularity of the fused MoE backend, and $M_{\mathrm{attn}}$ is the query granularity of the FlashInfer attention backend. For this evaluated backend configuration, the branch-validity bound satisfies \(\tau=E\), so the backend-aware form in Section~\ref{sec:nfp_law} reduces to the specialized formula used here.

As in the module-level MoE analysis, we use controlled routing patterns rather than input-dependent routing. The load-balanced pattern distributes tokens across experts to expose the aggregate expert-token padding slack, while the load-skewed pattern routes all tokens to the same selected experts to expose the lower-bound granularity-limited case. This keeps the expert-load distribution deterministic and makes the full-model boundary comparable with the module-level MoE principle.

We evaluate sequence lengths $L\in\{256,4096,16384,32768\}$. These settings cover short, medium, long, and very long contexts for the MoE model. The purpose of this sweep is to test whether the predicted limiting module changes with context length: at shorter contexts the boundary is expected to be dominated by MoE FFN granularity, whereas at longer contexts the increased attention workload can make the attention backend granularity the limiting factor.

For AR verification, we use Ling-2.0-mini with the same architecture and serving configuration as LLaDA-2.1-mini. The evaluation parameters (top-$k$ sweep, sequence lengths) are kept identical, and the experiment is conducted on NVIDIA H800.

\clearpage

\section{Model-Level Validation Results}
\label{app:model-level-results}

This section reports the full-model validation results of the NFP principle. The goal is to test whether the module-level constraints identified in the main text compose into accurate full-model predictions, and whether the principle can identify the active limiting module across sequence lengths, routing patterns, and GPU platforms.

\subsection{Dense Model Results}
\label{app:dense-model-results}

\begin{figure*}[htbp]
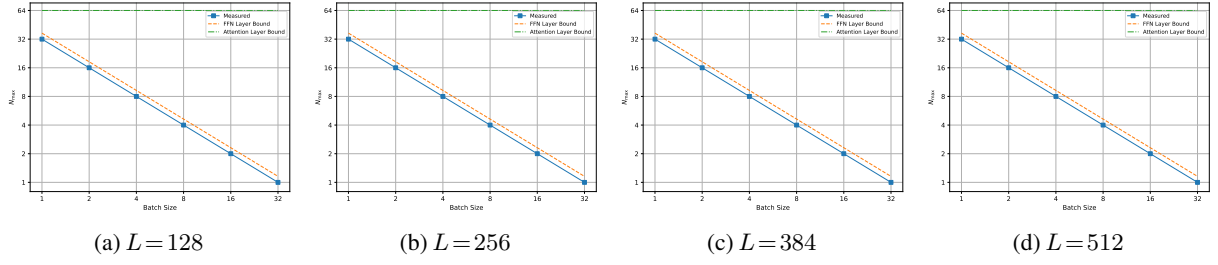

    \centering

    \begin{subfigure}[t]{0.24\textwidth}
        \centering
        \includegraphics[width=\linewidth]{figs/Dense_Model/H20/dense_seq128.pdf}
        \caption{$L=128$}
        \label{fig:dense-model-h20-seq128}
    \end{subfigure}
    \hfill
    \begin{subfigure}[t]{0.24\textwidth}
        \centering
        \includegraphics[width=\linewidth]{figs/Dense_Model/H20/dense_seq256.pdf}
        \caption{$L=256$}
        \label{fig:dense-model-h20-seq256}
    \end{subfigure}
    \hfill
    \begin{subfigure}[t]{0.24\textwidth}
        \centering
        \includegraphics[width=\linewidth]{figs/Dense_Model/H20/dense_seq384.pdf}
        \caption{$L=384$}
        \label{fig:dense-model-h20-seq384}
    \end{subfigure}
    \hfill
    \begin{subfigure}[t]{0.24\textwidth}
        \centering
        \includegraphics[width=\linewidth]{figs/Dense_Model/H20/dense_seq512.pdf}
        \caption{$L=512$}
        \label{fig:dense-model-h20-seq512}
    \end{subfigure}

    \caption{
    Dense model-level NFP principle validation results on \textbf{NVIDIA H20} (WeDLM-8B).
    }
    \label{fig:dense-model-h20}
\end{figure*}

\begin{figure*}[htbp]
    \centering

    \begin{subfigure}[t]{0.24\textwidth}
        \centering
        \includegraphics[width=\linewidth]{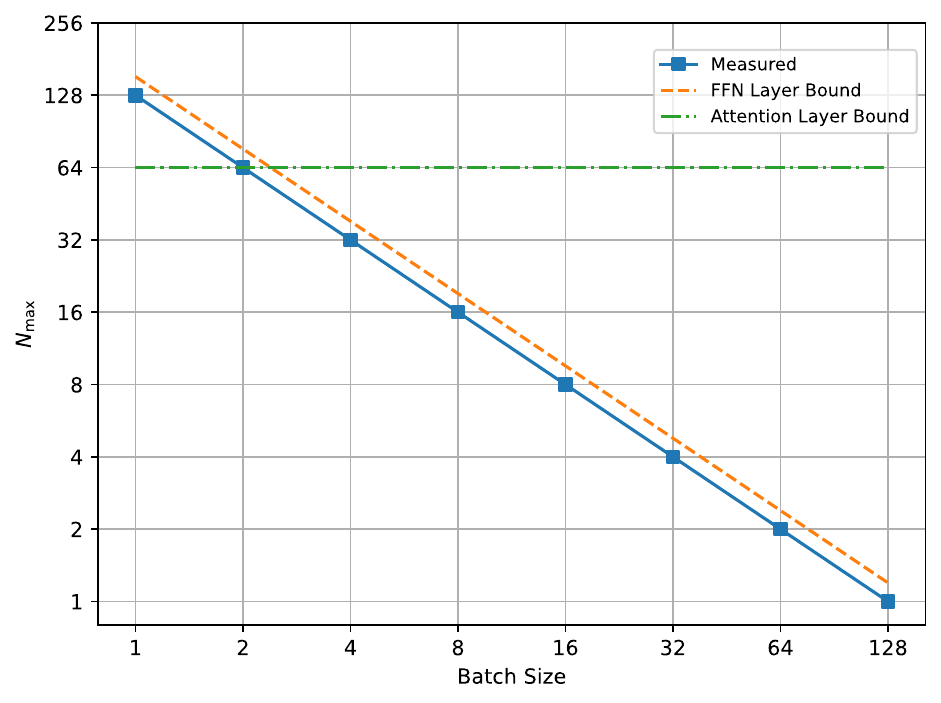}
        \caption{$L=128$}
        \label{fig:dense-model-a800-seq128}
    \end{subfigure}
    \hfill
    \begin{subfigure}[t]{0.24\textwidth}
        \centering
        \includegraphics[width=\linewidth]{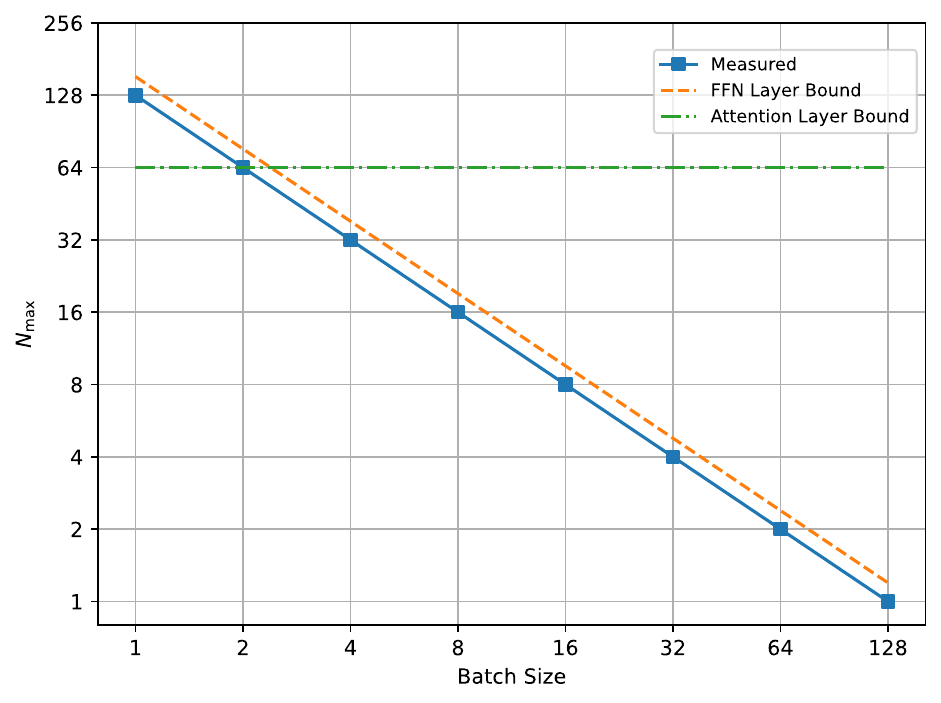}
        \caption{$L=256$}
        \label{fig:dense-model-a800-seq256}
    \end{subfigure}
    \hfill
    \begin{subfigure}[t]{0.24\textwidth}
        \centering
        \includegraphics[width=\linewidth]{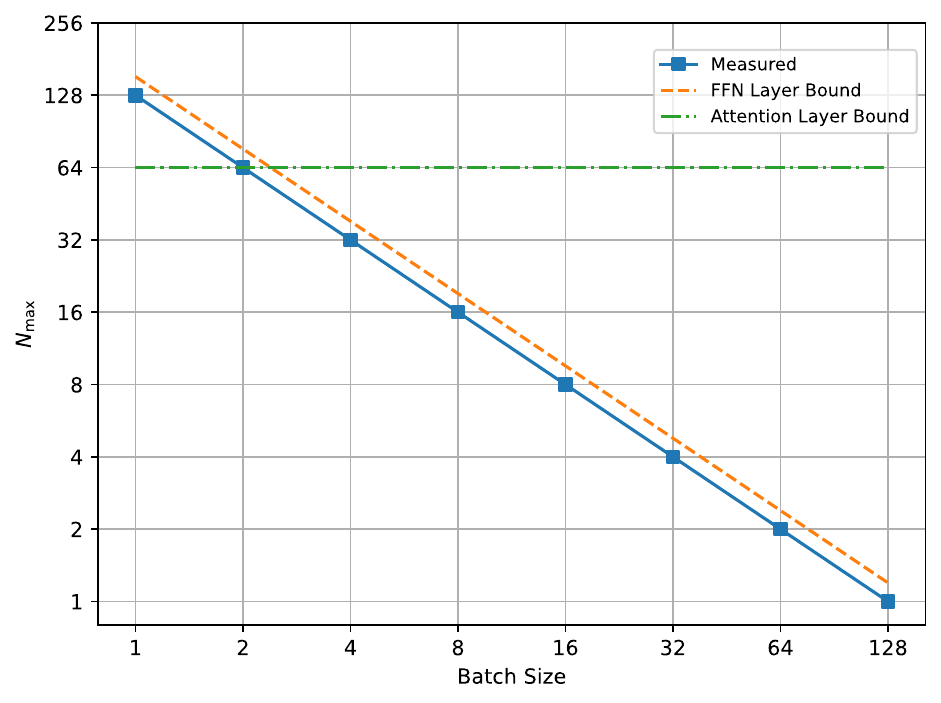}
        \caption{$L=384$}
        \label{fig:dense-model-a800-seq384}
    \end{subfigure}
    \hfill
    \begin{subfigure}[t]{0.24\textwidth}
        \centering
        \includegraphics[width=\linewidth]{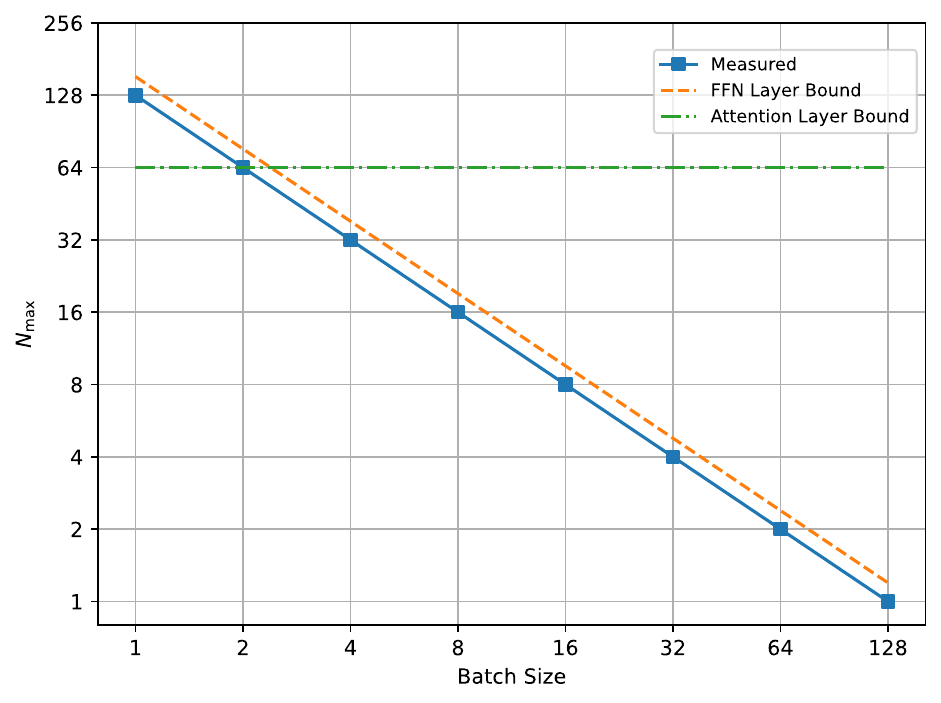}
        \caption{$L=512$}
        \label{fig:dense-model-a800-seq512}
    \end{subfigure}

    \caption{
    Dense model-level NFP principle validation results on \textbf{NVIDIA A800} (WeDLM-8B).
    }
    \label{fig:dense-model-a800}
\end{figure*}

\begin{figure*}[htbp]
    \centering

    \begin{subfigure}[t]{0.24\textwidth}
        \centering
        \includegraphics[width=\linewidth]{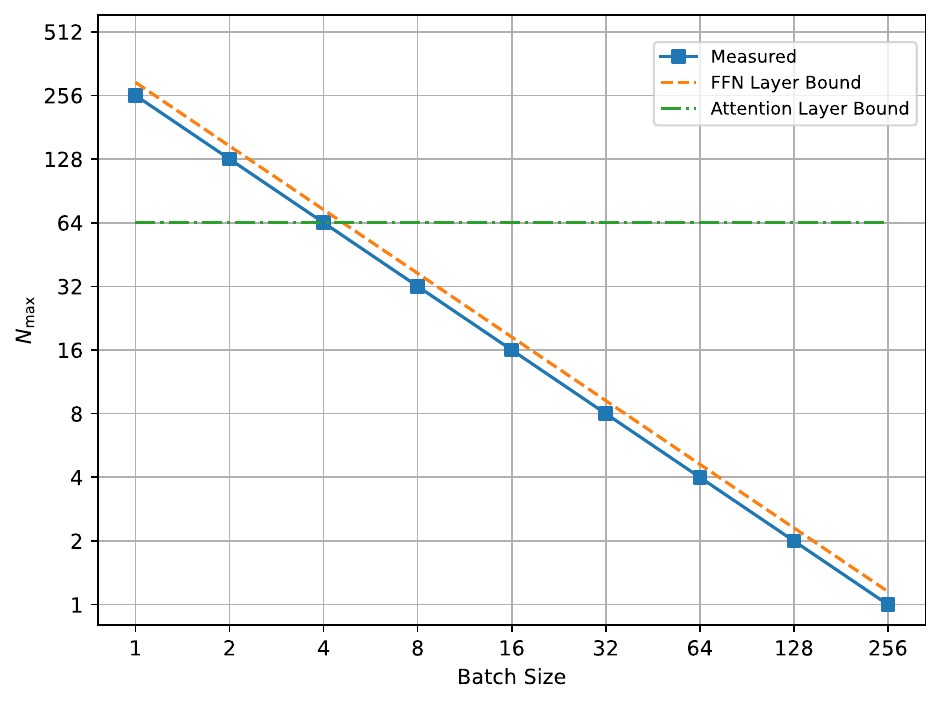}
        \caption{$L=128$}
        \label{fig:dense-model-h800-seq128}
    \end{subfigure}
    \hfill
    \begin{subfigure}[t]{0.24\textwidth}
        \centering
        \includegraphics[width=\linewidth]{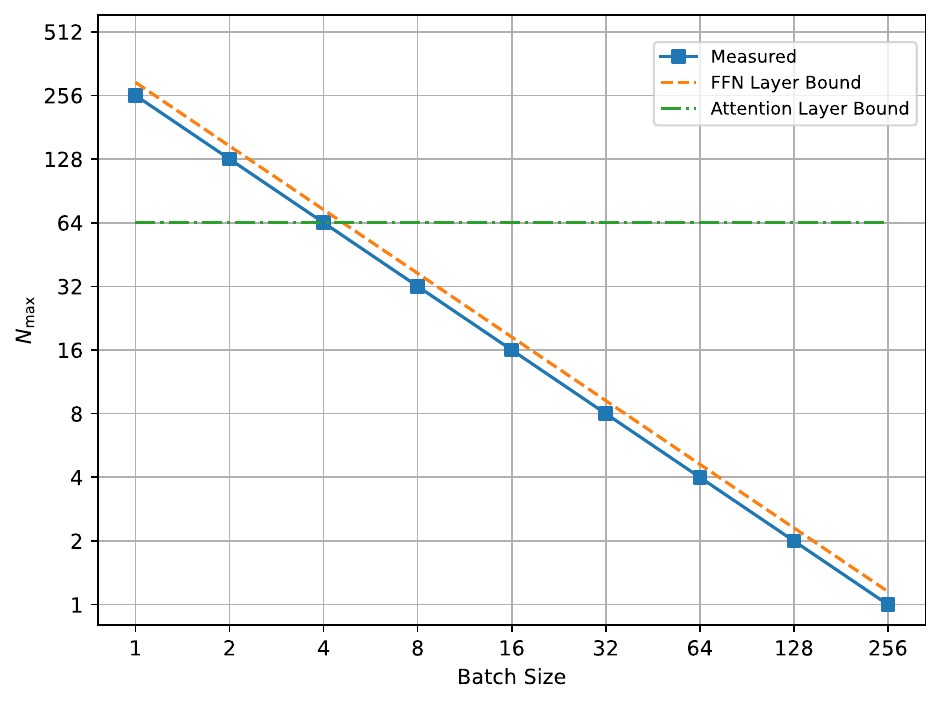}
        \caption{$L=256$}
        \label{fig:dense-model-h800-seq256}
    \end{subfigure}
    \hfill
    \begin{subfigure}[t]{0.24\textwidth}
        \centering
        \includegraphics[width=\linewidth]{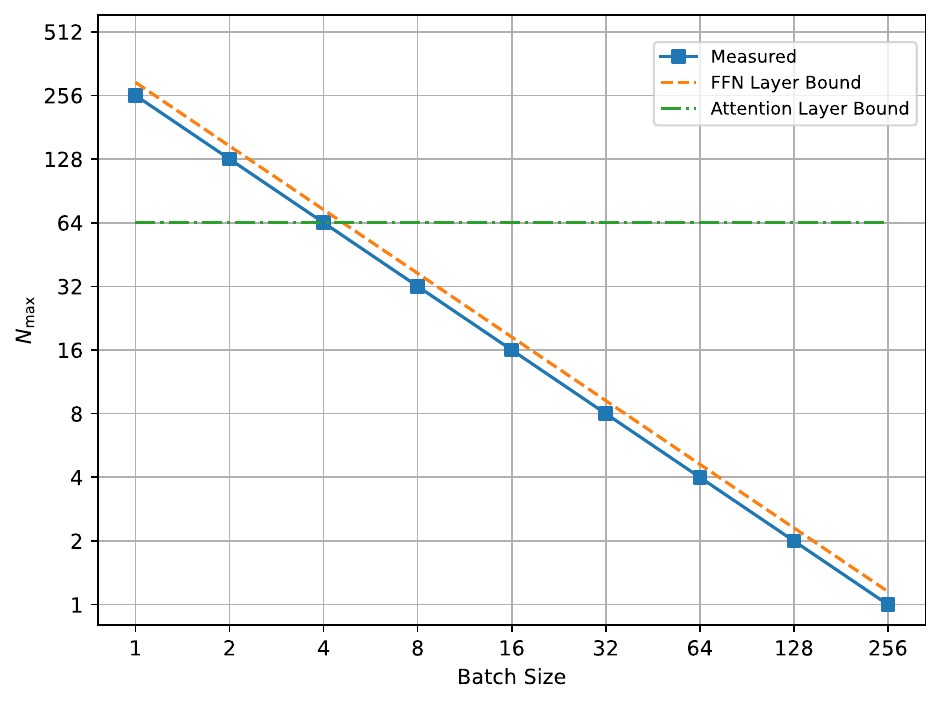}
        \caption{$L=384$}
        \label{fig:dense-model-h800-seq384}
    \end{subfigure}
    \hfill
    \begin{subfigure}[t]{0.24\textwidth}
        \centering
        \includegraphics[width=\linewidth]{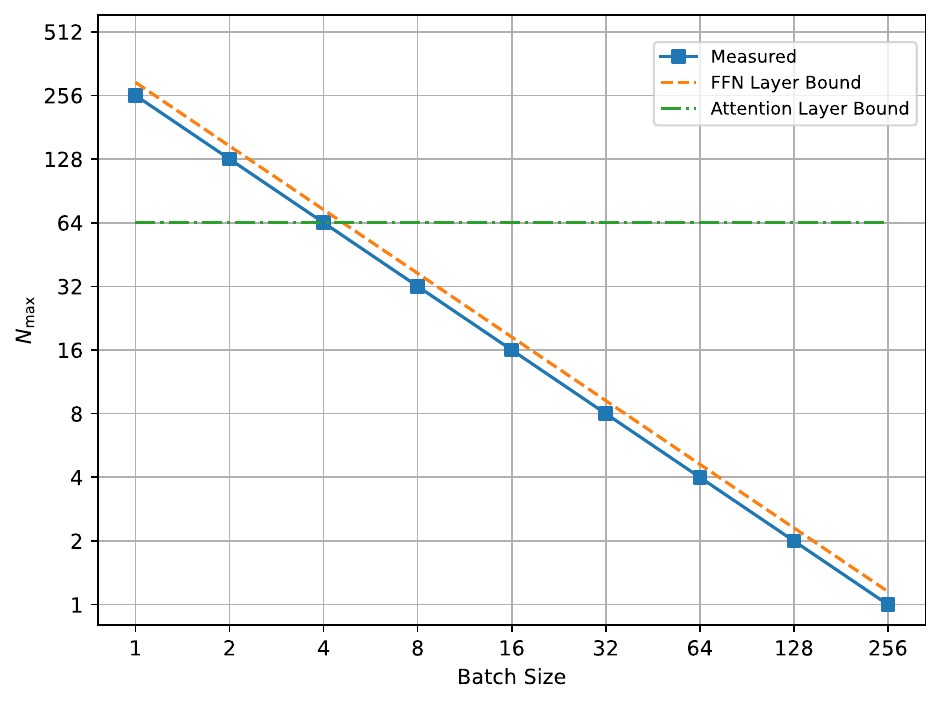}
        \caption{$L=512$}
        \label{fig:dense-model-h800-seq512}
    \end{subfigure}

    \caption{
    Dense model-level NFP principle validation results on \textbf{NVIDIA H800} (WeDLM-8B).
    }
    \label{fig:dense-model-h800}
\end{figure*}

Figures~\ref{fig:dense-model-h20}--\ref{fig:dense-model-h800} report the Dense model-level validation results on NVIDIA H20, A800, and H800. Across all evaluated GPU platforms, the measured NFP boundaries closely follow the principle-predicted boundaries. This agreement shows that the Dense FFN idle-compute constraint and the Attention query-granularity constraint compose into an accurate full-model predictor.

The same agreement holds across the evaluated sequence lengths. In this short-context validation range, changing the cached sequence length does not substantially change the measured boundary trend. The sequence-length robustness also supports the module-level Attention finding that the realized Attention boundary is determined mainly by backend query granularity.

Across hardware platforms, the measured boundary shifts consistently with the hardware-sensitive Dense FFN term. At the same batch size, GPUs with larger compute-to-memory balance points expose larger near-free capacity, while increasing the batch size reduces the boundary. Overall, the Dense model results confirm that the proposed principle transfers from isolated modules to full-model inference and preserves the expected hardware and batch-size trends.

\begin{figure*}[htbp]
    \centering

    \begin{subfigure}[t]{0.24\textwidth}
        \centering
        \includegraphics[width=\linewidth]{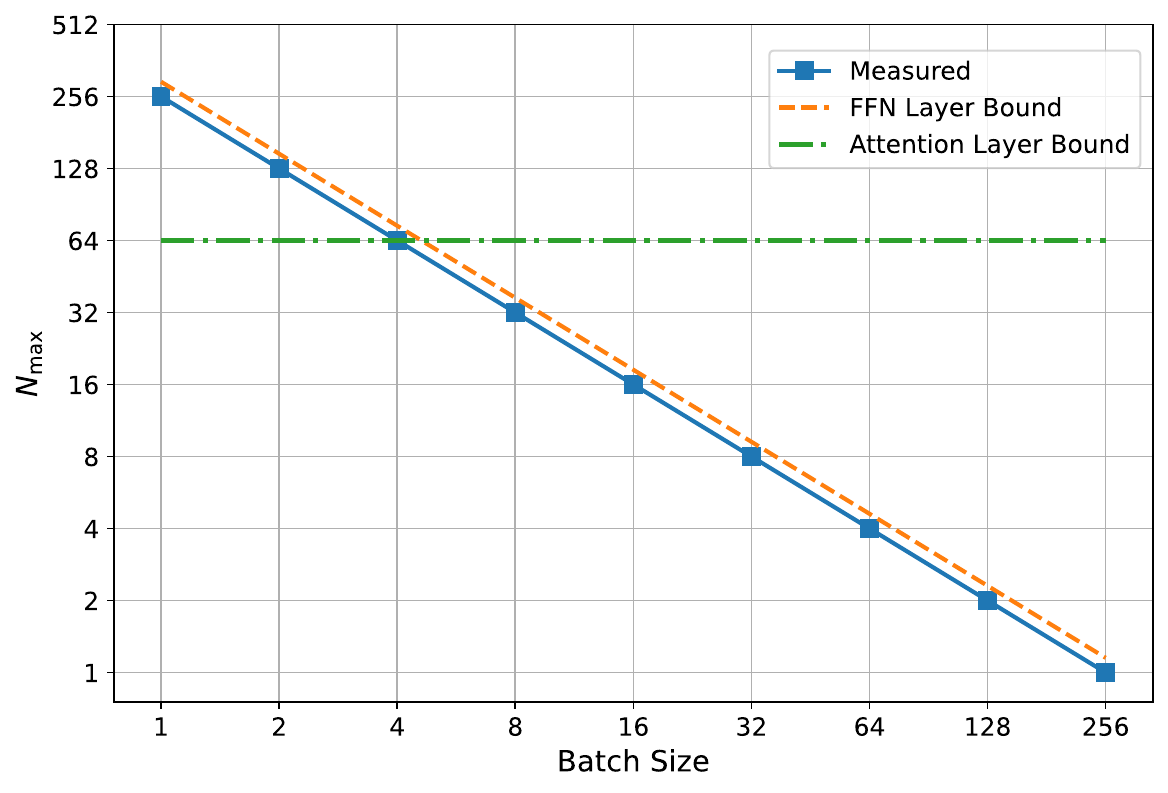}
        \caption{$L=128$}
        \label{fig:dense-ar-qwen3-seq128}
    \end{subfigure}
    \hfill
    \begin{subfigure}[t]{0.24\textwidth}
        \centering
        \includegraphics[width=\linewidth]{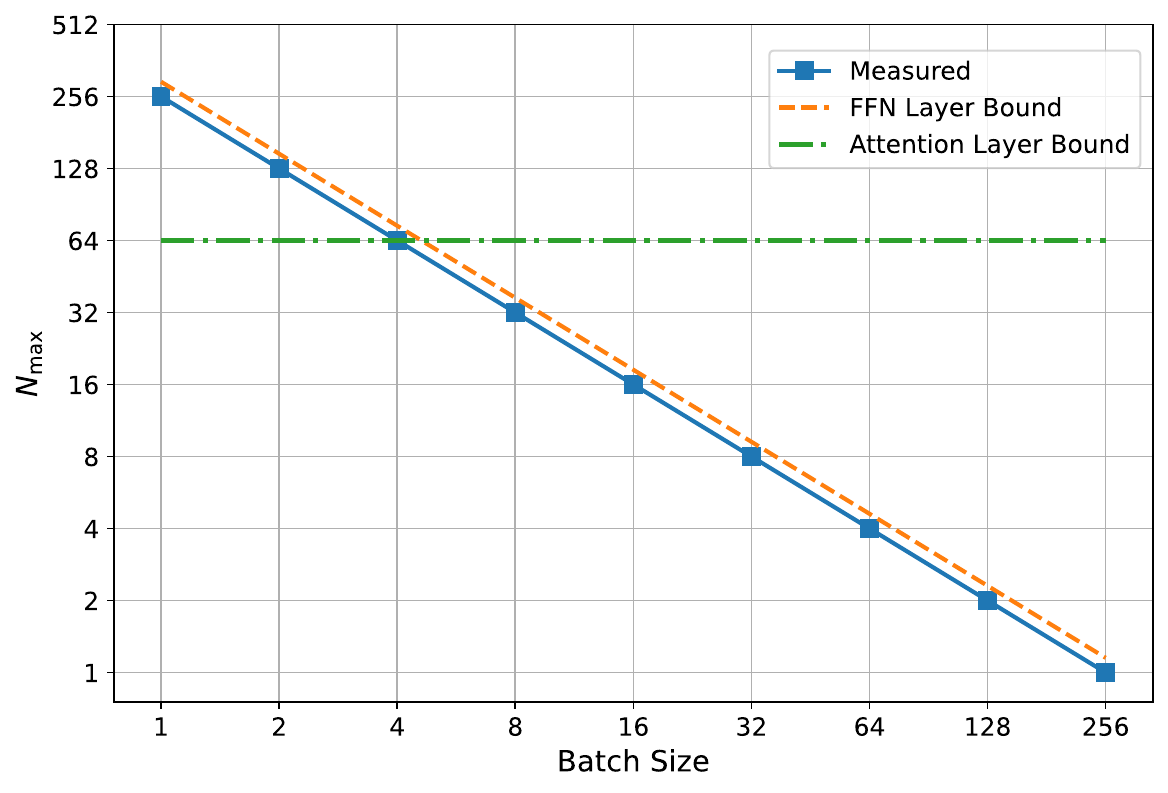}
        \caption{$L=256$}
        \label{fig:dense-ar-qwen3-seq256}
    \end{subfigure}
    \hfill
    \begin{subfigure}[t]{0.24\textwidth}
        \centering
        \includegraphics[width=\linewidth]{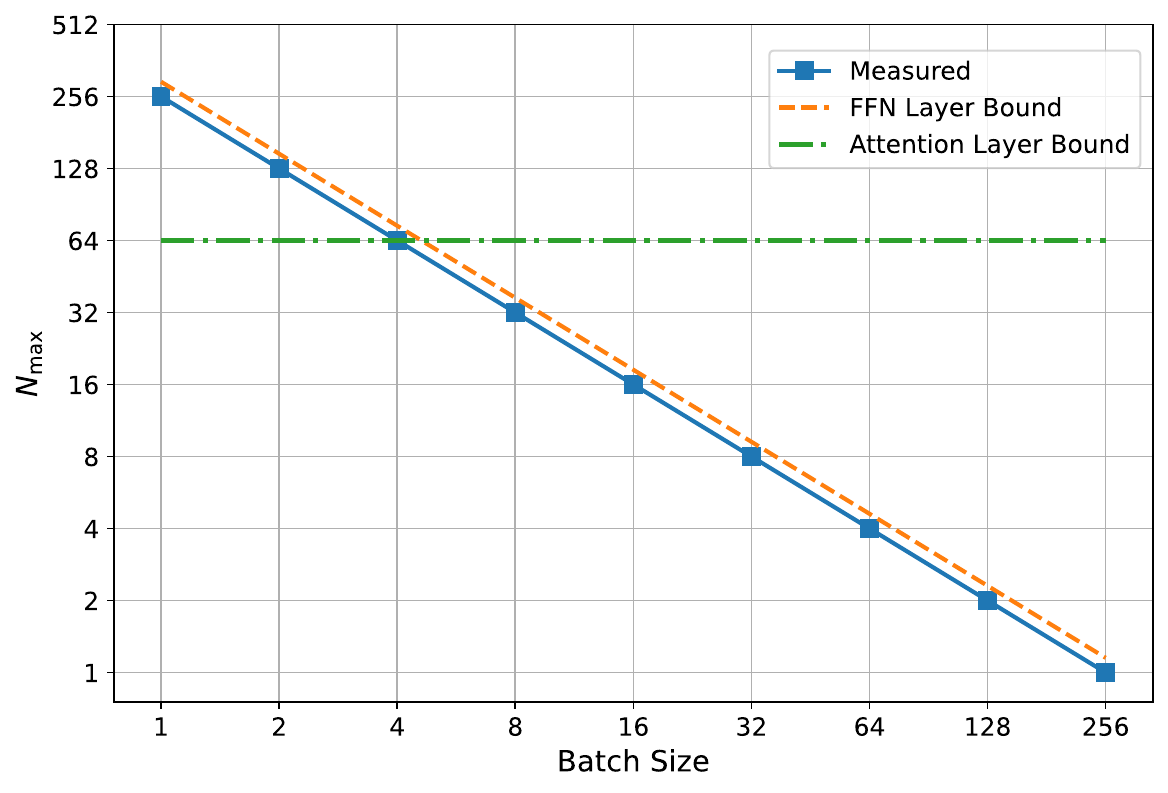}
        \caption{$L=384$}
        \label{fig:dense-ar-qwen3-seq384}
    \end{subfigure}
    \hfill
    \begin{subfigure}[t]{0.24\textwidth}
        \centering
        \includegraphics[width=\linewidth]{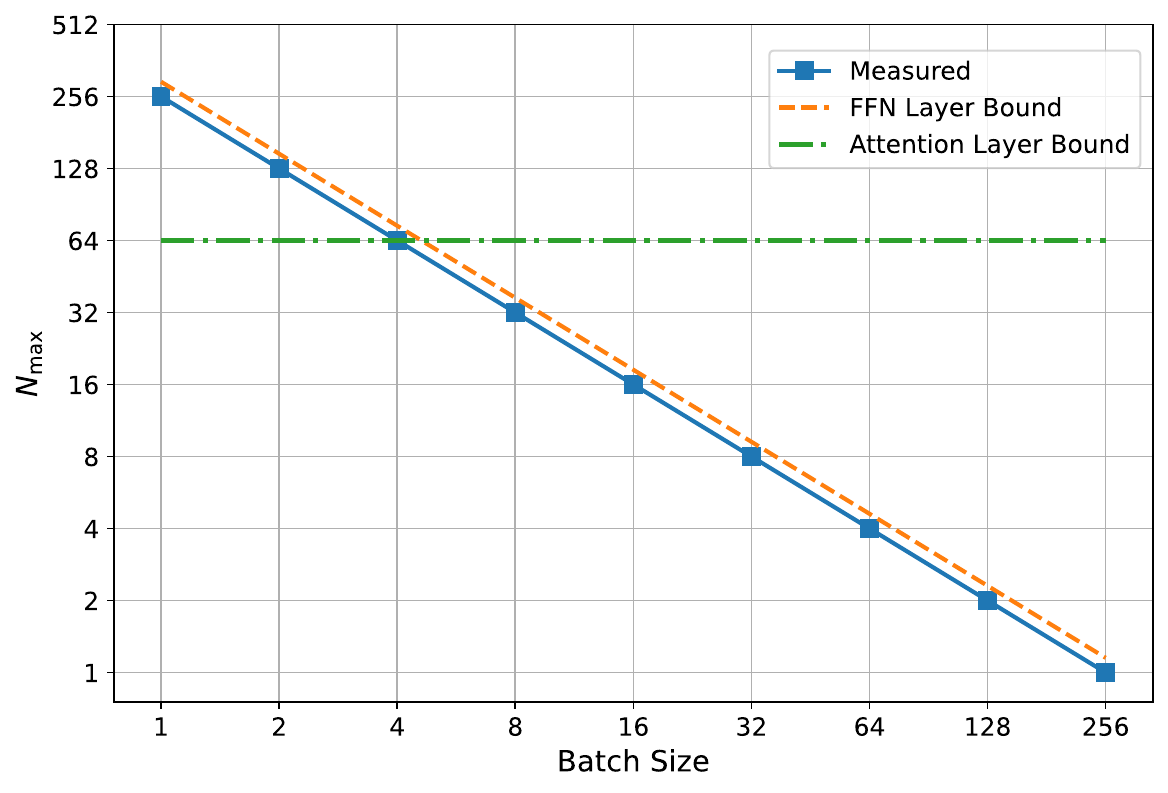}
        \caption{$L=512$}
        \label{fig:dense-ar-qwen3-seq512}
    \end{subfigure}

    \caption{
    Dense model-level NFP principle validation results on \textbf{NVIDIA H800} (Qwen3-8B).
    }
    \label{fig:dense-ar-qwen3-h800}
\end{figure*}

Figure~\ref{fig:dense-ar-qwen3-h800} extends the validation to an AR model (Qwen3-8B) on H800. The measured boundary exhibits the same $1/b$ scaling governed by idle-compute slack and the same Attention-limited plateau, confirming that the Dense NFP mechanism applies equally to the AR verification forward.

\clearpage

\clearpage

\subsection{MoE Model Results}
\label{app:moe-model-results}
\subsubsection{Load-balanced Routing as Upper Bound}

\begin{figure*}[htbp]
    \centering

    \begin{subfigure}[t]{0.24\textwidth}
        \centering
        \includegraphics[width=\linewidth]{figs/MoE_Upper/H20/moe_upper_seq256.pdf}
        \caption{$L=256$}
        \label{fig:moe-upper-h20-seq256}
    \end{subfigure}
    \hfill
    \begin{subfigure}[t]{0.24\textwidth}
        \centering
        \includegraphics[width=\linewidth]{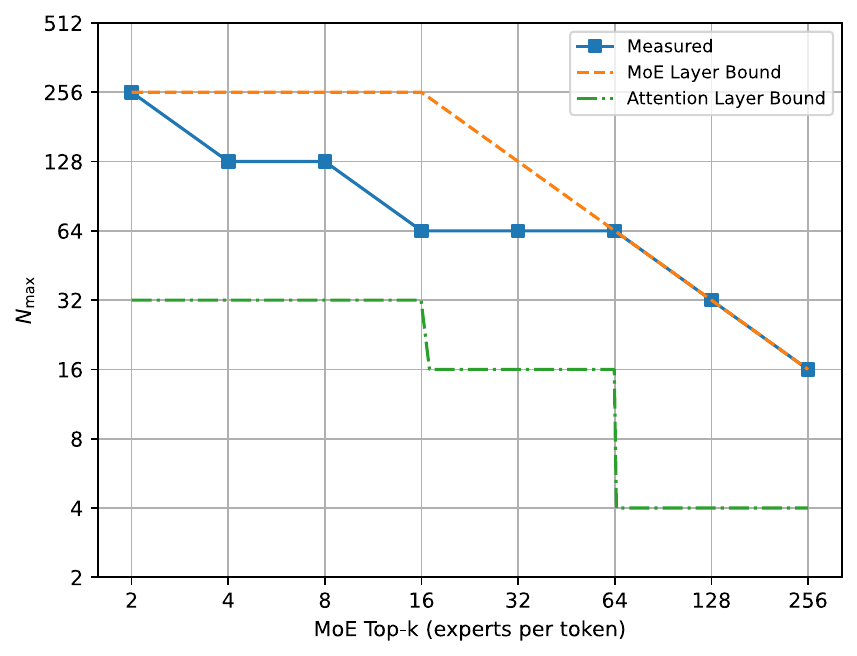}
        \caption{$L=4096$}
        \label{fig:moe-upper-h20-seq4096}
    \end{subfigure}
    \hfill
    \begin{subfigure}[t]{0.24\textwidth}
        \centering
        \includegraphics[width=\linewidth]{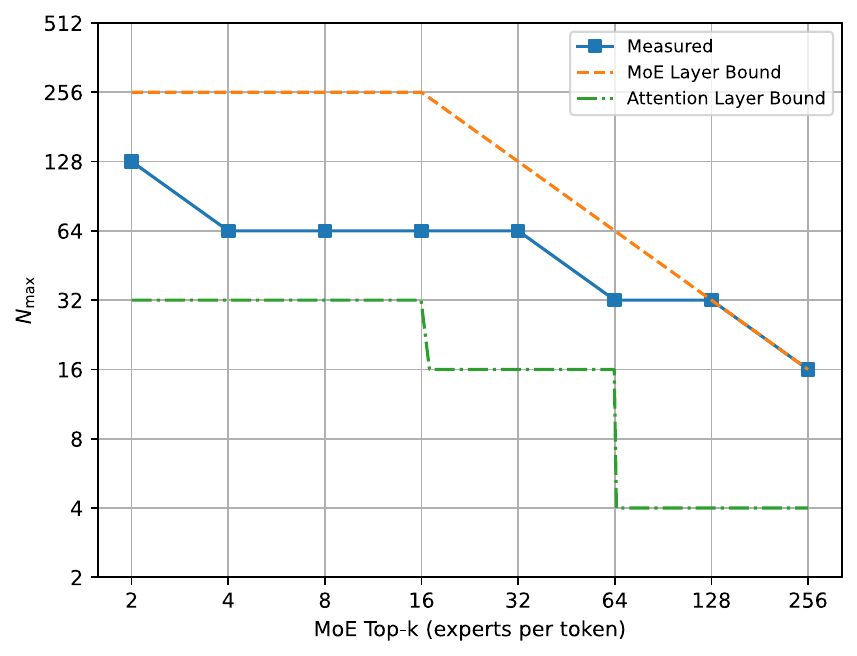}
        \caption{$L=16384$}
        \label{fig:moe-upper-h20-seq16384}
    \end{subfigure}
    \hfill
    \begin{subfigure}[t]{0.24\textwidth}
        \centering
        \includegraphics[width=\linewidth]{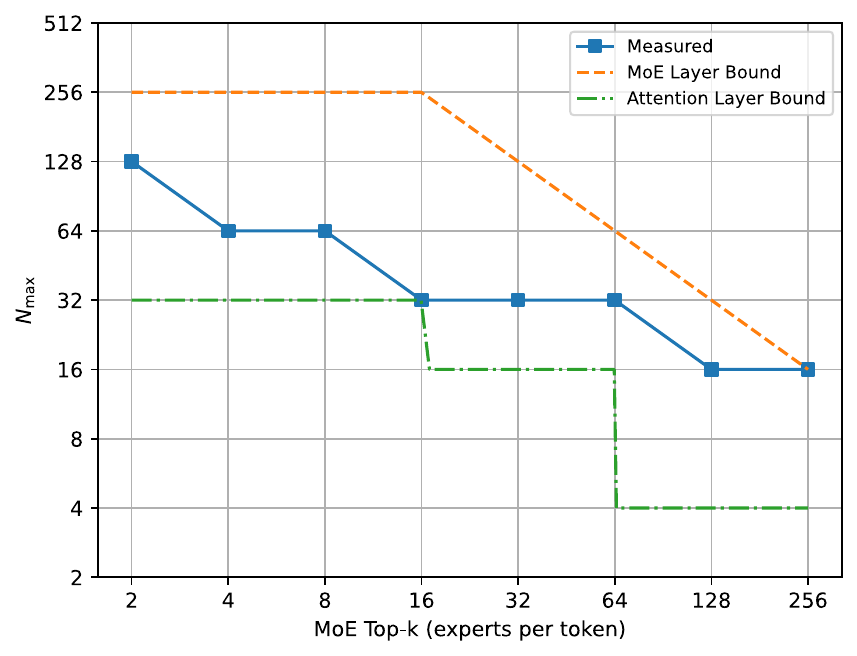}
        \caption{$L=32768$}
        \label{fig:moe-upper-h20-seq32768}
    \end{subfigure}

    \caption{
    MoE model-level NFP principle validation for load-balanced routing, the upper-bound case, on \textbf{NVIDIA H20} (LLaDA-2.1-mini).
    }
    \label{fig:moe-upper-h20}
\end{figure*}

\begin{figure*}[htbp]
    \centering

    \begin{subfigure}[t]{0.24\textwidth}
        \centering
        \includegraphics[width=\linewidth]{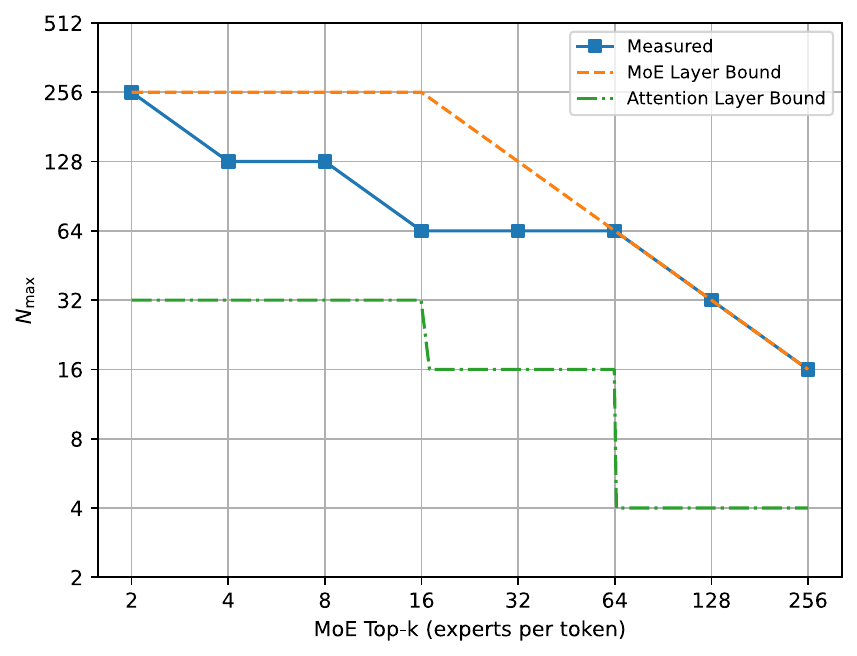}
        \caption{$L=256$}
        \label{fig:moe-upper-a800-seq256}
    \end{subfigure}
    \hfill
    \begin{subfigure}[t]{0.24\textwidth}
        \centering
        \includegraphics[width=\linewidth]{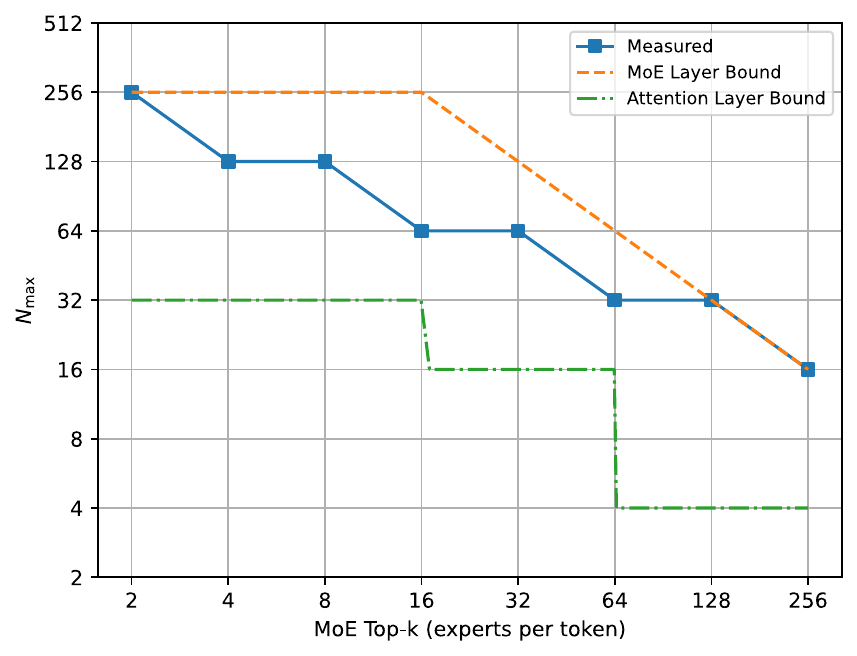}
        \caption{$L=4096$}
        \label{fig:moe-upper-a800-seq4096}
    \end{subfigure}
    \hfill
    \begin{subfigure}[t]{0.24\textwidth}
        \centering
        \includegraphics[width=\linewidth]{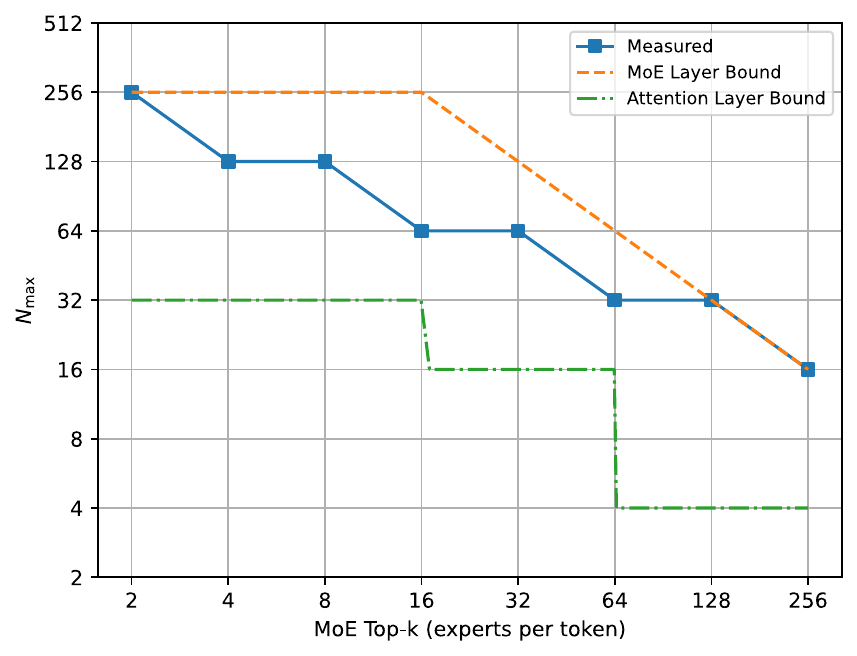}
        \caption{$L=16384$}
        \label{fig:moe-upper-a800-seq16384}
    \end{subfigure}
    \hfill
    \begin{subfigure}[t]{0.24\textwidth}
        \centering
        \includegraphics[width=\linewidth]{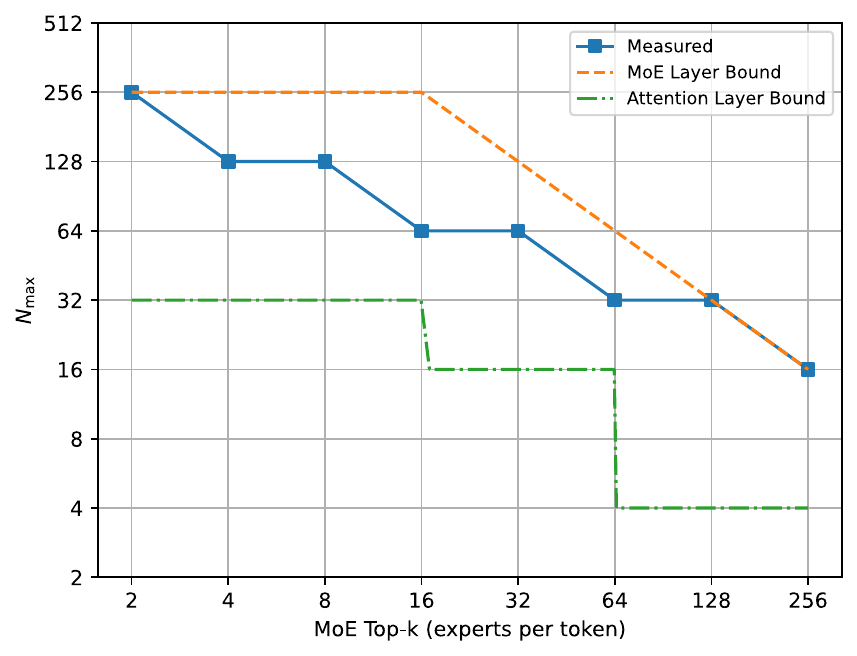}
        \caption{$L=32768$}
        \label{fig:moe-upper-a800-seq32768}
    \end{subfigure}

    \caption{
    MoE model-level NFP principle validation for load-balanced routing, the upper-bound case, on \textbf{NVIDIA A800} (LLaDA-2.1-mini).
    }
    \label{fig:moe-upper-a800}
\end{figure*}

\begin{figure*}[htbp]
    \centering

    \begin{subfigure}[t]{0.24\textwidth}
        \centering
        \includegraphics[width=\linewidth]{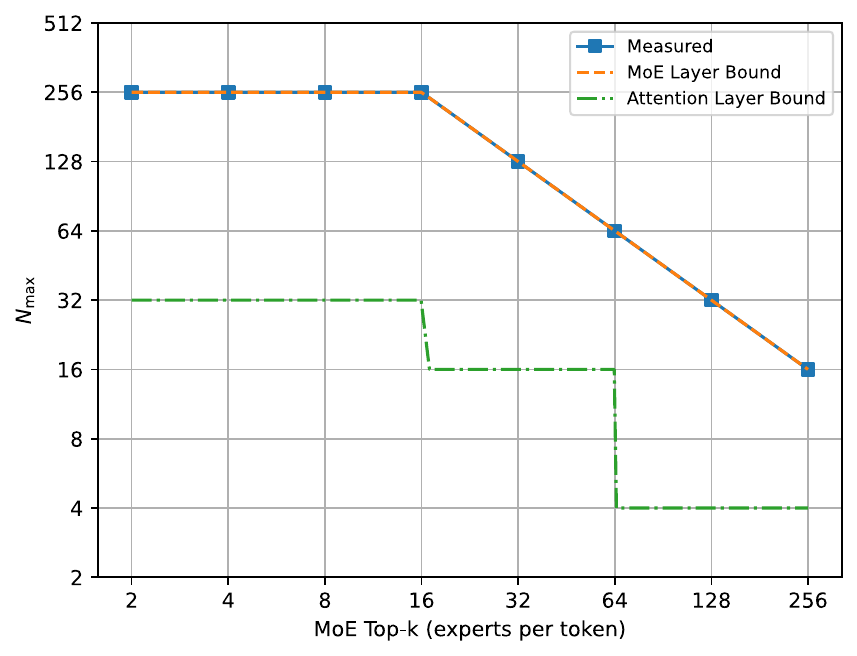}
        \caption{$L=256$}
        \label{fig:moe-upper-h800-seq256}
    \end{subfigure}
    \hfill
    \begin{subfigure}[t]{0.24\textwidth}
        \centering
        \includegraphics[width=\linewidth]{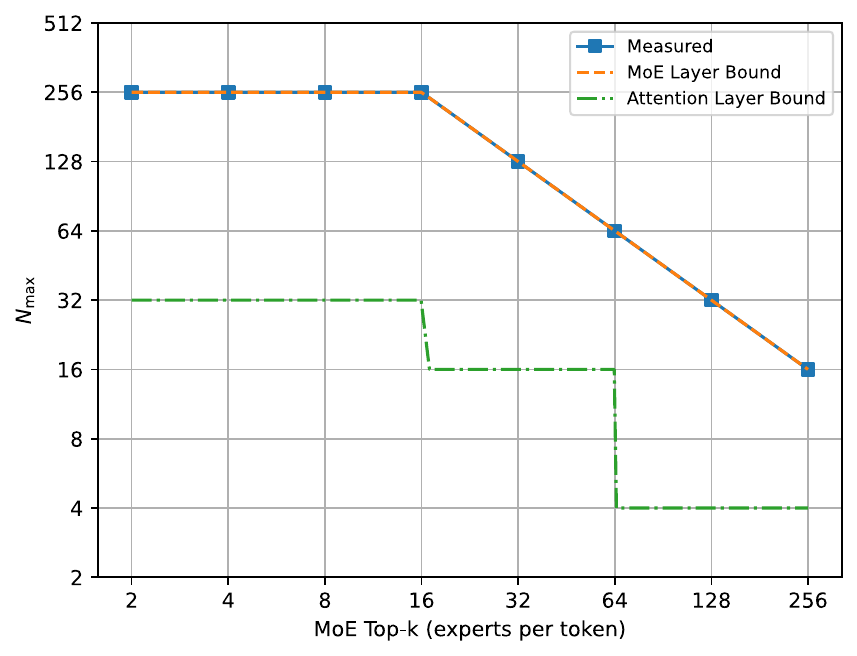}
        \caption{$L=4096$}
        \label{fig:moe-upper-h800-seq4096}
    \end{subfigure}
    \hfill
    \begin{subfigure}[t]{0.24\textwidth}
        \centering
        \includegraphics[width=\linewidth]{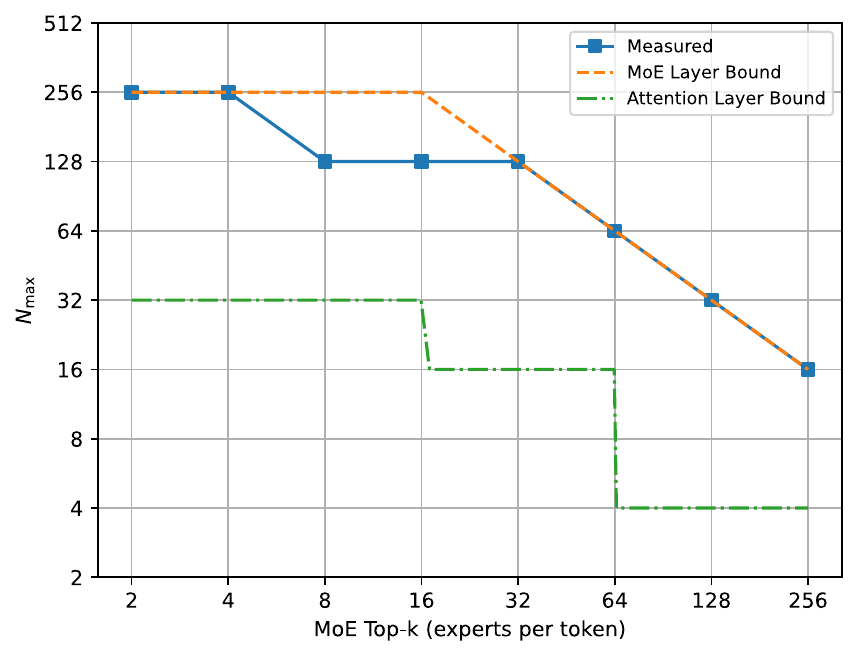}
        \caption{$L=16384$}
        \label{fig:moe-upper-h800-seq16384}
    \end{subfigure}
    \hfill
    \begin{subfigure}[t]{0.24\textwidth}
        \centering
        \includegraphics[width=\linewidth]{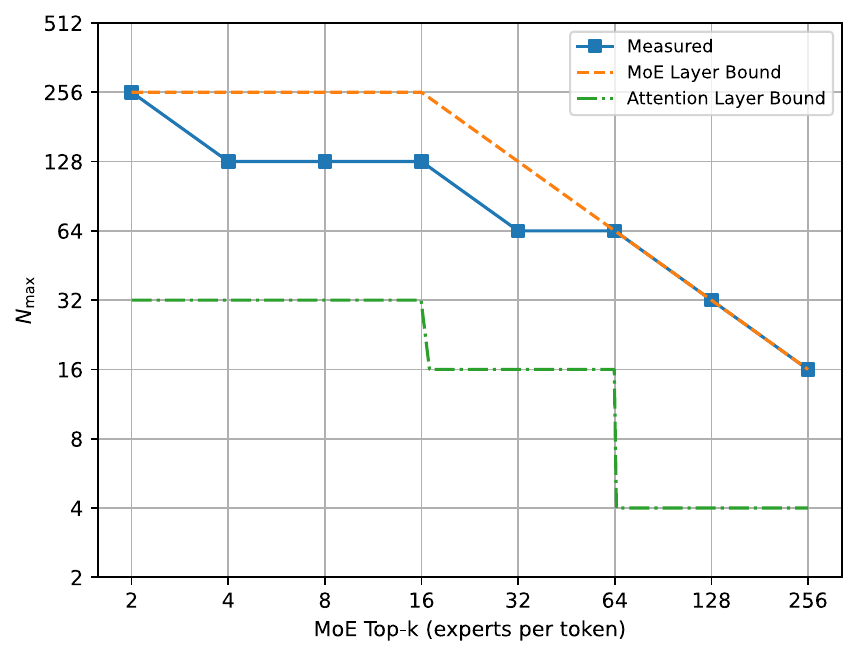}
        \caption{$L=32768$}
        \label{fig:moe-upper-h800-seq32768}
    \end{subfigure}

    \caption{
    MoE model-level NFP principle validation for load-balanced routing, the upper-bound case, on \textbf{NVIDIA H800} (LLaDA-2.1-mini).
    }
    \label{fig:moe-upper-h800}
\end{figure*}

Figures~\ref{fig:moe-upper-h20}--\ref{fig:moe-upper-h800} report the MoE model-level results under load-balanced routing, which represents the upper-bound case. The measured boundaries follow the predicted trend across GPU platforms and sequence lengths: smaller $k$ exposes more aggregate expert-token padding slack, while larger $k$ reduces the maximum near-free N. This validates that the load-balanced MoE FFN granularity term remains predictive at the full-model level.

The results also show a clear limiting-module transition. At short sequence lengths, the measured boundary is primarily controlled by the MoE FFN granularity term, since Attention does not yet impose the active constraint. As the cached sequence length increases, the Attention workload becomes more significant, and the boundary approaches the Attention-side limit. Thus, the principle captures not only the boundary value, but also when the active bottleneck shifts from MoE FFN to Attention.

\begin{figure*}[htbp]
    \centering

    \begin{subfigure}[t]{0.24\textwidth}
        \centering
        \includegraphics[width=\linewidth]{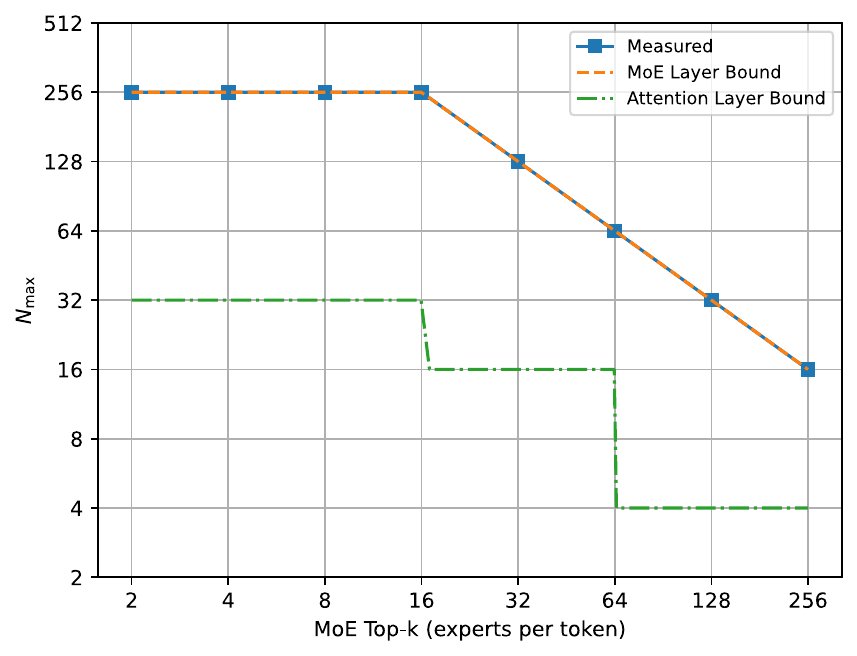}
        \caption{$L=256$}
        \label{fig:moe-upper-ar-ling2-seq256}
    \end{subfigure}
    \hfill
    \begin{subfigure}[t]{0.24\textwidth}
        \centering
        \includegraphics[width=\linewidth]{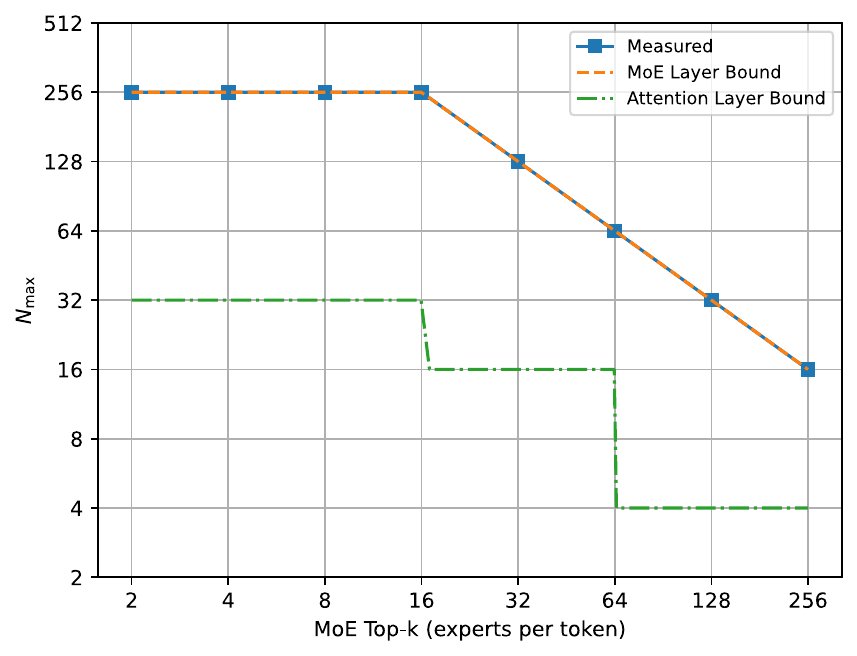}
        \caption{$L=4096$}
        \label{fig:moe-upper-ar-ling2-seq4096}
    \end{subfigure}
    \hfill
    \begin{subfigure}[t]{0.24\textwidth}
        \centering
        \includegraphics[width=\linewidth]{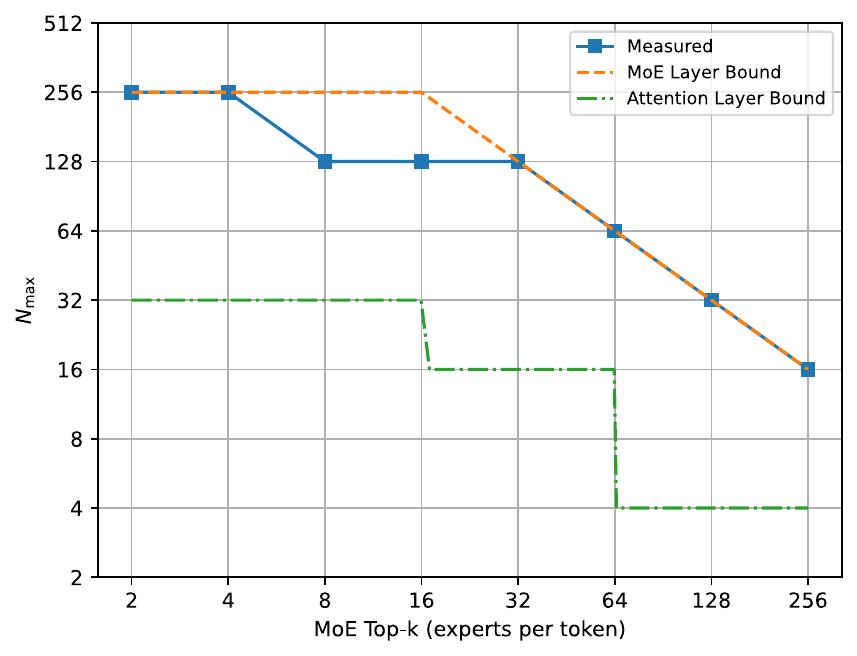}
        \caption{$L=16384$}
        \label{fig:moe-upper-ar-ling2-seq16384}
    \end{subfigure}
    \hfill
    \begin{subfigure}[t]{0.24\textwidth}
        \centering
        \includegraphics[width=\linewidth]{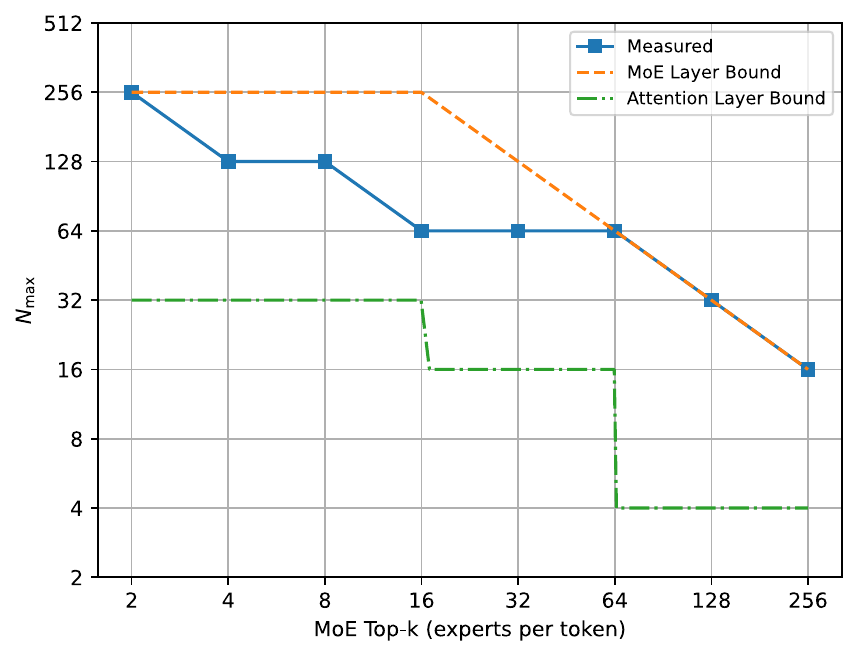}
        \caption{$L=32768$}
        \label{fig:moe-upper-ar-ling2-seq32768}
    \end{subfigure}

    \caption{
    MoE model-level NFP principle validation for load-balanced routing, the upper-bound case, on \textbf{NVIDIA H800} (Ling-2.0-mini).
    }
    \label{fig:moe-upper-ar-ling2-h800}
\end{figure*}

Figure~\ref{fig:moe-upper-ar-ling2-h800} extends the load-balanced validation to an AR model (Ling-2.0-mini) on H800. The boundary follows the same staircase decay with increasing $k$ and transitions to Attention-limited at long sequences, consistent with the DLLM results. This confirms that the aggregate expert-token padding slack mechanism governs the AR verification forward in the same manner.

\clearpage

\clearpage

\subsubsection{Load-skewed Routing as Lower Bound}
\begin{figure*}[htbp]
    \centering

    \begin{subfigure}[t]{0.24\textwidth}
        \centering
        \includegraphics[width=\linewidth]{figs/MoE_Lower/H20/moe_Lower_seq256.pdf}
        \caption{$L=256$}
        \label{fig:moe-lower-h20-seq256}
    \end{subfigure}
    \hfill
    \begin{subfigure}[t]{0.24\textwidth}
        \centering
        \includegraphics[width=\linewidth]{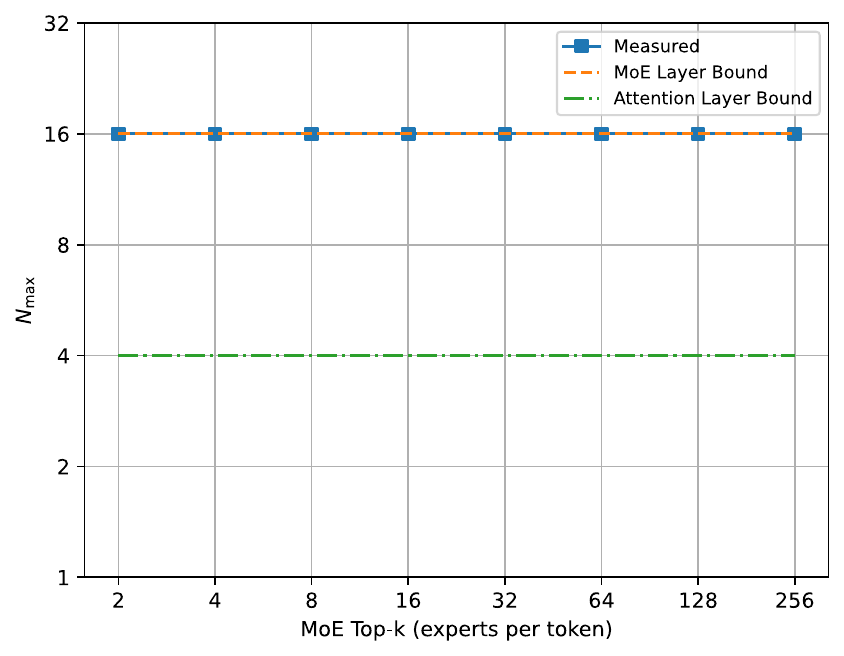}
        \caption{$L=4096$}
        \label{fig:moe-lower-h20-seq4096}
    \end{subfigure}
    \hfill
    \begin{subfigure}[t]{0.24\textwidth}
        \centering
        \includegraphics[width=\linewidth]{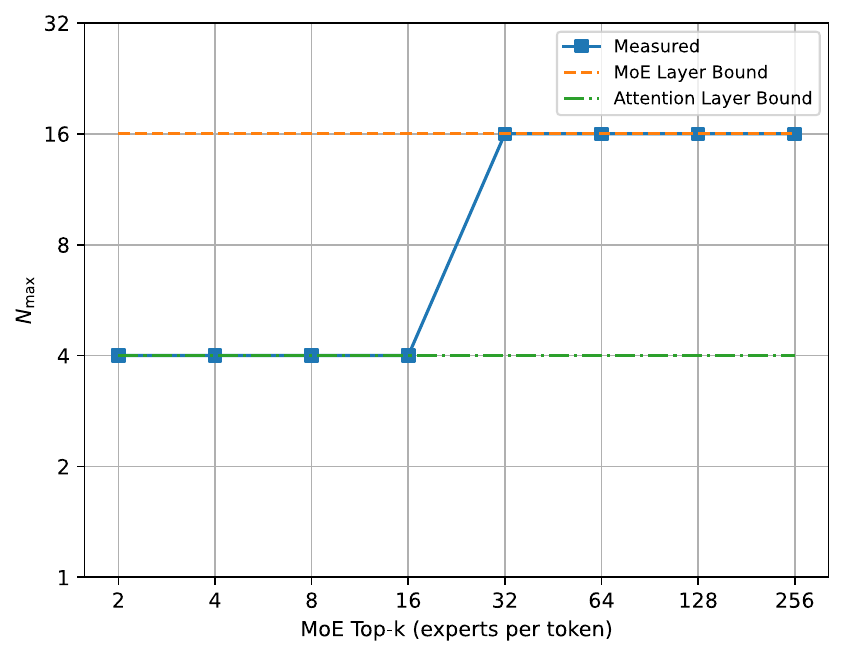}
        \caption{$L=16384$}
        \label{fig:moe-lower-h20-seq16384}
    \end{subfigure}
    \hfill
    \begin{subfigure}[t]{0.24\textwidth}
        \centering
        \includegraphics[width=\linewidth]{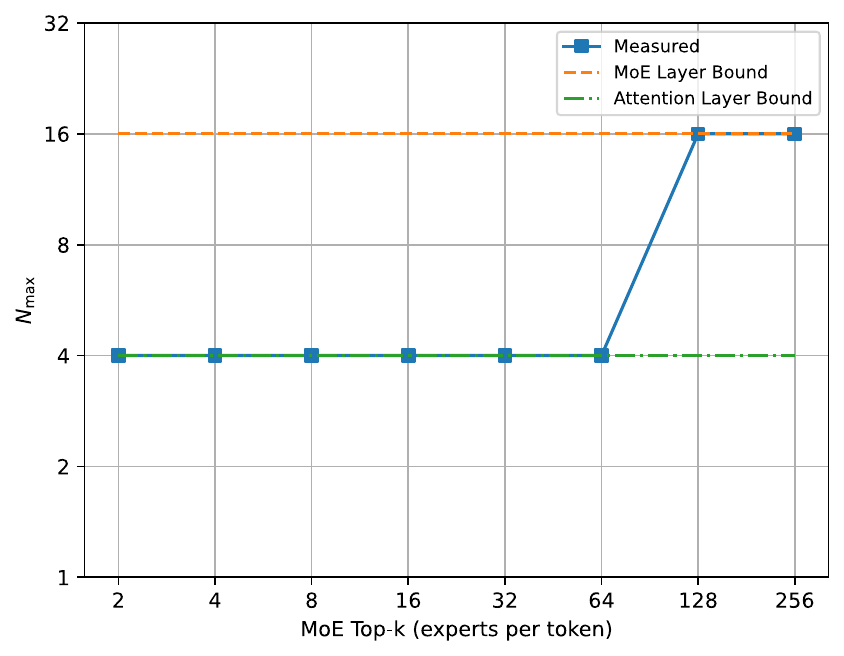}
        \caption{$L=32768$}
        \label{fig:moe-lower-h20-seq32768}
    \end{subfigure}

    \caption{
    MoE model-level NFP principle validation for load-skewed routing, the lower-bound case, on \textbf{NVIDIA H20} (LLaDA-2.1-mini).
    }
    \label{fig:moe-lower-h20}
\end{figure*}

\begin{figure*}[htbp]
    \centering

    \begin{subfigure}[t]{0.24\textwidth}
        \centering
        \includegraphics[width=\linewidth]{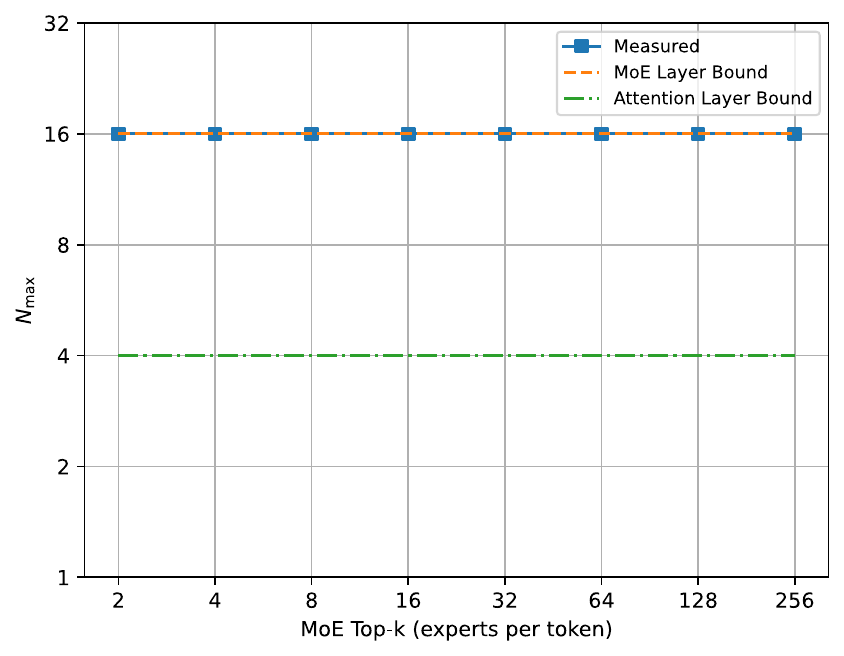}
        \caption{$L=256$}
        \label{fig:moe-lower-a800-seq256}
    \end{subfigure}
    \hfill
    \begin{subfigure}[t]{0.24\textwidth}
        \centering
        \includegraphics[width=\linewidth]{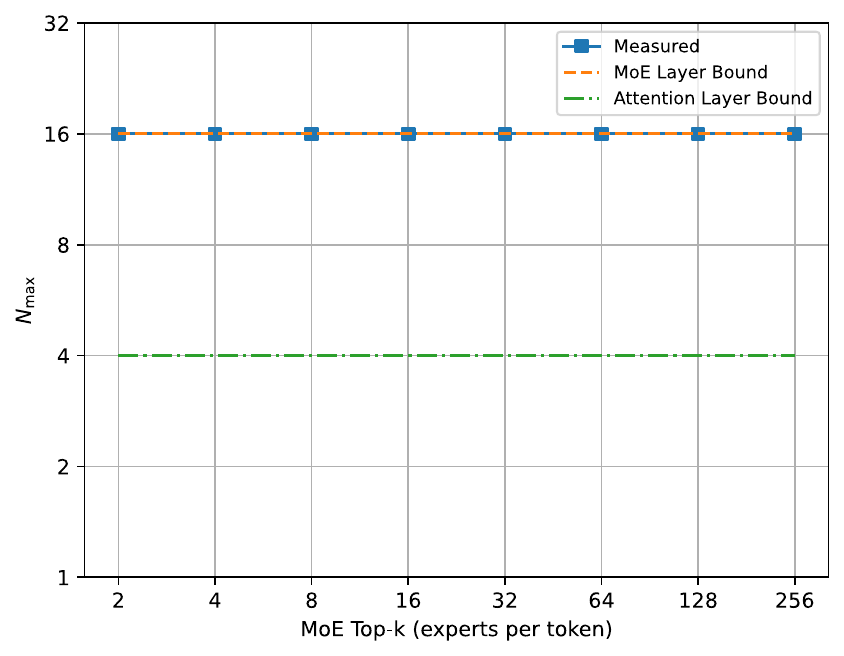}
        \caption{$L=4096$}
        \label{fig:moe-lower-a800-seq4096}
    \end{subfigure}
    \hfill
    \begin{subfigure}[t]{0.24\textwidth}
        \centering
        \includegraphics[width=\linewidth]{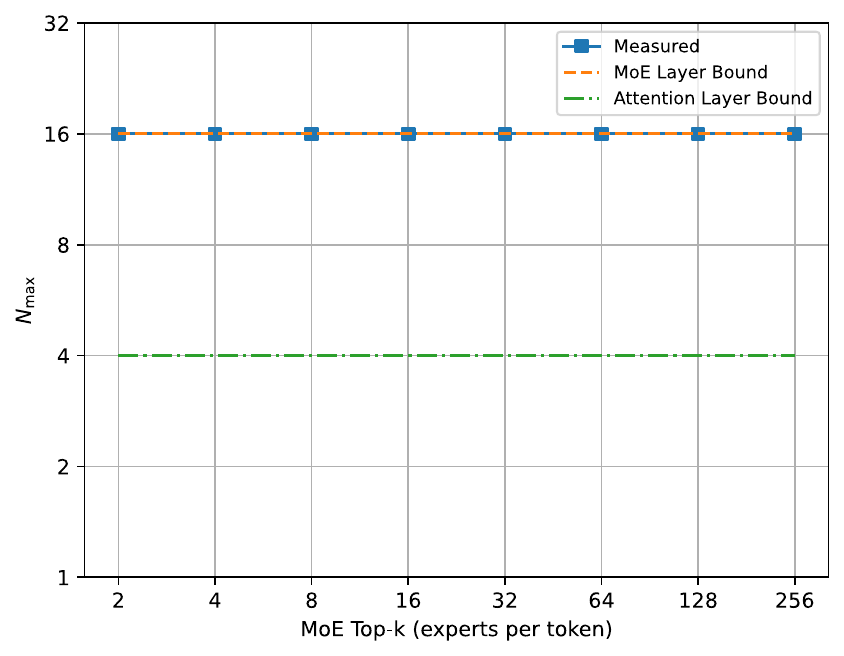}
        \caption{$L=16384$}
        \label{fig:moe-lower-a800-seq16384}
    \end{subfigure}
    \hfill
    \begin{subfigure}[t]{0.24\textwidth}
        \centering
        \includegraphics[width=\linewidth]{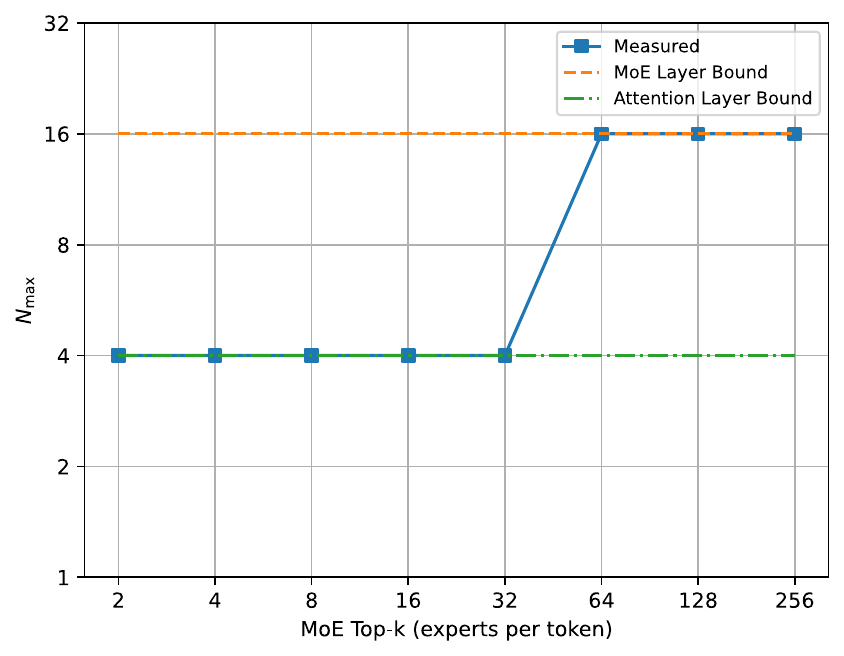}
        \caption{$L=32768$}
        \label{fig:moe-lower-a800-seq32768}
    \end{subfigure}

    \caption{
    MoE model-level NFP principle validation for load-skewed routing, the lower-bound case, on \textbf{NVIDIA A800} (LLaDA-2.1-mini).
    }
    \label{fig:moe-lower-a800}
\end{figure*}

\begin{figure*}[htbp]
    \centering

    \begin{subfigure}[t]{0.24\textwidth}
        \centering
        \includegraphics[width=\linewidth]{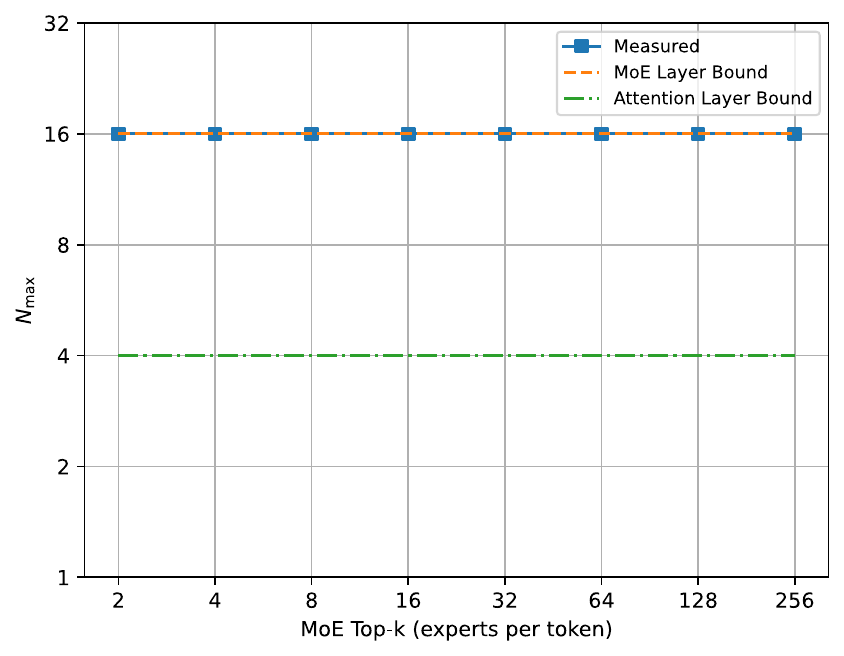}
        \caption{$L=256$}
        \label{fig:moe-lower-h800-seq256}
    \end{subfigure}
    \hfill
    \begin{subfigure}[t]{0.24\textwidth}
        \centering
        \includegraphics[width=\linewidth]{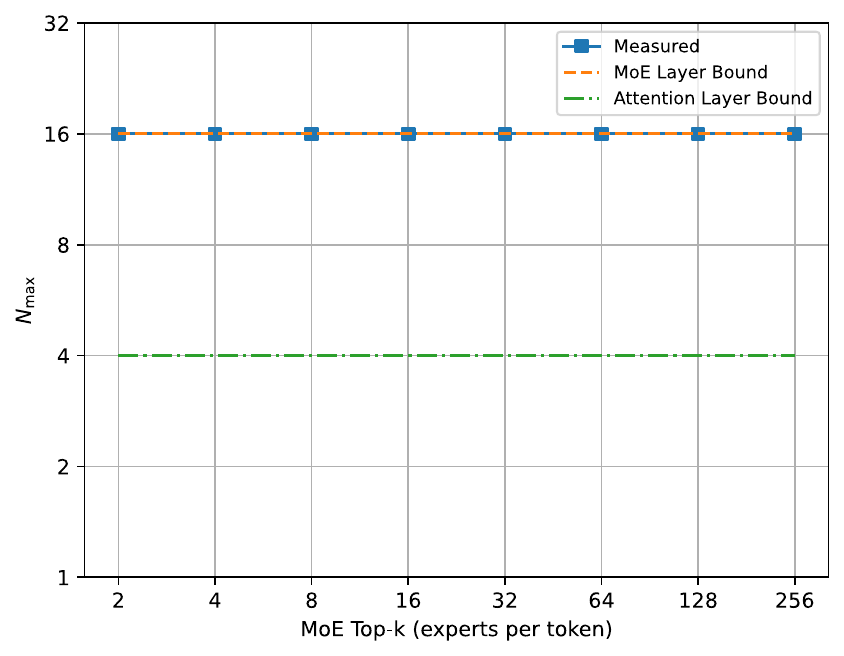}
        \caption{$L=4096$}
        \label{fig:moe-lower-h800-seq4096}
    \end{subfigure}
    \hfill
    \begin{subfigure}[t]{0.24\textwidth}
        \centering
        \includegraphics[width=\linewidth]{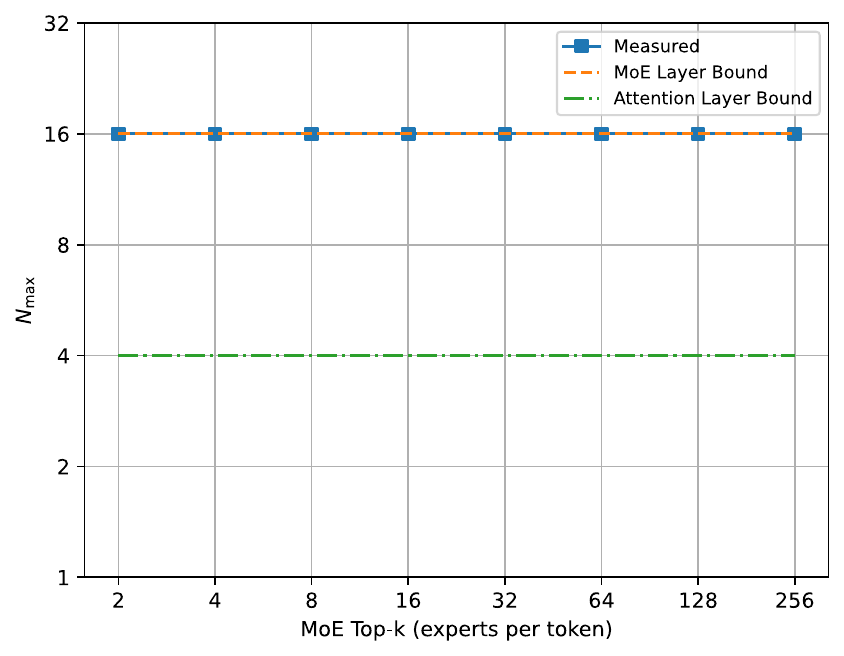}
        \caption{$L=16384$}
        \label{fig:moe-lower-h800-seq16384}
    \end{subfigure}
    \hfill
    \begin{subfigure}[t]{0.24\textwidth}
        \centering
        \includegraphics[width=\linewidth]{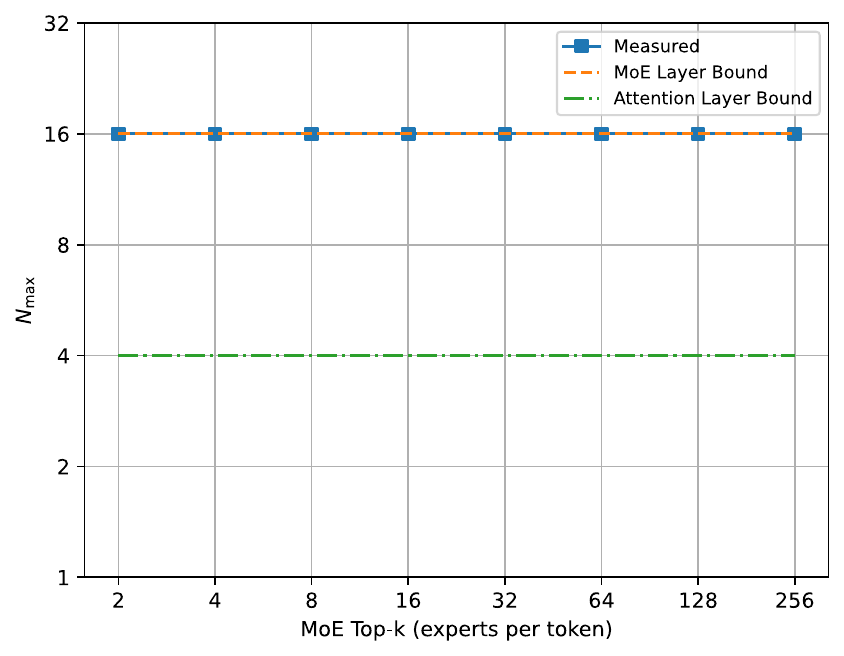}
        \caption{$L=32768$}
        \label{fig:moe-lower-h800-seq32768}
    \end{subfigure}

    \caption{
    MoE model-level NFP principle validation for load-skewed routing, the lower-bound case, on \textbf{NVIDIA H800} (LLaDA-2.1-mini).
    }
    \label{fig:moe-lower-h800}
\end{figure*}

Figures~\ref{fig:moe-lower-h20}--\ref{fig:moe-lower-h800} report the corresponding results under load-skewed routing, which represents the lower-bound case. When Attention is not the active limiter, the measured boundary remains nearly fixed across $k$, matching the prediction that load-skewed routing exposes only a local expert-token padding granularity rather than aggregate slack across experts. The contrast with the load-balanced case confirms that expert-load distribution remains a first-order factor in full-model MoE NFP.

The hardware trends further support the principle. On higher-performance GPU platforms, the MoE-side boundary remains active over a wider range of sequence lengths, while lower-capacity platforms become Attention-limited earlier at long contexts. This does not imply that $M_{\mathrm{attn}}$ itself is hardware-derived; rather, it shows that the point at which the Attention constraint becomes the active full-model bottleneck depends on the interaction between backend granularity, attention workload, and platform performance.

\begin{figure*}[htbp]
    \centering

    \begin{subfigure}[t]{0.24\textwidth}
        \centering
        \includegraphics[width=\linewidth]{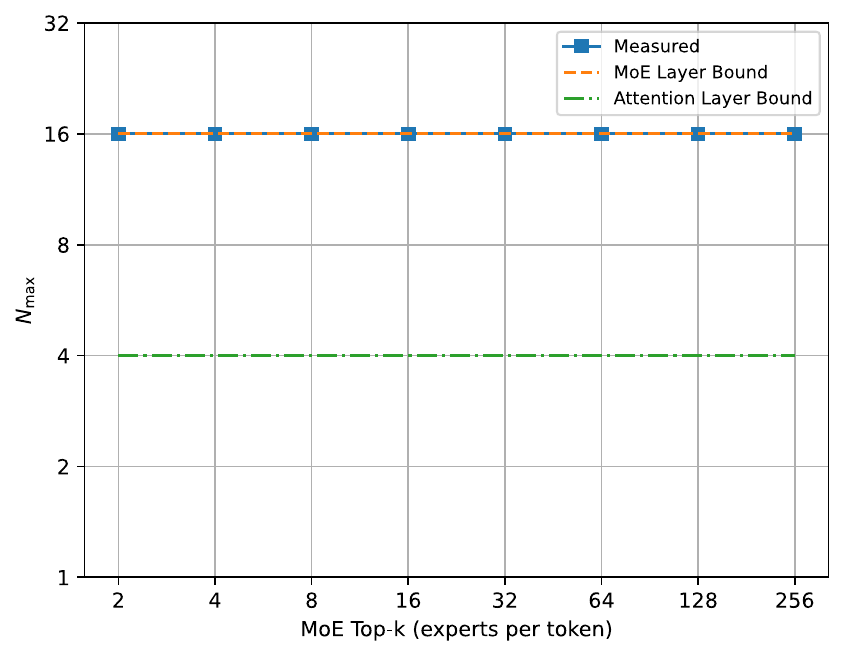}
        \caption{$L=256$}
        \label{fig:moe-lower-ar-ling2-seq256}
    \end{subfigure}
    \hfill
    \begin{subfigure}[t]{0.24\textwidth}
        \centering
        \includegraphics[width=\linewidth]{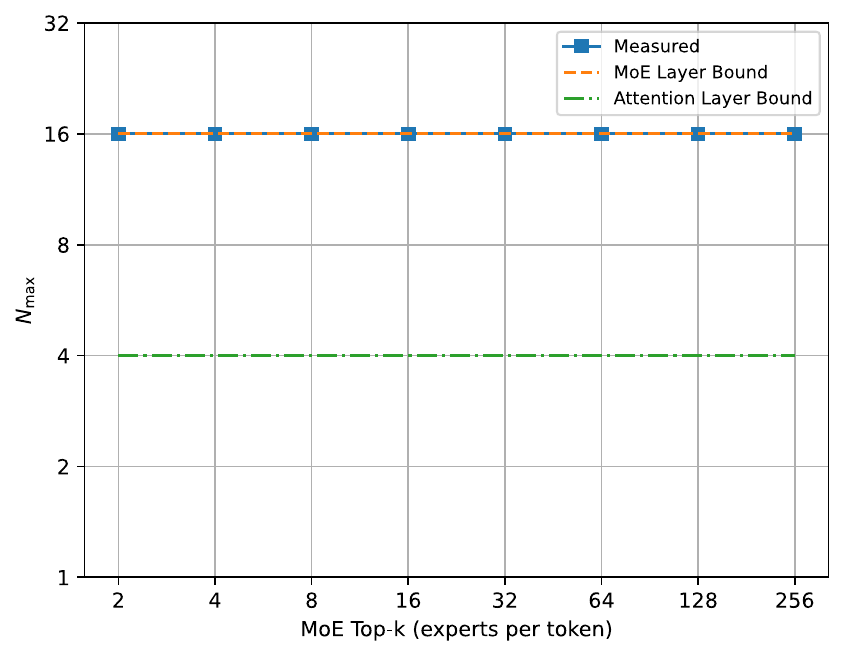}
        \caption{$L=4096$}
        \label{fig:moe-lower-ar-ling2-seq4096}
    \end{subfigure}
    \hfill
    \begin{subfigure}[t]{0.24\textwidth}
        \centering
        \includegraphics[width=\linewidth]{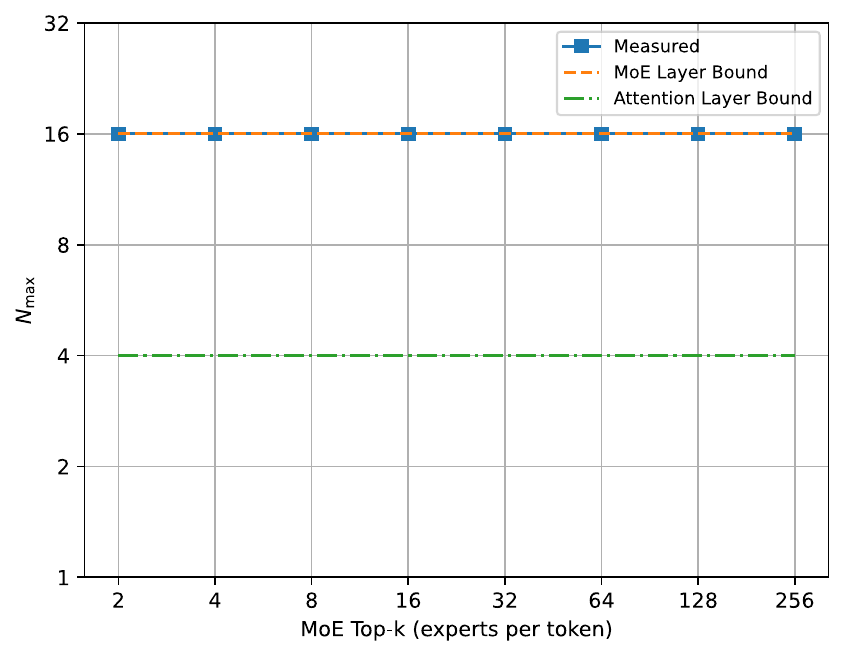}
        \caption{$L=16384$}
        \label{fig:moe-lower-ar-ling2-seq16384}
    \end{subfigure}
    \hfill
    \begin{subfigure}[t]{0.24\textwidth}
        \centering
        \includegraphics[width=\linewidth]{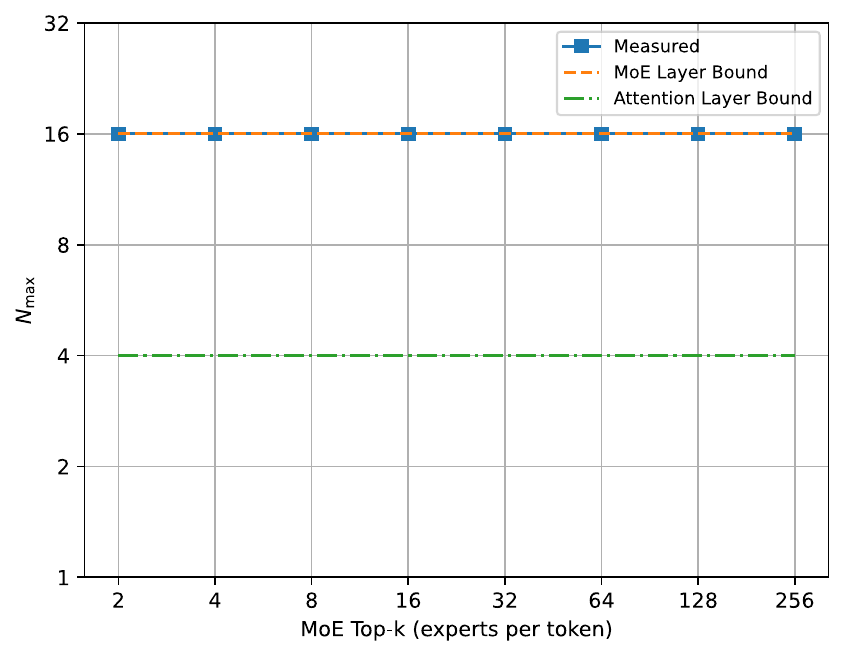}
        \caption{$L=32768$}
        \label{fig:moe-lower-ar-ling2-seq32768}
    \end{subfigure}

    \caption{
    MoE model-level NFP principle validation for load-skewed routing, the lower-bound case, on \textbf{NVIDIA H800} (Ling-2.0-mini).
    }
    \label{fig:moe-lower-ar-ling2-h800}
\end{figure*}

Figure~\ref{fig:moe-lower-ar-ling2-h800} extends the load-skewed validation to an AR model (Ling-2.0-mini) on H800. The boundary remains flat across all $k$ values, determined by local padding granularity, reproducing the DLLM observation. Together with the Dense and load-balanced results, these AR experiments confirm that the NFP principle is paradigm-independent: the system-level latency behavior is governed by the same granularity and resource-balance constraints regardless of whether decode positions originate from a diffusion schedule or a speculative draft.

Overall, the MoE model results confirm that the proposed NFP principle transfers to full-model routed inference. It predicts both upper- and lower-bound routing regimes, explains the transition from MoE-limited to Attention-limited behavior as sequence length increases, and identifies the active module constraint across hardware platforms.

\clearpage

\clearpage

\section{Sensitivity to Tolerance Threshold}
\label{app:sensitivity}

The NFP boundary is extracted using a fixed 20\% latency tolerance ($\epsilon=0.2$).
This section examines how the extracted boundary changes across tolerance values
$\epsilon \in \{0.05, 0.10, 0.15, 0.20, 0.30\}$ on NVIDIA H20, covering all
module classes and inference backends evaluated in this work.

For granularity-governed modules (MoE FFNs and Attention), $N_{\max}$ is nearly
invariant across the evaluated tolerance range under both vLLM/SGLang and
FlashAttention/FlashInfer backends
(Tables~\ref{tab:sensitivity-moe-upper-vllm}--\ref{tab:sensitivity-attn-fa}).
This robustness arises directly from the staircase latency pattern: within a
kernel-granularity block, all $N$ values produce nearly identical latency,
while crossing a block boundary causes a discrete jump well above any reasonable
tolerance. Any tolerance in $[0.05, 0.30]$ therefore yields the same extracted
boundary for these modules at sufficiently long sequence lengths.

For Dense FFNs (Table~\ref{tab:sensitivity-dense}), the boundary shifts by at most
one discrete $N$ step (e.g., from 16 to 32 at $b{=}1$) as $\epsilon$ varies, because
the latency transition is smoother. The $1/b$ scaling trend is preserved regardless
of threshold choice.

For the load-skewed MoE case (Tables~\ref{tab:sensitivity-moe-lower-vllm}--\ref{tab:sensitivity-moe-lower-sglang}),
a tighter tolerance ($\epsilon{=}0.05$) can yield a smaller boundary at some $k$ values.
This occurs because minor latency variation within a single padding block---well below
the magnitude of cross-block jumps---may exceed a very tight threshold. However, the
dominant staircase structure remains clearly identifiable regardless of $\epsilon$:
latency increases discretely at padding-block boundaries rather than smoothly with $N$.
The mechanism conclusion is therefore independent of the threshold.
For practical deployment guidance, $\epsilon \in [0.10, 0.30]$ produces consistent
boundaries across all evaluated modules and backends.

We also note that at short cached sequence lengths, the absolute kernel latency is very
small (sub-millisecond), making the percentage-based tolerance more susceptible to
GPU timing noise. This is visible in the FlashInfer results at $L \le 2048$
(Table~\ref{tab:sensitivity-attn-fi}), where the extracted boundary fluctuates at
tight tolerances. At longer sequence lengths where latency is larger and the
signal-to-noise ratio is higher, the boundary stabilizes and consistently reflects
the backend query-tile granularity identified in Section~4.

Overall, the mechanism identification and scaling trends reported in this work are
independent of the specific tolerance value.

\begin{table}[htbp]
\centering
\begin{tabularx}{\linewidth}{l|*{5}{>{\centering\arraybackslash}X}}
\toprule
$b$ & $\epsilon{=}0.05$ & $\epsilon{=}0.10$ & $\epsilon{=}0.15$ & $\epsilon{=}0.20$ & $\epsilon{=}0.30$ \\
\midrule
1  & 16 & 16 & 16 & 32 & 32 \\
2  & 8  & 8  & 8  & 16 & 16 \\
4  & 4  & 4  & 8  & 8  & 8  \\
8  & 2  & 2  & 4  & 4  & 4  \\
16 & 1  & 1  & 2  & 2  & 2  \\
32 & 1  & 1  & 1  & 1  & 1  \\
\bottomrule
\end{tabularx}
\caption{Sensitivity of Dense FFN $N_{\max}$ to tolerance threshold $\epsilon$ (H20).}
\label{tab:sensitivity-dense}
\end{table}

\clearpage
\begin{table}[htbp]
\centering
\begin{tabularx}{\linewidth}{l|*{5}{>{\centering\arraybackslash}X}}
\toprule
$L$ & $\epsilon{=}0.05$ & $\epsilon{=}0.10$ & $\epsilon{=}0.15$ & $\epsilon{=}0.20$ & $\epsilon{=}0.30$ \\
\midrule
256   & 64 & 64 & 64 & 64 & 256 \\
512   & 16 & 48 & 64 & 64 & 64  \\
1024  & 32 & 64 & 64 & 64 & 64  \\
2048  & 64 & 64 & 64 & 64 & 64  \\
4096  & 64 & 64 & 64 & 64 & 64  \\
8192  & 64 & 64 & 64 & 64 & 64  \\
16384 & 64 & 64 & 64 & 64 & 64  \\
32768 & 64 & 64 & 64 & 64 & 64  \\
\bottomrule
\end{tabularx}
\caption{Sensitivity of Attention $N_{\max}$ to tolerance threshold $\epsilon$ (FlashAttention, H20).}
\label{tab:sensitivity-attn-fa}
\end{table}

\begin{table}[htbp]
\centering
\begin{tabularx}{\linewidth}{l|*{5}{>{\centering\arraybackslash}X}}
\toprule
$L$ & $\epsilon{=}0.05$ & $\epsilon{=}0.10$ & $\epsilon{=}0.15$ & $\epsilon{=}0.20$ & $\epsilon{=}0.30$ \\
\midrule
256$^\dagger$   & 240 & 240 & 256 & 256 & 256 \\
512$^\dagger$   & 128 & 128 & 144 & 224 & 256 \\
1024$^\dagger$  & 1   & 2   & 16  & 16  & 16  \\
2048$^\dagger$  & 1   & 16  & 16  & 16  & 64  \\
4096  & 16  & 16  & 16  & 16  & 64  \\
8192  & 16  & 16  & 16  & 16  & 64  \\
16384 & 16  & 16  & 16  & 16  & 64  \\
32768 & 16  & 16  & 16  & 16  & 64  \\
\bottomrule
\end{tabularx}
\caption{Sensitivity of Attention $N_{\max}$ to tolerance threshold $\epsilon$ (FlashInfer, H20).
$^\dagger$At short sequence lengths ($L \le 2048$), kernel latency is very small (sub-ms), making the percentage-based tolerance susceptible to timing noise. Long-sequence results ($L \ge 4096$) reliably reflect the backend query-tile granularity.}
\label{tab:sensitivity-attn-fi}
\end{table}

\begin{table}[htbp]
\centering
\begin{tabularx}{\linewidth}{l|*{5}{>{\centering\arraybackslash}X}}
\toprule
$k$ & $\epsilon{=}0.05$ & $\epsilon{=}0.10$ & $\epsilon{=}0.15$ & $\epsilon{=}0.20$ & $\epsilon{=}0.30$ \\
\midrule
2   & 256 & 256 & 256 & 256 & 256 \\
4   & 256 & 256 & 256 & 256 & 256 \\
8   & 256 & 256 & 256 & 256 & 256 \\
16  & 256 & 256 & 256 & 256 & 256 \\
32  & 128 & 128 & 128 & 128 & 128 \\
64  & 64  & 64  & 64  & 64  & 64  \\
128 & 32  & 32  & 32  & 32  & 32  \\
256 & 16  & 16  & 16  & 16  & 16  \\
\bottomrule
\end{tabularx}
\caption{Sensitivity of MoE FFN $N_{\max}$ to tolerance threshold $\epsilon$ (load-balanced, vLLM, H20). Baseline is the smallest $N$ that activates all experts.}
\label{tab:sensitivity-moe-upper-vllm}
\end{table}

\begin{table}[htbp]
\centering
\begin{tabularx}{\linewidth}{l|*{5}{>{\centering\arraybackslash}X}}
\toprule
$k$ & $\epsilon{=}0.05$ & $\epsilon{=}0.10$ & $\epsilon{=}0.15$ & $\epsilon{=}0.20$ & $\epsilon{=}0.30$ \\
\midrule
2   & 256 & 256 & 256 & 256 & 256 \\
4   & 256 & 256 & 256 & 256 & 256 \\
8   & 256 & 256 & 256 & 256 & 256 \\
16  & 256 & 256 & 256 & 256 & 256 \\
32  & 128 & 128 & 128 & 128 & 128 \\
64  & 64  & 64  & 64  & 64  & 64  \\
128 & 32  & 32  & 32  & 32  & 32  \\
256 & 16  & 16  & 16  & 16  & 16  \\
\bottomrule
\end{tabularx}
\caption{Sensitivity of MoE FFN $N_{\max}$ to tolerance threshold $\epsilon$ (load-balanced, SGLang, H20). Baseline is the smallest $N$ that activates all experts.}
\label{tab:sensitivity-moe-upper-sglang}
\end{table}

\begin{table}[htbp]
\centering
\begin{tabularx}{\linewidth}{l|*{5}{>{\centering\arraybackslash}X}}
\toprule
$k$ & $\epsilon{=}0.05$ & $\epsilon{=}0.10$ & $\epsilon{=}0.15$ & $\epsilon{=}0.20$ & $\epsilon{=}0.30$ \\
\midrule
2   & 512 & 512 & 512 & 512 & 512 \\
4   & 128 & 128 & 128 & 128 & 128 \\
8   & 64  & 64  & 64  & 64  & 64  \\
16  & 16  & 32  & 32  & 32  & 32  \\
32  & 8   & 16  & 16  & 16  & 16  \\
64  & 4   & 16  & 16  & 16  & 16  \\
128 & 2   & 16  & 16  & 16  & 16  \\
256 & 16  & 16  & 16  & 16  & 16  \\
\bottomrule
\end{tabularx}
\caption{Sensitivity of MoE FFN $N_{\max}$ to tolerance threshold $\epsilon$ (load-skewed, vLLM, H20).}
\label{tab:sensitivity-moe-lower-vllm}
\end{table}

\begin{table}[htbp]
\centering
\begin{tabularx}{\linewidth}{l|*{5}{>{\centering\arraybackslash}X}}
\toprule
$k$ & $\epsilon{=}0.05$ & $\epsilon{=}0.10$ & $\epsilon{=}0.15$ & $\epsilon{=}0.20$ & $\epsilon{=}0.30$ \\
\midrule
2   & 128 & 128 & 128 & 128 & 512 \\
4   & 128 & 128 & 128 & 128 & 128 \\
8   & 64  & 64  & 64  & 64  & 64  \\
16  & 32  & 32  & 32  & 32  & 32  \\
32  & 4   & 16  & 16  & 16  & 16  \\
64  & 1   & 8   & 16  & 16  & 16  \\
128 & 4   & 4   & 16  & 16  & 16  \\
256 & 16  & 16  & 16  & 16  & 16  \\
\bottomrule
\end{tabularx}
\caption{Sensitivity of MoE FFN $N_{\max}$ to tolerance threshold $\epsilon$ (load-skewed, SGLang, H20).}
\label{tab:sensitivity-moe-lower-sglang}
\end{table}

\clearpage

\section{Discussion and Implications}

\subsection{Revisiting the Idle-Compute Baseline}

This section provides a unified retrospective interpretation of the per-module mechanisms identified in Section 3, explaining why the idle-compute baseline succeeds for one module class but not the others. The key point is that the idle-compute baseline is not only a resource-balance argument; it also assumes an idealized execution model in which logical decode positions are mapped to physical kernel work approximately continuously.

Dense FFNs are close to this idealized setting. For a fixed set of weights, increasing the number of processed positions smoothly increases the GEMM workload, while the dominant weight traffic is largely amortized across positions. Thus, the physical work executed by the system closely follows the logical workload implied by the number of positions. In this regime, increasing N primarily consumes otherwise idle tensor-core capacity in the memory-bound single-position case, so the NFP boundary is well described by the resource-balance condition. This also explains why increasing batch size reduces Dense FFN NFP: batching and increasing N consume the same compute slack.

MoE FFNs and Attention violate this smooth-scaling assumption. In practical fused MoE kernels, routed tokens are grouped by expert and padded or aligned to backend-defined expert-token blocks. Similarly, practical attention backends execute query positions in backend-defined query tiles. As a result, physical work is a discretized, rounded version of the logical workload. Larger N is near-free when it falls inside already allocated expert-token blocks or query tiles, but becomes expensive when it crosses a granularity boundary.

Therefore, the failure of the idle-compute baseline for MoE FFNs and Attention does not mean that resource balance is irrelevant. Rather, implementation granularity changes the mapping from logical parallelism to physical work, and this granularity-induced slack dominates the observed NFP boundary. The relevant distinction is thus continuous versus granular realization of logical parallelism: Dense FFNs are mainly governed by resource slack, whereas MoE FFNs and Attention are mainly governed by padding, tiling, routing distribution, and backend-specific execution granularity.

\clearpage

\subsection{Implications}
\label{app:implications}

The NFP principle turns near-free parallelism from an empirical latency observation into a system-side design signal. It does not predict the end-to-end speedup of a specific decoding algorithm. Instead, it estimates how much multi-position execution a given model-system stack can absorb at near-free latency, and which component limits this capacity.

\subsubsection{Deployment-Time Parallelism Budget}

For deployment, the NFP principle provides a latency-side budget for matching parallel decoding configurations to serving conditions. Given a model, hardware platform, batch size, and backend implementation, the principle predicts the range of decode positions that can be executed within the near-free regime. This is useful when the algorithmic parallelism level is fixed or only partially tunable. For example, an MTP model may expose a fixed prediction length, while a speculative decoding system may choose a verification length. The predicted NFP boundary indicates under which batch sizes, hardware platforms, and backend configurations these positions remain near-free, and when they begin to incur additional latency.
Table~\ref{tab:practical_lookup} instantiates this budget for representative dense and MoE serving configurations, comparing the NFP principle prediction with the idle-compute baseline.
The comparison shows that the standard resource-balance intuition over-predicts the realized boundary by $2.4$--$23\times$ in over half of the evaluated settings, confirming that kernel-granularity constraints must be accounted for when selecting parallelism levels.

\begin{table}[htbp]
\centering
\small
\begin{tabular}{llccc}
\toprule
Model type & Representative setting & Idle-compute baseline & NFP Principle & Baseline discrepancy \\
\midrule
Dense & H20, $b=1$ & $\sim 37$ & $\sim 37$ & — \\
Dense & H20, $b=4$ & $\sim 9$ & $\sim 9$ & — \\
Dense & A800, $b=1$ & $\sim 153$ & $\sim 64$ & $2.4\times$ over (Attn.\ tile) \\
Dense & H800, $b=1$ & $\sim 295$ & $\sim 64$ & $4.6\times$ over (Attn.\ tile) \\
Dense & H800, $b=8$ & $\sim 37$ & $\sim 37$ & — \\
\midrule
MoE, balanced & $E{=}256$, $k{=}8$ & $\sim 1453^\dagger$ & $\sim 64$ & $23\times$ over (granularity) \\
MoE, balanced & $E{=}256$, $k{=}32$ & $\sim 362^\dagger$ & $\sim 64$ & $5.7\times$ over (granularity) \\
MoE, balanced & $E{=}256$, $k{=}64$ & $\sim 181^\dagger$ & $\sim 64$ & $2.8\times$ over (granularity) \\
MoE, skewed & any $k$ & $\sim 45^\dagger$ & $\sim 16$ & $2.8\times$ over (MoE padding) \\
\bottomrule
\end{tabular}
\caption{
Deployment lookup comparing the idle-compute baseline with the NFP principle under representative settings ($M_{\mathrm{attn}}{=}64$).
The idle-compute baseline predicts the NFP boundary from resource balance alone, ignoring implementation granularity.
The baseline discrepancy column shows where and by how much this standard intuition over-predicts the realized boundary.
In over half of the evaluated settings, the idle-compute baseline over-predicts by $2.4$--$23\times$,
demonstrating the practical importance of accounting for kernel-granularity constraints.
$^\dagger$MoE idle-compute values use H20 ($\rho{=}37$) as a conservative lower bound; on higher-$\rho$ hardware (A800, H800) the over-prediction is larger or unbounded.
In the skewed case, $k$ appears in both numerator and denominator of $N_{\mathrm{idle}}^{\mathrm{moe}}$, yielding a near-constant prediction ($\sim\!45$) across all $k$.
}
\label{tab:practical_lookup}
\end{table}

\subsubsection{Parallel-Decoding-Aware Architecture Design}

NFP also provides a system-side criterion for model and architecture design. An architecture optimized for single-position autoregressive decoding is not necessarily optimal for parallel decoding, because parallel decoding requires residual execution headroom for multi-position execution. If the single-position path already consumes the available compute budget, larger N quickly becomes expensive. In contrast, architectures that preserve memory-bound resource slack or expose favorable implementation granularity can provide a larger near-free region for parallel decoding.

This perspective helps interpret architectural choices such as MoE sparsity, routing balance, attention design, and auxiliary prediction modules. Our MoE analysis shows that higher sparsity and balanced routing can enlarge the available NFP region by distributing expert-token granularity slack across more experts, whereas load skew collapses this aggregate slack. More broadly, NFP suggests that parallel-decoding-friendly architectures should not only optimize the cost of the single-position path, but also consider how much near-free capacity remains for candidate, predicted, or updated positions.

\clearpage

\subsubsection{Capacity-Normalized Evaluation}

NFP provides a way to separate system capacity from algorithmic utilization when evaluating parallel decoding methods. End-to-end speedup alone conflates two factors: how many positions the system can execute cheaply, and how many of those positions become useful generated tokens. By measuring NFP, one can ask a more diagnostic question: given the same near-free execution budget, how effectively does an algorithm convert these positions into useful outputs?

This enables capacity-normalized evaluation. A method with low speedup may be limited because the underlying model-system stack exposes little near-free capacity, or because the algorithm fails to utilize the available capacity through acceptance, prediction accuracy, or convergence efficiency. NFP separates these cases by providing an explicit system-side budget against which algorithm-side utilization can be interpreted.

\subsubsection{System and Backend Co-Design}

Finally, the NFP principle acts as a bottleneck diagnostic for system optimization. Since the model-level boundary is determined by the minimum of module-level constraints, the limiting term identifies which component restricts near-free parallelism. If the Dense FFN term is limiting, the relevant factors are hardware balance, batch size, and dense weight traffic. If the Attention term is limiting, further gains require changing the attention backend or its query-tile granularity. If the MoE term is limiting, the relevant knobs include expert-token padding, routing distribution, top-\(k\) sparsity, and expert grouping.

Thus, the principle supports model-system co-design: it indicates not only how much parallelism is near-free, but also where implementation or architectural changes can expand that region. In this sense, padding and tiling are not merely low-level implementation details; they directly shape the amount of usable parallelism available to parallel decoding algorithms.

\subsubsection{Implications for Future Hardware}

\begin{figure}[htbp]
\centerline{
\includegraphics[width=0.8\linewidth]{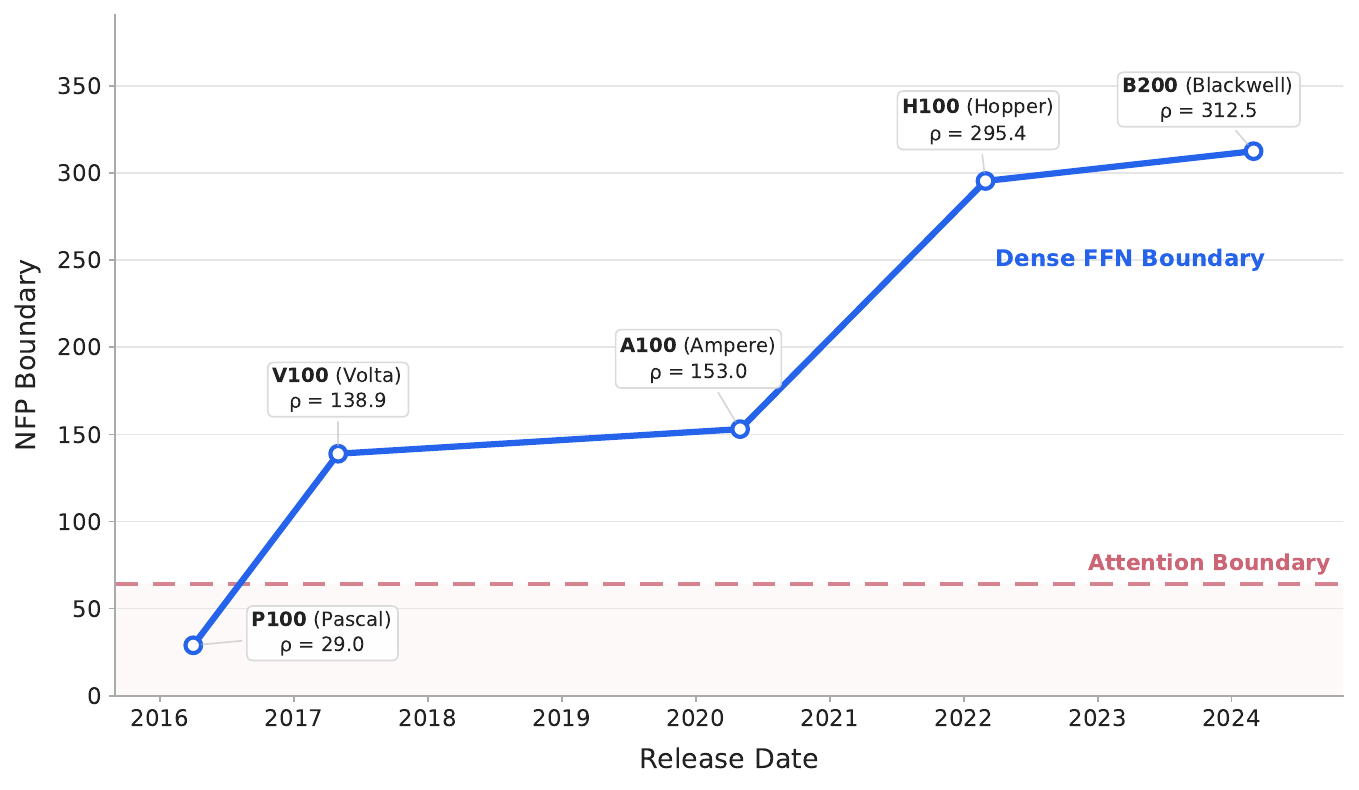}
}
\caption{
Hardware implication of the NFP principle for dense models. As the compute-to-memory balance point $\rho$ increases across GPU generations, the Dense FFN boundary $\rho s / 2b$ grows accordingly, while the Attention boundary $M_{\mathrm{attn}}$ remains an implementation-dependent reference set by backend-specific query granularity.
}
\label{fig:hardware-implication}
\end{figure}

The NFP principle links hardware evolution to dense-model parallel decoding capacity. 
For dense models, the hardware-sensitive boundary is the Dense FFN term \(\rho s / 2b\), while the Attention boundary \(M_{\mathrm{attn}}\) is mainly fixed by backend-specific query granularity. 
As shown in Figure~\ref{fig:hardware-implication}, \(\rho\) grows across GPU generations, leading to a larger Dense FFN boundary, with the Attention boundary shown as an implementation-dependent reference.
This trend suggests that future compute-heavy hardware can expand the NFP opportunity in short-context dense-model inference, where larger N can be absorbed by the increasing Dense FFN-side near-free capacity.

\end{document}